%% file: ttp96-18.tex
\documentstyle[12pt,psfig,epsf]{report}
\input{altmcom}

\begin{document}
\input{Title}
\tableofcontents
\include{intro}
\include{part1}

\include{part2}
\include{part3}

\include{part4}

\input{ack}

\input{main.bbl}
\end{document}

%% file: altmcom.tex
\renewcommand{\Re}{{\rm Re}}
\renewcommand{\Im}{{\rm Im}}

\newcommand{\bfp}{{\bf p}}
\newcommand{\rmp}{{\rm p}}
\newcommand{\bfk}{{\bf k}}
\newcommand{\rmk}{{\rm k}}

\def\be{\begin{eqnarray}}
\def\ee{\end{eqnarray}}
\def\ep{$e^+e^-$}
\def\eps{\epsilon}
\def\app#1#2#3{{\it Act. Phys. Pol. }{\bf B #1} (#2) #3}
\def\apa#1#2#3{{\it Act. Phys. Austr.}{\bf#1} (#2) #3}

\def\npb#1#2#3{{\it Nucl. Phys. }{\bf B #1} (#2) #3}
\def\plb#1#2#3{{\it Phys. Lett. }{\bf B #1} (#2) #3}
\def\prd#1#2#3{{\it Phys. Rev. }{\bf D #1} (#2) #3}
\def\prl#1#2#3{{\it Phys. Rev. Lett. }{\bf #1} (#2) #3}
\def\prc#1#2#3{{\it Phys. Reports }{\bf C #1} (#2) #3}
\def\cpc#1#2#3{{\it Comp. Phys. Commun. }{\bf #1} (#2) #3}

\def\pr#1#2#3{{\it Phys. Reports }{\bf #1} (#2) #3}
\def\sovnp#1#2#3{{\it Sov. J. Nucl. Phys. }{\bf #1} (#2) #3}
\def\jl#1#2#3{{\it JETP Lett. }{\bf #1} (#2) #3}
\def\jet#1#2#3{{\it JETP Lett. }{\bf #1} (#2) #3}
\def\zpc#1#2#3{{\it Z. Phys. }{\bf C #1} (#2) #3}

\newdimen\oldsize
\newdimen\skipsize
\newdimen\leng
\newdimen\vskipsize

\newbox\myown
\newbox\mycount
 
\newcommand{\ctowidth}[2]{ \setbox\mycount=\hbox{$#2$}
                          \hbox to \wd\mycount{$ \hss #1 \hss $} }
\newcommand{\ltowidth}[2]{ \setbox\mycount=\hbox{$#2$}
                          \hbox to \wd\mycount{$\hskip0pt plus0pt minus1fil
                           #1 \hfill $} }
\newcommand{\rtowidth}[2]{ \setbox\mycount=\hbox{$#2$}
                          \hbox to \wd\mycount{$\hfill #1
                          \hskip0pt plus0pt minus1fil$} }
\newcommand{\flowcap}[2]{\advance\skipsize by -1cm
                         \setbox\myown=\hbox{\parbox{\skipsize}{#2
                           $\vphantom{g}$}}
                         \leng=\ht\myown
                         \advance\leng by \dp\myown
                         \vskipsize=-#1\baselineskip
                         \advance\vskipsize by -\leng
                         \advance\vskipsize by -\parskip
                         \advance\vskipsize by -8mm
                         \vspace{\vskipsize}
                         \noindent \box\myown \par
                         \advance\vskipsize by \leng
                         \advance\vskipsize by \parskip
                         \multiply\vskipsize by -1
                         \vspace{\vskipsize}
                        }
\newcommand{\rflowcap}[2]{\advance\skipsize by -1cm
                         \setbox\myown=\hbox to \hsize{\hss
                           \parbox{\skipsize}{#2 $\vphantom{g}$}}
                         \leng=\ht\myown
                         \advance\leng by \dp\myown
                         \vskipsize=-#1\baselineskip
                         \advance\vskipsize by -\leng
                         \advance\vskipsize by -\parskip
                         \advance\vskipsize by -5mm
                         \vspace{\vskipsize}
                         \noindent \box\myown \par
                         \advance\vskipsize by \leng
                         \advance\vskipsize by \parskip
                         \multiply\vskipsize by -1
                         \vspace{\vskipsize}
                        }

\newbox{\k}
\newbox{\kk}
\newbox{\kkl}
\newbox{\kkk}
\newbox{\kkkk}
 
\sbox{\kkkk}{
\unitlength1ex
\begin{picture}(4,2)
\thinlines
\multiput(0,0.5)(1,0){4}{\framebox(1,1)[c]{}}
\end{picture}
}
 
\sbox{\kkk}{
\unitlength1ex
\begin{picture}(3,2)
\thinlines
\multiput(0,1)(1,0){3}{\framebox(1,1)[c]{}}
\put(0,0){\framebox(1,1)[c]{}}
\end{picture}
}
 
\sbox{\kk}{
\unitlength1ex
\begin{picture}(2,2)
\thinlines
\multiput(0,1)(1,0){2}{\framebox(1,1)[c]{}}
\multiput(0,0)(1,0){2}{\framebox(1,1)[c]{}}
\end{picture}
}
 
\sbox{\kkl}{
\unitlength1ex
\begin{picture}(2,3)
\thinlines
\multiput(0,1.5)(1,0){2}{\framebox(1,1)[c]{}}
\multiput(0,-0.5)(0,1){2}{\framebox(1,1)[c]{}}
\end{picture}
}
 
\sbox{\k}{
\unitlength1ex
\begin{picture}(1,4)
\thinlines
\multiput(0,-1)(0,1){4}{\framebox(1,1)[c]{}}
\end{picture}
}

\newcommand{\swin}{\sin^2\vartheta_W}

\newcommand{\epem}{$e^+ e^-$}

\newbox{\dra}
\sbox{\dra}{
\unitlength1ex
\begin{picture}(3,2)
\thinlines
\put(0,0.7){\line(0,1){1}}
\put(0,0.7){\vector(1,0){2}}
\end{picture}              }

\newcommand{\shatn}{\mbox{$\hat{s}$ }}
\newcommand{\ttb}{\mbox{$t\overline{t}$}}
\newcommand{\bbb}{\mbox{$b\overline{b}$}}
\newcommand{\ttbn}{\mbox{$t\overline{t}$ }}
\newcommand{\qqb}{\mbox{$q\overline{q}$}}
\newcommand{\ppb}{\mbox{$p\overline{p}$}}

\newcommand{\ra}{\mbox{$\rightarrow$ }}

\newcommand{\sm}{Standard Model}

\newcommand{\beq}{\begin{equation}}
\newcommand{\eeq}{\end{equation}}
\newcommand{\beqa}{\begin{eqnarray}}
\newcommand{\eeqa}{\end{eqnarray}}

\newcommand{\calckm}{$CKM$}

%% file: Title.tex
\setlength\topmargin{-0.5cm}
\setlength\textheight{23.0cm}
\setlength\textwidth{16cm}
\protect\setlength\oddsidemargin{-0.25cm}
\protect\setlength\evensidemargin{0.3cm}
\headsep 30pt


\begin{titlepage}
\noindent
\mbox{} 
\hfill {\small TTP96--18} \\
\mbox{}
May   1996   
\protect\vspace*{10mm}
%
%
%

\vspace*{20mm}

\begin{center}
  \begin{Large}
  \begin{bf}
Theory of Top Quark Production and Decay${}^{\dagger}$
  \\
  \end{bf}
\vspace{1.0cm}
  \end{Large}
%
%
  \vspace{1cm}
  \begin{large}
{J.H.~K\"uhn}
  \end{large}
  \vspace{0.5cm}

    Institut f\"ur Theoretische Teilchenphysik,
    Universit\"at Karlsruhe \\ 
    D-76128 Karlsruhe, Germany\\
email: johann.kuehn@physik.uni-karlsruhe.de

\vspace{30mm}

%
  \vspace{0.5cm}
  {\bf Abstract}
\end{center}
\begin{quotation}
\noindent
{
Direct and indirect information on the top quark mass and its decay
modes is reviewed.  The theory of top production in hadron- and
electron-positron-colliders is presented.}
\end{quotation}

\vfill

\vfill

\footnoterule
\noindent
$^{\dagger}${\footnotesize 
Lectures presented at XXIII SLAC Summer Institute on Particle Physics, 
``The Top Quark and the Electroweak Interaction,''  July 10--21, 1995, 
SLAC, Stanford, CA. \\
Supported by BMBF 057KA92P and Volkswagen-Stiftung grant I/70452.

}
\end{titlepage}


%% file: intro.tex
\addcontentsline{toc}{chapter}{Introduction}
\chapter*{Introduction}
An extensive search for top quarks has been performed at
electron-positron and hadron colliders for more than a decade.  First
evidence for top quark production in proton-antiproton collisions has
been announced by the CDF collaboration in the spring of 1994.  After
collecting  more luminosity subsequently both the CDF and the D0
experiments presented the definite analysis \cite{Sinerva}
 which demonstrated not only the existence of the
anti\-cipated quark but at the same time also provided a kinematic
determination of the top quark mass around 180 GeV and a production
cross section consistent with the QCD predictions.  The mass value is
in perfect agreement with the indirect mass determinations based on
precision measurements 
[2-7]
of the
electroweak 
parameters in \ep annihilation and in lepton-nucleon scattering.
Exploiting the quadratic top mass dependence of radiative corrections
an indirect mass measurement of 180 GeV with a present uncertainty of
roughly $20$  GeV  has been achieved.  

The top quark completes the fermionic spectrum of the Standard Model.
 All its properties are uniquely fixed after the mass has been
determined.  However, as a consequence of its large mass and decay
rate it will behave markedly different compared to the remaining five
lighter quarks.  

It is not just the obvious aim for completion which raises the
interest in the top quark.  With its mass comparable to the scale of
the electroweak symmetry breaking it is plausible that top quark
studies could provide  an important clue for the understanding of
fermion mass generation and the pattern of Yukawa couplings.  In fact,
it has been suggested that a top quark condensate could even be
responsible for the mechanism of spontaneous symmetry breaking
\cite{Hill}. 

These lectures will be mainly concerned with top quark phenomenology
within the framework of the Standard Model (SM).  
The precise understanding of its
production and decay constitutes the basis of any search for
deviations or physics beyond the SM. 

The properties of the top quark will be covered in chapter 1.  Direct
and indirect determinations of its decay rates, decay distributions
including QCD and electroweak corrections and decay modes predicted in
supersymmetric extensions will be discussed.  Top quark production at
hadron colliders will be the subject of chapter 2.  The production
cross section and momentum distributions are important ingredients in
any of the present analysis.   An alternative reaction, namely top
quark production through $W$-$b$-fusion allows to determine the
$W$-$b$-$\bar  t$ coupling and thus indirectly the top quark decay rate.

Perspectives for top studies at a future \ep collider will be
presented in chapter 3.  An accurate determination of the top quark
mass and its width to better than 1 GeV with a relative accuracy of
about 10\% seems feasible, and the electroweak couplings of the top
quark could be precisely measured with the help of polarized beams.
Of particular interest is the interplay between the large top quark
decay rate and the binding through the QCD potential which will be
also covered in chapter 3.
\setcounter{chapter}{0}

%% file: part1.tex
\chapter{The Profile of the Top Quark}
Hadron collider experiments at the TEVATRON have firmly established
the existence of the top quark and already provide a fairly accurate
determination of its mass.  The couplings of the top quark to the
gauge bosons are uniquely fixed by the SM. Thus all its properties
--- its production cross section and its decay rate and distributions
--- can be predicted unambiguously.  

The study of real top quarks at high energy colliders, in particular
the observation of a peak in the invariant mass of its decay products,
is certainly the most impressive proof of existence.  Nevertheless, the
indirect evidence for a top quark and the determination of its mass is
not only of historical interest.  The arguments which anticipated the
existence of the top quark and its mass around 180 GeV illustrate the
rigid structure of the SM, its selfconsistency and beauty.  They will be
presented in section \ref{indir}. 

These theoretical arguments have inspired the experimental searches.
The upper limit on the top mass around 200 GeV
deduced already relatively early from electroweak
precision studies has provided encouragement  that energies of
present colliders were suited to complete this enterprise.  The
agreement between the most recent indirect mass determinations both
through radiative corrections and through direct observation
strengthens  the present belief into the quantum field aspect of the
theory.  It furthermore justifies the corresponding line of reasoning
concerning the search for the last remaining ingredient of the SM, the
Higgs boson.  Section \ref{indir} of this first chapter will, with this
motivation in mind, be devoted to a discussion of the indirect
information on the top
quark, its existence and its mass.  Top decays, including aspects of
radiative corrections, polarisation effects and decays induced by
physics beyond the SM will be covered in section \ref{decays}.

\section{Indirect information}
\label{indir}
\subsection{Indirect evidence for the top quark}
Several experimental results already prior to its discovery did
provide strong evidence that the
fermion spec\-trum of the \sm\
\[
   { \left[ \nu_e    \atop    e^- \right]_L } \quad { \atop    e^-_R } \qquad
   { \left[ \nu_\mu  \atop  \mu^- \right]_L } \quad { \atop  \mu^-_R } \qquad
   { \left[ \nu_\tau \atop \tau^- \right]_L } \quad { \atop \tau^-_R }
\]
\nopagebreak
\[ \!
   { \left[ {\;    u \;    } \atop d \right]_L }
     \quad { u_R \atop d_R } \qquad
   { \left[ {\, \, c \, \, } \atop s \right]_L }
     \quad { c_R \atop s_R } \qquad
   \,
   { \left[ {\, \, t \, \, } \atop b \right]_L }
     \quad { t_R \atop b_R }
\]
does include the top quark, imprinting the same multiplet structure on
the third family as the first two families. The evidence is
based on theoretical selfconsistency (absence of anomalies),  the
absence of flavour changing neutral currents (FCNC) and
measurements of the weak isospin of the $b$ quark which has been
proved to be non-zero, $I_3 = -1/2$, thus demanding an $I_3 = +1/2$
partner in this isospin multiplet.
\subsubsection{Absence of triangle anomalies}
A compelling argument for the existence of top quarks follows
from a theoretical consistency requirement. The renormalizability of the
\sm\ demands the absence of triangle anomalies.
Triangular fermion loops built-up by an axialvector charge $I_{3A} = -I_{3L}$
combined with two electric vector charges $Q$ would spoil the renormalizability
of the gauge theory. Since the anomalies do not depend on the masses of the
fermions
circulating in the loops, it is sufficient to demand that the sum
\begin{eqnarray*}
\hspace*{5mm}
\psfig{figure=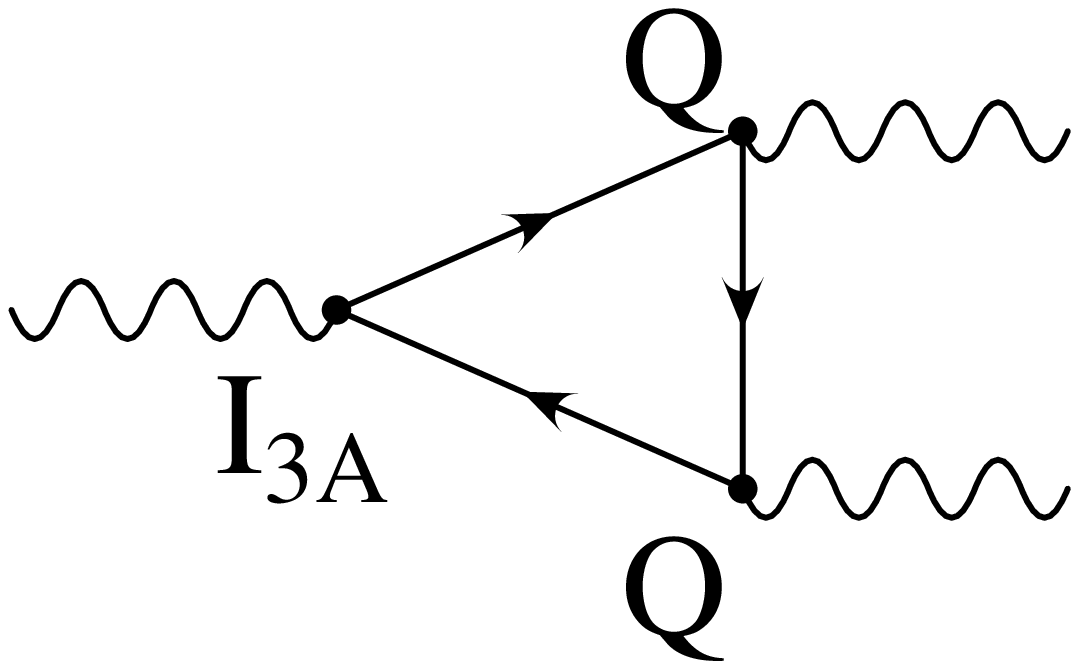,width=50mm,bbllx=210pt,bblly=490pt,%
bburx=630pt,bbury=550pt}  
\hspace*{-5mm}
   & & \\
   \hspace{4cm} & & \sim \sum_L I_{3A} Q^2 = - \sum_L I_3 \left[ I_3 +
     \frac{1}{2} Y \right]^2 \\
   \hspace{4cm} & & \sim \sum_L Y \sim \sum_L Q \\
   & &
\end{eqnarray*}
of all contributions be zero. Such a requirement can be translated into a
condition on the electric charges of all the left-handed fermions
\beq
   \sum_L Q = 0.
\eeq
This condition is met in a complete standard family in which the
electric charges of the
lepton plus those of all color components of the up and down quarks add up
to zero,
\[
   \sum_L Q = - 1 + 3 \times \left[ \left( + \frac{2}{3} \right) + \left(
   - \frac{1}{3} \right) \right] = 0.
\]
If the top quark were absent from the third family, the condition would be
violated and the \sm\ would be theoretically inconsistent.

\subsubsection{Absence of FCNC decays}
Mixing between quarks which belong to different isospin multiplets
\[
   { \left[ c \atop s^{'} \right]_L } \quad { \atop b^{'}_L }
   \hspace{2cm} {
     \! s^{'}_L = \: s_L \cos \vartheta^{'} + b_L \sin \vartheta^{'} \atop
     \: b_L^{'} =   -s_L \sin \vartheta^{'} + b_L \cos \vartheta^{'} }
\]
generates non-diagonal neutral current couplings, i.e. the breaking of the
GIM mechanism

\begin{eqnarray*}
   <I_3> & = & + \frac{1}{2} \left( \bar c_L,c_L \right)
               - \frac{1}{2} \left( \bar s^{'}_L, s^{'}_L \right) \\
         & = & {1\over 2}\left( \bar c_L,c_L \right)-
-{1\over 2}\left( \bar s_L,s_L \right) \cos^2\vartheta^\prime
-{1\over 2}\left( \bar b_L,b_L \right) \sin^2\vartheta^\prime
\\
&& - \frac{1}{2} \sin \vartheta^{'} \cos \vartheta^{'}
              \left( \left( \bar s_L, b_L \right)
                   + \left( \bar b_L, s_L \right)\right).
\end{eqnarray*}
The non-diagonal current induces flavor-changing neutral lepton pair decays
$b \ra s + l^+l^-$ which have been estimated to be a
substantial fraction of all semileptonic $B$ meson decays.
The relative strenth of neutral versus charged current induced rate is
essentially given by
\be
{\Gamma_{\rm NC} \over \Gamma_{\rm CC} }\sim
{1\over 2} \left({M_W^2\over M_Z^2}\right)^2 
{(v_b^2+a_b^2)(v_e^2+a_e^2)\over (1+1) (1+1)}
\sim 0.06.
\ee
Taking the proper momentum dependence of the matrix element and the
phase space into account one finds \cite{kaue1}
\beq
   \frac{ \mbox{BR} \left( B \ra l^+l^-X \right) }
        { \mbox{BR} \left( B \ra l^+ \nu_l X \right) } \ge 0.12.
\eeq
This ratio is four orders of magnitude larger than a bound set  by
the UA1 Collaboration \cite{albaj3,PDG}
\beq
   \frac{ \mbox{BR} \left( B \ra \mu^+ \mu^-X    \right) }
        { \mbox{BR} \left( B \ra \mu   \nu_\mu X \right) } <
   \frac{ 5.0 \times 10^{-5} }{0.103 \pm 0.005 }.
\eeq
so that the working hypothesis of an isosinglet $b$ quark is clearly ruled out
experimentally also by this method.

\subsubsection{Partial width $\Gamma ( Z \ra \bbb )$ and
                        forward-backward asymmetry of $b$ quarks}
The $Z$ boson couples to quarks through vector and axial--vector
charges with the well--known strength
 
{\hangindent=4cm \hangafter=-5
\[
\hspace*{-20mm}
\psfig{figure=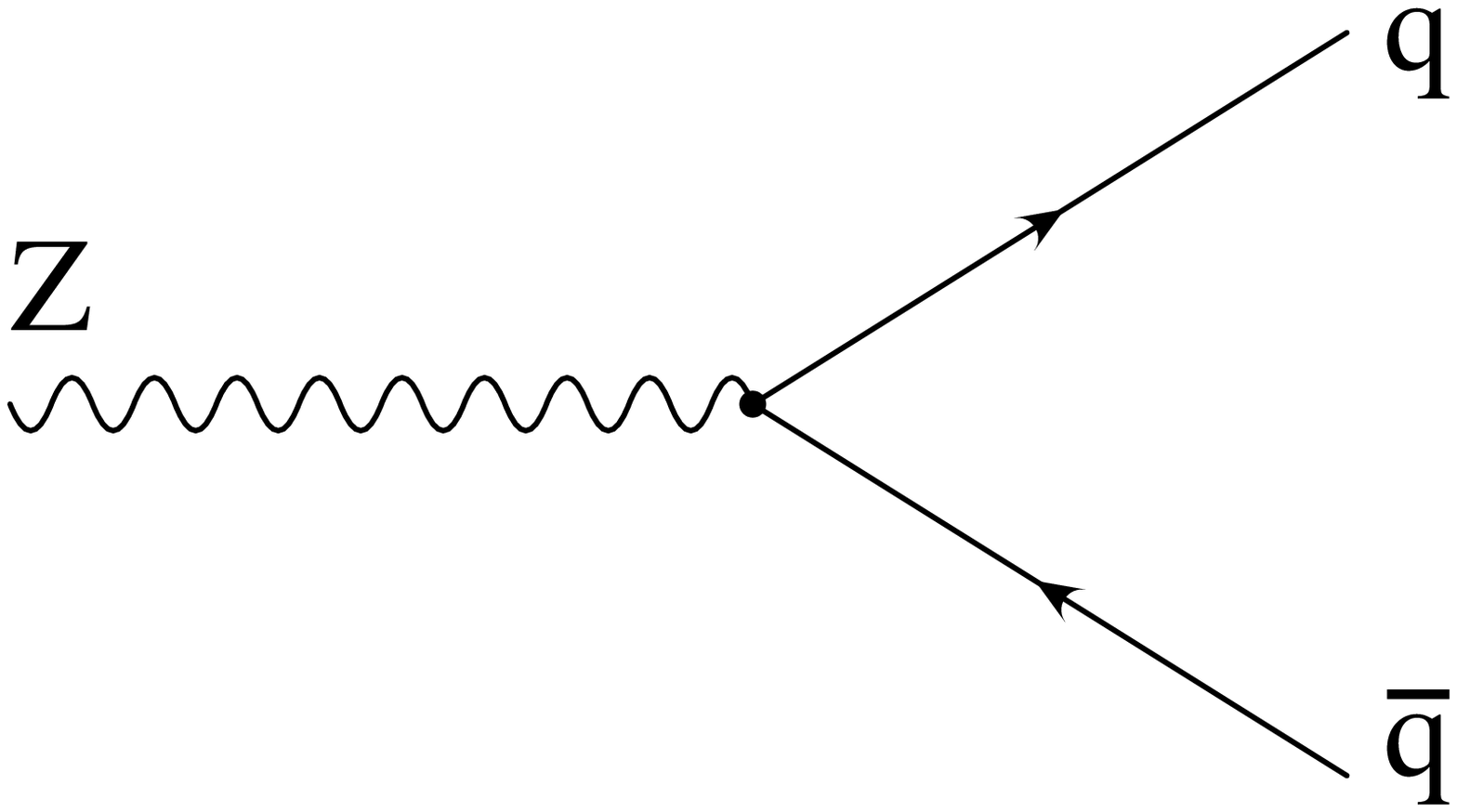,width=35mm,bbllx=210pt,bblly=410pt,%
bburx=630pt,bbury=550pt}  
=\sqrt{\frac{G_F m_Z^2}{2\sqrt{2}}}\gamma_\mu \left[v_q - a_q \gamma_5
\right].
\]
}

\noindent Depending on the isospin assignment of righthanded and
lefthanded quark fields these charges are defined as
\be
v_q&=&2 (I_q^{3L}+I_q^{3R})-4e_q\swin \\
a_q&=&2(I_q^{3L} -I_q^{3R})
\ee
 
For the present application the {\em Born approximation} in the
massless limit provides an adequate
representation of the partial $Z$ decay rate
\begin{equation}
\label{eqrate}
\Gamma_B(Z \rightarrow b \overline{b}) \approx
 \frac{G_F m_Z^3}{8\sqrt{2}\pi}\beta
\left( v_b^2 + a_b^2 \right).
\end{equation}
In the Standard Model $2I_q^{3R}=0$ and $2I_q^{3L}=\pm 1$ for up/down
quarks respectively.

\begin{figure}
\psfig{figure=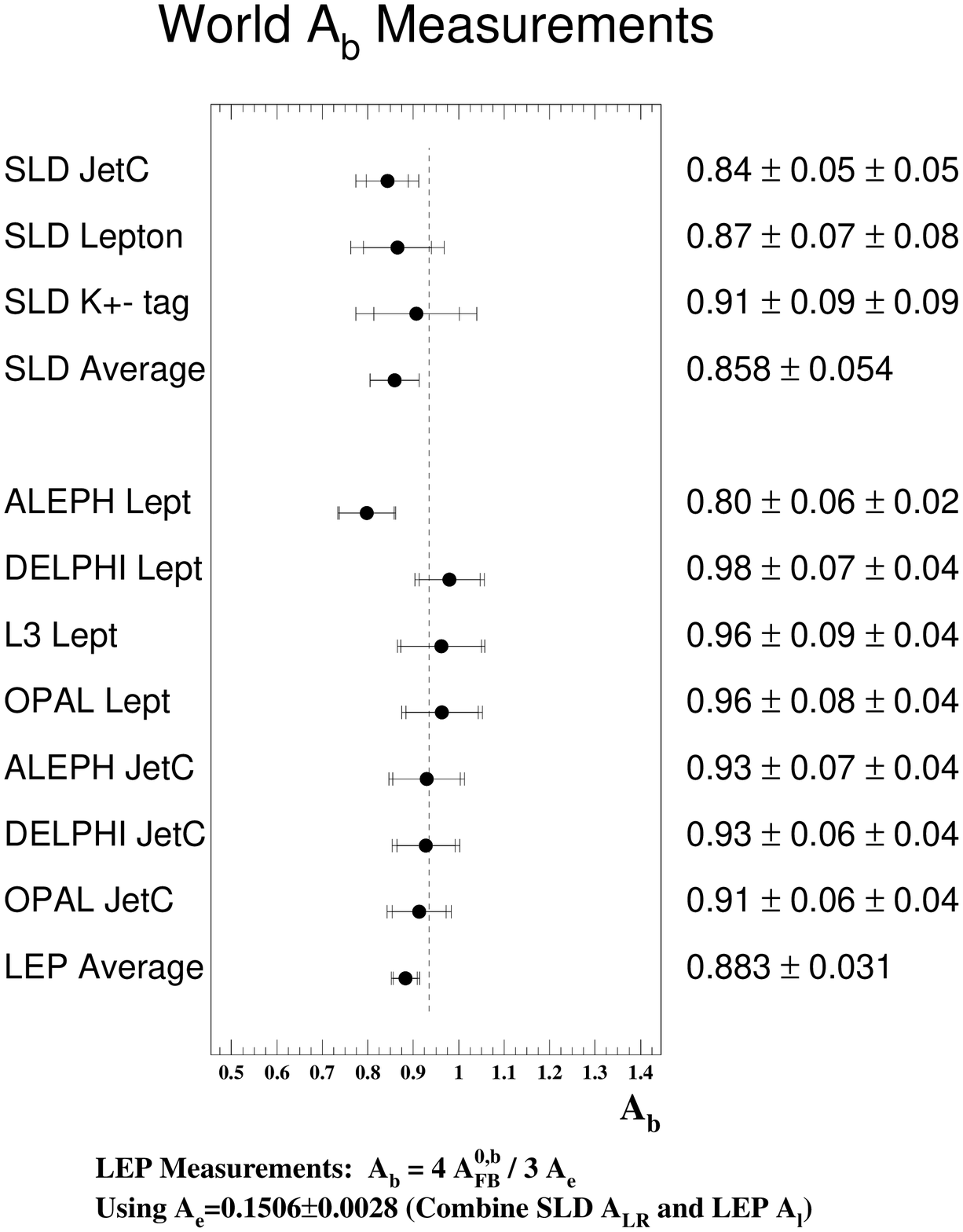,width=150mm,bbllx=-40pt,bblly=50pt,%
bburx=590pt,bbury=600pt}  
\caption{Experimental results for the asymmetry parameter $A_b$
\protect\cite{SuDong}.} 
\label{fig:sudong}
\end{figure}

The ratio between the predictions in the context of a topless model
and the SM amounts to 
\be
{\Gamma^{\rm topless}\over \Gamma^{\rm SM}} 
= { (4Q_b \sin^2\theta_W)^2 \over (1+4Q_b \sin^2\theta_W)^2 + 1}
\approx 1/13
\ee
whereas theory and  LEP experiments are well consistent
\be
{ \Gamma(Z\to b\bar b) \over \Gamma(Z\to {\rm had})} =
\left\{
\begin{array}{lll}
0.2155\pm 0.0004 & {\rm theory} & \cite{HollikBrus} \\
0.2219\pm 0.0017 & {\rm experiment} & \cite{HollikBrus}
\end{array}
\right.
\ee
ruling out the $I=0$ assignement for the $b$-quark.
The forward-backward asymmetry at the $Z$ resonance 
\be
A^{\rm FB} = {3\over 4} A_eA_b
\ee
with 
\be 
A_f \equiv {2v_f a_f \over v_f^2 + a_f^2}
\ee
is sensitive toward the relative size of vector and axial $b$ quark
couplings. Up to a sign, the first of these factors, $A_e\approx 0.15$,
can be interpreted as the degree of longitudinal $Z$ polarisation,
$P_Z= - A_e$, which is induced by the electron coupling even for
unpolarized beams. For longitudinally polarized beams it can be
replaced by unity.  The second factor represents essentially the
analyzing power of $b$ quarks. With a predicted value of 0.93 it is
close to its maximum in the SM.
In fact, this high analysing power is the reason for the large
sensitivity of $A^{FB}$ toward $\sin^2\theta_W$ \cite{djoua1}.
  For a fictitious topless model $A_b$ is
zero. The most recent experimental results from LEP and SLC are
displayed in Fig.~\ref{fig:sudong}.  

A remaining sign ambiguity is finally resolved by the interference
between NC and electromagnetic amplitude.  It leads to a forward
backward asymmetry at low energies
\be
A_{\rm FB} = -{3G_F s\over 16\sqrt{2} \pi\alpha} {a_e a_b\over Q_eQ_b}
\ee
which has been studied in particular at PEP, PETRA, and most recently
with highest precision at Tristan at a cm energy of 58 GeV \cite{Trist}.   
 Using the data available
shortly after the turnon of LEP and  combining $\Gamma_{b\bar
b}$,  $A_b$, and $A_{\rm FB}$
\begin{eqnarray}
\ltowidth{\{ I_3^L(b) \}_{exp} = -0.490^{+0.015}_{-0.012}}{\{ I_3^R(b) \}_{exp}
 = -0.028 {\pm 0.056}  }
 & \qquad \ra \quad & I_3^L(b) = -1/2 \nonumber \\
\{ I_3^R(b) \}_{exp} = -0.028 {\pm 0.056} & \qquad \ra \quad& I_3^R(b) = 0
\nonumber
\end{eqnarray}
has been obtained already some time ago \cite{SchailZ}.
As shown in Fig.~\ref{FSchailZ} all
measurements are nicely consistent with the predictions  of the 
SM\footnote{For a
discussion of the most recent results for $R_b$, 
however, see \protect\cite{HollikBrus}.}.
The isospin assignment of the
Standard Model is thus well confirmed.

\begin{figure}[hbt]

\vspace{11.5cm}

\caption[]{\label{FSchailZ}%
The weak isospins $I_3^L(b)$ and $I_3^R(b)$ of the left-- and
right--handed $b$ quark components, extracted from the data on $\Gamma(Z\ra
b\overline{b})$ and $A_{FB}(b)$ at LEP, and PETRA/PEP and TRISTAN,
Ref.\cite{SchailZ}.}
\end{figure}

\newpage

\subsection{Mass limits and indirect mass determinations}

\subsubsection{Theoretical constraints}

Present theoretical analyses of the \sm\ are based
almost exclusively on 
perturbation theory. If this method is assumed to apply also to the
top-quark sector, in particular when linked to the Higgs sector,
the top mass must be bounded as the strength of the Yukawa-Higgs-top
coupling is determined by this parameter. The following consistency
conditions must be met:

\noindent
\underline{Perturbative Yukawa coupling $g_Y \left( ttH \right)$}
\nopagebreak
\vspace{-\parskip}
\vspace{2mm}

\nopagebreak
Defining the Yukawa coupling in the  \sm\  through
\be
 {\cal L}_Y = g_Y \left({v+H\over \sqrt{2}}\right)
\left( \bar t_L t_R+ \mbox{h.c.} \right)
\ee
the coupling constant $g_Y$
is related to the top mass by
\beq
   g_Y \left( ttH \right) = m_t \sqrt{2\sqrt{2} G_F}.
\eeq
Demanding the effective expansion parameter $g^2_Y/4 \pi$
to be smaller than 1, the top mass is bounded to
\beq
   m_t < \sqrt{ \frac{\sqrt{2}\pi}{ G_F} } \approx 620 \mbox{GeV}.
\eeq
For a top mass of 180 GeV the coupling $g^2_Y / 4 \pi \approx 0.085$ is
comfortably small so that perturbation theory can safely be applied in this
region.

\noindent
\underline{Unitarity bound} \\[0.2cm]
At asymptotic energies the amplitude of the zeroth partial wave for elastic
\ttb\ scattering in the color singlet same-helicity channel \cite{chano1}
\begin{eqnarray}
   a_0 \left(\ttb \ra \ttb \right) & = & - \frac{3 g^2_Y}{8 \pi} \\
                                   & = & - \frac{3 G_F m^2_t}{4 \sqrt{2} \pi}
                                         \nonumber
\end{eqnarray}
grows quadratically with the top mass. Unitarity however demands this real
amplitude to be bounded by $\left| \mbox{Re} \, a_0 \right| \le 1/2$.
This condition translates to
\beq
   m_t < \sqrt{\frac{2 \sqrt{2} \pi}{3 G_F}} \approx 500 \, \mbox{GeV}.
\eeq
The bound improves by taking into account the running of the
Yukawa coupling \cite{cabib1}.
These arguments are equally applicable for any additional species of
chiral fermions with mass induced via spontaneous symmetry breaking. 

\newpage
\noindent
\underline{Stability of the Higgs system: top-Higgs bound} \\[0.2cm]
The quartic coupling
$\lambda$
in the effective Higgs potential
\[
V = \mu^2 \left| \phi \right|^2 + \frac{\lambda}{2} \left| \phi \right|^4
\]
depends on the scale at which the system is interacting. The running of
$\lambda$
is induced by higher-order loops built-up by the Higgs particles themselves,
the vector bosons and the fermions in the \sm\ \cite{cabib1,lindn1}.
For moderate values of the top mass,
$m_t \le 77$ GeV,
these radiative
corrections would have generated a lower bound on the Higgs mass
of 7 GeV.
With the present experimental lower limit $m_H> 65$ GeV and 
the top quark mass determined around 180 GeV this bound is of no practical
relevance any more.
At high energies the radiative corrections make
$\lambda$
rise up to the Landau pole at the cut-off parameter
$\Lambda$
beyond which the \sm\ in the present formulation cannot be continued
[``triviality bound'', as this bound could formally be misinterpreted
as requiring the low energy
coupling to vanish]. If for a fixed Higgs mass the top mass is increased,
the top loop radiative corrections lead to negative values of the quartic
coupling
$\lambda$
\beq
   \frac{\partial \lambda}{\partial \log s} = \frac{3}{8 \pi^2} \left[
      \lambda^2 - 4g^4_{Y} + \mbox{gauge couplings} \right]
\eeq
Since the potential is unbounded from below in this case, the Higgs
system becomes instable. Thus the stability requirement defines an upper
value of the top mass
$m_t$
for a given Higgs mass
$m_H$
and a cut-off scale
$\Lambda$.
The result of such an analysis is presented
in Fig.\ref{FLindner}. Depending on the cut-off scale $\Lambda$
where new physics may set in, 
\begin{figure}[hbt]
      \vspace{10cm}
      \caption[]{Bounds on the Higgs and top masses
      following from triviality of the Higgs's quartic self-coupling and the
      stability of the Higgs system; from \cite{lindn1}.}
\label{FLindner}
\end{figure}
the top mass is bounded to $m_t \le 200$ GeV
if $\Lambda$
exceeds the Planck scale but it rises up to 400 to 500 GeV if the
cut-off is reached at a level of 1 TeV and below. The estimates are similar
to the unitarity analysis in the preceeding
subsection. Lattice simulations of the Yukawa model have arrived at
qualitatively similar results (see e.g. \cite{Munster94}).

These theoretical analyses have shown that for the top
mass around 180 GeV
the \sm\ may be valid up to a cut-off at the Planck scale. [The hierarchy
problem, that is not touched in the present discussion,
may enforce nevertheless new additional physical phenomena already
in the TeV range.]

In the context of the SM the top Yukawa coupling is simply present as
a free parameter. In the minimal supersymmetric model, however, the
picture is changed completely.  A large Yukawa coupling may play the
role of a driving term for the spontaneous breaking of $SU(2)\times
U(1)$, as discussed in \cite{HallThis} and in fact the large mass of the
top quark had been predicted on the basis of these arguments prior to
its experimental discovery.

\subsubsection{Mass estimates from radiative corrections}
First indications of a high top quark mass were derived from the rapid
$(B - \overline{B})$ oscillations observed by ARGUS \cite{albre1}.
However, due to the
uncertainties of the \calckm\ matrix element $V_{td}$ and of the
$(b \bar{d})$
wave function, not more than qualitative conclusions can be drawn from such an
analysis as the oscillation frequency $\Delta m \sim \left| V_{td} \right|^2
f^2_B m^2_t$ depends on three [unknown] parameters.

The analysis of the radiative corrections to high precision electroweak
observables is much more advanced [2-7].
Since Higgs mass effects are
weak as a result of the screening theorem \cite{veltm1},
the top mass is the dominant
unknown  in the framework of the \sm.
Combining the high precision
measurements of the $Z$ mass with $\sin^2 \theta_W$ from the $Z$ decay
rate, from the forward-backward
asymmetry and  from
LR polarization measurements, the top quark mass 
has been
determined up to a residual uncertainty of less than 10 GeV
plus an additional uncertainty of  about 20 GeV
induced through the variation of $m_H$
\be
 m_t= 178\pm 8 ^{+17}_{-20}\; {\rm GeV}. 
\ee

The close agreement between direct and indirect top mass determination
can be considered a triumph of the Standard Model. Its predictions are
not only valid in Born approximation, as expected for any effective
theory, also quantum corrections play an important role, and are
indirectly confirmed.  Encouraged by this success and in view of the
improved accuracy of theory and experiment it is conceivable that the
same strategy can lead to a rough determination of the Higgs mass, or,
at least, to a phenomenologically relevant upper limit.

\subsection{The quadratic top mass dependence of $\delta\rho$}
The quadratic top mass dependence of $\delta\rho$ is a cornerstone of
the present 
precision measurements \cite{veltm2}.  
In view of its importance and the pedagogical
character of these lectures it is perhaps worthwhile to present a
fairly pedestrian derivation of this result.  

\begin{figure}[htb]
\hspace*{60mm}
\begin{tabular}{c}
\psfig{figure=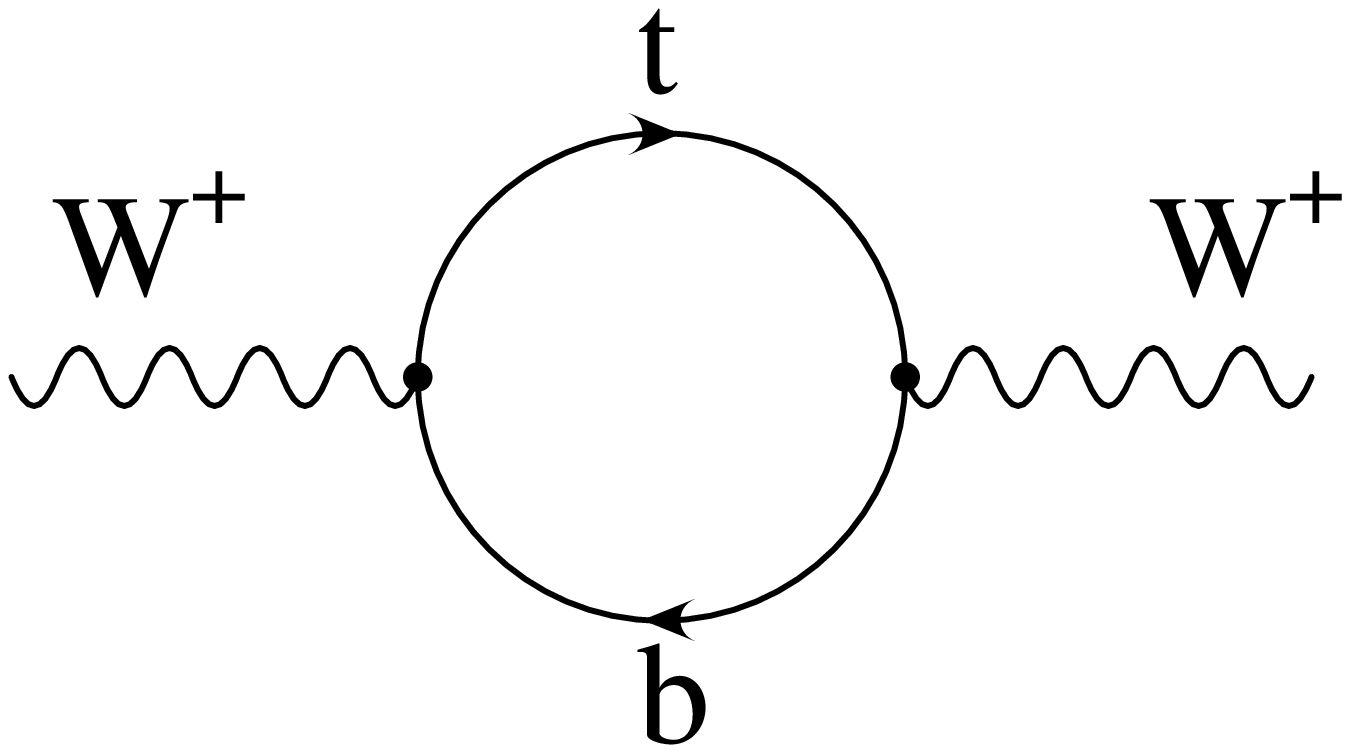,width=50mm,bbllx=210pt,bblly=410pt,%
bburx=630pt,bbury=550pt}
\end{tabular}\\[10mm]
\hspace*{35mm}
\begin{tabular}{cc}
\psfig{figure=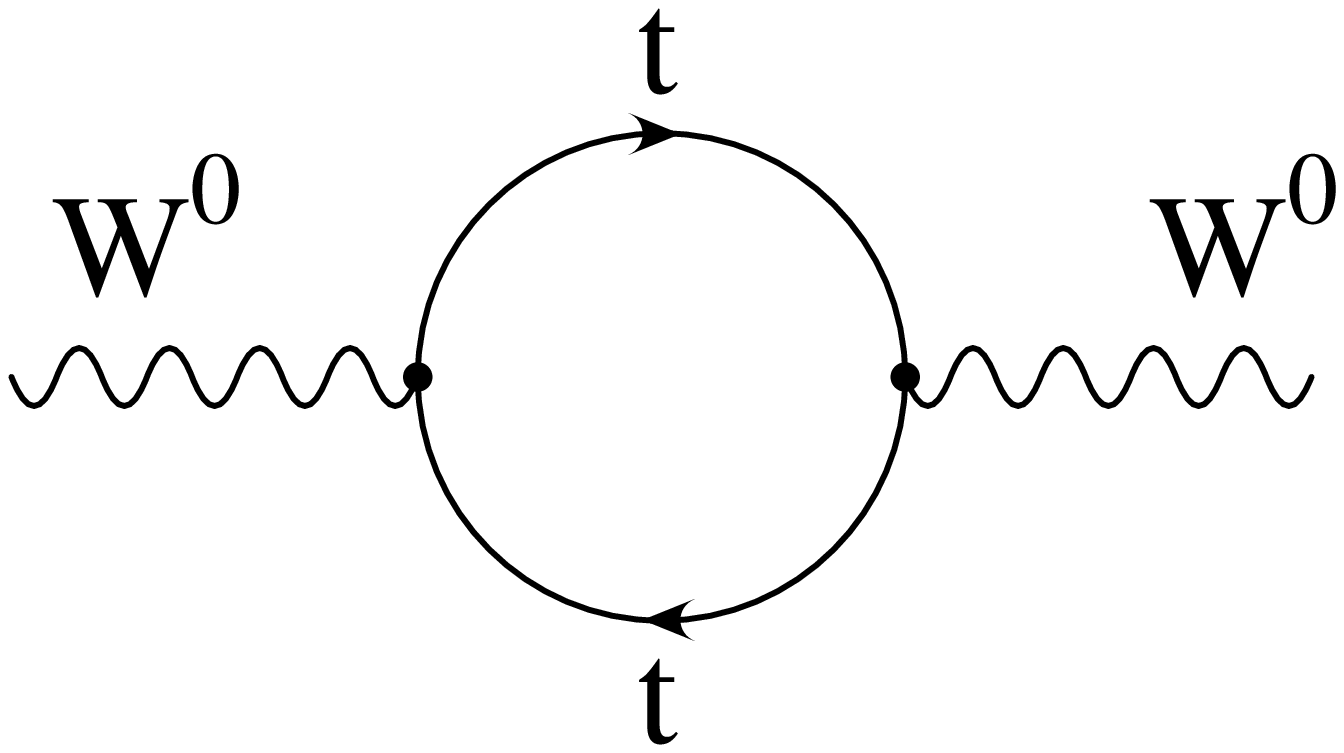,width=50mm,bbllx=210pt,bblly=410pt,%
bburx=630pt,bbury=550pt} 
& 
\psfig{figure=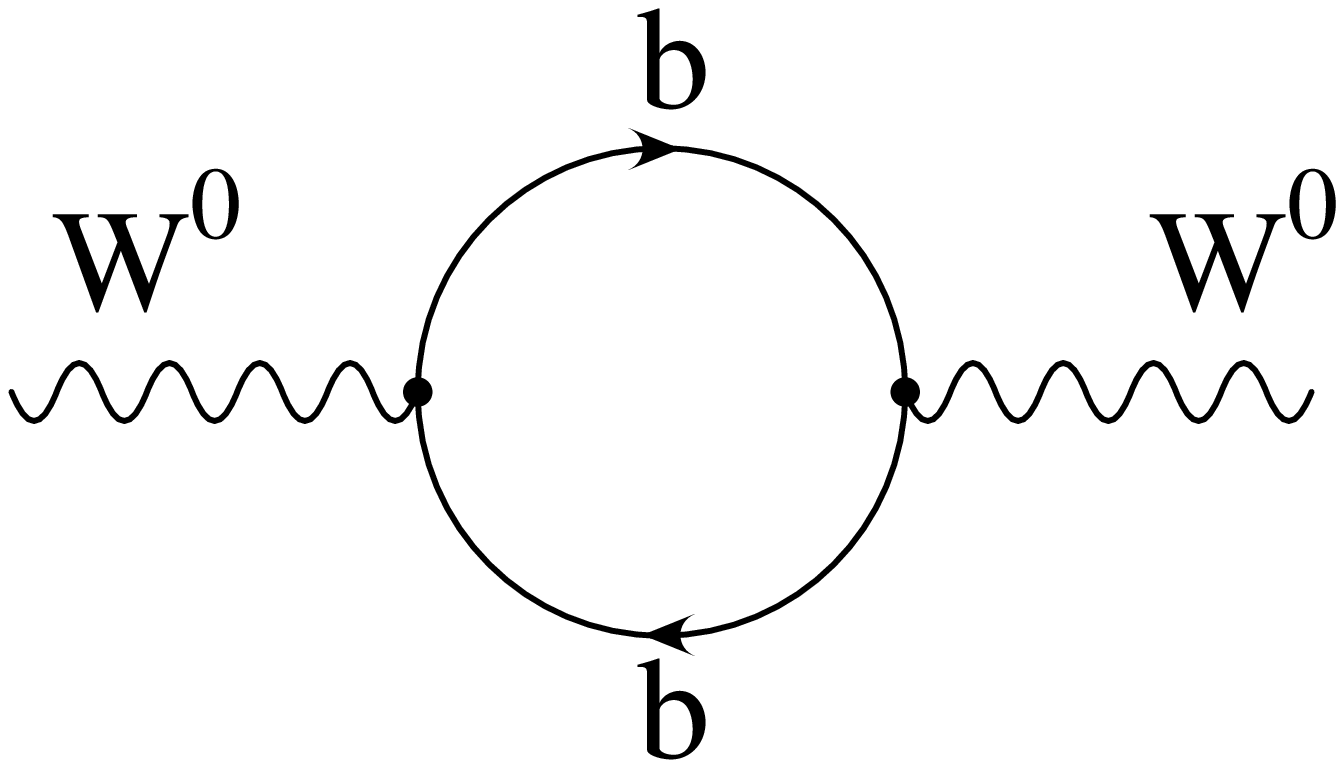,width=50mm,bbllx=210pt,bblly=410pt,%
bburx=630pt,bbury=550pt}    
\end{tabular}\\[15mm]
\caption{Self energy diagrams which induce the mass splitting between
$W^\pm$ and $W^0$  and a $\rho$ parameter different from one.}
\label{fig:self}
\end{figure}

Let us first consider the definition of the weak mixing angle in Born
approximation.  It can be fixed through the relative strength of
charged vs.~neutral current couplings:
\be
\hspace*{10mm}
\psfig{figure=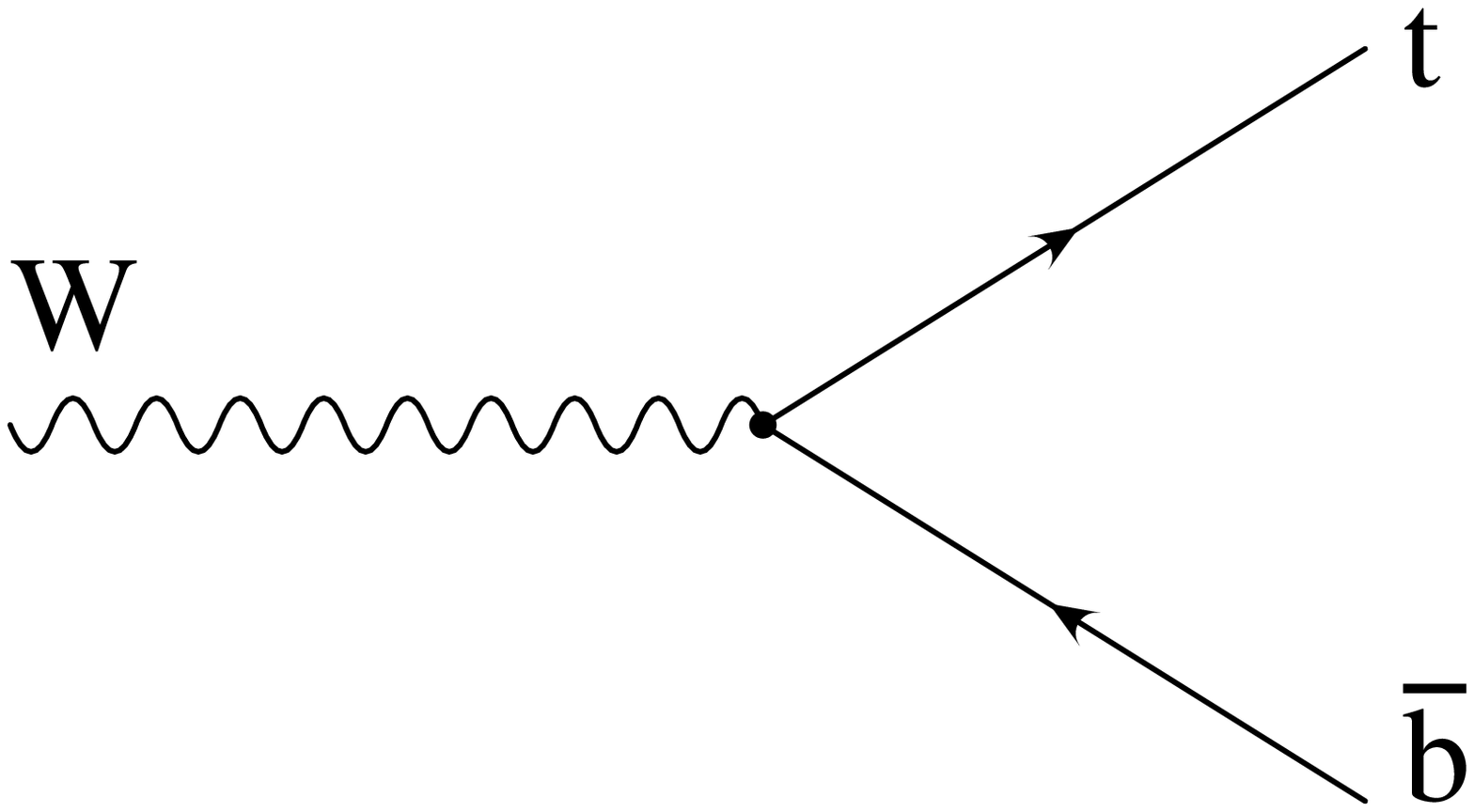,width=35mm,bbllx=210pt,bblly=410pt,%
bburx=630pt,bbury=550pt}  
& =&  {g\over 2\sqrt{2}} \gamma_\mu (1-\gamma_5)
\\[8mm]
\hspace*{10mm}
\psfig{figure=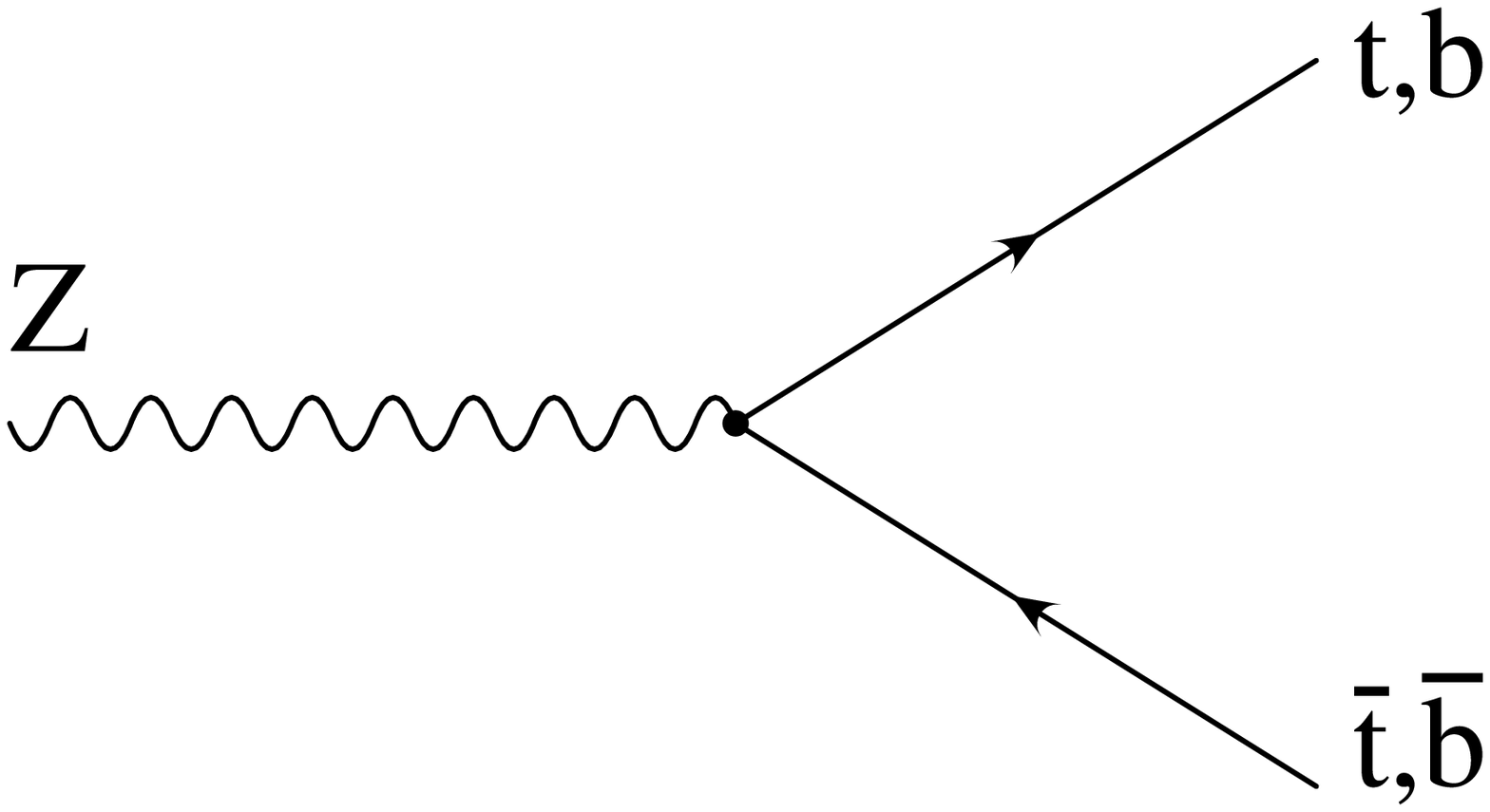,width=35mm,bbllx=210pt,bblly=410pt,%
bburx=630pt,bbury=550pt}  
& =&  {g\over 4\cos\bar \theta_W} 
\left[\left(2I_3^f-4Q_f\sin^2\bar\theta_W\right)\gamma_\mu 
+2I_3^f\gamma_\mu \gamma_5
\right]
\\[4mm]\nonumber
\ee
with the SU(2) coupling related to the electromagnetic coupling
through $g=e/\sin\bar \theta_W$.  Alternatively $\sin^2\theta_W$ is
defined through the mass ratio
\be
M_W^2/M_Z^2 = 1 - \sin^2\theta_W
\ee
These two definitions coincide 
in Born approximation 
\be
\theta_W = \left. \bar \theta_W \right|_{\rm Born}
\ee
However, the self energy diagrams depicted 
in Fig.~\ref{fig:self} lead to marked differences between the two
options, in particular if $m_t \gg M_{W,Z} \gg m_b$. 

This difference can be traced to a difference in the mass shift for
the $W$ and the $Z$ boson.  For a simplified discussion consider, in a
{\em first step},  $\sin^2\theta_W=0$ and hence the $SU(2)$ part
of the theory only.  The neutral boson will be denoted by $W^0$. 
In the lowest order this implies
\be
M_- = M_0 \equiv M
\ee
and the couplings simplify to
\be
\psfig{figure=Wbb.ps,width=35mm,bbllx=210pt,bblly=410pt,%
bburx=630pt,bbury=550pt}  
& =&  {g\over 2\sqrt{2}} \gamma_\mu (1-\gamma_5)
\\[8mm]
\psfig{figure=Ztb.ps,width=35mm,bbllx=210pt,bblly=410pt,%
bburx=630pt,bbury=550pt}  
 & =&  {g\over 4}  2I_3^f \gamma_\mu (1-\gamma_5)
\\[8mm]\nonumber
\ee
In order $g^2$ the propagators of charged and neutral bosons  are
modified by the self energies
\be
W^+:\qquad\qquad 
{1\over M^2 - s} & \Rightarrow & {1 \over M^2 - s - \Pi_+(s)}
\nonumber \\
W^0: \qquad\qquad 
{1\over M^2 - s} & \Rightarrow & {1 \over M^2 - s - \Pi_0(s)}
\ee
The mass shifts individually are given by
\be
\delta M_{+,0}^2 = M_{+,0}^2 - M^2 = - Re\, \Pi_{+,0}(M^2).
\ee
They are most easily calculated through dispersion relations from
their respective imaginary parts. These can be interpreted as the
``decay rate'' of a fictitious virtual boson of mass $\kappa$:
\be
\kappa \Gamma(W_{+}^*\to t + \bar b) = Im\,\Pi_{+}(\kappa^2)
\nonumber\\
\delta M^2_{+} = {1\over \pi} \int {d\kappa^2
\over \kappa^2-M^2} Im\,\Pi_{+}(\kappa^2).
\label{stern}
\ee
The decay rates of the virtual bosons are easily calculated
($m\equiv m_{top}$)
\be
\kappa \Gamma(W_{+}^* \to t\bar b) = 
{3\over 12\pi} \left( g\over 2\sqrt{2}\right)^2
\kappa^2 2 \left(1-{m^2\over 2\kappa^2} -{m^4\over
2\kappa^4}\right) 
\left(1-{m^2\over \kappa^2}\right) 
\Theta(\kappa^2-m^2).
\label{eq:rates}
\ee
The factor 3 originates from color, the factor 2 from the identical
vector and axial contributions, the  squared
matrix element and the  phase space 
 are responsible for the second and third factors in
brackets respectively.

Similarly one finds
\be  
\lefteqn{
\kappa \Gamma(W_{0}^* \to b\bar b) +
\kappa \Gamma(W_{0}^* \to t\bar t) =
}
\nonumber \\
&&
{3\over 12\pi} \left( g\over 4\right)^2
\kappa^2
\left\{
2\Theta(\kappa^2) + 2
\left(1-{m^2\over \kappa^2}\right)
\left(1-{4m^2\over \kappa^2}\right)^{1/2}  
\Theta(\kappa^2-4m^2)\right\}
\nonumber\\
\lefteqn{
\delta M^2_0 =  {1\over \pi} \int {d\kappa^2
\over \kappa^2-M^2} Im\,\Pi_{0}(\kappa^2).
}
\label{eq:simil}
\ee

With the large $\kappa^2$ behaviour of ${\rm Im}\,\Pi$ given by ${\rm Im}\,\Pi\sim
\kappa^2$ the 
dispersive integral eq.~(\ref{stern}) is  evidently quadratically
divergent.  In the limit of large $\kappa$ the leading ($\sim
\kappa^2$) and next-to-leading ($\to$ const) terms of
eqs. (\ref{eq:rates}) and (\ref{eq:simil}) coincide.  The leading and
next to leading divergences can therefore be absorbed in  a $SU(2)$
invariant 
mass renormalization.  The relative mass shift, however, the only
quantity accessible to experiment, remains finite and is given by
\be
\delta M^2 &=& \delta M^2_+ - \delta M^2_0 
\nonumber\\
&=& {3\over 12 \pi} {g^2\over 8} {1\over \pi} 
  \int_0^\infty {d\kappa^2\over \kappa^2-M^2}\kappa^2
\nonumber\\
&& \times \left\{ 
\begin{array}{ll}
-2\left(1-{m^2\over \kappa^2}\right) 
   \left( 1 -{m^2\over 2\kappa^2} - {m^4\over 2 \kappa^4} \right)
&
\times \Theta(\kappa^2-m^2) \\
+1 & \times \Theta(\kappa^2) \\
\left(1-{m^2\over \kappa^2}\right) 
   \left( 1 -{4m^2\over \kappa^2} \right)^{1/2}
&
\times \Theta(\kappa^2-4m^2) 
\end{array}
\right.
\ee
We are only interested in the leading top mass dependence: $m^2 \gg
M^2$.  The leading term is obtained by simply setting $M^2\to 0$ in
the integrand.  Introducing a cutoff $\Lambda^2 \gg m^2$ the leading
contributions to the three integrals are given by 
\be
\left\{ \rule{-1mm}{5mm} \right.
&&-2\Lambda^2 + 3m^2 \ln{\Lambda^2\over m^2} + {3\over 2}m^2+ \ldots
\nonumber\\
&& +\Lambda^2
\nonumber\\
&& +\Lambda^2 -  3m^2 \ln{\Lambda^2\over m^2}
 \left.  \rule{-1mm}{5mm}
\right\}
\ee
and hence
\be
\delta M^2 = {3\over 12 \pi^2} {g^2\over 8} {3\over 2}m^2
\ee
Up to the proportionality constant this result could have been guessed
on dimensional grounds from the very beginning.

 It has become customary to express the $SU(2)$ coupling in terms of
$G_F$ and the $W$ mass
\be
{g^2\over 8} = {G_F\over \sqrt{2}} M_W^2
\ee
such that
\be
{\delta M^2 \over M^2} = {3\over 16 \pi^2} \sqrt{2}G_Fm^2.
\label{eq:deltaMM}
\ee

The ratio of neutral versus charged current induced amplitude at small
momentum transfers is thus corrected by a factor
\be
{M_+^2\over M_0^2}  \equiv 1 + \delta\rho 
=\left(1 + {\delta M^2\over M^2} \right)
\ee
with $\delta\rho$ given in eq. (\ref{eq:deltaMM}).

To discuss the phenomenological implications of this result it is now
necessary to reintroduce the weak mixing between the neutral $SU(2)$
and $U(1)$ gauge bosons.  The gauge boson masses are induced by the
squared covariant derivative acting on the Higgs field
\be
D_\mu \phi \to -i\left( g\vec W_\mu {\vec \tau\over 2} + g^\prime
BY\right) 
\left( \begin{array}{c} 0\\ v/\sqrt{2} \end{array} \right)
\ee
giving rise to the following mass terms in the Lagrangian
\be
{\cal L}_M &=& {v^2\over 8} (g^2+ g^{\prime 2} )
\left({g \over \sqrt{g^2 + g^{\prime 2} }} W_3
-{g^\prime \over \sqrt{g^2 + g^{\prime 2} }} B\right)^2 
\nonumber \\
&& + {v^2\over 8} g^2 (W_1^2 + W_2^2) 
(1+\delta\rho)
\label{lagr}
\ee
The last term has been added to represent a contribution from a non
vanishing $\delta \rho$, induced e.g.~by the large top mass.  The
finite mass shift has been without loss of generality entirely
attributed to the charged $W$ boson.

The mass eigenstates are easily identified from eq.~(\ref{lagr})
\be
W^\pm &=& (W_1 \mp iW_2)/\sqrt{2}
\nonumber \\
Z &=& (\bar c W_3 - \bar s B)
\nonumber \\
A &=& (\bar c B + \bar s W_3)
\ee
with the weak mixing angle $\bar\theta_W$
\be
\bar c \equiv \cos \bar \theta_W 
       \equiv {g \over \sqrt{g^2 + g^{\prime 2} }} 
\nonumber \\
\bar s \equiv \sin \bar \theta_W
       \equiv {g^\prime \over \sqrt{g^2 + g^{\prime 2} }} 
\label{couplconst}
\ee
defined through the {\em couplings}.  This definition is, of
course, very convenient for measurements at LEP, where couplings are
determined most precisely.  The couplings of the photon and the $Z$
boson are thus also given in terms of $\bar s$.  The masses are read
off from eq.~(\ref{lagr})
\be
m_A^2 &=& 0
\nonumber\\
m_Z^2 &=& {v^2\over 4} (g^2 + g^{\prime2})
\nonumber\\
m_W^2 &=& {v^2\over 4} g^2(1+\delta\rho)
\ee
and 
\be
{m_W^2\over m_Z^2} = \bar c^2(1+\delta\rho)
\ee
which constitutes the standard definition of the $\rho$ parameter.
Alternatively one may define the mixing angle directly through the
mass ratio
\be
s^2 \equiv \sin^2\theta_W \equiv 1-M_W^2/M_Z^2
\label{massratio}
\ee
The two definitions coincide in the Born approximation; they differ,
however, for $\delta\rho \neq 0$:
\be
s^2 = 1-\bar c^2(1+\delta\rho) = \bar s^2 - \bar c^2\delta \rho
\ee
It is, of course, a matter of convention and convenience, which of the
two definitions (or their variants) are adopted.  The choice of input
parameters  and observables
will affect the sensitivity towards $\delta\rho$ --- and hence
towards $m_t$.  The observables which are measured with the highest
precision at present and in the forseable future are the
fine-structure constant $\alpha$, the muon life time which provides a
value for $G_F$ and the $Z$ boson mass.  To obtain the dependence of
$\sin^2\theta_W$ on $\delta\rho$ we predict $M_W^2$ from these
observables.  We start from
\be
{g^2\over 8} {1\over M_W^2} = {G_F\over \sqrt{2}}
\ee
and express $g^2$ through $G_F$ and $M_Z^2$ 
\be
g^2 = 4\sqrt{2} G_F M_Z^2 c^2
\ee
Alternatively $g$ can be related to the fine structure constant
\be
g^2 = e^2/\bar s^2
\ee
Note the appearance of $c^2$ in eq.~(\ref{massratio}) and of $\bar
s^2$ in eq.~(\ref{couplconst}).

One thus arrives at
\be
e^2 = 4\sqrt{2}  G_F^2 M_Z^2 c^2 (s^2 + c^2 \delta\rho)
\ee
or, equivalently, at
\be
{4\pi \alpha\over 4 \sqrt{2} G_F M_Z^2} {1\over 1+\cot^2\theta_W
\delta\rho} = s^2 c^2.
\ee
Solving for $\sin^2\theta_W$ (defined through the mass ratio) one
obtains on one hand
\be
\sin^2\theta_W \approx {1\over 2} \left[ 1 - \sqrt{1-{4\pi \alpha\over
\sqrt{2}G_F M_Z^2}}\; \right] -
{c^4\over c^2-s^2}\delta\rho
\label{eq:thet}
\ee
where the Born values for $c^2$ and $s^2$ can be taken in the
correction term. 
The definition of $\sin^2\bar \theta_W$ through the relative strength of
$SU(2)$ and $U(1)$ couplings leads on the other hand to
\be
\sin^2\bar\theta_W \approx {1\over 2} \left[ 1 - \sqrt{1-{4\pi \alpha\over
\sqrt{2}G_F M_Z^2}}\; \right] -
{s^2c^2\over c^2-s^2}\delta\rho
\label{eq:thetbar}
\ee

For the actual evaluation the running coupling $\alpha(M_Z)^{-1}
\approx 129$ must be employed \cite{Swartz}.  
Eq.~(\ref{eq:thet})  and eq.~(\ref{eq:thetbar})  exhibit rather
different sensitivity towards a 
variation of $\delta \rho$ and hence of $m_t$, with a ratio between the
two coefficients of $c^2/s^2 \approx 3.3$.  
 For a precise comparison between theory and
experiment subleading one-loop corrections must be included, and subtle
differences between variants of  $\bar\theta_W$ must be taken into
consideration, with  $\theta_{\overline{\rm MS}}$ and $\theta_{\rm eff}^{\rm
lept}$ as most frequently used options \cite{bardin}.  

With increasing experimental accuracy numerous improvements must be
and have been
incorporated into the theoretical predictions.

\begin{itemize}
\item
The full one loop corrections including all subleading terms are know
since long and are certainly indispensible (see
\cite{Swartz,bardin,avdeev,chetyr} and references therein).  
\item
 Two loop
purely weak corrections increase proportional $(G_Fm_t^2/16\pi^2)^2$.
A detailed discussion can be found in \cite{degrassi}.
\item
QCD corrections are available at the two- and even three-loop level 
\cite{Kniehl,avdeev,chetyr}.
\end{itemize}

With $\alpha$, $G_F$, and $M_Z$ fixed one may determine $m_t$ either
from $M_W$ or alternatively  from $\sin^2\bar\theta_W$ 
(corresponding to a measurement
of the left right asymmetry with polarised beams, the $\tau$
polarisation or forward backward asymmetries of unpolarised beams).
These measurements are in beautiful agreement with the determination
of $m_t$ in production experiments at the TEVATRON
(Fig.~\ref{fig:logo}).
\begin{figure}[htb]
\vspace*{10cm}
\caption{ }
\label{fig:logo}
\end{figure}

\section{Top Decays}
\label{decays}

Various aspects of top decays have been scrutinized in the literature.
The large top decay rate predicted in the SM governs top quark
physics.  Radiative correctons from QCD and electroweak interactions
have been calculated for the decay rate and for differential
distributions of the decay products.  Non-standard top decays are
predicted in SUSY extensions of the SM, with $t\to Hb$ and $t\to
\tilde t\tilde \gamma$ as most promising and characteristic
signatures.  Born predictions and radiative corrections (at least in
part) have been worked out also for these decay modes.  Beyond that a
number of even more exotic decay modes, in particular FCNC decays, have
been suggested.

\subsection{Qualitative aspects -- Born approximation}
\label{sect:qualit}
\begin{figure}[h]
\begin{minipage}{16.cm}
\hspace*{1cm}
$
\mbox{
\hspace*{18mm}
\begin{tabular}{|cll|}
\hline
 & & $S^t = +1/2$  \\
\psfig{figure=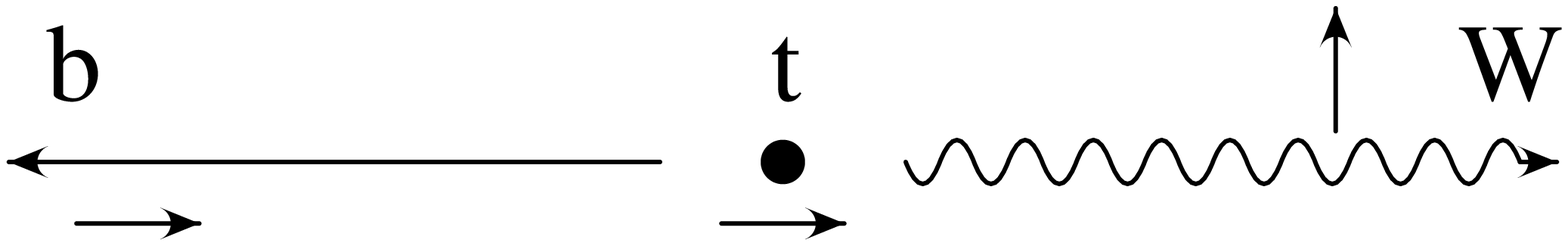,width=50mm,bbllx=0pt,bblly=300pt,%
bburx=630pt,bbury=420pt}
& $S_z^b = +1/2$ & $h^W = 0$
\\[2mm]
\hline
 & & $S^t = -1/2$  \\
\psfig{figure=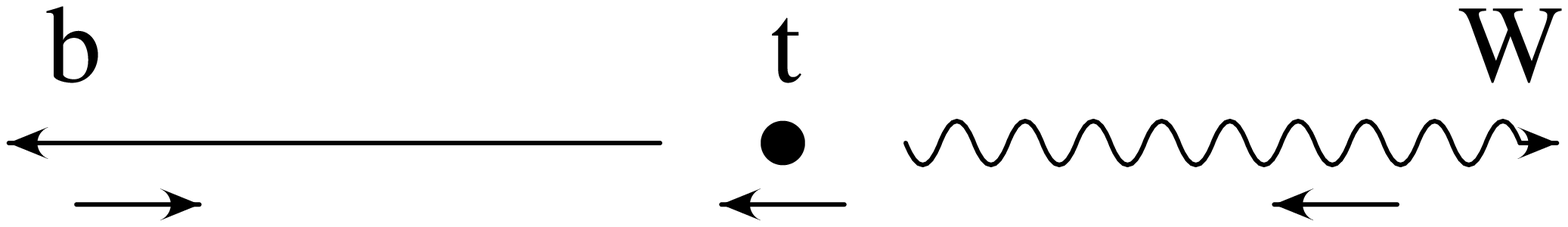,width=50mm,bbllx=0pt,bblly=300pt,%
bburx=630pt,bbury=420pt}
& $S_z^b = +1/2$ & $h^W = -1$
\\
\hline
 & & $S^t = \pm1/2$  \\
\psfig{figure=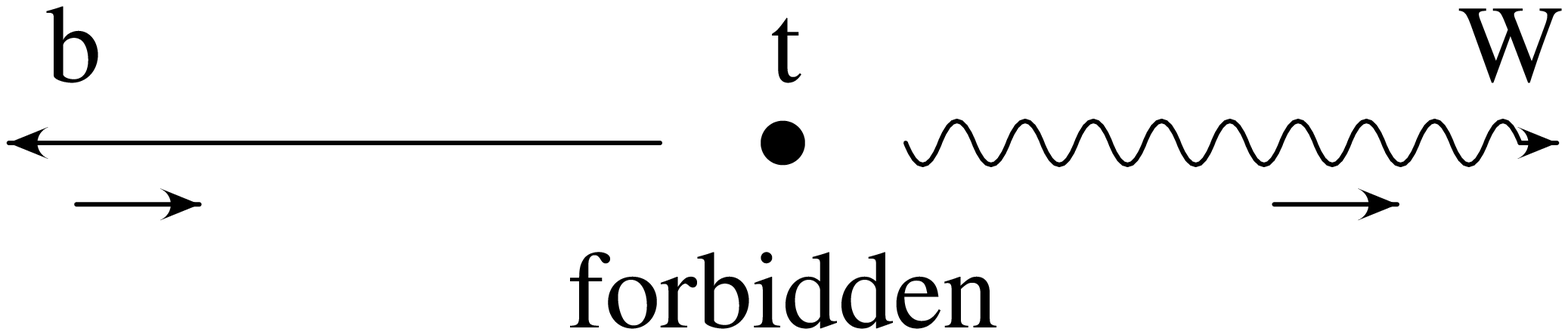,width=50mm,bbllx=0pt,bblly=300pt,%
bburx=630pt,bbury=420pt}
& $S_z^b = +1/2$ & $h^W = +1$
\\
\hline
\end{tabular}
}
$
\end{minipage}
\caption{Top decays: angular momentum conservation}
\label{fig:TopDec}
\end{figure}

The decay of the top quark into $b+W$ is governed by the following
amplitude
\be
{\cal M}(t\to bW) = {ig\over \sqrt{2}} \bar b \not \varepsilon^W
{1-\gamma_5\over 2} t
\ee
Adopting the high energy limit ($m_t^2>M_W^2$) for the polarisation
vector $\epsilon_L$ of the longitudinal $W$ (corresponding to helicity
$h^W=0$)  
\be
\varepsilon^W_L = 
\left(
\begin{array}{c}
p_3^W\\
0\\
0\\
p_0^W
\end{array}
\right)
{1\over M_W}
=
\left(
\begin{array}{c}
p_0^W\\
0\\
0\\
p_3^W
\end{array}
\right)
{1\over M_W}
+
{\cal O}(M_W/m_t)
\ee
the amplitude is dominated by contribution from longitudinal $W$'s
\be
{\cal M}_L &=& {ig\over \sqrt{2}} \bar b \not \varepsilon_L^W
{1-\gamma_5\over 2} t 
\approx 
{ig\over \sqrt{2}} {m_t\over M_W} 
\bar b  {1+\gamma_5\over 2} t 
\nonumber \\
&=& i\sqrt{2} {m_t\over v} 
\bar b  (1+\gamma_5) t 
\ee
This  part is thus 
proportional to
the Yukawa coupling 
\be
g_Y = \sqrt{2} {m_t\over v} 
\ee
with a rate growing proportional $m_t^3$.  In contrast, 
the amplitude for the decay
into transverse $W$'s, is obtained with the polarisation vectors
\be
\varepsilon^\pm_T = 
{1\over \sqrt{2}}
\left(
\begin{array}{c}
0\\
1\\
\pm i\\
0\\
\end{array}
\right)
\ee
and remains constant in the high mass limit.
The rate is governed by the gauge coupling $g$ and increases only
linearly with $m_t$.
The longitudinal or transversal 
$W$ is produced in conjunction with a lefthanded $b$ quark.  The
production of $W$'s with helicity $h^W=+1$ is thus forbidden by angular
momentum conservation (see fig.~\ref{fig:TopDec}).  

In total one finds 
\be
(h^W=-1) : (h^W=0) : (h^W=+1) = 1:{m_t^2\over 2 M_W^2}:0.
\ee
The implications for the angular distributions of the decay products
will be discussed below.  The decay rate
\be
\Gamma &=& {G_Fm_t^3\over 8 \sqrt{2}\pi} 
\left(1-{M_W^2\over m_t^2}\right)^2
\left(1+2{M_W^2\over m_t^2}\right)
\nonumber \\
&\approx& 175\, \mbox{MeV} 
\left({m_t\over M_W}\right)^3
\ee
increases with the third power of the quark mass and, for a
{\em realistic} top mass around 180 GeV amounts to more than 1.5 GeV,
exceeding significantly all hadronic scales.  Before we discuss the
implications of this fact let us briefly pursue the close similarity
between the coupling of the longitudinal $W$ to the $tb$ system and
the decay into a charged Higgs boson in a two Higgs doublet model
(THDM).  The decay rate is given by (see also section \ref{nonstan}) 
\be
\Gamma(t\to H^+b)= {G_Fm_t^3\over 8 \sqrt{2}\pi} 
\left(1-{m_H^2\over m_t^2}\right)^2
\left[
\left({m_b\over m_t}\right)^2\tan^2\beta + \cot^2\beta
\right]
\ee
The similarity between this rate and the rate for the decay into
longitudinal $W$'s is manifest  from the cubic top mass
dependence.  The minimal value of the term in brackets is assumed for
$\tan\beta = \sqrt{m_t/m_b}$.  Adopting $m_b(\mbox{running})\approx 3$
GeV, the minimal value of the last factor amounts to about $1/30$. On
the other hand, in any plausible THDM  the value of 
$\tan\beta$ should not exceed
$m_t/m_b$.  The $W$ decay mode will hence never be swamped by the
Higgs channel. (This fact is of course also implied by the actual
observation of the top quark at the TEVATRON.)  Up to this point we
have, tacitely, assumed the CKM matrix element $V_{tb}$ to be close to
one.  In fact, in the three generation SM one predicts (90\% CL)
\begin{eqnarray}
V_{tb} = 0.9990 \pm 0.0004 & \Rightarrow & BR(b) \approx 1\nonumber\\
V_{ts} = 0.044 \pm 0.010 & \Rightarrow & BR(s) \approx0.2\%\nonumber\\
V_{td} = 0.011 \pm 0.009 & \Rightarrow & BR(d) \approx 0.01\%
\label{eq:2}
\end{eqnarray}
on the basis of CKM unitarity.  In a four generation model, however,
sizeable mixing between third and fourth generation could arise.
Methods to determine the strength of $V_{tb}$ either through single
top production at a hadron collider or through a direct measurement of
$\Gamma_t$ in $e^+e^-$ colliders 
will be discussed below in sects.~\ref{singletop} and 3.2.

The large top decay rate provides a cutoff for the long distance QCD
dynamics.  The implications can be summarised in the statement: ``$t$
quarks are produced and decay like free quarks'' \cite{acta}. In
particular the angular distributions of their decay products follow
the spin 1/2 predictions.  This is in marked contrast to the situation
for $b$ quarks, with $B$ mesons decaying isotropically.  The arguments
for this  claim are either based on a comparison of energy scales, or,
alternatively, on a comparison of the relevant time scales.

  Let us
start with the first of these two equivalent viewpoints:
The mass difference between $B^{**}$ and $B$ mesons amounts to 450
MeV.  In the nonrelativistic quark model the $B^{**}$ is interpreted
as orbitally excited $b\bar q$ state.  With increasing mass of the
heavy quark this splitting remains approximately constant: it is
essentially governed by light quark dynamics. The hyperfine splitting
between $B^*$ and $B$, in contrast, is proportional to the color
magnetic moment and hence decreases $\sim 1/m_Q$.  Given a decay rate
of about 1.5 GeV it is clear that $T$-, $T^*$-, and $T^{**}$- mesons
merge and act coherently, rendering any distinction between individual
mesons meaningless.  In fact even individual toponium states cease to
exist.  From the perturbative QCD potential an energy difference
between $1S$ and $2S$ states around 1.2--1.5 GeV is predicted.  This
has to be contrasted with the toponium decay rate $\Gamma_{t\bar
t}\sim 2\Gamma_t \sim 3$ GeV.  All resonances merge and result in an
excitation curve which will be discussed in chapter 3.

A similar line of reasoning is based on the comparison between
different characteristic time scales:  The formation time of a hadron
from a locally produced $t$ quark is governed by its size which is
significantly larger than its lifetime
\be
\tau_{\mbox{\small Formation}} \approx \mbox{size} \approx 1/0.5 \mbox{GeV}
\gg \tau_{\mbox{\small Decay}} \approx 1/\Gamma_t
\ee
Top quarks decay before they have time to communicate hadronically
with light quarks and dilute their spin orientation.
For sufficiently rapid top quark decay even $t\bar t$ bound states
cease to exist.  The classical time of revolution $T_{\rm rev}$ 
for a Coulombic
bound state is given by $\left(\alpha_{\rm eff} = {4\over
3}\alpha_s\right)$
\be
T_{\rm rev} = \pi \alpha_{\rm eff} \sqrt{{m_{\rm red}\over 2|E|^3}}
\ee
With $E=-\alpha_{eff}^2 m_{\rm red}/2$ for the ground state
\be
T_{\rm rev} = \pi/|E| \gg 1/\Gamma_{t\bar t}
\ee
is obtained. The lifetime of the $t\bar t$ system is too small to
allow for the proper definition of a bound state with sharp binding
energy.

\subsection{Radiative corrections to the rate}

Perturbative corrections to the lowest-order result affect the total decay
rate as well as differential distributions.
Their inclusion is a necessary
prerequisite for any analysis that attempts
a precision analysis of top decays.
Both QCD and electroweak corrections are well under control and will be
discussed in the following.

\subsubsection{QCD corrections}

\begin{figure}

\psfig{figure=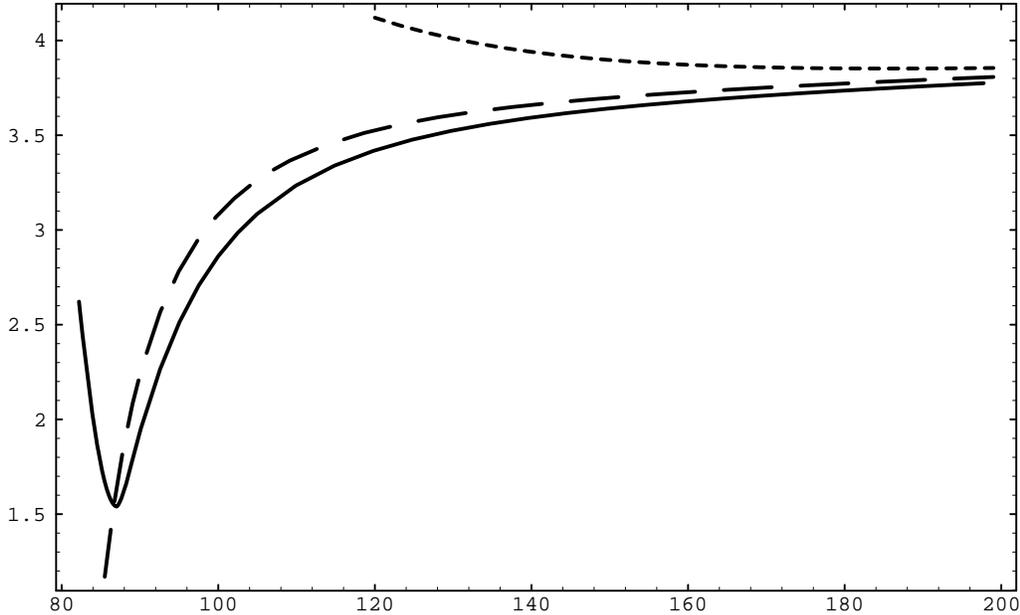,width=180mm,bbllx=40pt,bblly=230pt,%
bburx=670pt,bbury=580pt}

\caption{\label{fig:2.3}QCD correction function $f$ for the top quark
decay rate.  Solid line:  exact form; dashes:  $m_b = 0$, dotted:
approximate form for large $m_t/M_W$.}
\end{figure}

The correction to the {\em decay rate} is usually written in the form
\begin{equation}
\Gamma = \Gamma _{\rm Born}\left(1+\delta _{QCD} \right)
 =\Gamma _{\rm Born} 
\left( 1 - \frac{2}{3} \frac{\alpha _s}{\pi} f \right)
\label{eq:20}
\end{equation}
The correction function $f$ has been calculated in \cite{jk3}
for nonvanishing and vanishing
$b$ mass.
 In the limit $m^2_b /
m^2_t \to
0$ the result simplifies considerably, but remains a valid approximation
(Fig.\ref{fig:2.3}):
\begin{eqnarray}
f & = & {\cal F}_1 / {\cal F}_0 \nonumber \\
{\cal F}_0 & = & 2(1-y)^2 (1+2y) \nonumber \\
{\cal F}_1 & = & {\cal F}_0 \left[ \pi ^2 + 2 Li_2 (y) - 2Li_2 (1-y)\right] \nonumber \\
 & & + 4y(1-y-2y^2) \ln y + 2(1-y)^2 (5+4y) \ln(1-y) \nonumber \\
 & & - (1-y)(5+9y-6y^2),
\label{eq:21}
\end{eqnarray}
where $y=m^2_W / m^2_t$.
  In the limit $m^2_t \gg m^2_W$ $f$ is well approximated by
$f(y) = 2/3 \pi ^2 - 5/2 -3y +9y^2/2-3y^2\ln y \approx 4$.
For $m_t \approx 180~GeV$
the QCD correction amounts to
\begin{equation}
\delta _{QCD} \approx -3.7 \alpha _s / \pi
\label{eq:22}
\end{equation}
and lowers the decay rate by about 10\%.  This has a non-negligible impact
on the height and width of a toponium resonance or its remnant.

The $\alpha_s^2$ corrections are presently unknown, and the scale
$\mu$ in $\alpha_s(\mu^2)$ is uncertain.  Indications for a
surprisingly large correction of order $\alpha_s^2$, corresponding to
a rather small scale have been obtained recently.  Diagrams with
light fermion insertions into the gluon propagator have been
calculated numerically \cite{smith} and analytically \cite{ac95a} in
the limit $m_t\gg M_W$
\be
\Gamma &=& \Gamma_{\rm Born} \left[
1-{2\over 3} {\alpha_{\overline{\rm MS}}(m^2_t)\over \pi}
\left(4\zeta_2-{5\over 2}\right)
\right.
\nonumber\\
&&+
\left.
\left( {\alpha_{\overline{\rm MS}}\over \pi} \right)^2 
\left( -{2n_f\over 3} \right)
\left( {4\over 9} -{23\over 18}\zeta_2 - \zeta_3\right)
\right]
\ee
The BLM prescription \cite{BLM} suggests that the dominant
$\alpha_s^2$ coefficients can be estimated through the replacement
\be
-{2n_f\over 3} \to \left( 11 -{2n_f\over 3}\right)
\ee
and absorbed through a change in the scale.  For the problem at hand
this corresponds to a scale $\mu = 0.122m_t$ resulting in a fairly
large effective value of $\alpha_s$  of $0.15$  instead of
0.11 for $\mu = m_t$. 

\subsubsection{Electroweak corrections}

Electroweak corrections to the top quark decay rate can be found in
\cite{den1,eil}. They involve a large number of diagrams.
For asymptotically  large top masses the Higgs exchange diagram
 provides the dominant contribution.
Defining the Born term by means of the Fermi coupling $G_F$,
one derives in this limit
\begin{eqnarray}
\Gamma & = & \Gamma(G_F)_{\rm Born}\left[ 1 + \delta_{EW} \right], \\
 \delta_{EW} &=& \frac{G_F m_t^2}{4\sqrt2\pi^2} \left[ \frac{17}{4} + \log
\frac{m_H^2}{m_t^2}  \right] + \mbox{subleading terms}. \nonumber
\end{eqnarray}
While the Higgs--top coupling is the origin of the strong quadratic dependence
 on the
top mass, the Higgs itself is logarithmically screened in this limit.
However, the detailed
analysis reveals that the subleading terms are as important as the leading
terms so that finally one observes only a very weak
dependence of $\delta_{EW}$
on the top and the Higgs masses, Fig.\ref{FDenner}. The numerical value
of the corrections turns out to be small, $\delta_{EW} \approx +2 \%$.
Electroweak corrections in the context of the two Higgs doublet model
can be found in \cite{DennerHoang} and are of comparable magnitude.
\begin{figure}[hbt]

\vspace{9cm}

\caption[]{\label{FDenner}QCD and electroweak radiative
corrections to the top decay width;
adapted from Ref.\cite{den1}.}
\end{figure}

The positive correction $\delta_{EW}$ is nearly cancelled by the
negative correction $\delta_\Gamma$ of -1.5\% from the nonvanishing
finite width of the $W$.  The complete prediction  taken from
\cite{JezK} is displayed in table~\ref{compilation} for the choice
$\alpha_s(\mu^2=m_t^2)$. 
\begin{table}[h]
\begin{tabular}{|c|c|c|c|c|c|c|}
\hline
$m_t$ [GeV] & $\alpha_s(m_t)$ & $\Gamma^{\rm Born}_{\rm nar.\,w.}$ [GeV]&
$\delta^{(0)}_\Gamma$ [\%] & 
$\delta^{\rm nw}_{\rm QCD}$ [\%] & 
$\delta_{\rm EW}$ [\%] & 
$\Gamma$ [GeV] \\
\hline
170 & .108 & 1.41 & -1.52 & -8.34 & 1.67 & 1.29 \\
\hline
180 & .107 & 1.71 & -1.45 & -8.35 & 1.70 & 1.57 \\
\hline
190 & .106 & 2.06 & -1.39 & -8.36 & 1.73 & 1.89 \\
\hline
\end{tabular}
\caption{Top width as a function of top mass and the comparison of the
different approximations. }
\label{compilation}
\end{table}
For $\mu = 0.112 m_t$ the QCD correction amounts to -11.6 \% instead
of -8.3\%. This variation characterises the present theoretical
uncertainty, which could be removed by a full $\alpha_s^2$ calculation
only.  Additional uncertainties, e.g. from the input value of $\alpha_s$
($\sim 1\%$) or from the fundamental uncertainty in the relation
between the pole mass $m_t$ and the experimentally measured excitation
curve (assuming perfect data) of perhaps 0.5 GeV can be neglected in the
forseable future. 

Hence, it appears that the top quark width (and similarly the spectra
to be discussed below) are
well under theoretical control, including QCD and electroweak corrections.
The remaining uncertainties are clearly smaller than the experimental error
in $\Gamma_t$, which will amount to 5-10\% even at a linear collider
\cite{Fujii}. 

\subsection{Decay spectra and angular distributions}

\subsubsection{Born predictions}
Arising from a two body decay, the energy of the $W$ and of the hadronic
system ($\equiv b$ jet) are fixed to
\begin{eqnarray}
E_W & = & \frac{m^2_t + m^2_W - m^2_b}{2m_t} \nonumber\\
E_h & = & \frac{m^2_t + m^2_b - m^2_W}{2m_t}
\label{eq:5}
\end{eqnarray}
as long as gluon radiation is ignored.  The smearing of this $\delta$
spike 
by the combined effects of perturbative QCD and from the finite width
of the $W$ will be treated below.

Top quarks will in general be polarized through their electroweak
production mechanism.  For unpolarized beams and close to threshold
their polarization is given by the right/left asymmetry which would
be measured with longitudinally polarized beams \cite{stre}:
\begin{equation}
P_t = A_{RL}
\label{eq:11}
\end{equation}

For fully longitudinally polarized electron (and unpolarized positron)
beams the spin of both $t$ and $\overline{t}$ is aligned with the spin
of the $e^-$.  Quark polarization then leads to angular distributions of
the decay products which allow for various tests of the chirality of the
$tbW$ vertex.

The angular distribution of the longitudinal
and transverse W's is analogous
to those of $\rho$ mesons from $\tau$ decay $(m_{\tau} \to m_t,
m_{\rho} \to m_W)$
\begin{eqnarray}
\frac{dN_{T/L}}{d \cos\theta} = \frac{1}{2}
(1 \mp P_t \cos\theta )\quad  f\!or \quad
h_W = \left\{ {+1 \choose 0 } \right.
\label{eq:12}
\end{eqnarray}
and, after summation over the W polarizations
\begin{eqnarray}
\frac{dN}{d \cos\theta} = \frac{1}{2} \left( 1 +
\frac{m^2_t - 2m^2_W}{m^2_t + 2m^2_W} P_t \cos\theta \right)
\label{eq:13}
\end{eqnarray}
The angle between top quark spin and direction of the W
is denoted by $\theta$.
In the limit of $m_t \gg M_W$ the coefficient of the $P_t \cos\theta$
term rises to 1, for $ m_t= 180$ GeV, however, it amounts to 0.43
only. 

The angular distribution of leptons from the chain
$t \to b + W(\to \ell ^+\nu )$
will in general follow a complicated pattern with an energy
dependent angular distribution
\begin{equation}
\frac{dN}{dx d \cos\theta} = f(x) + g(x) P_t \cos\theta.
\label{eq:15}
\end{equation}

In the SM, however, a remarkable simplification arises.
Energy and angular distribution factorize \cite{stre,jk1}
\begin{equation}
\frac{dN}{dx d \cos\theta} = f(x) (1 + P_t \cos\theta )/2.
\label{eq:16}
\end{equation}
 
This factorisation holds true for arbitrary $m_t$ and even including
the effect of the nonvanishing $b$-quark mass \cite{JezK}.

\subsubsection{QCD corrections}
The $\delta$ spike in the energy distribution of the hadrons from the
decay $t \to b+W$ is smeared by quark fragmentation (not treated in this
context).

Hard gluon radiation leads to a slight shift and distortion of the energy
spectra with a tail extending from the lower limit given by two-body kinematics
upwards to $m_t - m_W$
\begin{equation}
\frac{m^2_t + m^2_b - m^2_W}{2m_t} \leq E_{had} < m_t - m_W
\label{eq:23}
\end{equation}

\begin{figure}

\psfig{figure=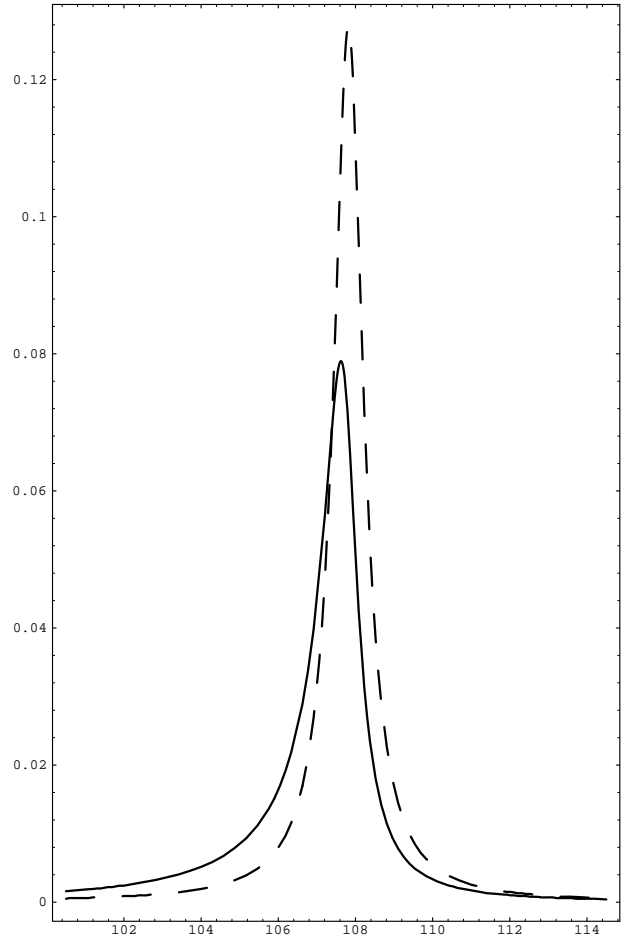,width=120mm,bbllx=-100pt,bblly=50pt,bburx=530pt,bbury=800pt}
\caption[]{\label{fig:2.4}Distribution of
the W energy for $m_t = 180~GeV$
without (dashed) and with (solid curve) QCD corrections.}
\end{figure}

Including finite W-width effects and $m_b \not= 0$ the differential hadron
energy distribution has been calculated in \cite{cjk}. The hadron energy
distribution is shown in Fig.\ref{fig:2.4} for $m_t = 180$ GeV.

The lepton spectrum (as well as the neutrino spectrum)
receives its main
correction close to the end points where the counting rates are fairly
low. 

Including QCD corrections \cite{jk1,czajk91,cjkk94,cj94} the spectrum
of both charged lepton and neutrino can be cast into the form
\be
{ {\rm d} N\over {\rm d}x{\rm d}\cos\theta}
= A(x) + B(x) \cos\theta
\ee
The shape of the charged lepton spectrum is hardly different from the
lowest order result \cite{jk1} with main corrections towards the end
point.
 $B_e(x)\approx A_e(x)$ remains
valid to extremely high precision \cite{czajk91}.  The charged lepton
direction is thus a perfect analyser of the top spin, even after
inclusion of QCD corrections.  A small admixture of $V+A$ couplings
will affect spectrum and angular distributions of electron and
neutrino as well.  Assuming a $V+A$ admixture of relative rate
$\kappa^2 = 0.1$, the functions $A_e$, $B_e$ and $A_\nu$ are only marginally
modified (Fig.~\ref{fig-1} and \ref{fig-2}).  The angular dependence
part of the 
neutrino spectrum $B_\nu$, however, is changed significantly 
(Fig.~\ref{fig-2}).  This observation could provide a useful tool in
the search for new couplings.
\begin{figure}
\begin{center}
\leavevmode
\epsffile[70 320 550 500]{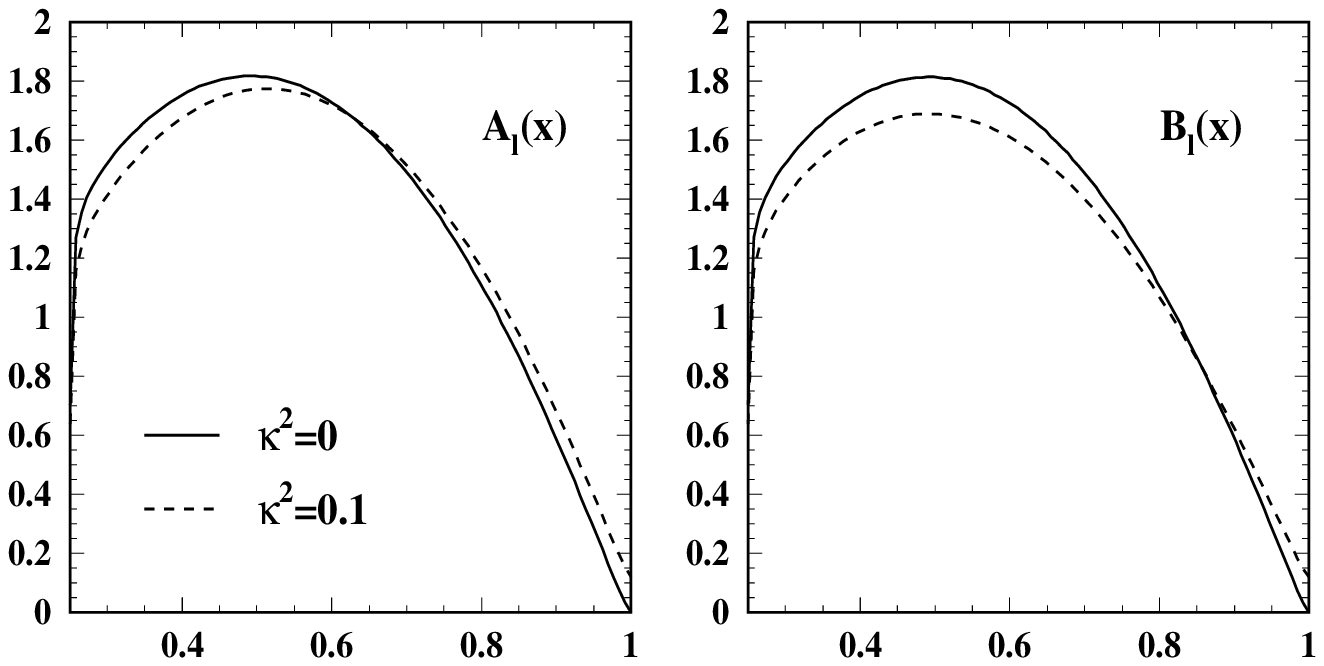}
\caption{The coefficient functions a) ${\rm A}_\ell(x)$ and
b) ${\rm B}_\ell(x)$ defining the charged lepton angular-energy
distribution for $y=0.25$ and $\alpha_s(m_t)=0.11$ : \ \
$\kappa^2=0$ -- solid lines and $\kappa^2=0.1$ -- dashed lines.
\label{fig-1} }
\end{center}
\begin{center}
\leavevmode
\epsffile[70 320 550 500]{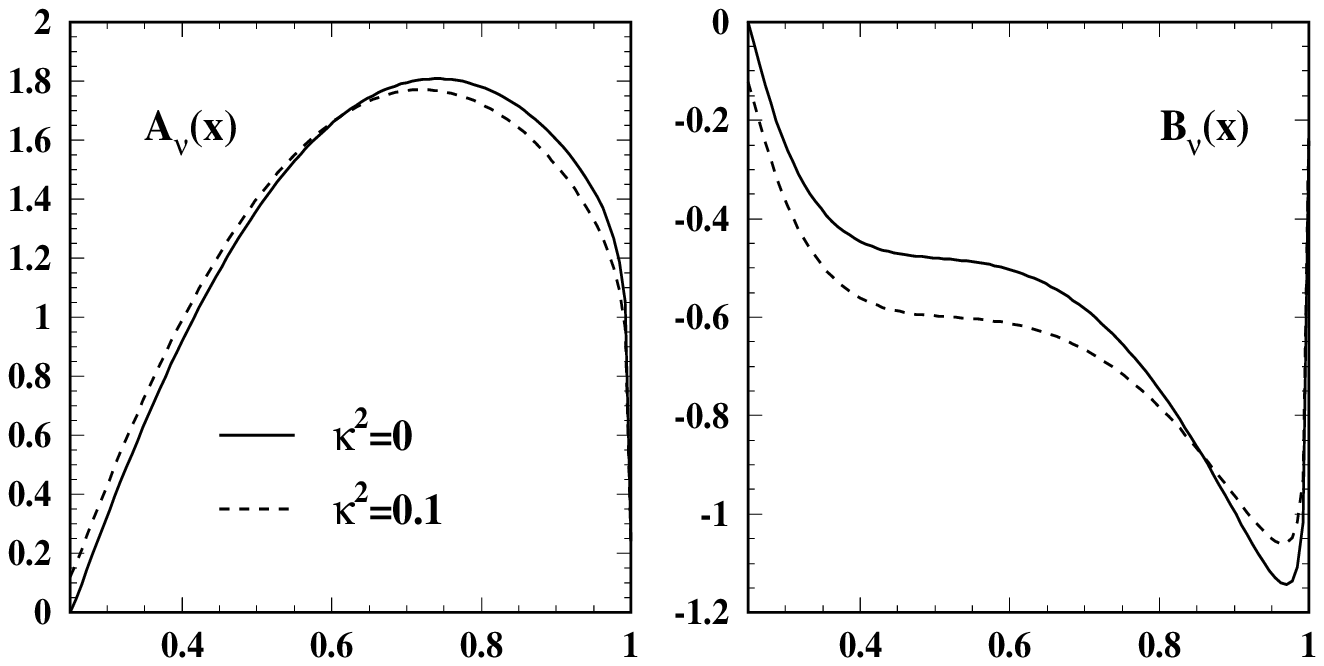}
\caption{The coefficient functions a) ${\rm A}_\nu(x)$ and
b) ${\rm B}_\nu(x)$ defining the neutrino angular-energy
distribution for $y=0.25$ and $\alpha_s(m_t)=0.11$ : \ \
$\kappa^2=0$ -- solid lines and $\kappa^2=0.1$ -- dashed lines.
\label{fig-2}  }
\end{center}
\end{figure}

\subsection{Non--standard top decays}
\label{nonstan}
The theoretical study of non--standard top decays is motivated by
the large top quark mass which could allow for exciting novel
decay modes, even at the Born level.
A few illustrative, but characteristic examples will be discussed
in some detail in the following section.

\subsubsection{Charged Higgs decays}

Charged Higgs states $H^\pm$ appear in 2--doublet Higgs models in
which out of the eight degrees of freedom three Goldstone bosons build up the
longitudinal states of the vector bosons and three neutral and two charged
states correspond to real physical particles. A strong motivation for this
extended Higgs sector is provided by supersymmetry
which requires the \sm\
Higgs sector to be doubled in order to generate masses for the up and
down--type fermions. In the minimal version of that
model the masses of the charged Higgs particles
are predicted to be larger than the $W$ mass, mod.\ radiative corrections,
\[
m(H^\pm) > m(W^+) \mbox{ [mod. rad. corr.]}
\]
We shall adopt this specific model for the more detailed discussion in the
following paragraphs.

If the charged Higgs mass is lighter than the top mass, the top quark may
decay into $H^+$ plus a $b$ quark \cite{ns30},
\[
t \ra b + H^+
\]
The  coupling of the charged Higgs to the scalar $(t,b)$ current is defined by
the quark masses and the parameter $\tan \beta$,
\begin{equation}\label{gl10}
J(b,t) = {1\over \sqrt{2}v} \left[ \left( m_b \tan \beta + m_t \cot
 \beta
\right) - \gamma_5 \left( m_b \tan \beta - m_t \cot \beta  \right) \right].
\end{equation}
The parameter $\tan\beta$ is the ratio of the vacuum expectation values of
the
Higgs fields giving masses to up and down--type fermions, respectively. For
the sake of
consistency, related to grand unification, we shall assume
$\tan\beta$ to be bounded by
\begin{equation}\label{gl11}
1 < \tan\beta = \frac{v_2}{v_1}  < \frac{m_t}{m_b} \sim 60
\end{equation}
\noindent with $v = \sqrt{v_1^2 + v_2^2}=(\sqrt{2}G_F)^{-1/2}$ corresponding
to the ground state of the \sm\  Higgs field.
The width following from the coupling (\ref{gl10}) has a form quite
similar to the \sm\  decay mode [see e.g.\ \cite{bigi2}],
\begin{equation}
\Gamma(t\ra b + H^+) = \frac{G_F m_t^3}{8\sqrt{2}\pi} \left[ 1 -
\frac{m_H^2}{m_t^2} \right]^2 \left[ \left( \frac{m_b}{m_t} \right)^2
\tan^2\beta + \cot^2\beta \right]
\end{equation}

\begin{figure}[p]

\vspace{8.5cm}

\caption[]{{\label{FHiggs} \bf (a)} Branching ratios for the decays $t
\ra b W^+$ and $t \ra b  H^+$ in two--doublet Higgs models \cite{reya1}.}
\addtocounter{figure}{-1}

\vspace{10cm}

\caption[]{{\bf (b)} Branching ratios for charged Higgs decays to
$\tau$ leptons  and quarks \cite{reya1}.}
\end{figure}

The branching ratio of this novel Higgs decay mode is
compared with the $W$ decay mode in Fig.\ref{FHiggs}a
(The behaviour is qualitatively similar for $m_t=180$ GeV.)
In the parameter range  eq.~(\ref{gl11}) the $W$ decay mode is
dominant; the Higgs decay branching ratio is in general small, yet large
enough to be clearly observable \cite{PIK}.
The Higgs branching ratio is
minimal at $\tan \beta = \sqrt{m_t / m_b} \sim 6-8$. QCD corrections
to the $t\to Hb$ mode have been calculated in \cite{CzarDav} and
electroweak corrections in \cite{EWHiggs}.

The detection of this scalar decay channel is facilitated by the
characteristic decay pattern of the charged Higgs bosons
\[
H^+ \ra \tau^+ + \nu_\tau \quad \mbox{ and }\quad c + \overline{s}
\]
Since $H^\pm$ bosons couple preferentially to down--type fermions
\cite{ns38} for $\tan\beta > 1$,
\begin{eqnarray}
\Gamma(H^+ \ra \tau^+ \nu_\tau) & = &    \frac{G_F m^2_\tau}{\sqrt{2}}
\frac{m_H}{4\pi} \tan^2\beta \\
\Gamma(H^+ \ra c \overline{s} ) & = &   \frac{3G_Fm^2_c}{\sqrt{2}}
\frac{m_H}{4\pi} \left[ \left( \frac{m_s}{m_c} \right)^2 \tan^2 \beta
+ \cot^2 \beta \right]
\end{eqnarray}
\noindent the $\tau$ decay mode wins over the quark decay
mode [Fig.\ref{FHiggs}b], thus
providing a clear experimental signature. A first signal of top decays into
charged Higgs particles would therefore be the breakdown
of $\mu, e$ {\it vs.}\ $\tau$
universality in semileptonic top decays.

An interesting method for a determination of $\tan\beta$ is based on
an analysis of the angular distribution of Higgs bosons in the decay
of polarized top quarks
\be
{ {\rm d} N \over  {\rm d} \cos\theta}
\sim
m_t^2 \cot^2\beta (1+\cos\theta_H)
+m_b^2 \tan^2\beta (1-\cos\theta_H),
\ee
an immediate consequence of the couplings given in (\ref{gl10}).

\subsubsection{Top decay to stop}

Another exciting decay mode in supersymmetry models is the decay of
the top to 
the SUSY scalar partner stop plus neutralinos, mixtures of neutral
gauginos and higgsinos
\cite{ns31,ns39}. This possibility is intimately
related to the large top mass  which leads to novel phenomena induced
by the strong Yukawa interactions. These effects do not occur in light--quark
systems but are special to the top.

The mass matrix of the scalar SUSY partners $(\tilde{t}_L, \tilde{t}_R)$ to
the left-- and right--handed top--quark components $(t_L, t_R)$ is
built--up by 
the following elements \cite{ns40}
\[
{\cal{M}}^2 = \left\| \begin{array}{cc}
          m_{\tilde{t}_L}^2 + m_t^2 & \delta  \tilde{m}_{LR}^2 \\
          \delta \tilde{m}_{LR}^2 & m_{\tilde{t}_R}^2 + m_t^2
        \end{array}
\right\|
\]
Large Yukawa interactions lower the diagonal matrix elements $\sim
-m_t^2$ with respect to the common squark mass value in supergravity
models, and they mix the $\tilde{t}_L$ and $\tilde{t}_R$
states with the strength $\sim m_t$ to form the
mass eigenstates $\tilde{t}_1, \tilde{t}_2$. Unlike the five light quark
species, these Yukawa interactions of ${\cal{O}}(m_t)$
can be so large in the top sector that after
diagonalizing the mass matrix, the smaller eigenvalue may fall below the top
quark mass,
\[
m_{\tilde{t}_1} < m_t \mbox{\ \ [ : possible ]}.
\]
The decay modes
\[
t \ra \tilde{t} + \mbox{neutralinos}
\]
\noindent then compete with the ordinary $W$ decay mode. Identifying
the lightest SUSY particle with the photino $\tilde{\gamma}$ 
(the mass of which we neglect in this estimate) one finds
\begin{equation}
\frac{\Gamma(t \ra \tilde{t} \tilde{\gamma})}{\Gamma(t \ra bW)}  \approx
\frac{8
\sqrt{2}\pi \alpha}{9 G_F m_t^2}  \frac{[1 - m_{\tilde{t}}^2/m_t^2 ]^2}{[1 -
m_W^2 /m_t^2 ]^2 [1 + 2m_W^2/m_t^2] }
\end{equation}
\noindent This ratio is in general less than 10\%. The subsequent
$\tilde{t}$ decays
\begin{eqnarray}
\tilde{t} & \ra & b \tilde{W}, \tilde{W} \ra W \tilde{\gamma} \mbox{\
\ or\ \ }
l\tilde{\nu} \mbox{\ \ {\it etc.}\ \ } \nonumber \\
\tilde{t} & \ra & c \tilde{\gamma} \nonumber
\end{eqnarray}
\noindent lead to an overall softer charged lepton spectrum and,
as a result of the escaping photinos, to an increase of the  missing energy, the
characteristic signature for SUSY induced phenomena.

Depending on the SUSY parameters however, stop decays could even be
more enhanced if the top is heavy. Decays into strongly coupled,
fairly light higgsinos could thus occur frequently.

\subsubsection{FCNC decays}

Within the Standard Model, FCNC decays like $t\ra c\gamma$ are forbidden at
the tree level by the GIM mechanism. However, they do occur in
principle at the one--loop
level, though strongly suppressed. The suppression is particularly severe for
top decays since the quarks building up the loops, must be down--type quarks
with $m_b^2$ setting the scale of the decay amplitude, $\Gamma_{FCNC} \sim
\alpha G_F^2 m_b^4 m_t$. A sample of branching ratios is given below
\cite{ns41}:

\[
\begin{array}{c}
\psfig{figure=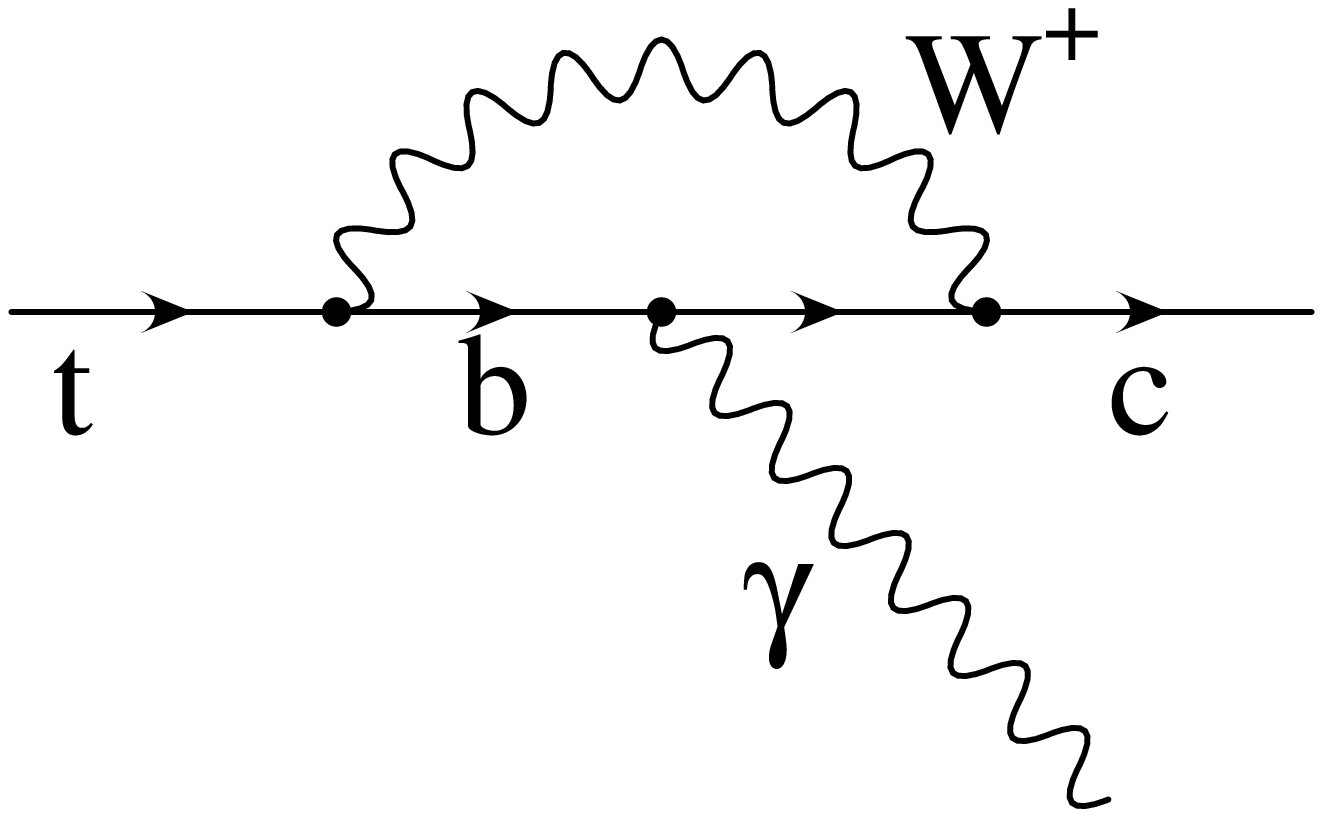,width=60mm,bbllx=0pt,bblly=300pt,%
bburx=630pt,bbury=420pt}
\end{array}\\
\hspace*{-5mm} \begin{array}{lclclcl}
              BR (t \ra cg) & \sim & 10^{-10} & \quad & BR(t \ra cZ) & \sim &
              10^{-12} \\
              BR(t \ra c\gamma) & \sim & 10^{-12} & & BR (t \ra cH) & \sim &
              10^{-7}
             \end{array}
\]

\vspace*{1cm}

At this level, no \sm\  generated $t$ decays can be observed, even given
 millions
of top quarks in proton colliders. On the other hand, if these decay modes
were detected, they would be an undisputed signal of new physics beyond the
\sm. From such options we select one illustrative, though very speculative
example for brutal GIM breaking. It is tied to the large top mass and holds out
faint hopes to be observable even in low rate $e^+e^-$ colliders.

The GIM mechanism requires all $L$ and $R$ quark
components of the same electric charge
in different families to carry identical isospin quantum numbers,
respectively. This rule is
broken by adding quarks in LR symmetric vector
representations \cite{ns32} to the
``light'' chiral representations or mirror quarks \cite{ns33}:
\[
 \begin{array}{lccccc}
   \mbox{\underline{vector quarks}: } & \quad \cdots \quad &
         \left[ \begin{array}{c} t \\ b \end{array} \right]_L &
         \begin{array}{c} t_R \\ b_R \end{array} & \qquad
         \left[ \begin{array}{c} U \\ D \end{array} \right]_L &
         \left[ \begin{array}{c} U \\ D \end{array} \right]_R  \\
\\
   \mbox{\underline{mirror quarks}: } & \quad \cdots \quad &
         \left[ \begin{array}{c} t \\ b \end{array} \right]_L &
         \begin{array}{c} t_R \\ b_R \end{array} & \qquad
         \begin{array}{c} U_L \\ D_L \end{array} &
         \left[ \begin{array}{c} U \\ D \end{array} \right]_R
 \end{array}
\]
Low energy phenomenology requires the masses $M$ of the new $U,D$ quarks to be
larger than 300 GeV.

Depending on the specific form of the mass matrix,
mixing between the normal chiral states and the new states
may occur at the level $\sim\sqrt{m/M}$, so
that FCNC $(t,c)$ couplings of the order $\sim\sqrt{m_tm_c/M^2}$ can be induced.
FCNC decays of top quarks, for example,
\[
BR(t \ra cZ) \sim \mbox{fraction of }\%
\]
\noindent are therefore not excluded. Such branching ratios would be at the
lower edge of the range accessible at $e^+e^-$ colliders.

%% file: part2.tex
\chapter{Top quarks at hadron colliders}

The search for new quarks and the exploration of their properties
has been a most important task at hadron colliders in the past.
The recent observation of top quarks with a mass of around 180 GeV at
the TEVATRON has demonstrated again the discovery power of hadron
colliders in the high energy region.  Several ten's of top quarks have
been observed up to now.  The significant increase of luminosity
toward the end of this decade will sharpen the picture.  The branching
ratios of the dominant decay modes will be determined and the
uncertainty in the top mass reduced significantly.  For
a detailed study of the top quark properties the
high energy collider LHC will  provide the required large
number of top events [order $10^7$].

\begin{figure}[htb]
\vspace*{10mm}
\begin{tabular}{cc}
\psfig{figure=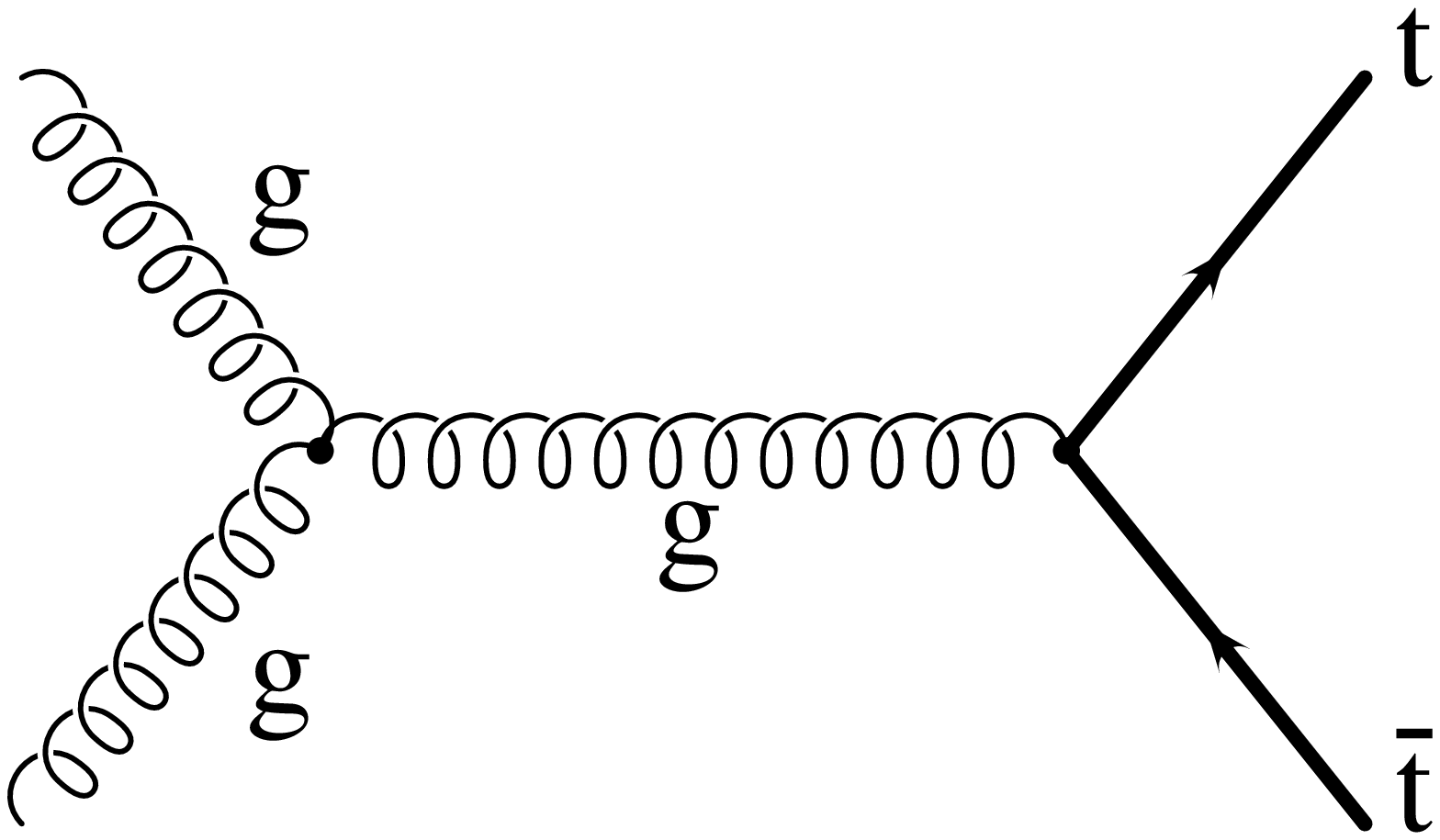,width=80mm,bbllx=0pt,bblly=300pt,%
bburx=630pt,bbury=420pt}
&
\psfig{figure=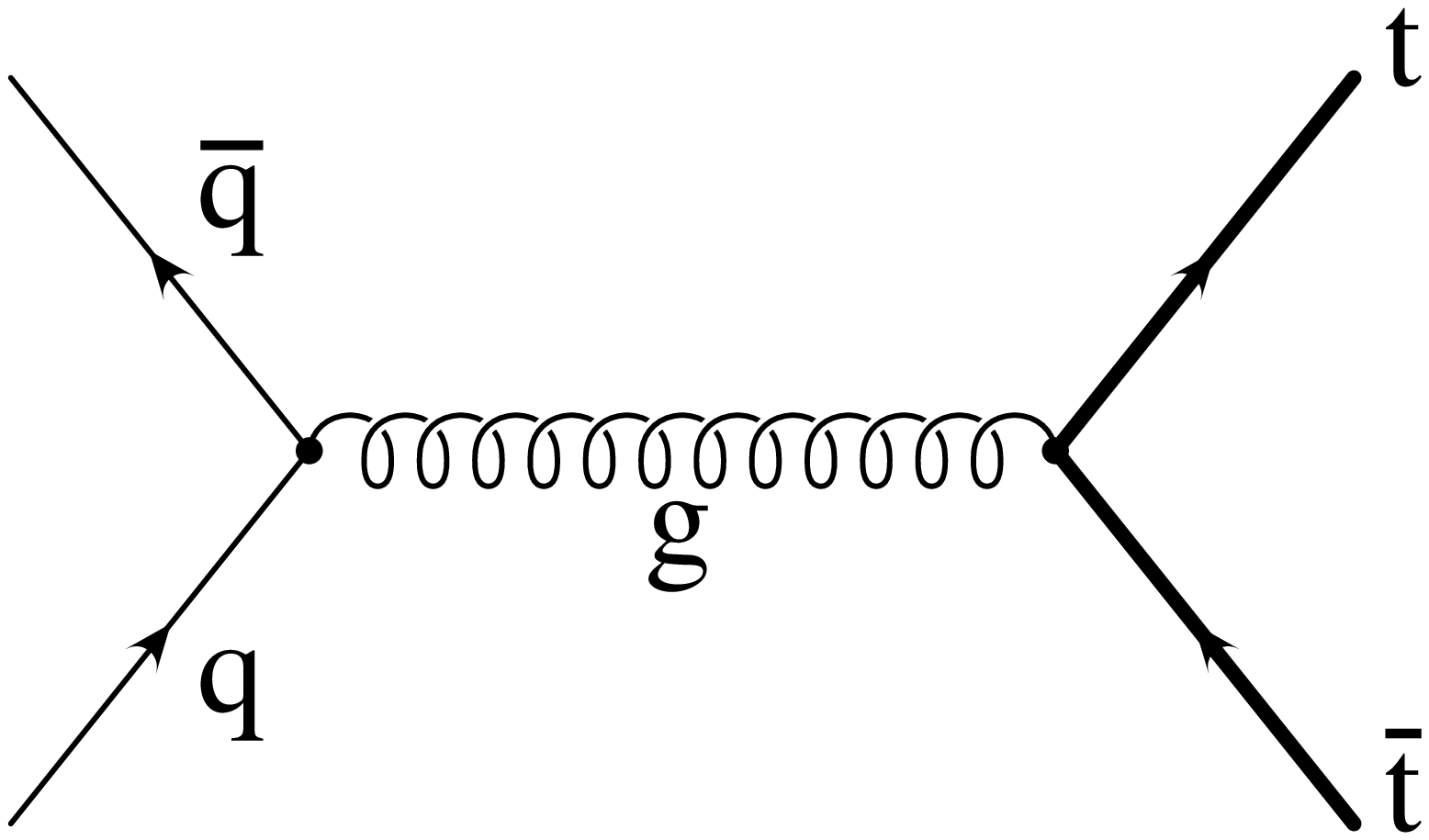,width=80mm,bbllx=0pt,bblly=300pt,%
bburx=630pt,bbury=420pt}
\end{tabular}
\\[25mm]
\hspace*{30mm}
\begin{tabular}{c}
\psfig{figure=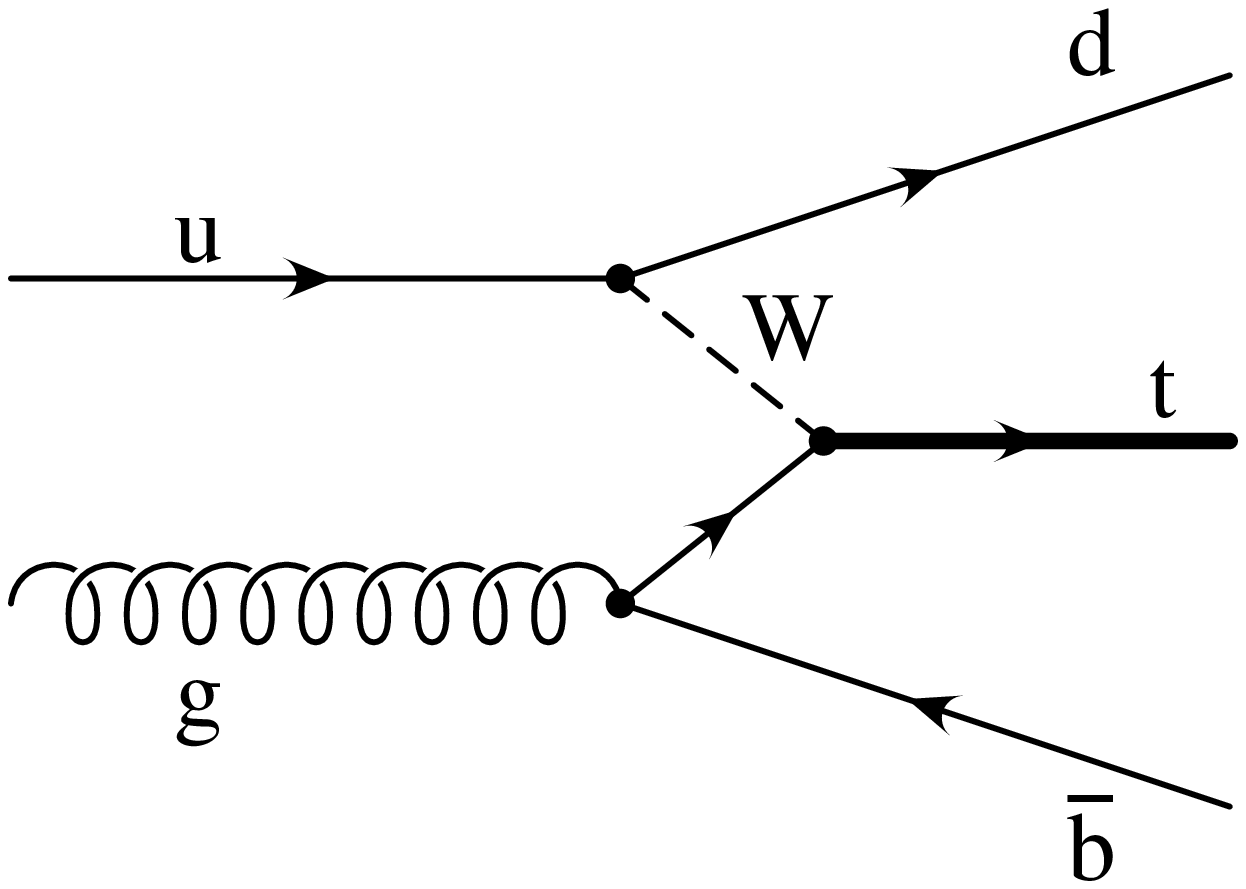,width=120mm,bbllx=0pt,bblly=300pt,%
bburx=630pt,bbury=420pt}
\end{tabular}
\vspace*{5mm}
   \caption[dummi]{\label{Fhaddia} The main production
   mechanisms for top quarks in \ppb\ and $pp$ colliders
   [generic diagrams].}
\end{figure}

The main production mechanisms
for top quarks in proton-antiproton collisions, Fig.\ref{Fhaddia},
are  quark-antiquark fusion  supplemented by a small
admixture of gluon-gluon fusion \cite{gluec1}.
$$
gg \;\;\mbox{and} \;\;q\overline{q} \rightarrow t\overline{t}
$$
Top production at the LHC is of course dominated by the second
reaction. The  $W$-gluon fusion process \cite{wille1}
$$ Wg \rightarrow t\overline{b} $$
is interesting  on its own. It is about a factor 0.1 -- 0.2 below
the dominant reaction and thus well accessible
at the high energy $pp$ colliders --- and perhaps even at the TEVATRON.

\section{Lowest order predictions and qualitative features}
The dominant Born terms for the {\em total top cross section} in
$gg/\qqb \ra \ttb$ fusion are well-known to be of the form \cite{gluec1}
\stepcounter{equation}
$$ \label{eq29a}
   \sigma_{gg}(\hat{s})   =   \frac{4\pi\alpha^2_s}{12\hat{s}}
   \; \left[\;(1+\rho+\frac{\rho^2}{16})\;\ln\frac{1+\beta}{1-\beta} -
   \beta\;(\frac{7}{4}+\frac{31}{16}\rho)\;\right] \\
   \eqno{(\arabic{equation}a)}
$$
$$\hspace{-5.8cm}
   \sigma_{q\overline{q}}(\hat{s})   =
   \frac{8\pi\alpha^2_s}{27\hat{s}} \;\beta\;[1+\frac{\rho}{2}]
   \eqno{(\arabic{equation}b)}
$$
with $\rho$ = 4$m_t^2/\hat{s}\;$ and
$\beta = \sqrt{1 - \rho}$ being the velocity of the $t$ quarks in
the $t\overline{t}$ cm frame with invariant energy $\sqrt{\hat{s}}$.
The total $p\bar p$ cross
sections then follow by averaging the partonic cross sections over the
$q\overline{q}$ and $gg$ luminosities in $p\bar p$ (and similarly in $pp$) 
collisions,
\beq
   \sigma \left( p\bar p \ra \ttb \right) = \int^1_{4m^2_t / s}
   d\tau \; \frac{ d {\cal L} \left( gg \right) }{d \tau}
   \sigma_{gg} \left( \tau s \right) + \left[ \qqb \right]
\eeq
The relative enhancement  of the $q\bar q$ cross section by about a
factor 3, as evident from the threshold behaviour
\be
\sigma_{q\bar q} &\approx& {4\over 9} {\pi\alpha_s^2\over \hat s}\beta
\nonumber\\
\sigma_{gg} &\approx& {7\over 48} {\pi\alpha_s^2\over \hat s}\beta
\ee
has to be combined with the prominent $q\bar q$ luminosity at the TEVATRON.
\begin{figure}
\psfig{figure=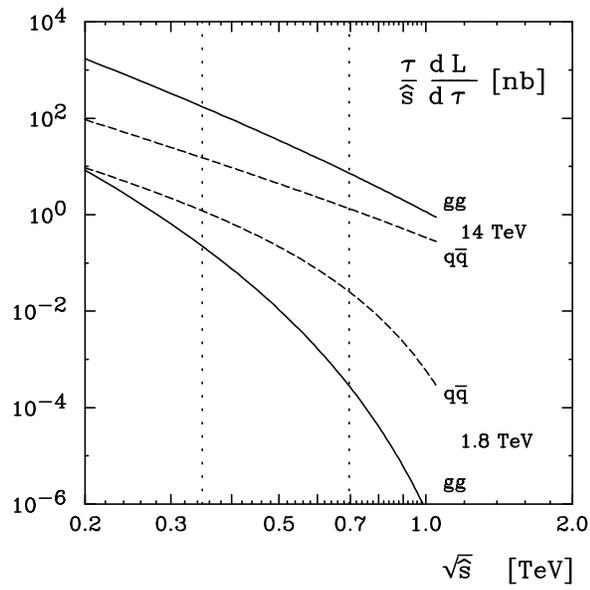,width=160mm,bbllx=-100pt,bblly=240pt,%
bburx=630pt,bbury=520pt}
\caption{Parton luminosities for TEVATRON and LHC energies.}
\label{Lumi}
\end{figure}
\begin{figure}
   \vspace{12.5cm}
   \caption[dummi]{\label{Fhadsig} Production cross section
   for $\ttb\ $
   pairs in $\ppb$ and $pp$ colliders: Tevatron (1.8 GeV);
   Tevatron II (3.6 TeV); LHC (16 TeV); SSC (40 TeV). Ref.\cite{reya1}.}
\label{fig213}
\end{figure}
As  shown in Fig.~\ref{Lumi}
\be
\left. {\rm d}{\cal L}_{q\bar q} : {\rm d}{\cal L}_{gg} 
\right|_{\sqrt{\hat s}=2m}
\approx
\left\{
\begin{array}{rl}
5 & {\rm TEVATRON} \\
0.1 & {\rm LHC}
\end{array}
\right.
\ee
which implies the dominance of $q\bar q$ annihilation, in contrast to
the situation at the LHC, where gluon fusion is the dominant reaction. 

A number of important features can be read off from this lowest order
result:  
\begin{itemize}
\item
Since the parton luminosities rise steeply with decreasing $\tau$, the
production cross sections increase dramatically with the energy
(Fig.~\ref{fig213}) 
\item
Structure functions and quark-antiquark luminosities in the region of
interest for the TEVATRON, i.e. for $\sqrt{\tau}$ between 0.2 and 0.4
are fairly well known from experimental measurements at lower energies
(combined with evolution equations) and collider studies.  The
predictions are therefore quite stable with respect to variations
between different sets of phenomenologically acceptable 
parton distributions.  The near tenfold increase of the  energy at the
LHC and the corresponding decrease of $x$ and $\sqrt{\tau}$
 by nearly a factor of ten
leads to the dominance of gluon-gluon fusion   and results in a
significantly 
enhanced uncertainty in the production cross section.
\item
With the cross section proportional to $\alpha_s^2$ and uncertainties
in $\alpha_s(M_Z)$ which may be stretched up to $\pm 10\%$ one might
naively expect a resulting uncertainty in the predicted cross section.
However, the increase in the parton cross section with increasing
$\alpha_s$  is, to some extent,
compensated by a decrease in the parton luminosity (with increasing
$\alpha_s$) for the kinematical
region of interest at the TEVATRON.  This compensation mechanism has
been studied in \cite{Martin95} for inclusive jet
production (Fig.~\ref{Jet}) and applies equally well for top quark 
production.
\item
At the TEVATRON the rapidity distribution is strongly dominated by central
production, $|y| \leq 1$, a consequence of the balance between the
steeply falling proton and antiproton parton distributions. 
At the LHC, however, a rapidity plateau develops gradually and the
distribution spans nearly four units in rapidity (Fig.~\ref{transpS8}).
\begin{figure}
\begin{center}
    \leavevmode
    \epsfxsize=16cm
    \epsffile[18 180 577 819]{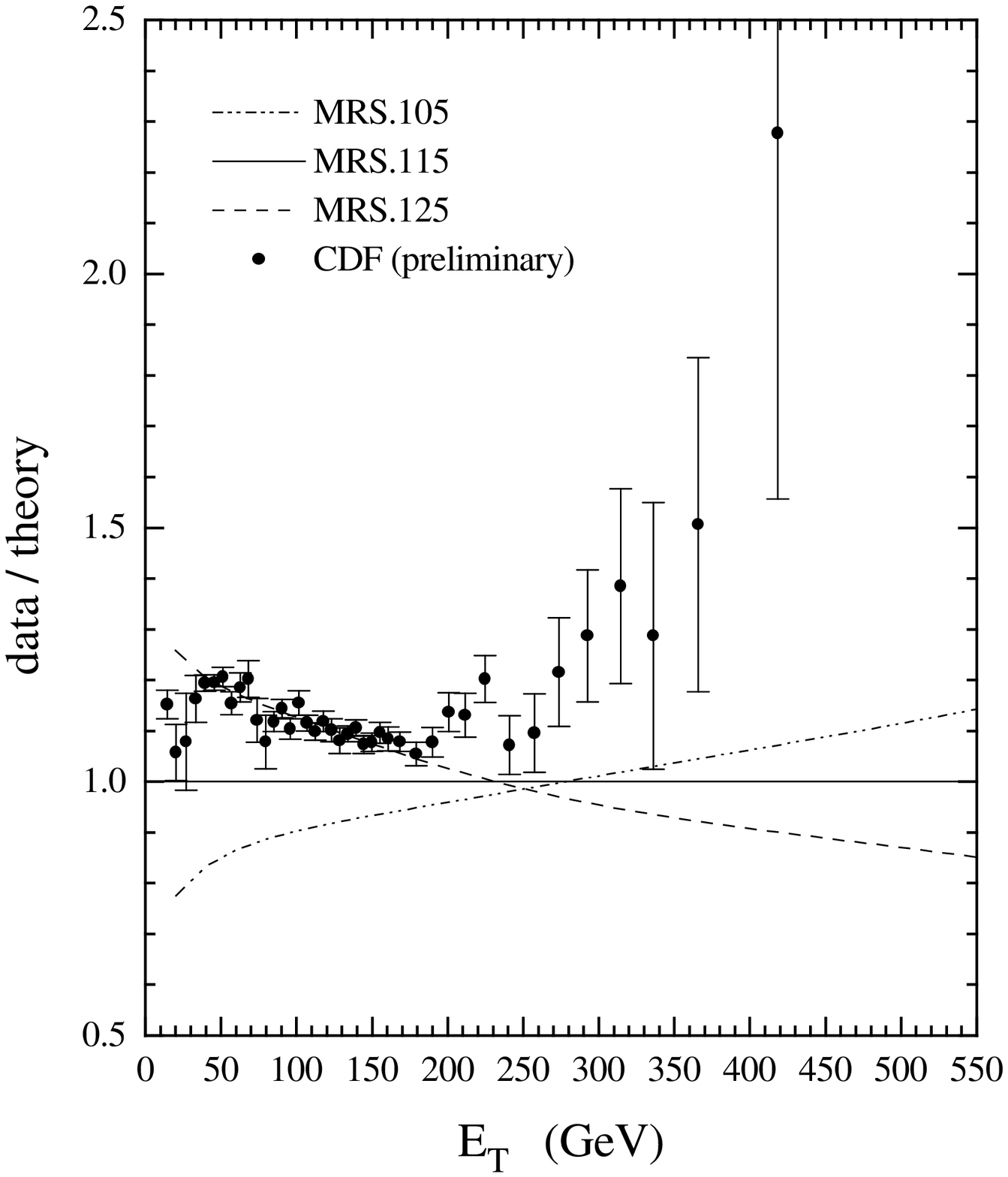}
\end{center}
\caption{The $p\bar p$--initiated jet $E_T$ distribution at
$\protect\sqrt{s}=1.8$ TeV normalized to the prediction from partons
with $\alpha_S = 0.115$ (i.e. MRS.115).  The data are the CDF
measurements of $d^2 \sigma/dE_T d \eta$ averaged over the
rapidity interval $0.1 < |\eta| < 0.7$.  The curves
are obtained from a leading-order calculation evaluated at $\eta =
0.4$.  The data are preliminary and only the statistical errors
are shown.  The systematic errors are approximately 25\% and are
correlated between different $E_T$ points.  
(From \protect\cite{Martin95}).}
\label{Jet}
\end{figure}

\begin{figure}[htb]
\vspace*{110mm}
\caption{Rapidity distribution of top quarks in $gg/q\bar q$ fusion at
$\protect\sqrt{s}=16$ TeV
\protect\cite{reya1}. }
\label{transpS8}
\end{figure}
\item
The transverse momentum distribution is relatively flat, dropping down
to half its peak value at around $p_t \approx  m_t/2$, again a
consequence of the competition between the increase of the phase space
factor $\propto \beta$ in the parton cross section and the steeply
decreasing parton luminosity (Fig.~\ref{FigS8Beenaker}). 
\begin{figure}[htb]
\vspace*{105mm}
\caption{The differential cross section for $p+\bar p\to t(\bar t) +X$
with $m_t=120$ GeV/c$^2$ and $\mu_R=Q=\protect\sqrt{m_t^2+p_t^2}$ at
$\protect\sqrt{s}=1.8$ TeV.  The cross section is shown at different
values of rapidity for (1) dashed lines: lowest order contribution
scaled by an arbitrary factor (2) solid lines: full order $\alpha_s^2$
calculation. (From \protect\cite{Been91}.)}
\label{FigS8Beenaker}
\end{figure}
At the LHC the distribution
will be even flatter and $p_t$ values around $2m_t$ are well
within reach (Fig.~\ref{fig:transverse}).
This corresponds to  CMS energies between 0.5 and 1
TeV in the parton subsystem and extremely large subenergies
 are therefore accessible.  This opens the
possibility to search for the radiation of $W$, $Z$, or Higgs bosons
in this reaction.  For high energies the suppression of the cross
section through
electroweak virtual corrections (cf.~sect.~\ref{ssec:weak})
is, at least partially, compensated by the large
logarithm $\ln \hat s/M^2_{W,Z,H}$.
\end{itemize}
\begin{figure}[htb]
\vspace{110mm}
\caption{Transverse momentum distribution of top quarks in $gg/q\bar
q$ at $\protect\sqrt{s}=16$ TeV. (From \protect\cite{reya1}.)}
\label{fig:transverse}
\end{figure}  

\newpage

\section{QCD and electroweak corrections}
The observation of top quarks has been well established during the
last year.  One of the tools to study its properties, in particular
its mass and its decay modes, is a precise experimental determination
of its production cross section and subsequent decay
in the $t\to b+W$ channel.  A large
deficiency in the comparison between theory and experiment would
signal the presence of new decay modes which escape the canonical
experimental cuts; with $t\to bH^+$ as most prominent example.
However, the early round of experiments had indicated even an excess
of top events when compared to the theoretical prediction for
$m_t\approx 180$ GeV.  This observation
 was difficult to interpret and the original
calculations were  scrutinized again by various authors.  In particular,
the resummation of leading logarithms and the influence of the Coulomb
threshold enhancement was investigated --- in the end, however, the
prediction remained fairly stable.

In these lectures we will, therefore, in a first step, present the
results from a complete NLO calculation (sect.~\ref{ssec:NLO}). This
is supplemented by a qualitative discussion
of the resummation of higher order leading logarithmic terms.  The
influence of the Coulomb enhancement is studied in section
\ref{ssec:Coul}, electroweak corrections are presented in
sect.~\ref{ssec:weak}.  Radiation of gluons may have a sizeable effect
on the aparent mass of top quarks as observed in the experiment
(sect.~\ref{ssec:exper}), with distinct differences between initial
and final state radiation. 

\subsection{Next to leading order (NLO) corrections and resummation of
large logarithms}
\label{ssec:NLO}

\begin{figure}[h]
\begin{minipage}{16.cm}
\hspace*{40mm}
\[
\mbox{
\hspace*{15mm}
\begin{tabular}{cc}
\psfig{figure=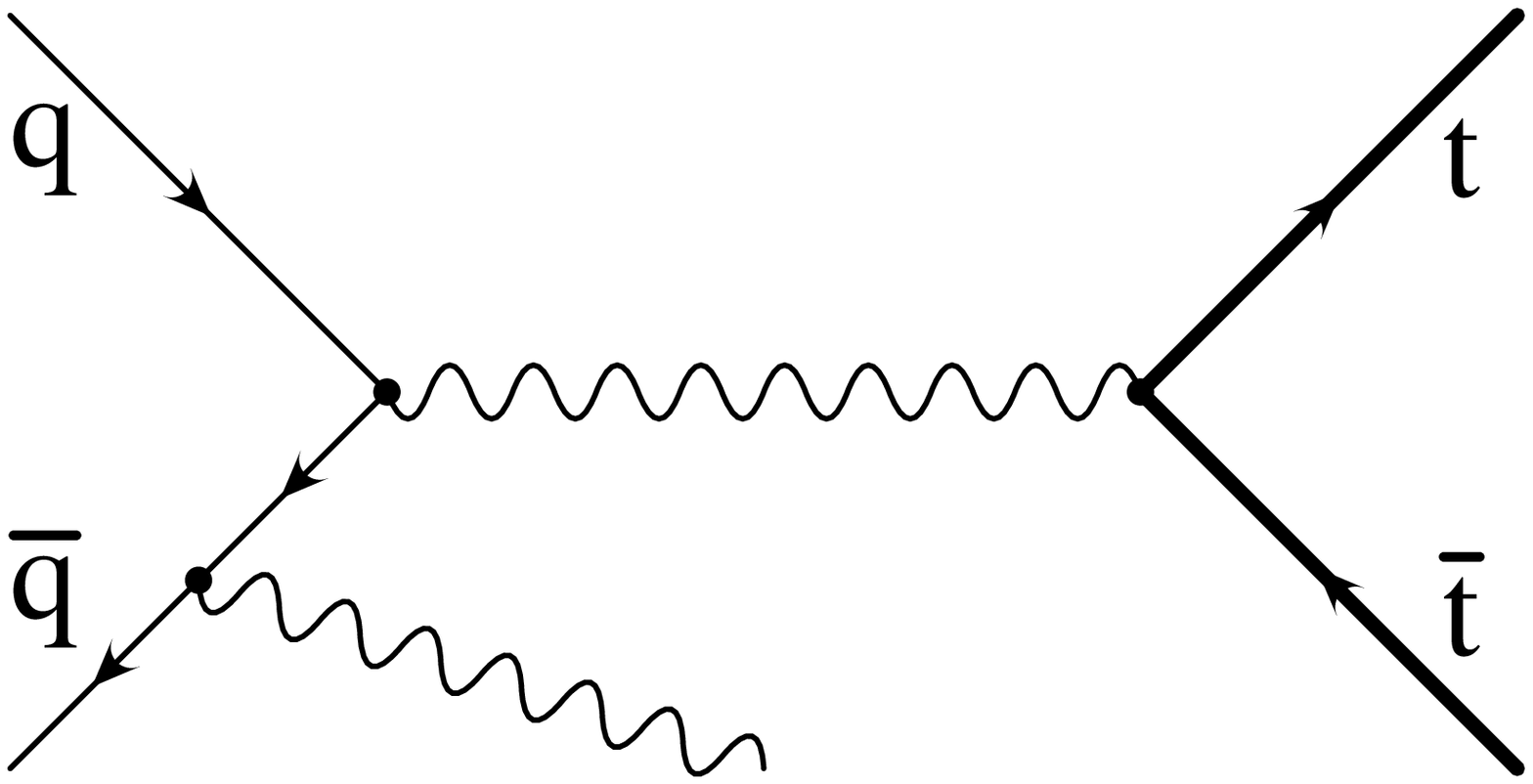,width=40mm,bbllx=210pt,bblly=410pt,bburx=630pt,bbury=550pt} 
&\hspace*{15mm}
\psfig{figure=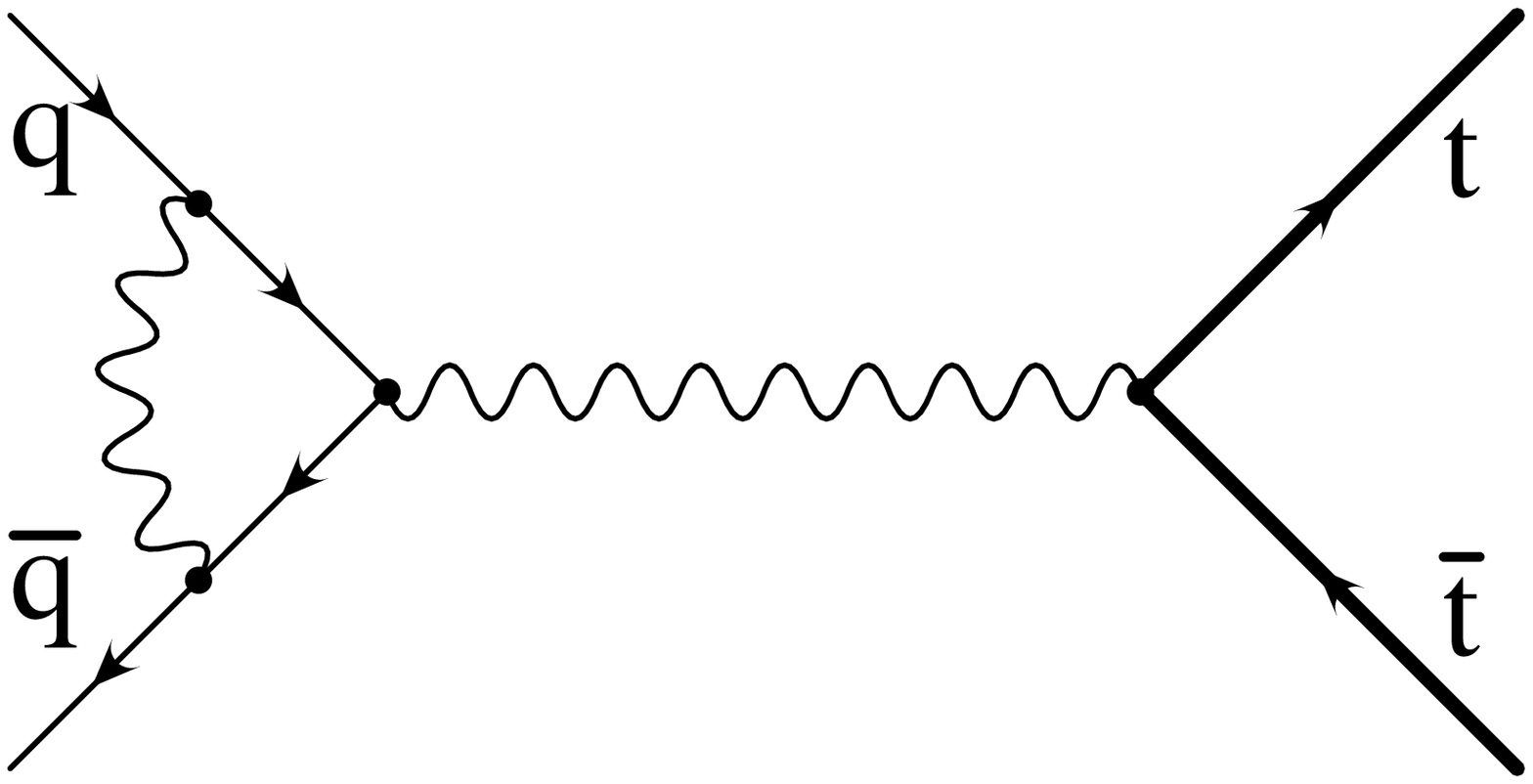,width=40mm,bbllx=210pt,bblly=410pt,bburx=630pt,bbury=550pt}
\\[15mm]
\rule{-25mm}{0mm} (a) &\hspace*{-.1cm} \rule{-4mm}{0mm}(b) 
\\
\psfig{figure=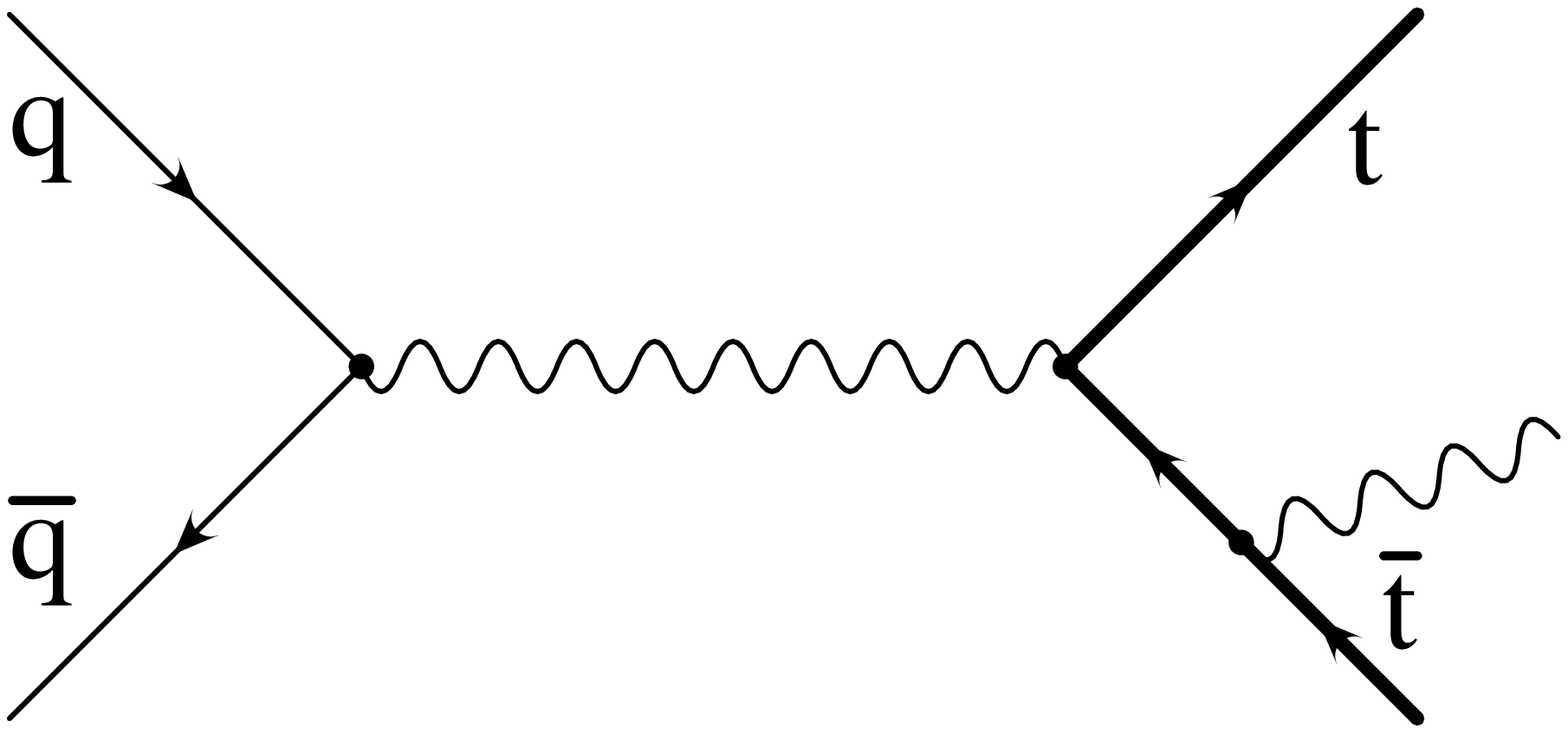,width=40mm,bbllx=210pt,bblly=410pt,bburx=630pt,bbury=550pt} 
&\hspace*{15mm}
\psfig{figure=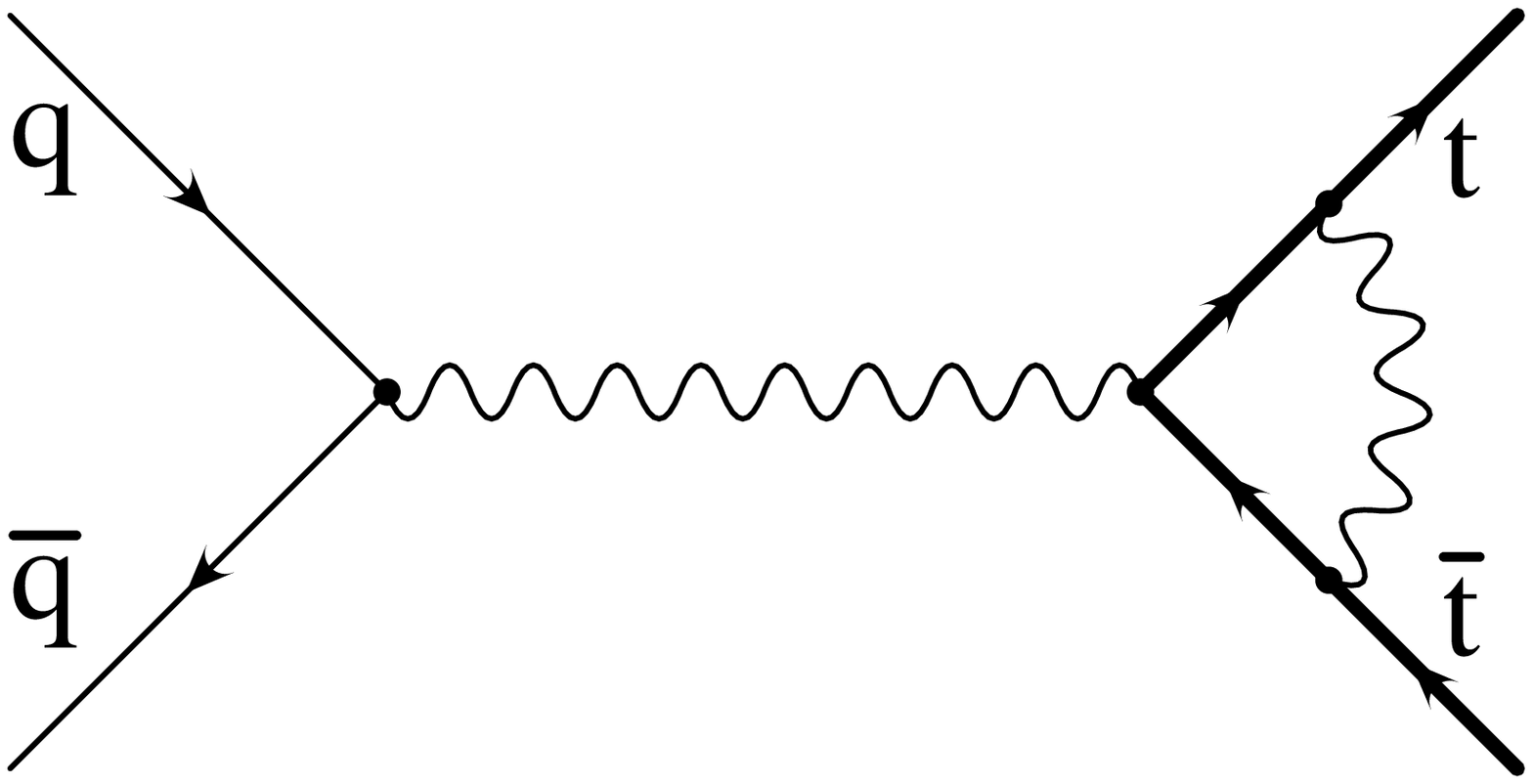,width=40mm,bbllx=210pt,bblly=410pt,bburx=630pt,bbury=550pt}
\\[15mm]
\rule{-25mm}{0mm} (c) &\hspace*{-.1cm} \rule{-4mm}{0mm}(d) 
\end{tabular}}
\]
\end{minipage}
\caption{Initial and finite state radiation in the reaction $q\bar
q\to t\bar t$}
\label{fig:ifsr}
\end{figure}

Higher-order QCD corrections \cite{nason1,beena1,Been91,meng1} 
include loop corrections to
the Born terms and $2\to 3$ contributions like $gg\to t\bar t g$,
$q\bar q\to t\bar t g$ etc.  For $q\bar q$ annihilation a few
characteristic diagrams are displayed in Fig.~\ref{fig:ifsr}.  
Real and virtual
\begin{figure}[h]
\begin{minipage}{16.cm}
\hspace*{40mm}
\[
\mbox{
\hspace*{25mm}
\begin{tabular}{cc}
\psfig{figure=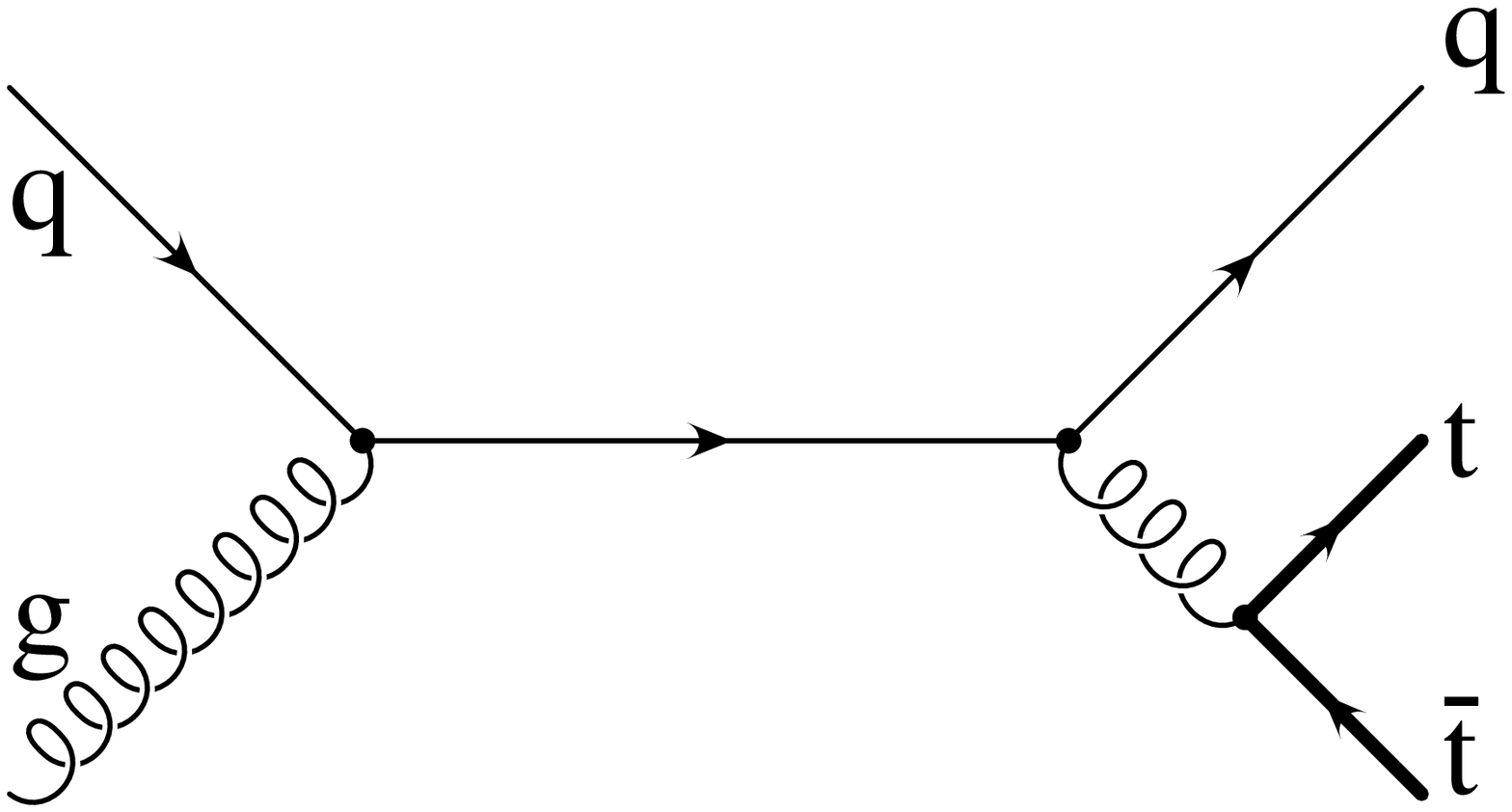,width=40mm,bbllx=210pt,bblly=410pt,bburx=630pt,bbury=550pt} 
&
$+\ldots$ 
\\[15mm]
\end{tabular}}
\]
\end{minipage}
\caption{Reaction $qg\to t\bar t q$}
\label{fig:quarkgluon}
\end{figure}
initial state radiation (Fig.~\ref{fig:ifsr}a,b) dominate, final state radiation
from the slow top quarks (Fig.~\ref{fig:ifsr}c) is
unimportant, 
virtual gluon exchange at the $t\bar t$ vertex (Fig.~\ref{fig:ifsr}d) 
leads to the Coulomb
enhancement and will be discussed in sect.~\ref{ssec:Coul}.  The
separation between $q\bar q$ annihilation and $qg$ reactions
(Fig.~\ref{fig:quarkgluon}) 
depends on the choice of the so-called factorisation scale $\mu^2$
which effectively enters the definition of the structure functions.  

The differential as well as the total production cross section can be
cast into the following form
\be
{\rm d}\sigma = \sum_{i,j=partons} \int {\rm d}x_1 {\rm d}x_2
F_i^p(x_1,\mu^2)
F_j^{\bar p}(x_2,\mu^2)
{\rm d}\sigma_{ij}\left( \hat s, \ldots, \mu^2\right)
\ee
The renormalization scale $\mu_R$ and the factorisation scale $\mu_F$
are in general identified, $\mu_R = \mu_F \to \mu$,
a matter of convention and convenience more than a matter of
necessity.  The parton distributions are extracted from structure
functions as measured in deep inelastic scattering, and the analysis
has to be taylored to the order of the calculation, i.e. to the NLO in
the present case.  The integrated expressions for the total cross
sections can still be cast into a simple form
\begin{equation}
   \sigma_{ij}(\hat{s},m_{t}^2,\mu^{2}) =
   \frac{\alpha_{s}^2(\mu^{2})}{m_{t}^{2}} \left[ f_{ij}^{(0)}(\rho)
   + 4 \pi \alpha_{s}(\mu^{2}) \left( f_{ij}^{(1)}(\rho) +
   \bar{f}_{ij}^{(1)}(\rho) \ln \frac{\mu^{2}}{m_{t}^{2}} \right) \right]
\end{equation}
where $\hat{s} = x_1 x_2 s$ and the dominant lowest-order contributions
$f^{(0)}_{ij} \left( \rho \right)$ are given by the parton cross sections
above;
in addition
$f^{(0)}_{gq} = f^{(0)}_{g \bar{q}} = 0$. The subleading higher-order
expressions for $f^{(1)}_{ij}$ and $\bar{f}^{(1)}_{ij}$ are given in
Refs.\cite{nason1}, \cite{beena1}. The heavy quarks are treated within the on-shell
renormalization scheme with $m_t$ being the ''physical'' mass of the
top quark. Outside the heavy quark sector, the $\overline{MS}$ scheme
has been employed. These higher-order terms have to be used in
conjunction with the running coupling $\alpha_s \left( \mu^2 \right)$
and the gluon/light-quark parton densities evolved in 2-loop evolution
equations. $\mu$ is the renormalization scale, identified here
also with the factorization scale; typical scales that have been chosen
are $\mu = m_t$ and $\sqrt{m^2_t + p^2_T}$. More technical details are
discussed in Ref.\cite{reya1}.

\begin{sloppypar}
The lowest-  and higher-order predictions are compared with each other
in Fig.\ref{Fhadsig}.
In \cite{reya1} it has been argued that the subdominant $2 \ra 3$
contributions add up to less 
than $10 \%$ of the dominant lowest-order results.
The theoretical uncertainties of the predictions for the LHC due to
different parton distributions \cite{diemo1} were estimated  about
$\pm 10 \%$ plus 
a $\pm 10 \%$ variation due to the scale ambiguity $\mu$.
The impact of the additional shift from the resummation of large logs
arising in higher orders
will be discussed below.
Note that the ''K factor'', defined formally by the higher order
corrections to the LO parton cross section, but the parton
distributions and $\alpha_s$ kept fixed, amounts to an $\approx 50 \%$
correction of the Born terms.
\end{sloppypar}

\begin{figure}
\hspace*{10mm}
\begin{tabular}{cccc}
\psfig{figure=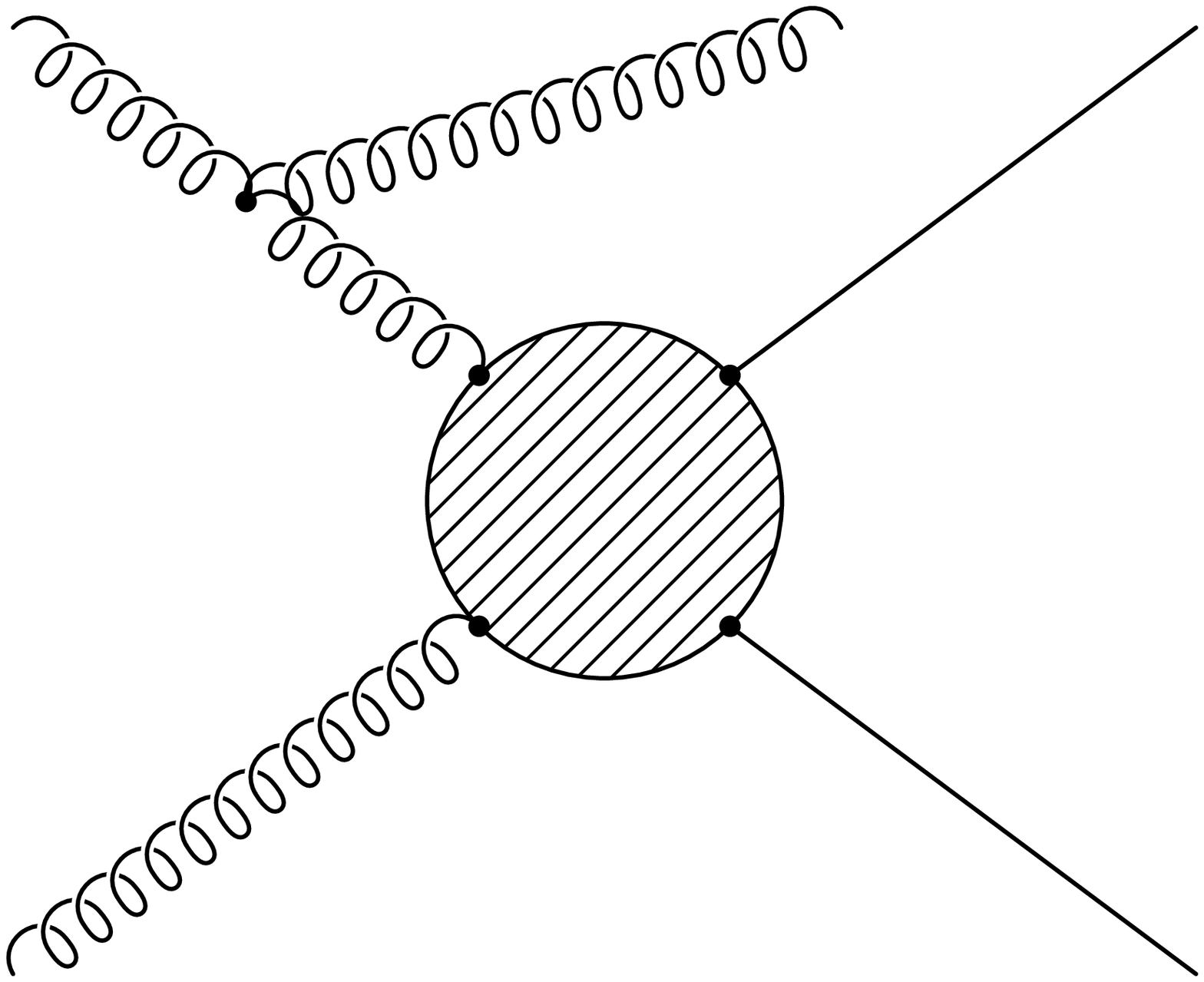,width=30mm,bbllx=210pt,bblly=410pt,%
bburx=630pt,bbury=550pt}
&
\hspace*{5mm}
\psfig{figure=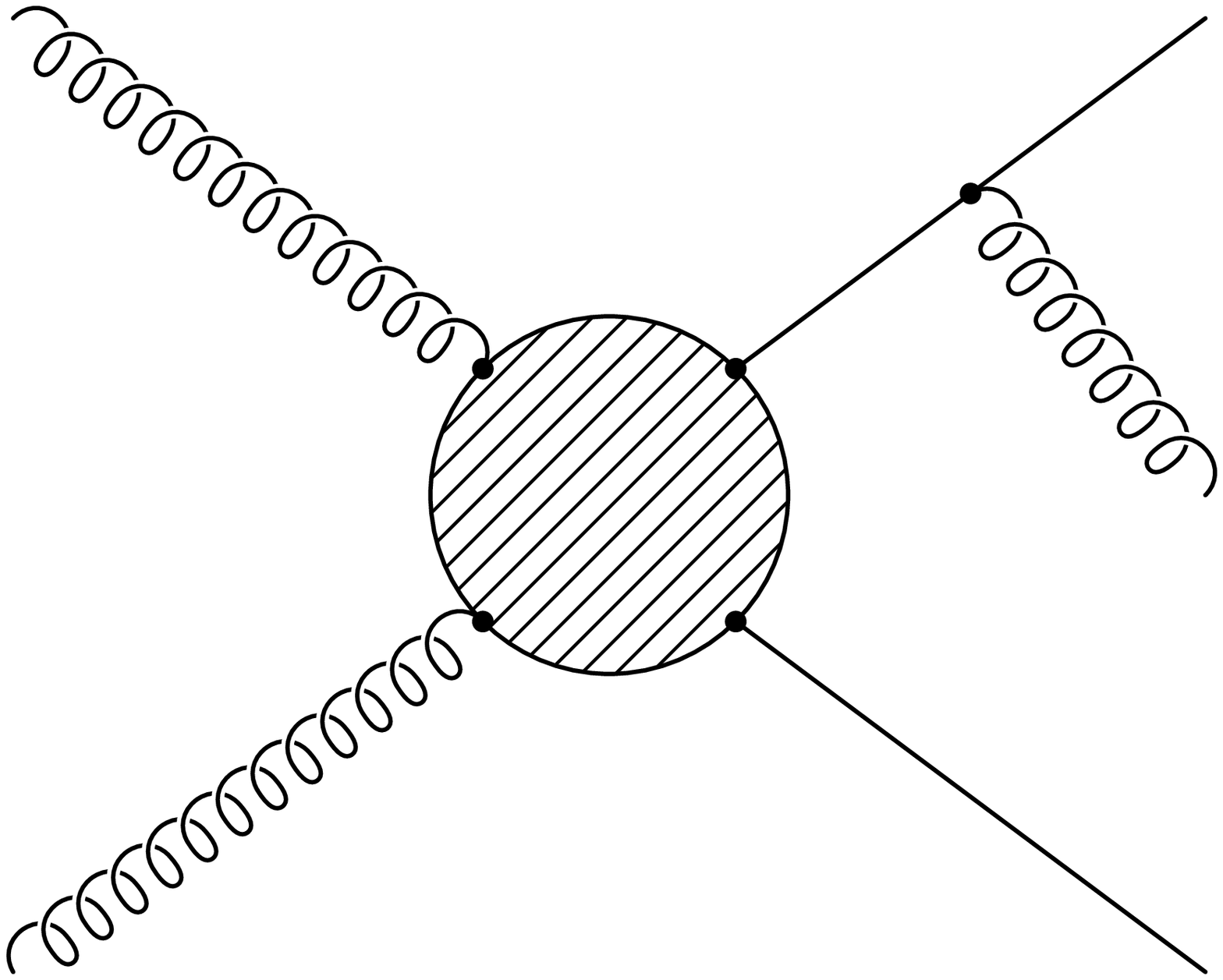,width=30mm,bbllx=210pt,bblly=410pt,%
bburx=630pt,bbury=550pt}
&
\hspace*{5mm}
\psfig{figure=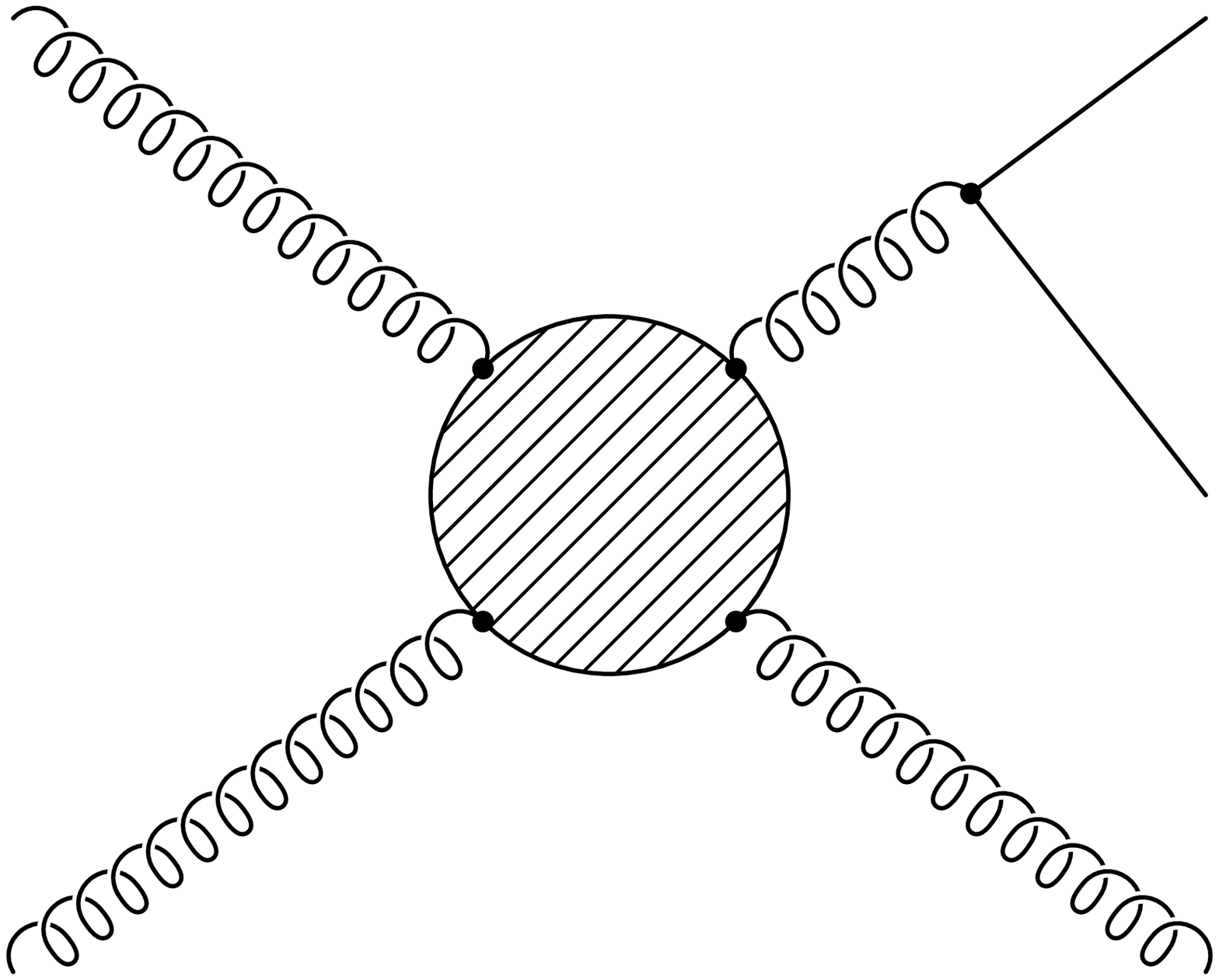,width=30mm,bbllx=210pt,bblly=410pt,%
bburx=630pt,bbury=550pt}
&
\hspace*{5mm}
\psfig{figure=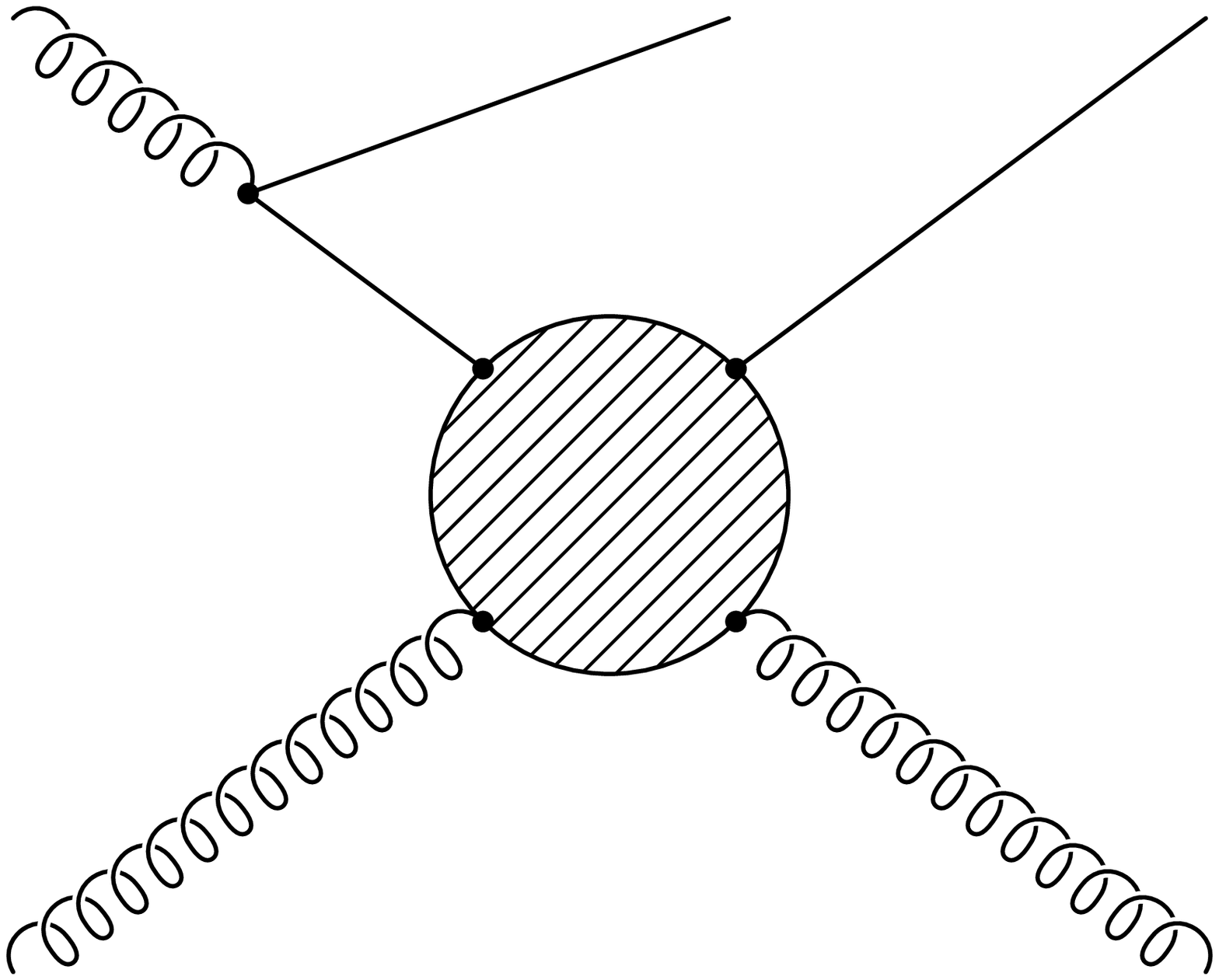,width=30mm,bbllx=210pt,bblly=410pt,%
bburx=630pt,bbury=550pt}
\end{tabular}
\\[15mm]
\begin{tabular}{cccc}
\hspace*{12mm} ISGB &\hspace*{24mm} FSGB
 & \hspace*{26mm} GS & \hspace*{30mm} FE
\end{tabular}
\caption{Generic QCD diagrams of the dominant higher order mechanisms.}
\label{Fhaddi2}
\end{figure}

It is also instructive to study separate, physically distinct components of
the $\alpha_s^3$ results
\cite{meng1}.  The initial  state bremsstrahlung (ISGB)
processes, illustrated for the gluon initiated reactions
in Fig.\ref{Fhaddi2}, dominate
around threshold [$\sqrt{\shatn}\ge 2m_t$ or $p_t<m_t)$], the case of
relevance at the TEVATRON. 
The gluon splitting (GS)
and the flavor excitation (FE) contributions become  
increasingly important for
$\sqrt{\shatn}\gg 2m_t$, the situation  anticipated for
the LHC.

\begin{figure}[h]
\begin{minipage}{16.cm}
\hspace*{40mm}
\[
\mbox{
\hspace*{15mm}
\begin{tabular}{cc}
\psfig{figure=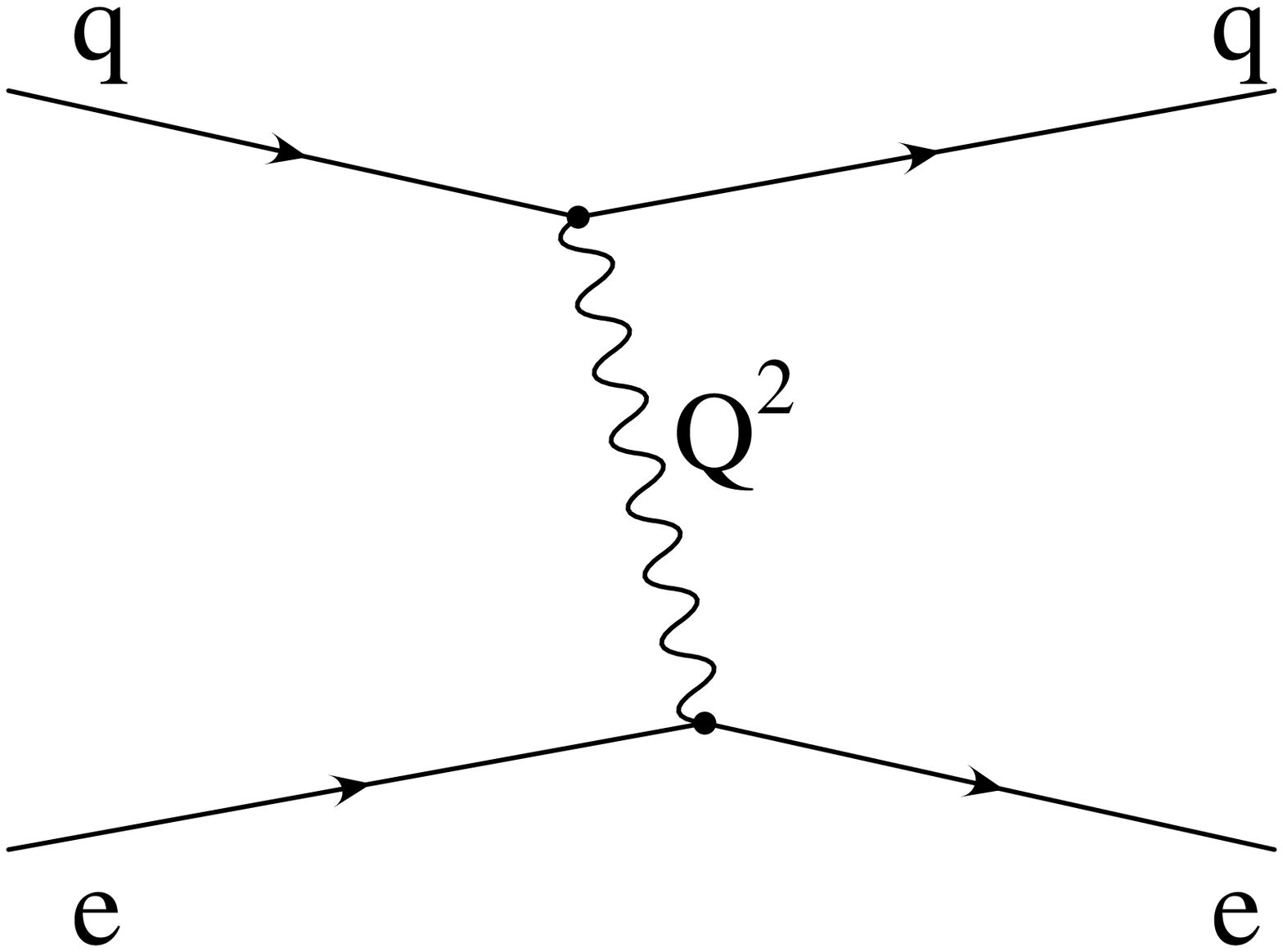,width=40mm,bbllx=210pt,bblly=410pt,bburx=630pt,bbury=550pt} 
&\hspace*{15mm}
\psfig{figure=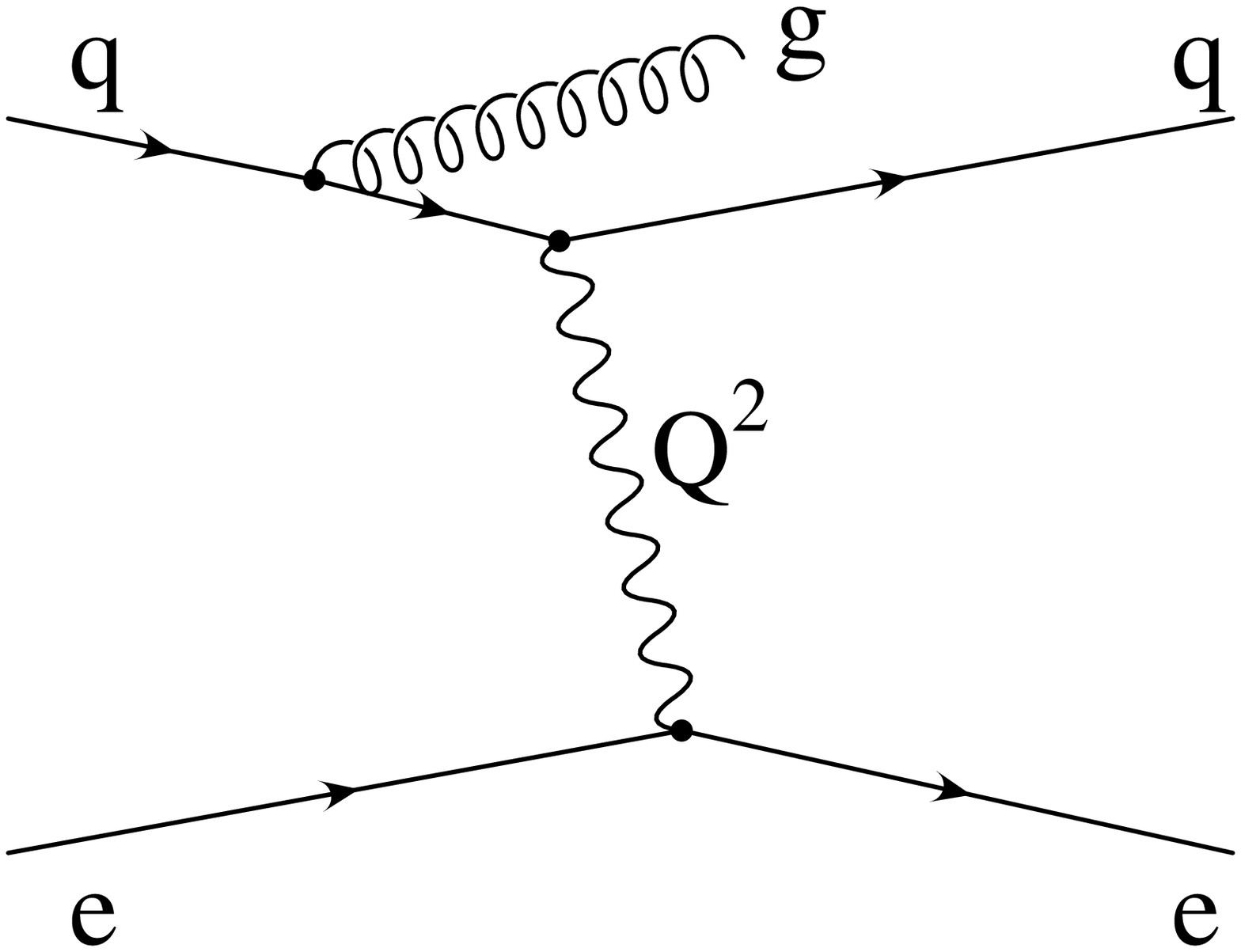,width=40mm,bbllx=210pt,bblly=410pt,bburx=630pt,bbury=550pt}
\\[15mm]
\rule{-25mm}{0mm} (a) &\hspace*{-.1cm} \rule{-4mm}{0mm}(b) 
\end{tabular}}
\]
\end{minipage}
\caption{Definition of quark distributions, including NLO.}
\label{fig:sei13}
\end{figure}

Let us concentrate in the remainder of this section on the predictions
for TEVATRON energies.  Initial state radiation reduces the effective
energy in the partonic subsystem, requiring larger initial parton
energies to reach the threshold for top pair production.  Considering
the steeply falling parton distributions $F_j(x)$ one might, therefore,
expect a reduction of $\sigma$ through NLO contributions.  However,
the same effect is operative in the very definition of $F_j$
(Fig.~\ref{fig:sei13})
through deep inelastic scattering, including NLO corrections.  In fact,
without this compensation mechanism the result would not even be
finite.  However, the magnitude or even the sign of the correction
cannot be guessed on an intuitive basis, and, not quite unexpected,
even the precise form of $f^{(1)}$ and $\bar f^{(1)}$ depends on the
definition of the structure functions.  
The most prominent examples are the $\overline{MS}$ scheme where
$1/\epsilon$ poles (plus $\ln(4\pi) -\gamma$) 
from collinear singularities are simply dropped [more precisely, they
are combined with the corresponding singular terms which arise in the
NLO definition of the structure function] and
finite corrections have to be applied when comparing to deep inelastic
scattering, and the DIS scheme, where $F_j$ are defined through deep
inelastic scattering to all orders.

Let us illustrate the qualitative aspects in the simpler example of NLO
contributions to the Drell Yan process.  
The dominance of initial state radiation in the corrections to $t\bar
t$ production will allow to apply the same reasoning to the case of
interest in these lectures.  
Including NLO corrections one
obtains 
\be
{ {\rm d} \sigma \over {\rm d} Q^2 }
&=& \sigma_0 \sum_{i,j} \int_0^1 { {\rm d} x_i\over x_i} 
{ {\rm d} x_j\over x_j} F_i^p(x_i,Q^2) F_j^{\bar p}(x_j,Q^2)
\cdot 
\omega_{ij}\left({\tau\over x_ix_j},\alpha_s\right)
\nonumber \\
&=& \sigma_0 \sum_{i,j} \int_\tau^1 { {\rm d} z\over z} 
{\cal F}_{ij}(\tau/z)
\cdot 
\omega_{ij}(z,\alpha_s)
\label{eq:27a}
\ee
with
\be
\sigma_0 &\equiv & {4\pi\alpha^2\over 9Q^2s}\qquad \qquad \qquad 
\tau \equiv
{Q^2\over s}
\nonumber\\
\omega^{(0)}_{q\bar q} (z,\alpha_s) &=& e_q^2\delta(1-z)
\nonumber\\
\omega^{(1)}_{q\bar q} (z,\alpha_s) &=& e_q^2{\alpha_s\over 2\pi}C_F
 \left[
2(1+z^2)\left( \ln(1-z)\over 1-z\right)_+ +3\left({1\over
1-z}\right)_+ \right.
\nonumber\\
&& \left. \qquad
+\left(1+{4\pi^2\over 3}\right)\delta(1-z)+\mbox{regular functions}
\right]
\label{eq:DY}
\ee
(The quark-gluon induced reactions will not be discussed in this
connection.) 
The plus prescription which regulates the singularity of the
distributions  at $z=1$ arises from the subtraction of collinear
singularities.  It can be understood by considering the limit 
\be
\left( { \ln^n(1-z)\over 1-z} \right)_+ \equiv \lim_{\epsilon\to 0}
\left( { \ln^n(1-z)\over 1-z} \Theta(1-\eps-z) + \delta(1-z)
{\ln^{n+1}\eps \over n+1} 
\right)
\ee
with the coefficient of the $\delta$ function adjusted such that the
integral from zero to one vanishes.

Equivalently  the plus-distribution 
can be defined through an integral with test functions
$f(z)$.  If $f(z)$ vanishes outside the interval $\langle
\tau,1\rangle$ a convenient formula which will be of use below 
reads as follows
\be
\int_\tau^1 {\rm d}z f(z)
\left( { \ln^n(1-z)\over 1-z} \right)_+ =
\int_\tau^1 {\rm d}z
{ f(z) -f(1) \over 1-z} \ln^n(1-z) - f(1)
\int_0^\tau
dz\, { \ln^n(1-z)\over 1-z}
\ee
The Born term $\omega^{(0)}$ is simply given by a
 $\delta$ function peak at $z=1$,
corresponding to the requirement that the squared energy of the
partonic system $\tau/z\times s$ and the squared mass of the muon pair
$\tau\cdot s= Q^2$ be equal.  ${\cal O} (\alpha_s)$ corrections
contribute to the $\delta$ function through vertex corrections and a
continuous part from initial state radiation extending through the
range 
\be
\tau \leq z \leq 1
\ee
The upper limit $z=1$ corresponds to the kinematic endpoint without
radiation, the requirement $\tau\leq z$ originates from the fact that
the parton luminoisities ${\cal F}_{ij}(\tau/z)$ 
vanish for $\tau/z >1$. The
regular and the subleading pieces of $\omega$
are process dependent, the leading
singularity $\sim (1+z^2)\left( { \ln(1-z)\over 1-z} \right)_+ $ is
universal (and closely related to the $q\to qg$ splitting function) and
equally present in $t\bar t$ production. 

The suppression of final state radiation in top pair production allows
to extend the analogy to the Drell Yan process and to employ
resummation techniques that were successfully developed and applied
for muon pair production \cite{Appel}.
A complete treatment of this resummation is outside the scope of these
lectures.  Nevertheless we shall try to present at least qualitative
arguments which allow to understand the origin of these large
logarithms.  (For a similar line of argument see \cite{Appel}.)
 With the energies of the partonic reaction
$\sqrt{\hat s}={\cal O}(2-4m_t)\approx 350 - 700$ GeV and the CMS energy
$\sqrt{s}=1800$ GeV of comparable magnitude it is clear that the ratio
$\hat s/s$ will not give rise to large logs.  However, large
logarithms can be traced to the interplay between the collinear
singularity in the subprocess and the rapidly falling parton
luminosity ${\cal F}_{ij}$ (cf.~eq.~(\ref{eq:27a})). This rapid decrease
leads to a reduction 
in the effective range of integration.  Let us, for the sake of
argument assume a range reduced from
\be
\tau \leq z \leq 1
\ee
to 
\be
1-\delta \leq z \leq 1
\ee
and evaluate the leading term.  For a constant luminosity ${\cal
F}(\tau/z)/z={\cal F}$ one would obtain
\be
\int_\tau^1 {\rm d}z\; 2 \left( { \ln (1-z)\over 1-z} \right)_+ 
{\cal F}
= \ln^2(1-\tau) 
{\cal F}
\ee
If the region of integration extended through the full kinematic range
and with $\tau = 4m_t^2/s=0.04$ there would be no large log. 
For the restricted range of integration, however, one finds
\be
\int_{1-\delta}^1 {\rm d}z \;2 \left( { \ln (1-z)\over 1-z} \right)_+ 
{\cal F}
= \ln^2(\delta) 
{\cal F}
\ee
For small $\delta$, corresponding in practice to steeply falling
luminosities one thus obtains large, positive (!) corrections from the
interplay between ${\cal F}_{ij}(\tau/z)$ and $\omega_{q\bar q}$. 

To arrive at a reliable prediction the leading terms of the form
$(\alpha_s\ln^2\ldots)^n$ thus have to be included.  The results are based
on alternatively momentum space or impact parameter techniques which
were originally developed for the Drell Yan process
and applied to top pair production in \cite{Laenen}. An additional
complication arises from the blow up of the coupling constant
associated with the radiation of soft gluons for $\hat s\to 4m^2_t$.
This has been interpreted in \cite{Laenen} as a breakdown of
perturbation theory.  Different regulator prescriptions have been
advertised. In \cite{Laenen}
 a cutoff $\hat s-4m^2_t >\mu_0^2$ with $\mu^2_0 \ll
4m^2_t$ was introduced to exclude a small fraction
of the phase space.

The result is fairly stable for $q\bar q$ induced reactions with
$\mu_0/m_t$ chosen between 0.05 and 0.2.  The small contribution from
gluon fusion, however, is  sensitive towards $\mu_0$ which had
to be chosen in the range between $0.2m_t$ and $0.3m_t$, a consequence
of the enhancement of radiation from gluons.
A slightly different approach (``principal value resummation'')
 has been advocated in \cite{Berger}
which circumvents the explicit $\mu_0$ dependence of the result, but
leads essentially to the same final answer (Table \ref{table:cross}).
\begin{table}[htb]
\begin{center}
\begin{tabular}{|l|l|l|}
\hline
$m_t$ [GeV] & 175 & 180 \\ \hline \hline
$\mu_0$ (min) &4.72 & 3.86  \\ \hline
$\mu_0$ (centr) & 4.95 &  4.21\\ \hline
$\mu_0$ (max) & 5.65 & 4.78  \\ \hline
``principal value'' & 5.6 & 4.8  
\\ \hline
\end{tabular}
\end{center}
\caption{Top production cross section (in pb)
for $\protect\sqrt s$ =1800 GeV for
different values of the cutoff $\mu_0$ \protect\cite{Laenen} (first
three lines) and 
for the  ``principal value'' prescription \protect\cite{Berger}
(fourth line).
} 
\label{table:cross}
\end{table}
\begin{table}[htb]
\begin{center}
\begin{tabular}{|l|l|}
\hline
  &  $\protect\sigma$ [pb] \\ \hline
Altarelli et al. \protect\cite{nason1}  & $3.52$ (DFLM) 
\\
                                          & $4.10$ (ELHQ)
\\ \hline
Laenen et al. \protect\cite{Laenen} & 
           $ \left.\begin{array}{ll} 3.5 &\quad(\mu^2=4m^2)\\
                                   3.8  &\quad(\mu^2=m^2)\\
                                   4.05 &\quad(\mu^2=m^2/4)
                  \end{array}
\right\}$ MRSD
\\ \hline
\hline
Resummation &   \\ \hline
Laenen et al. \cite{Laenen}   &  $
            \left.\begin{array}{c} 3.86\\4.21\\4.78\end{array}
\right\} $ vary $\mu_0$
\\ \hline
Berends et al. \cite{Berends} & 4.8 central value
\\ \hline
Berger et al. \cite{Berger} & 4.8 ``principal value res.''
\\ \hline
\end{tabular}
\end{center}
\caption{History of predictions for the production cross section for
$\protect\sqrt{s} =1.8$ TeV and $m_t = 180$ GeV.}
\label{table:history}
\end{table}

The result of the improved prediction (central value) is compared to
the fixed order calculation (with $\mu^2=4m^2,m^2,m^2/4$) in
fig.~\ref{fig:seite15}. 
\begin{figure}[htb]
\vspace*{13cm}
\caption{The NLO exact cross section as a function of the top quark
  mass for  three choices of scale: 
$\mu = m/2$ (upper solid line),
$\mu = m$ (central solid line)
and $\mu = 2m$ (lower solid line), and the NLO exact cross section
plus the ${\cal O}(\alpha^4_s)$ contribution at $\mu=m$ (dashed
line) (from \protect\cite{Laenen}).
}
\label{fig:seite15}
\end{figure}
Resummation evidently increases the cross section slightly above the
previously considered range.  The history of predictions is shown in
table~\ref{table:history}, with $\sqrt{s}=1.8$ TeV and $m_t=180$ GeV
as reference values.  The table demonstrates that the spread of
predictions through a (fairly extreme) variation of structure
functions (DFLM vs. ELHQ) 
and through a variation of the renormalisation and factorisation are
comparable --- typically around $\pm 10\%$. Leading log resummation
increases the cross sections by $10-15\%$, with a sizeable sensitivity
towards the cutoff prescription.  A reduction in $m_t$ by 5 GeV leads
to an increase of $\sigma$ by about 0.8 pb. 
Theory and experiment,
with its present result of $7.6^{+2.4}_{-2.0}$ pb and $6.3\pm 2.2$ pb
from CDF and D0 respectively are thus well compatible
(Fig.~\ref{fig:seite17}).  
\begin{figure}[htb]
\vspace*{13cm}
\caption{Comparison of experimental results for the top quark mass and
production cross section with the theoretical prediction.}
\label{fig:seite17}
\end{figure}

\subsection{Threshold behaviour}
\label{ssec:Coul}

\begin{figure}
\hspace*{20mm}
\begin{tabular}{cl}
\psfig{figure=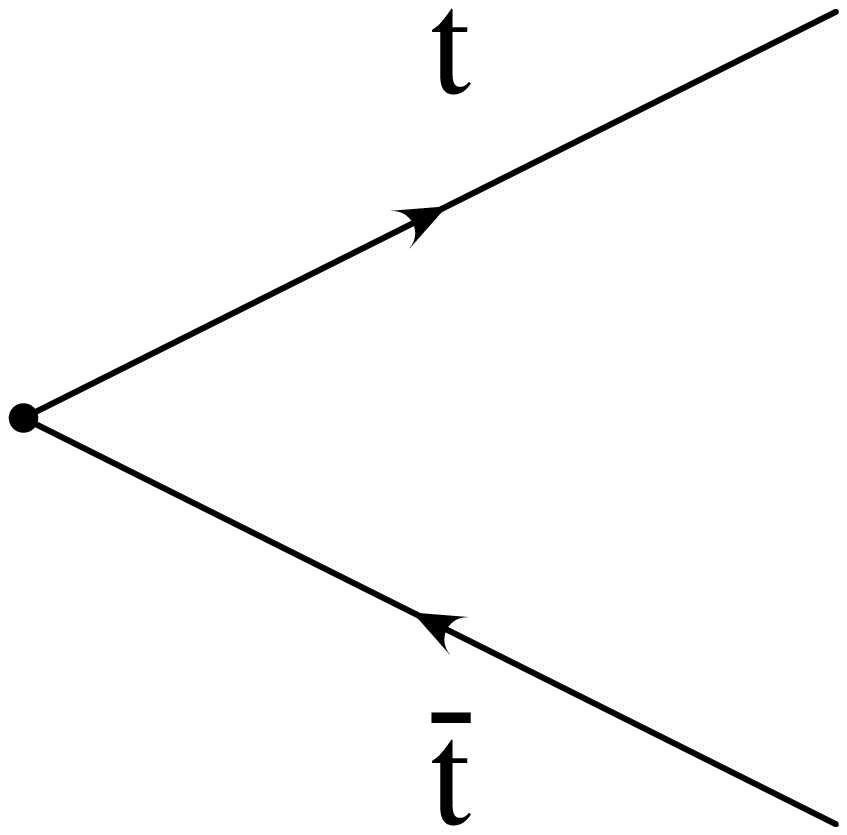,width=35mm,bbllx=210pt,%
bblly=410pt,bburx=630pt,bbury=550pt} & $\sim \beta_t$
\\[12mm]
\psfig{figure=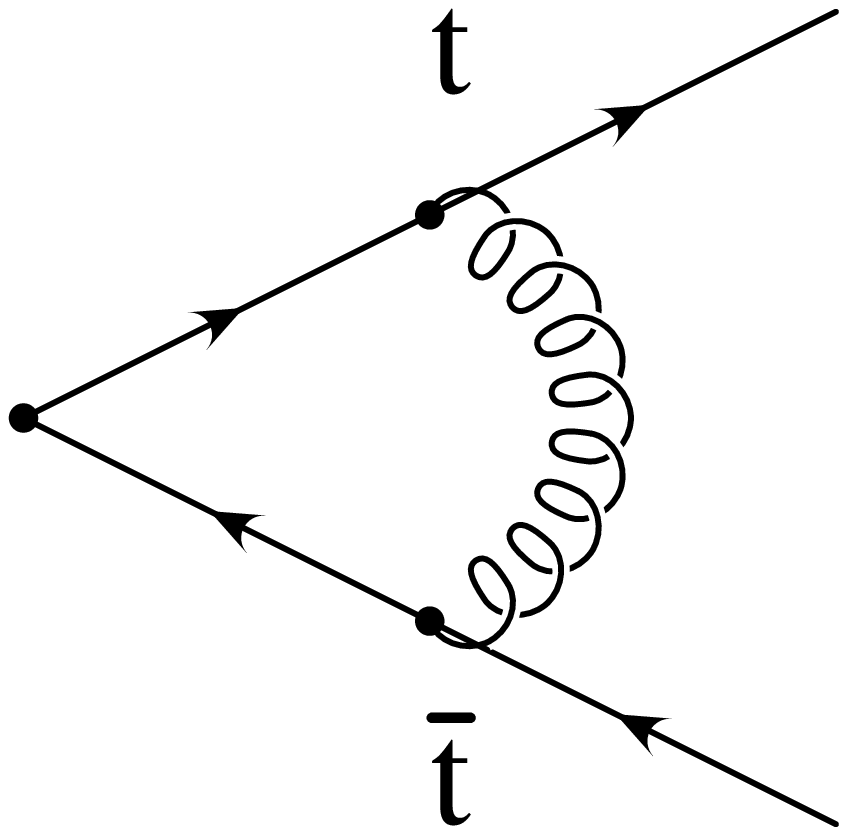,width=35mm,bbllx=210pt,%
bblly=410pt,bburx=630pt,bbury=550pt} & $\sim \beta_t\cdot
{\pi\alpha_s\over
2\beta_t} \left\{\begin{array}{ll}
4/3 & {\rm singlet}\\
-1/6 & {\rm octet}
\end{array}\right.
$
\end{tabular}
\vspace*{15mm}
\caption{Vertex corrections from gluon exchange in the threshold
region.}
\label{fig:coul}
\end{figure}

Near the $\ttb$ threshold the cross sections are affected by resonance
production and Coulomb rescattering forces \cite{bigi2}, \cite{guesk1},
\cite{fadin1}. These
corrections can be estimated in a simplified potential picture. The
driving one-gluon exchange potential is attractive if the $\ttb$ is in
color-singlet state and repulsive in a color-octet state \cite{fadin1},
\begin{eqnarray}
\sigma^{(1)}(gg\ra t\overline{t}) & = & \frac{2}{7}
       \;\sigma_B(gg\ra t\overline{t}) \mid\!\Psi^{(1)}\!\mid^2 \nonumber \\
\sigma^{(8)}(gg\ra t\overline{t}) & = & \frac{5}{7}
       \;\sigma_B(gg\ra t\overline{t}) \mid\!\Psi^{(8)}\!\mid^2 \\
\sigma^{(8)}(q\overline{q}\ra t\overline{t}) & = & \phantom{\frac{5}{7}}
       \;\sigma_B(q\overline{q}\ra t\overline{t})
       \mid\!\Psi^{(8)}\!\mid^2 \nonumber
\end{eqnarray}

with the correction factors (see fig.~\ref{fig:coul})
given in NLO by 
\be
\beta |\psi|^2 = \beta \left(
1+{\pi\alpha_s\over 2\beta_t}\left\{
\begin{array}{cc}
4/3 & \mbox{singlet} \\
-1/6 & \mbox{octet} 
\end{array}
\right.
\right)
\label{eq:star1}
\ee
The summation of the leading $\pi\alpha_s/\beta$ terms to all orders
results in the familiar Sommerfeld correction factor
\be
\beta|\psi|^2 = \beta {x\over 1-e^{-x}}
\label{eq:18a}
\ee
For $t\bar t$ in the singlet configuration
$x=x^{(1)}\equiv {4\over 3} {\pi \alpha_s\over \beta}$, for octet states
 $x=x^{(8)}\equiv -{1\over 6} {\pi \alpha_s\over \beta}$.

The Coulombic attraction thus leads to a sharp rise of the cross section
at the threshold in the singlet channel, even if no resonance can be
formed anymore, since the phase space suppression of the Born term
$\sigma_B \propto \beta_t$
is neutralized by the Coulomb enhancement of the wave function
$\left| \Psi_1 \right|^2 \propto \alpha_s / \beta$.
In the octet channel (dominant for $q\bar q$ annihilation), by
contrast, the cross sections are strongly 
reduced by the Coulombic repulsion which leads effectively to an
exponential fall-off of the cross sections
$\sigma_8 \propto \exp \left[ - \pi \alpha_s / 6 \beta_t \right]$
at the threshold \cite{fadin1}.
Due to the
averaging over parton luminosities the effects are less
spectacular in $p\bar p$ or $pp$ than in $e^+e^-$ collisions.

The enhancement and suppression factors are compared to simple phase
space $\sim \beta$ in Fig.~\ref{fig:seite20}.
The dotted line corresponds to the phase space factor $\beta$, the
dashed line to the perturbative NLO calculation
(\ref{eq:star1}),
the solid line to the Coulomb enhancement given in eq.~(\ref{eq:18a}).
The predictions for the singlet, octet
($\equiv q\bar q$), and properly weighted gluon fusion channel are
displayed in Figs.~\ref{fig:seite20} a, b, and c, respectively.
\begin{figure}[htbp]
  \begin{center}
    \leavevmode
\psfig{figure=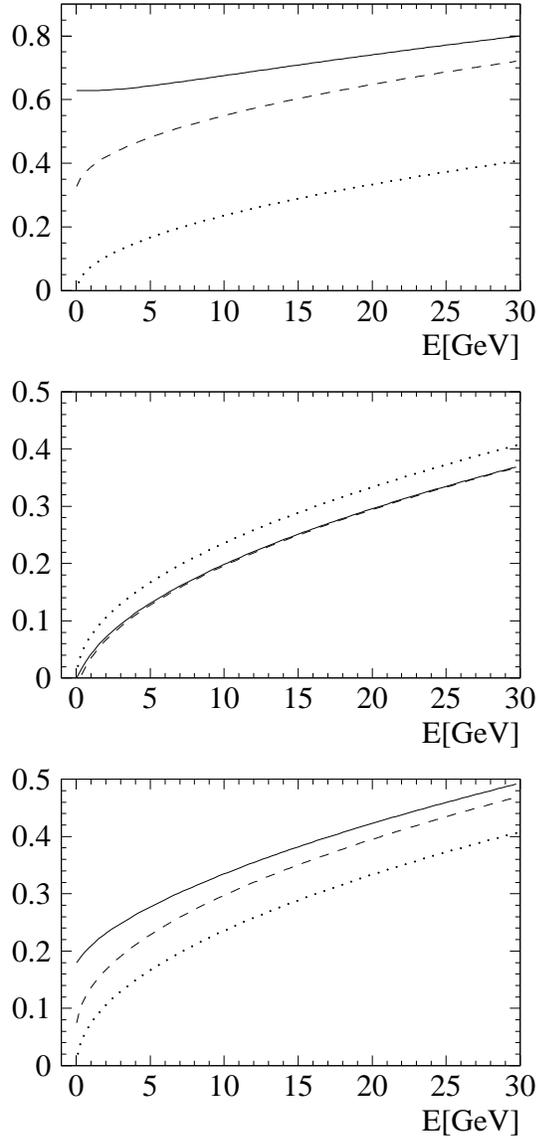,width=150mm,bbllx=0pt,bblly=115pt,%
bburx=630pt,bbury=699pt}
  \end{center}
  \caption{Threshold enhancement and suppression factors for singlet
(upper), octet (middle), and properly weighted gluon fusion (lower
figure) for $\alpha_s=0.15$. Dotted line: Born approximation; dashed
line: NLO approximation; solid line: Coulomb enhancement.}
  \label{fig:threshold}
\end{figure}
\subsection{Electroweak corrections}
\label{ssec:weak}
Another potentially important modification which is closely tied to
the Coulomb enhancement originates from vertex corrections induced by
light Higgs boson exchange.  In a simplified treatment these are
lumped into a Yukawa potential 
\be
V(r) = { \sqrt{2} G_F m_t^2\over 4\pi} {e^{-m_Hr}\over r}
\ee
resulting in a reduction $\delta E$ of the apparent threshold, with
$\delta E = -200$ MeV for $m_H = 100$ GeV as characteristic example.
The change in the normalisation by $+ 10\%$ could become
relevant for precision measurements.  The situation is quite similar
to the one discussed for $e^+e^-$ colliders in section 3.2.

\begin{figure}
\hspace*{20mm}
\begin{tabular}{ccc}
\psfig{figure=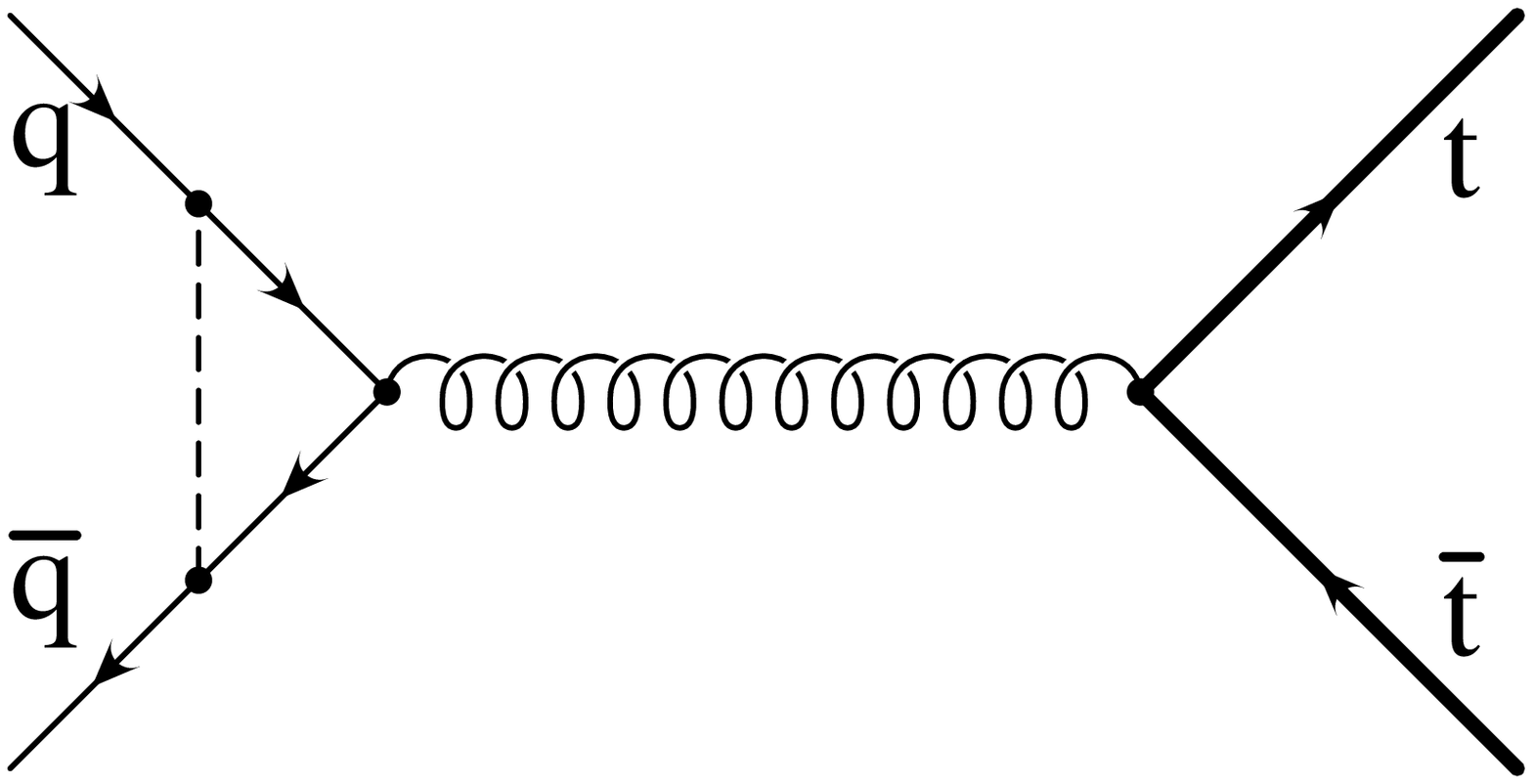,width=35mm,bbllx=210pt,%
bblly=410pt,bburx=630pt,bbury=550pt}
&
\hspace*{5mm}
\psfig{figure=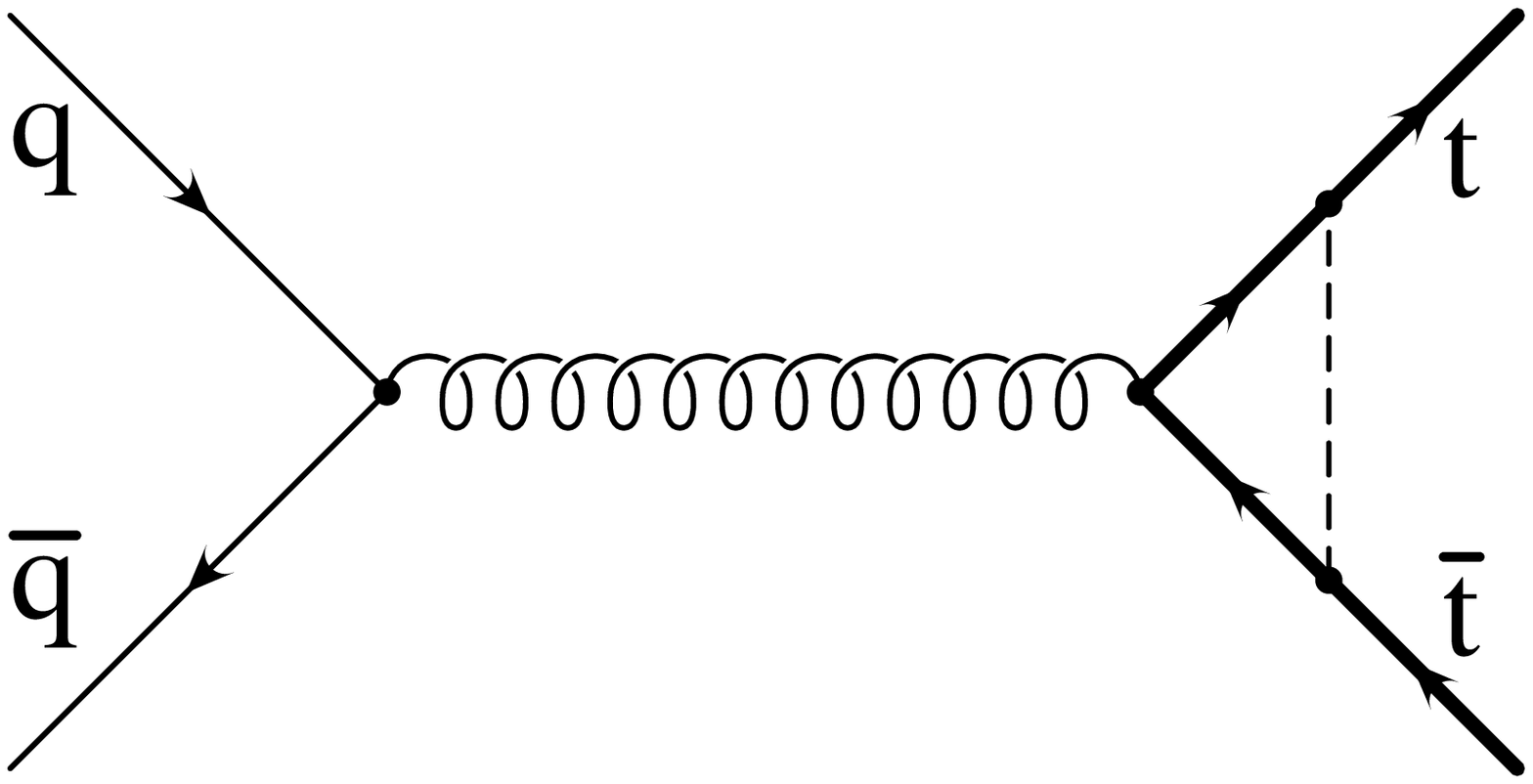,width=35mm,bbllx=210pt,%
bblly=410pt,bburx=630pt,bbury=550pt}
&
\hspace*{5mm}
\psfig{figure=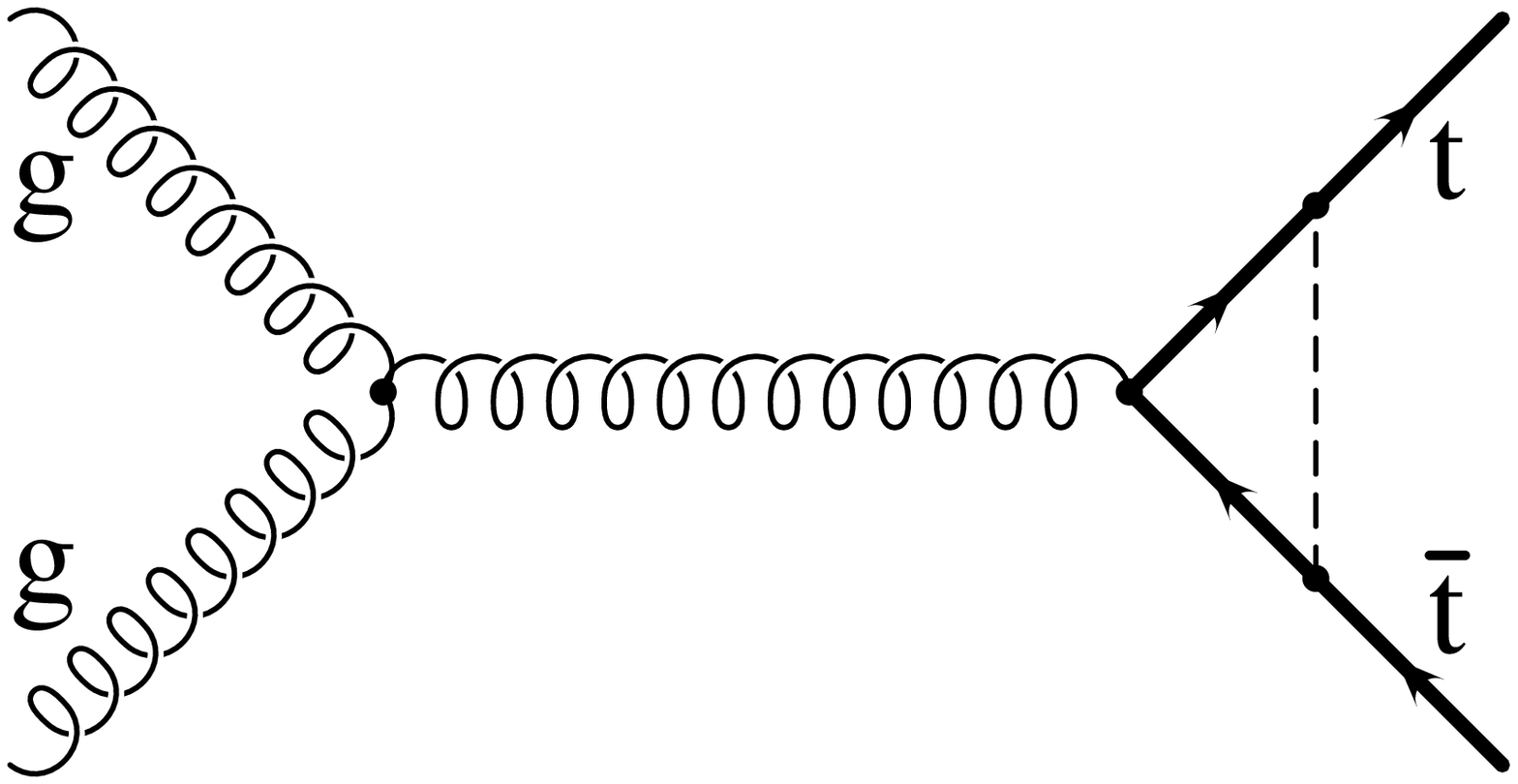,width=35mm,bbllx=210pt,%
bblly=410pt,bburx=630pt,bbury=550pt}
\\[20mm]
\psfig{figure=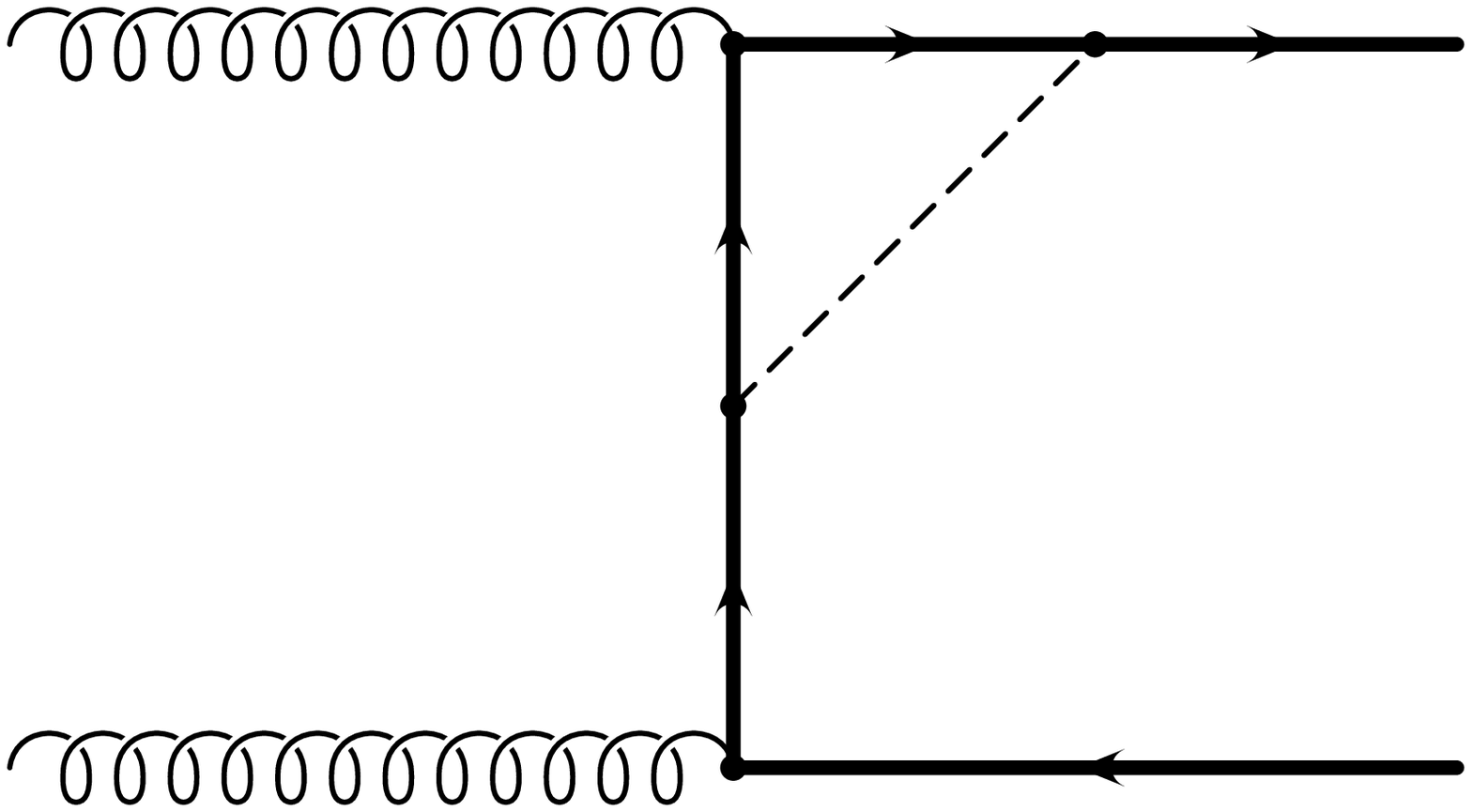,width=35mm,bbllx=210pt,%
bblly=410pt,bburx=630pt,bbury=550pt}
&
\hspace*{5mm}
\psfig{figure=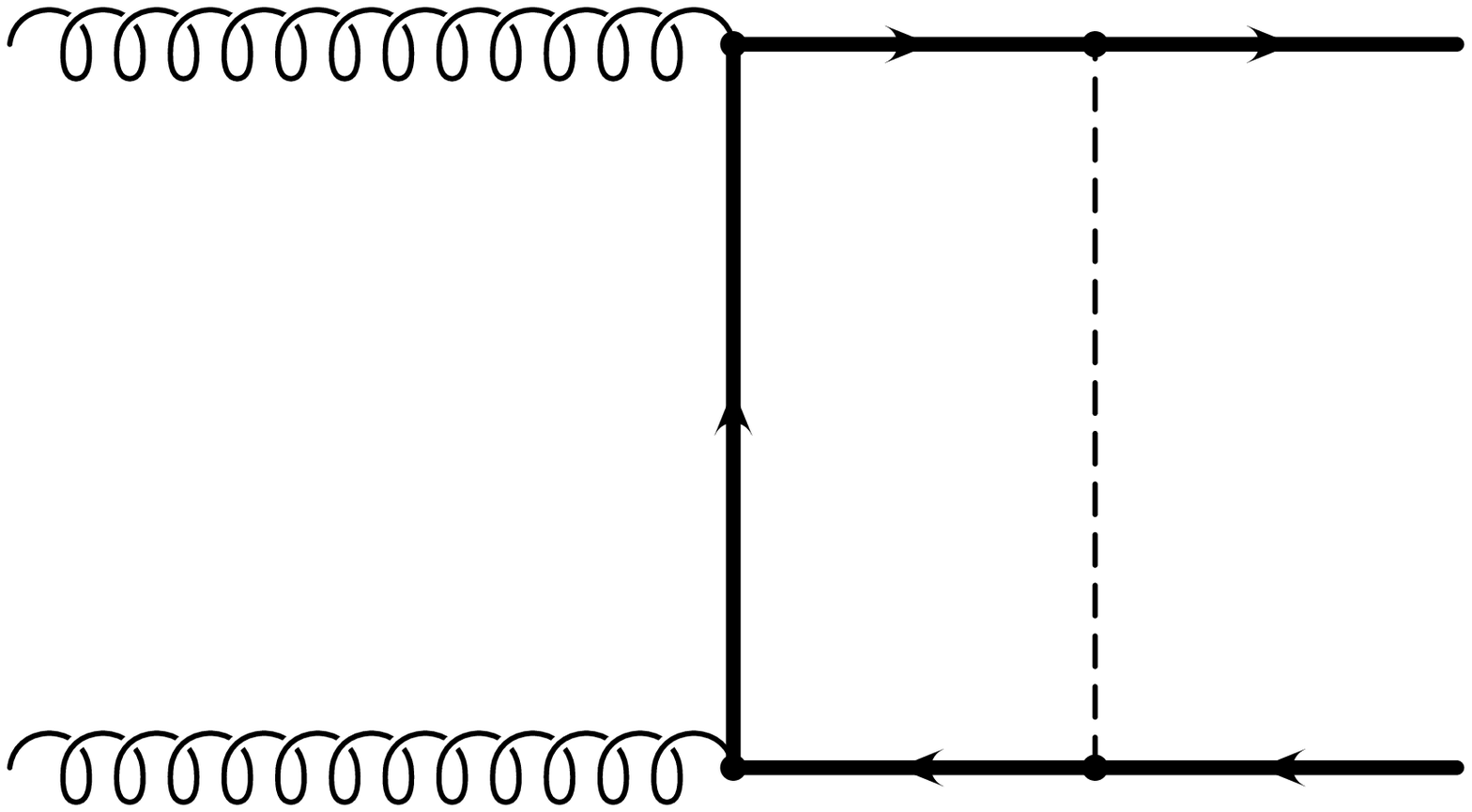,width=35mm,bbllx=210pt,%
bblly=410pt,bburx=630pt,bbury=550pt}
&
\hspace*{5mm}
\psfig{figure=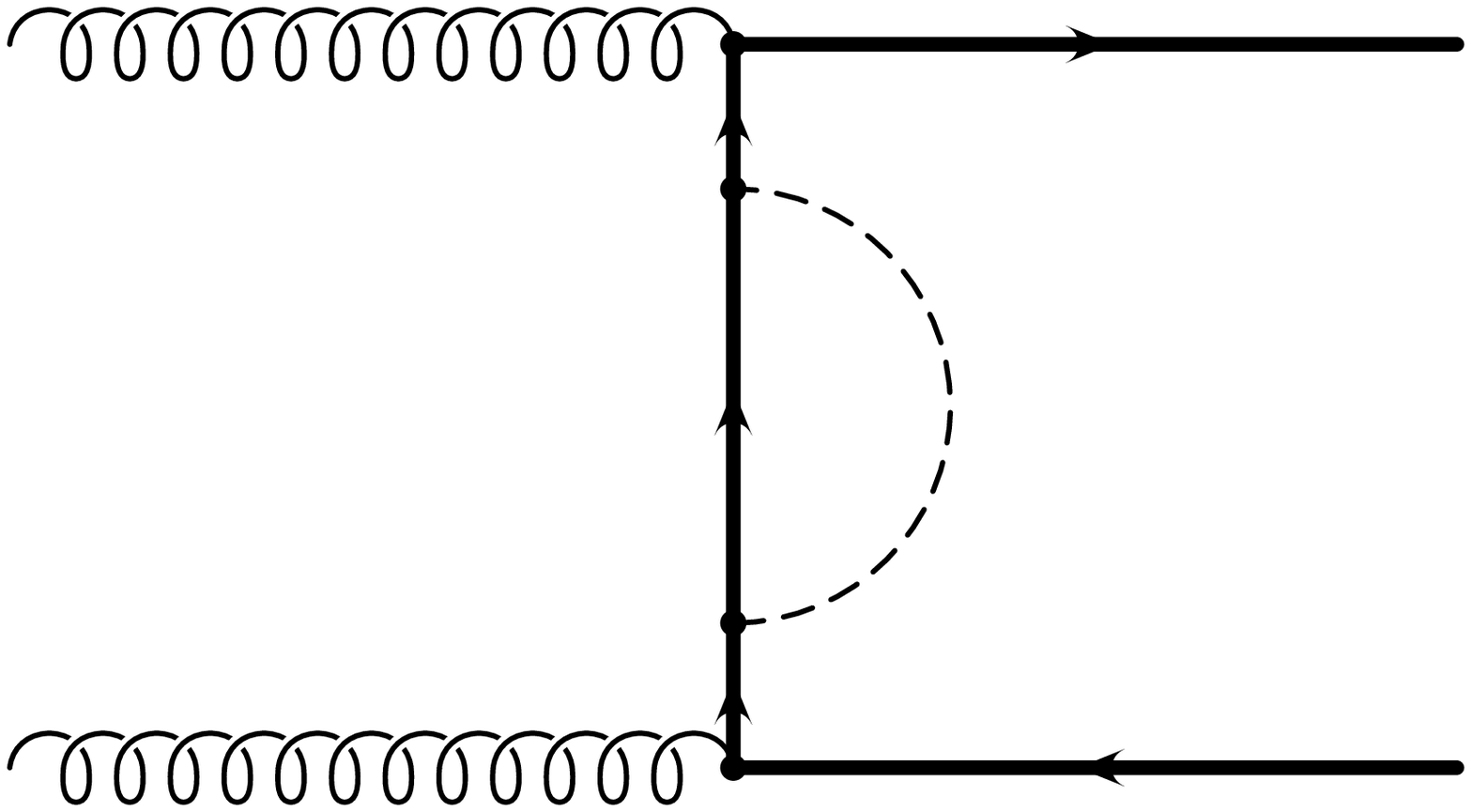,width=35mm,bbllx=210pt,%
bblly=410pt,bburx=630pt,bbury=550pt}
\end{tabular}
\vspace*{15mm}
\caption{Feynman diagrams contributing to the electroweak radiative
  corrections.  Dashed lines stand for $H$, $Z$, $\gamma$, or $W$.}
\label{fig:seite21}
\end{figure}

Genuine electroweak contributions of
${\cal O} \left( \alpha \alpha^2_s \right)$
have been calculated to both the
$\qqb$ and $gg \ra \ttb$
subprocesses \cite{holli2}. The corrections include vertex corrections and
box diagrams built-up by vector bosons and the Higgs boson
(Fig.~\ref{fig:seite21}). 

\begin{figure}[htbp]
  \begin{center}
    \leavevmode
    \vspace*{15cm}
  \end{center}
  \caption{Relative correction to the parton cross section for
$m_t=250$ GeV (upper figure: $q\bar q\to t\bar t$, lower figure:
$gg\to t\bar t$) (from \protect\cite{holli2}).}
  \label{fig:seite20}
\end{figure}

Except for a small region close to the production threshold, which is
dominated by the Yukawa potential, the
corrections are always negative; they can become sizeably large,
in particular
if the top is very heavy
and if the energy of the subsystem exceeds 1 TeV, not uncommon for
$t\bar t$ production at the LHC.  In this case however, the large
negative corrections are compensated by positive contributions from
real radiation of $W$, $Z$, or $H$.  The corrections for the $q\bar q$
and $gg$ subprocess as functions of the parton energies are shown in
Fig.~\ref{fig:seite20}.  The sharp increase of the corrections
close to threshold for a light Higgs is clearly visible and,
similarly, the large negative correction for large parton energies. 
 After convoluting
the cross sections of the subprocesses with the parton distributions,
a reduction of the Born
cross section at a level of a few percent is observed
(Fig.~\ref{fig:been17}). 
\begin{figure}
\vspace*{90mm}
\caption{Relative correction to the hadronic cross section for $s=(16
  $ TeV)$^2$ (from \protect\cite{holli2})} 
\label{fig:been17}
\end{figure}

\subsection{Gluon radiation}
\label{ssec:exper}

Up to this point the discussion has centered around the predictions
for inclusive top quark production.  Additional ingredients for the
experimental analysis are the detailed topological structure of the
signal, the number of jets,  the characteristics of the underlying
event, and, of course, predictions for the background.  This
information allows to adjust in an optimal way experimental cuts and
to measure the top quark through a kinematic analysis of its decay
products.  As a typical example the impact of gluon radiation on the
top mass determination has been analysed recently.  An idealized study
has been performed e.g. in \cite{Lampe}.  Radiated gluons are merged
with the $b$ jet from top decay or with the quark jet from $W\to q\bar q$ if
they are found within a cone of opening angle
\be
R=\sqrt{(\Delta \eta)^2 + (\Delta \phi)^2} < R_{cut}
\ee
with respect to $b$, $q$ or $\bar q$ and if their rapidity is below
$|\eta| < 2$. In this case the gluon is considered as top decay
product and hence contributes to its invariant mass.  If the gluon jet
falls outside the cuts, it is assigned the rest of the event.

\begin{figure}[htb]
\hspace*{25mm}
\begin{tabular}{cc}
\psfig{figure=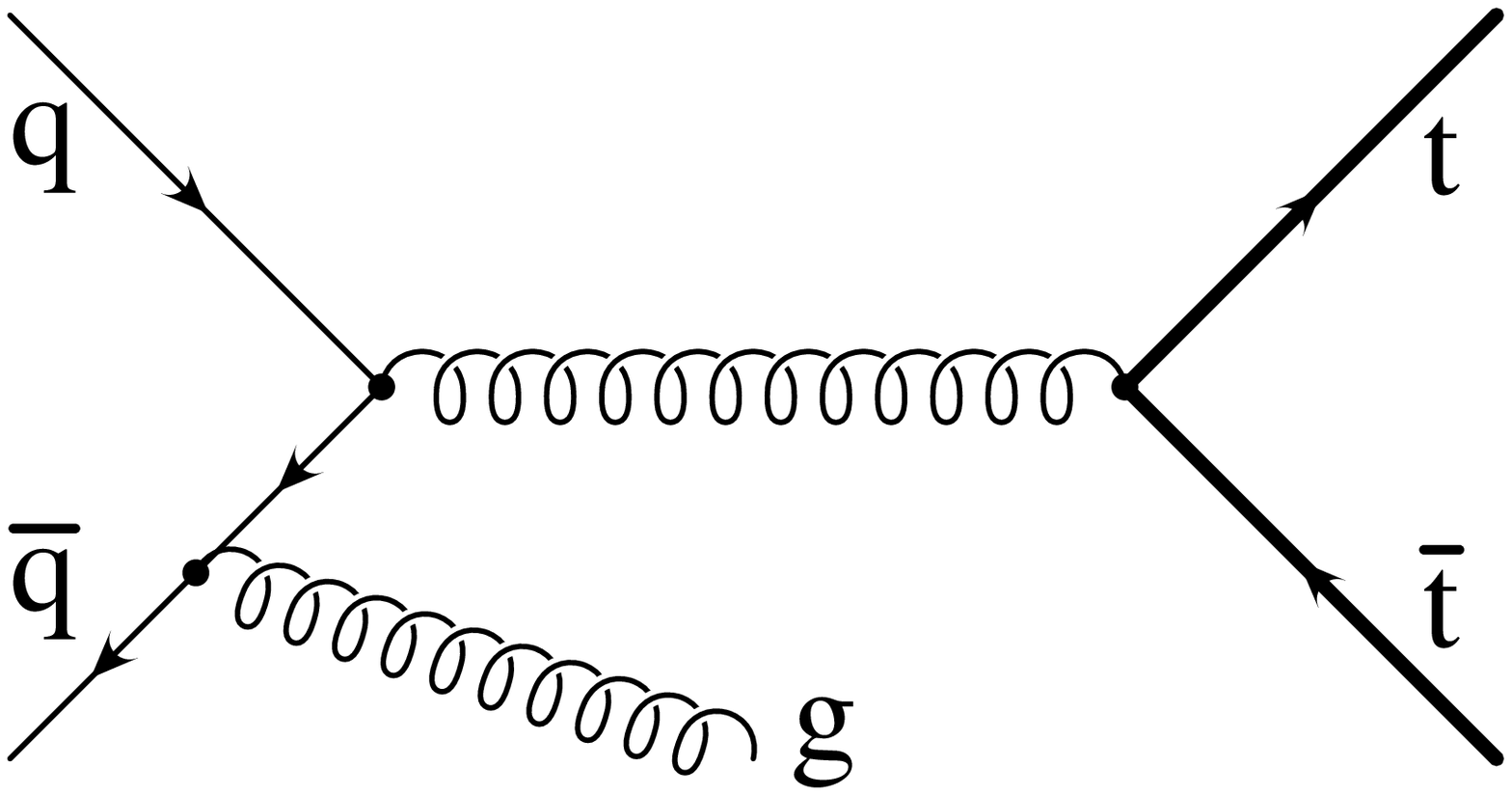,width=50mm,bbllx=210pt,%
bblly=410pt,bburx=630pt,bbury=550pt}
&
\hspace*{20mm}
\psfig{figure=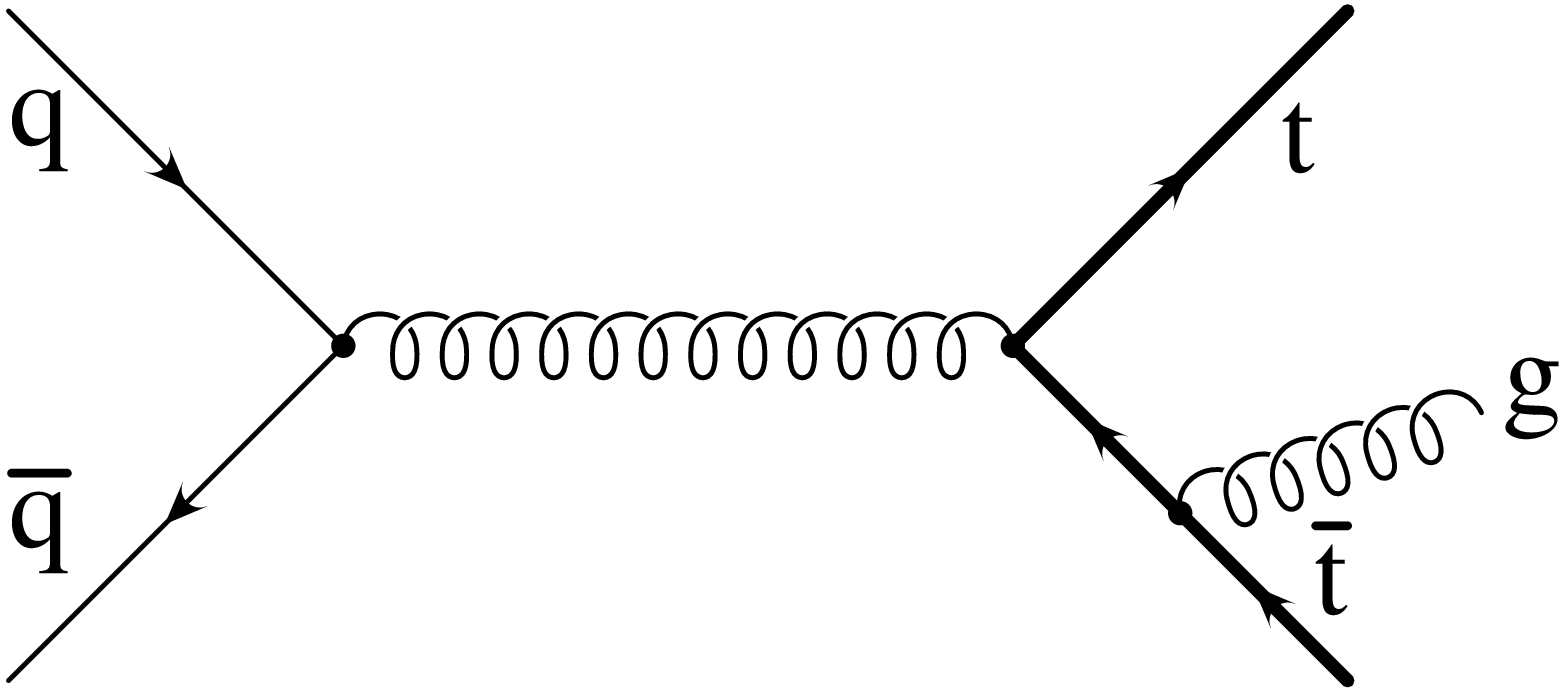,width=50mm,bbllx=210pt,%
bblly=410pt,bburx=630pt,bbury=550pt}
\end{tabular}
\\[30mm]
\hspace*{65mm}
\psfig{figure=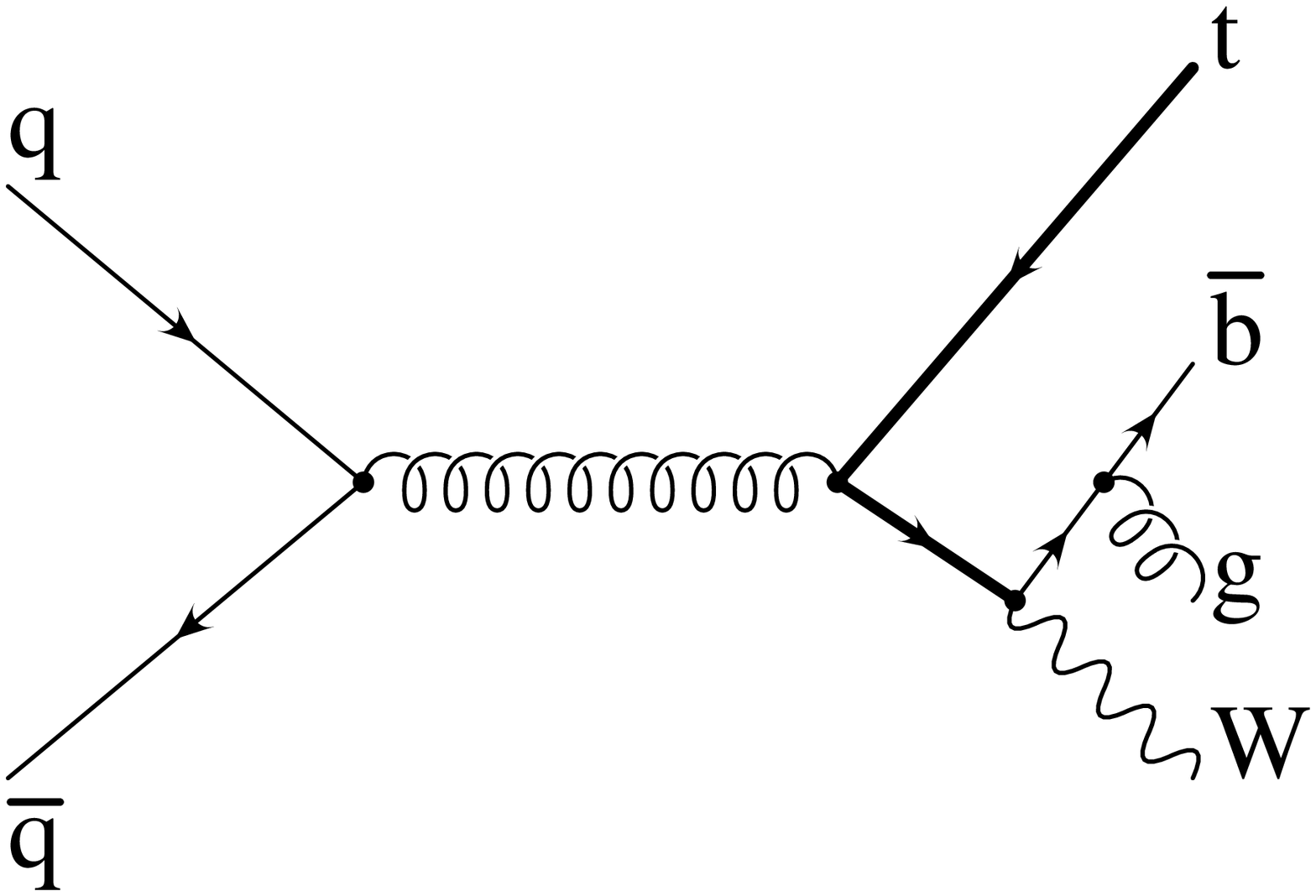,width=50mm,bbllx=210pt,%
bblly=410pt,bburx=630pt,bbury=550pt}
\\
\vspace*{15mm}
\caption{Gluon radiation from top production (upper) and decay (lower
  diagrams).}
\label{fig:seite22}
\end{figure}
Gluon radiation associated with top quark production, if erroneously
associated with top decay, will thus increase the apparent $m_t$.
Radiation from top decay, if outside the forementioned cuts, will
however, decrease the measured mass of the quark.  The interplay
between the two compensating effects is displayed in
Fig.~\ref{fig:seite23}. \begin{figure}[htbp]
\begin{center}
  \leavevmode
  \vspace*{10cm}
\end{center}
\caption{Shift in the apparent top mass as a function of $R$ cut (from
  \protect\cite{Lampe}).}
\label{fig:seite23}
\end{figure}
For a realistic $R_{cut} \approx 0.6$ a reduction $\Delta m$ of around
2 GeV is predicted.

\section{Single top production}
\label{singletop}
\begin{figure}
\begin{minipage}{16.cm}
\hspace*{40mm}
\[
\mbox{
\hspace*{15mm}
\begin{tabular}{c}
\psfig{figure=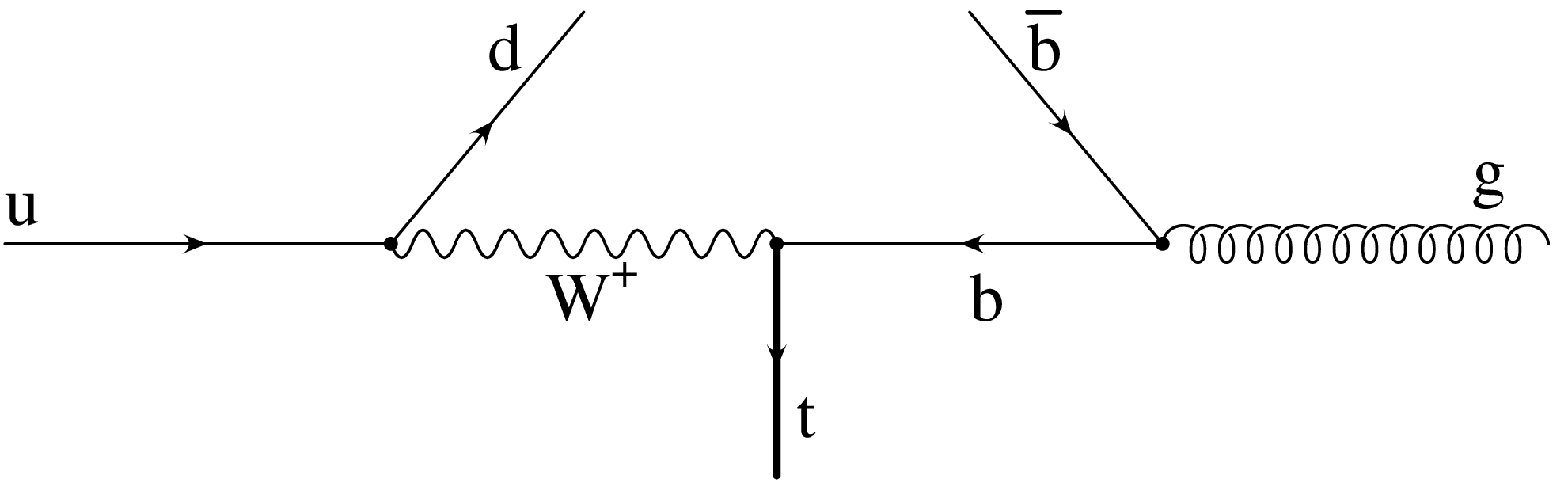,width=40mm,bbllx=210pt,bblly=410pt,bburx=630pt,bbury=550pt}
\\[15mm]
\end{tabular}}
\]
\end{minipage}
\caption{Characteristic diagram for single top production. }
\label{fig:seite25}
\end{figure}

Virtual $W$ bosons, originating from $u\to dW$ splitting, can merge
with bottom quarks from gluon splitting $g\to b\bar b$ to produce
single top quarks in association with fairly collinear $d$- and $\bar
b$- jets (Fig. \ref{fig:seite25}).
The interaction radius in the QCD $gg$ fusion process shrinks with
rising energy so that the cross section $\sigma(gg \rightarrow \ttbn) \sim
\alpha_s^2 / \shatn$ [mod.~log's] vanishes asymptotically. By contrast, the
interaction radius in the weak fusion process \cite{wille1} is set by the
Compton wave length of the $W$ boson and therefore asymptotically non-zero,
$\sigma \rightarrow G_F^2 m_W^2 / 2\pi$. The subprocess has to be
folded
 with the
quark-gluon luminosities
\be
\sigma(p\bar p\to t + X) = \int_\tau^1 {{\rm d}z\over z} {\cal
L}_{ug}(\tau/z)
\sigma(ug\to t+\bar b+d)
\ee
plus a similar contribution from $\bar d g\to t + \bar b + \bar u$.  
The fall-off of the total cross section
$\sigma(pp \rightarrow t\overline{b})$ is less steep than for the QCD fusion
processes. As a result, the $Wg$ fusion process
would have dominanted for large top
quark masses $\ge$ 250 GeV at the LHC (Fig.~\ref{Fhads2}).

\begin{figure}[htb]
\vspace*{5mm}
\begin{tabular}{cc}
\hspace*{25mm}
\psfig{figure=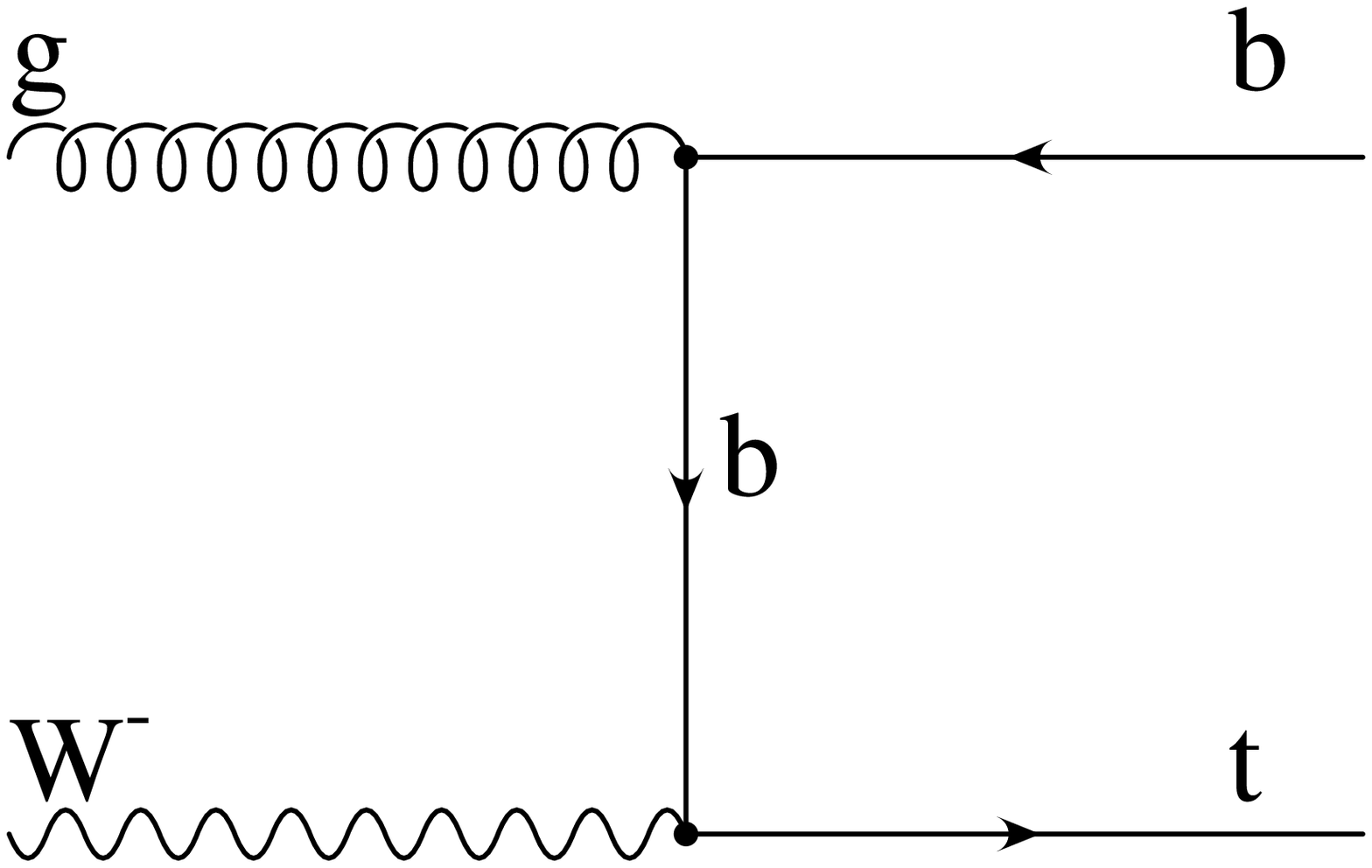,width=50mm,bbllx=210pt,%
bblly=410pt,bburx=630pt,bbury=550pt}
&
\hspace*{20mm}
\psfig{figure=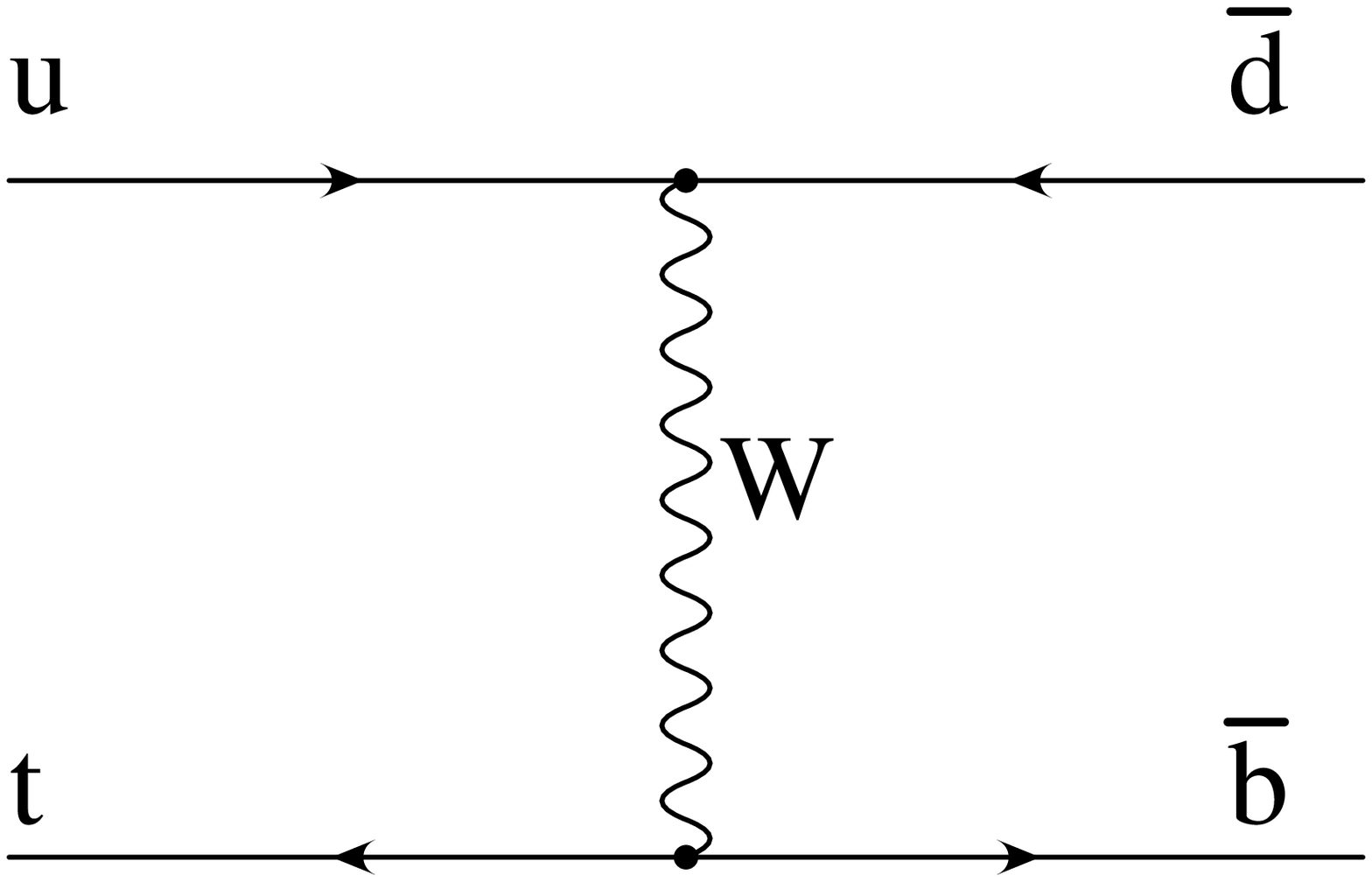,width=50mm,bbllx=210pt,%
bblly=410pt,bburx=630pt,bbury=550pt}
\\[18mm]
(a) & (b)
\end{tabular}
\vspace*{5mm}
\caption{Characteristic diagrams for top production via Compton
  scattering (a) and Drell Yan process (b)}
\label{fig:Wg}
\end{figure}

For $m_t =$ 180 GeV, the case of practical interest, single top
production is about a factor 5 below the QCD reaction.
Nevertheless, as shown in Fig.\ref{Fhads2},
about 10$^6$ top quarks will be produced at the LHC by this mechanism
at an integrated luminosity of $\int\cal L$ = 10$^4
pb^{-1}$. Also at the TEVATRON this process should be accessible with
the anticipated luminosity.
\begin{figure}[hbt]
\vspace*{115mm}
\caption{Cross sec\-ti\-ons for several
      mechanisms of top quark production \protect\cite{reya1}. Various
                   pa\-ra\-me\-tri\-za\-ti\-ons and models of the parton
          densities \protect\cite{diemo1} have been used.}
\label{Fhads2}
\end{figure}

A close inspection of the various contributions to the subprocess
$ug\to t+\bar b+d$ 
reveals immediately that the
by far dominant part of the cross section is due to  $b$ exchange, with the
b quark being near its mass shell. Since the $b$
quark is almost collinear to the incoming gluon,
this cross section is logarithmically enhanced $\sim\ln(m_t^2/m_b^2)$
over other mechanisms.
This naturally suggests to approximate
the process by the subprocess $u$ + $b$
$\stackrel{\mbox{\scriptsize W}}{\rightarrow}$ $d$ + $t$ with the $b$-quark
distribution generated perturbatively
by gluon splitting based on
massless evolution equations.
The weak cross sections can be presented in a compact form,
\begin{eqnarray}
\sigma(ub\stackrel{\mbox{\scriptsize W}}{\rightarrow}dt) & = &
\frac{G_F^2 m_W^2}{2\pi}\;\frac{(\shatn-m_t^2)^2}
{\shatn(\shatn+m_W^2-m_t^2)}
\mbox{\hspace{5.00cm} }\rightarrow\frac{G_F^2 m_W^2}{2\pi}
\nonumber \\
\sigma(d\overline{b}\stackrel{\mbox{\scriptsize W}}{\rightarrow}
u\overline{t})
&\rule[-1mm]{0mm}{10mm}= &
\frac{G_F^2 m_W^2}{2\pi}\;
\left[\;1+\frac{\shatn(2m_W^2-m_t^2)-2m_W^2m_t^2}{\shatn^2} \right.
\nonumber \\
& &    \mbox{\hspace{1.6cm}} \left.
-\;\frac{(2\shatn+2m_W^2-m_t^2)m_t^2}{\shatn^2}\;\log
\frac{\shatn+m_W^2-m_t^2}
{m_W^2}\;\right] \mbox{\hspace{0.00cm}}\rightarrow\frac{G_F^2 m_W^2}
{2\pi} \nonumber
\end{eqnarray}
and identically the same expressions for the $\cal C$-conjugate reactions.

Top quarks are created in $u+g$ collisions, anti-top quarks in
$d+g$ collisions where the absorption of a $W^-$ transforms a
$\overline{b}$ quark to a $\overline{t}$ quark. The
na$\ddot{\mbox{\i}}$ve expectation from valence quark counting
for the ratio of $t / \overline{t}$ cross sections,
$\sigma(u\rightarrow t):\sigma(d\rightarrow \overline{t})$ $\sim$ 2 : 1 is
corroborated by a detailed analysis; in fact, the ratio
turns out to be 2.1 for top quark
masses of about 150 GeV.

The remaining possibilities for single top production are Compton
scattering (Fig.~\ref{fig:Wg}(a))
\be 
g+b\to W+t
\ee
and the Drell Yan process (Fig.~\ref{fig:Wg}(b))
\be
u+\bar d \to t+ \bar b
\ee
The predicted cross sections are too small to be of practical
interest.  Single top quark production via the dominant mechanism
(Fig.~\ref{fig:seite25})  offers a unique way for a
measurement of the CKM matrix element $V_{tb}$ and thus, indirectly of
the top quark life time.  As discussed in section \ref{sect:qualit}
 $V_{tb}$ is strongly constrained to be very close to one for three 
generations --- in a four generation
model $V_{tb}$ may be quite different from these expectations. 

\section{Quarkonium production}

Both charm and bottom quarks have been discovered at hadron colliders
in the form of quarkonium resonances $J/\psi$ and $\Upsilon$ through
their distinct signals in the $\mu^+\mu^-$ channel. The search for
toponium at a hadron collider is, however, entirely useless.  The
broad ($\Gamma\sim$ 2 GeV) resonances decay with an overwhelming
probability through single quark decay and are therefore
indistinguishable from open top quarks produced close to threshold.  

The situation could be different in extensions of the SM. Decays of a
fourth generation $b^\prime$ 
\be
b^\prime \to t + W
\ee
are suppressed by small mixing angles.  Alternatively, if
$m_{b'}<m_t+m_W$, the $b'\to c+W$ mode would have to compete with loop
induced FCNC decays --- leaving ample room for narrow quarkonium
states. Another example would be the production of weak isosinglet
quarks which are predicted in Grand Unified Theories.  The decay of
these quarks would again be inhibited by small mixing angles.  

Of particular interest is the search for $\eta_{b'}$, the $^1S_0$ state
composed of $b'$ and $\bar b'$ \cite{Barger,KM,KP}. It is produced
with appreciable cross section. Its dominant decay mode
\be
\eta_{b'} \to H+Z
\ee
is enhanced by the large Yukawa coupling, governing the coupling of
the heavy quark to the Higgs and the longitudinal $Z$.  For large
$m_{b'}$ one obtains 
\be
{\Gamma(\eta_{b'}\to H+Z)\over \Gamma(\eta_{b'}\to \gamma\gamma)}
\sim
{m_{b'}^4\over M_W^4}
\ee
The branching ratios as functions of $M_\eta$ are displayed in
Fig.~\ref{fig:seite30oben}.
\begin{figure}
\vspace*{105mm}
\caption{Branching ratios of $\eta_{b'}$ for the dominant decay modes
  as functions of the bound state mass $M$.}
\label{fig:seite30oben}
\end{figure}
The complete set of QCD corrections for leading and subleading
annihilation decay modes
can be found in \cite{KP}.  They do not alter the picture
significantly.

It should be emphasized that the decay $\eta_{b'}\to H+Z$ proceeds
through the axial part of the neutral current coupling which, in turn,
is proportional to the third component of the weak isospin.  
Bound states of isosinglet quarks would, therefore, decay dominantly
into two gluon jets.

The cross section for {\em open} $b\bar b'$ production at the
LHC (with $m_{b'}=300$ GeV) amounts to about 100 pb.  The fraction of
the phase space where bound states can be formed, i.e. for relative
quark velocity $\beta<0.1$, covers around $10^{-2}$ of the relevant
region
\be
{\int_0^{0.1} {\rm d}\beta \beta 
\over
\int_0^{1} {\rm d}\beta \beta }
=10^{-2}
\ee
and indeed one predicts a production cross section somewhat less than
1 pb from a full calculation.
 
For a  detailed calculation of the production cross section a proper
treatment of the QCD potential is required to obtain a reliable
prediction for the bound state wave function at the origin.  The
structure of the 
NLO corrections for the production cross section, in particular of the
dominant terms, bears many similarities with the result for open
production and for the Drell Yan process (eq.~\ref{eq:DY}).  For gluon
fusion the partonic cross section is 
(in the $\overline{MS}$ scheme) given by \cite{Letter,Paper}
\be
\lefteqn{\hat \sigma^{gg}  =  {1\over s}{\pi^2\over 3}{R^2(0)\over M^3}
\alpha^2_{\overline{\rm MS}} (M^2)
\left\{ \rule{0mm}{5mm} \delta(1-z) \right.}
\nonumber
\\
&&\left. +{\alpha_{\overline{\rm MS}} (M^2) \over \pi}
\left\langle
\delta(1-z)\left( N_C\left(1+{\pi^2\over 12}\right)
C_F\left({\pi^2\over 4}-5\right)
-{4\over 3}T_f \ln(2) \right) +N_C F(z)\right\rangle \right\}
\nonumber\\
\label{eq:ttp9235}
\ee
where
\be
F(z)&=&\Theta(1-z)\left[
{11z^5+11z^4+13z^3+19z^2+6z-12\over 6z(1+z)^2}\right.
\nonumber\\
&&+4\left( {1\over z} +z(1-z) -2\right) \ln(1-z) +4\left(
{\ln(1-z)\over 1-z} \right)_+
\nonumber\\
&&\left.
+\left( {2(z^3-2z^2-3z-2)(z^3-z+2)z\ln(z)\over
(1+z)^3(1-z)}-3\right)
{1\over 1-z}\right]
\ee
and $z=M^2/{\hat s}$.  Both Born term and the virtual correction are
proportional to $\delta(1-z)$, the structure of the dominant term due
to gluon splitting $\sim \ln(1-z)/(1-z)$ is again
universal. Quark-gluon and quark-antiquark initiated subprocesses of
order $\alpha^3_s$ can be found in \cite{Letter,Paper}.  It may
be worth mentioning that the structure of QCD corrections to light
Higgs production \cite{lighthiggs} is nearly identical to
eq.~\ref{eq:ttp9235}. 
From Fig.~\ref{fig:prod} 
it is evident that $\eta_{b'}$ states with masses up to 1
TeV are produced at the LHC with sizeable rates.  The fairly clean
signature of the $Z+H$ decay mode might allow to discover these exotic
quarkonia and the Higgs boson at the same time.

\begin{figure}
\vspace*{105mm}
\caption{Production cross section for $\eta_{b'}$ including QCD
  corrections as function of the bound state mass $M$ for
  $\protect\sqrt{s}=16$ TeV and 40 TeV.}
\label{fig:prod} 
\end{figure}

%% file: part3.tex
\chapter{Top quarks in $e^+e^-$ annihilation}

A variety of reactions is conceivable for top quark production
at an electron positron collider.  Characteristic Feynman
diagrams are shown in
Fig.\ref{fig:3.1}.
$e^+ e^-$ annihilation through the virtual photon and
$Z$ (Fig.\ref{fig:3.1}a)
dominates and constitutes the reaction of interest for the currently
envisaged energy region.
\begin{figure}
\vspace{10cm}
\caption{Feynman diagrams for
 $t\bar{t}$ or $t\bar{b}$ production.}

\label{fig:3.1}
\end{figure}

In addition one may also consider \cite{fusi} a variety of
gauge boson fusion reactions (Fig.\ref{fig:3.1}b-d) that
are in close analogy to $\gamma \gamma$
fusion into hadrons at $e^+e^-$ machines of lower energy.
Specifically these are single top production,
\begin{equation}
e^+e^- \to \bar{\nu} e^- t \bar{b}
\label{eq:26}
\end{equation}
or its charge conjugate and top pair production through neutral or charged
gauge boson fusion
\begin{eqnarray}
e^+e^- & \to & e^+e^- t \bar{t} \nonumber\\
e^+e^- & \to & \bar{\nu} \nu t \bar{t}
\label{eq:27}
\end{eqnarray}

The experimental observation of these reactions would allow
to determine the coupling of top quarks to gauge bosons,
in particular also to longitudinal $W$ bosons and $Z$ bosons,
in the space-like region and eventually at large momentum transfers.
This would constitute a nontrivial test of the mechanism
of spontaneous symmetry breaking.

The various cross sections increase with energy in close analogy
to $\gamma \gamma$ reactions, and eventually even exceed $e^+e^-$
annihilation rates.  However, at energies accessible in the
foreseeable future these reactions are completely
negligible:  for an integrated luminosity of $10^{40}$ cm$^{-2}$,
at $E_{cm} = 500~GeV$ and for $m_t = 150~GeV$ one expects about
one $e^+e^ - t\bar{t}$ event (still dominated by
$\gamma \gamma$ fusion). At that same energy the cross sections for
$e^+ \nu \bar{t} b + c.c.$ and $\nu \bar{\nu} t \bar{t}$ final states
are still one to two orders of magnitude smaller.

Another interesting class of reactions is $e^+e^-$ annihilation
into heavy quarks in association with gauge or Higgs bosons:
\begin{eqnarray}
e^+ e^- & \to & t \bar{t} Z \\
e^+ e^- & \to & t \bar{b} W^- \\
e^+ e^- & \to & t \bar{t} H \\
e^+ e^- & \to & t \bar{b} H^-
\label{eq:31}
\end{eqnarray}

Two amplitudes contribute to the first reaction \cite{hagi}:
The $t \bar{t}$ system may be produced through a virtual Higgs
boson which by itself was radiated from a $Z$ (Fig.\ref{fig:3.2}).
The corresponding amplitude dominates the rate and
provides a direct measurement of the Yukawa coupling.
The radiation of longitudinal $Z$'s from the quark line
in principle also carries information on the symmetry
breaking mechanism of the theory.
\begin{figure}
\vspace{50mm}
\caption{\label{fig:3.2}
Amplitudes relevant for
$e^+e^- \to t\bar{t}Z$ and for $e^+e^- \to t\bar{t}H$.}
\end{figure}
The transverse part of the $t \bar{t} Z$ coupling,
i.e. the gauge part, can
be measured directly through the cross section or
various asymmetries in
$e^+e^- \to t \bar{t}$.  The longitudinal part, however,
could only be isolated
with $t \bar{t} Z$ final states.  For an integrated
luminosity of $10^{40}$ cm$^{-2}$
one expects only about 40 events (see sect. 3.1.7)
and it is therefore not
clear whether these can be filtered from the huge background
and eventually used for a detailed analysis.

Light Higgs bosons may be produced in conjunction with
$t \bar{t}$ \cite{djo}. They are radiated either from
the virtual $Z$ with an amplitude that is present
also for massless fermions or directly from heavy quarks as a consequence of
the large Yukawa coupling (Fig.\ref{fig:3.2}).
The latter dominates by far and may therefore be tested
specifically with heavy quark final states.  The predictions for the
rate will be discussed in sect.~3.1.7.  Depending on the mass of the
Higgs and the top 
quark, the reaction could perhaps be detected with an integrated luminosity
of $10^{40}cm^{-2}$.

Top quark production in $\gamma \gamma$ collisions is conceivable at a
``Compton collider''.  It requires special experimental provisions for the
conversion of electron beams into well-focused beams of energetic photons
through rescattering of laser light.  A detailed discussion can be
found in \cite{DESY}. 

Chapter 3 will be entirely devoted to $t\bar t $ 
production in $e^+e^-$ annihilation.  Section 3.1 will be 
concerned with the energy region far above threshold --- with electroweak 
aspects as well as with specific aspects of top hadronisation. The emphasis 
of section 3.2 will be on the threshold region which is governed by the 
interplay between bound state formation and the rapid top decay.

\section{Top production above threshold}
\subsection{Born predictions}
From the preceding discussion it is evident that the bulk of top studies
at an $e^+ e^-$ collider will rely on quarks produced in $e^+ e^-$ annihilation
through the virtual $\gamma$ and Z, with a production cross section of the
order of $\sigma _{point}$.  For quarks tagged at an angle $\vartheta $, the
differential cross section in Born approximation is a binomial in $\cos \vartheta $
\begin{equation}
\frac{d\sigma}{d \cos\vartheta} = \frac{3}{8} \left( 1 + \cos^2 \vartheta \right) \sigma _U + \frac{3}{4} \sin^2 \vartheta \sigma _L + \frac{3}{4} \cos \vartheta \sigma _F
\label{eq:32}
\end{equation}
$U$ and $L$ denote the contributions of unpolarized and longitudinally polarized
gauge bosons along the $\vartheta$ axis, and $F$ denotes the difference
between right and left polarizations.  The total cross section is the sum
of $U$ and $L$,
\begin{equation}
\sigma = \sigma _U + \sigma _L
\label{eq:33}
\end{equation}
the forward/backward asymmetry is given by the ratio
\begin{equation}
A^{FB} = \frac{3}{4} \frac{\sigma _F}{\sigma}
\label{eq:34}
\end{equation}

The $\sigma^i$ can be expressed in terms of the cross sections for the
massless case in Born approximation,
\begin{eqnarray}
\sigma ^U_B & = & \beta \sigma^{VV} + \beta ^3 \sigma ^{AA} \nonumber \\
\sigma ^L_B & = & \frac{1}{2} \left( 1 - \beta ^2 \right) \beta \sigma^{VV} \nonumber \\
\sigma ^F_B & = & \beta ^2 \sigma^{VA}
\label{eq:35}
\end{eqnarray}
with
\begin{eqnarray}
\sigma ^{VV} & = & \frac{4 \pi \alpha^2 (s) e^2_e e^2_Q}{s} \nonumber \\
& & + \frac{G_F (s)}{\sqrt{2}} e_e e_Q (\upsilon _e + \rho a_e )
\upsilon _Q \frac{m^2_Z \left( s-m^2_Z \right)}
{\left( s-m^2_Z \right)^2 + \left(\frac{s}{m_Z}
\Gamma _Z \right)^2} \nonumber \\
& & + \frac{G^2_F}{32 \pi} \left( \upsilon ^2_e
+ a^2_e + 2 \rho \upsilon _e a_e \right) \upsilon ^2_Q \frac{m^4_Z s}
{\left( s-m^2_Z \right)^2
+ \left(\frac{s}{m_Z} \Gamma _Z \right)^2} \nonumber \\
\sigma ^{AA} & = & \frac{G^2_F}{32 \pi}
\left( \upsilon ^2_e + a^2_e + 2 \rho \upsilon _e a_e \right)
a^2_Q \frac{m^4_Z s}{\left( s-m^2_Z \right)^2
+ \left(\frac{s}{m_Z} \Gamma _Z \right)^2} \nonumber \\
\sigma ^{VA} & = & \frac{G_F \alpha (s)}{\sqrt{2}} e_e
(\rho \upsilon _e + a_e )e_Q a_Q \frac{m^2_Z
\left( s-m^2_Z \right)}{\left( s-m^2_Z \right)^2
+ \left(\frac{s}{m_Z} \Gamma _Z \right)^2} \nonumber\\
& & + \frac{G^2_F}{16 \pi} \left( 2 \upsilon _e a_e
+ \rho (\upsilon ^2_e + a^2_e)\right) \upsilon_Q a_Q
\frac{m^4_Z s}{\left( s-m^2_Z\right)^2
+ \left(\frac{s}{m_Z}\Gamma _Z \right)^2}
\label{eq:36}
\end{eqnarray}

The fermion couplings are given by
\begin{eqnarray}
\upsilon_f = 2I^f_3 - 4e_f \sin ^2 \theta _w \quad , \qquad a_f = 2I^f_3
\label{eq:37}
\end{eqnarray}
and the possibility of longitudinal electron polarization
($\rho = -1;+1;0$ for
right\-handed; lefthanded; unpolarized electrons) has been included.
Alternatively one may replace $G_F m^2_Z$ by
\begin{equation}
G_F m^2_Z = \frac{\pi \alpha (s)}{\sqrt{2} \sin^2 \theta_W \cos^2 \theta _W}
\label{eq:38}
\end{equation}
With $\sin^2\theta_W$ ($\approx$0.23)
interpreted as $\sin^2 \theta_{eff}$ \cite{con}
this formula accommodates the leading logarithms from the running coupling
constant as well as the quadratic top mass terms in the threshold region.  

\subsection{Radiative corrections}

\underline{QCD corrections}
to this formula are available for arbitrary $m^2 / s$ up
to first order in $\alpha _s$:
\begin{eqnarray}
\sigma & = & \frac{(3 - \beta ^2)}{2} \beta \sigma^{VV} \left( 1 + \frac{4}{3} \frac{\alpha _s}{\pi} K_V \right) \nonumber \\
& & + \beta ^3 \sigma ^{AA} \left( 1 + \frac{4}{3} \frac{\alpha _s}{\pi} K_A \right)
\label{eq:40}
\end{eqnarray}
The exact result \cite{exac} for $K_{V,A}$ can be found in \cite{KueZer}.
These  QCD enhancement factors are
well approximated by \cite{appr}
\begin{eqnarray}
K_V & = & 1 + \frac{4}{3} \alpha_s
\left[ \frac{\pi}{2\beta} - \frac{3+\beta}{4}
\left( \frac{\pi}{2} - \frac{3}{4\pi} \right) \right] \nonumber \\
K_A & = & 1 + \frac{4}{3} \alpha_s
\left[ \frac{\pi}{2\beta} -
\left(\frac{19}{10} - \frac{22}{5}\beta +\frac{7}{2}\beta^2\right)
\left( \frac{\pi}{2} - \frac{3}{4\pi} \right) \right] \nonumber \\
\alpha_s & = & \frac{12\pi}{25 \log (4p^2_t / \Lambda ^2)}
\label{eq:43}
\end{eqnarray}
The next to leading order corrections 
to $K_V$ were calculated only recently \cite{HKT,CKS}.
The scale in $\alpha _s$ chosen
above was guessed on the basis of general
arguments \cite{guesk1} which were confirmed by the forementioned
complete calculations.
\par\noindent
For small $\beta$ these factors develop the familiar Couloumb
enhancement $\sim{2\pi\alpha_s\over 3\beta}$, compensating the
phase space $\sim \beta$.  This leads to a nonvanishing cross
section which smoothly joins the resonance region.  Details of
this transition will be treated in section 3.2.

To prepare this discussion, let us briefly study the limit of applicability 
of fixed order perturbation theory.  The leading terms in the perturbative 
expansion close to threshold are obtained from Sommerfeld's rescattering 
formula $\left(x\equiv {4\over 3}{\pi\alpha\over \beta}\right)$
\be
K_V^{\rm Som} = {x\over 1 - e^{-x}} &=& 1 +{x\over 2} + {B_1 x^2\over 2!} 
- {B_2 x^4\over 4!}+  {B_3 x^6\over 6!}\pm\ldots
\nonumber \\
&=& 1+{x\over 2} + { x^2\over 12} - {x^4\over 720} + {x^6\over 
5040}\pm\ldots
\ee
with $B_i$ being the Bernoulli numbers.  At first glance one might require 
$x\le 1$ for the perturbative expansion to be valid.  However, 
significantly larger values of $x$ are acceptable. The full  Sommerfeld 
factor $K_V^{\rm Som}$ is remarkably well approximated by the first three 
terms of the series for surprisingly large $x$ (only 6\% deviation for 
$x=4!$).  For top quarks this corresponds to $\beta\approx 0.13-0.14$ 
and hence down to about 3 GeV above the nominal threshold.  Upon closer 
inspection one also observes that the formula given in
eq.~(\ref{eq:43}) (a result  
of order $\alpha_s$) coincides numerically well with the correction factor 
$K_V^{\rm Som} \left(1-{16\over 3}{\alpha_s\over \pi}\right)$ which 
incorporates rescattering and hard gluon vertex corrections.  The results 
presented in these lectures are based on the Born predictions plus ${\cal 
O} (\alpha_s)$ corrections.

\begin{figure}[ht]
  \begin{center}
    \leavevmode
    \epsfxsize=12.cm
    \epsffile[65 220 410 450]{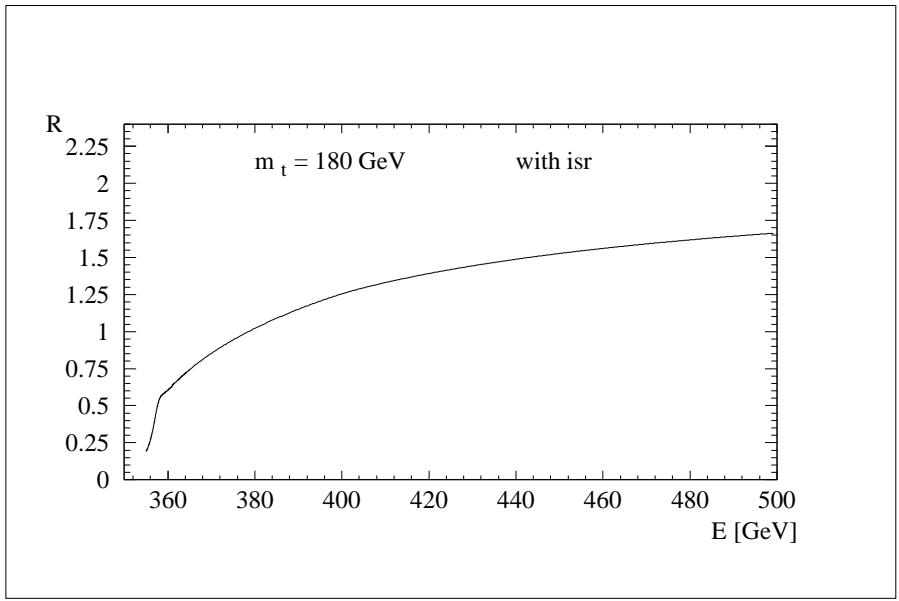}
    \hfill
\caption{Cross section for $t\bar{t}$ production,
including resonances, QCD corrections
and initial state radiation in units of $\sigma_{point}$.}
\label{fig:3.5}
  \end{center}
\end{figure}

\underline{Initial state radiation}
has an important influence on the magnitude
of the cross section.  $\sigma(s_{eff})$ is folded with
the Bonneau Martin structure function, supplemented by the
summation of large logarithms.  A convenient formula for
the non-singlet structure function in the leading logarithmic
approximation has been obtained in \cite{jez},
which is a natural extension of a formula
proposed in \cite{jad}.
This leads to a significant suppression by about a factor
\begin{equation}
\left( \frac{\delta W}{m_t} \right) ^{ {2\alpha\over \pi}\left(
\ln{m_t^2\over m_e^2}-1\right)}
 \approx 0.5-0.6
\label{eq:47}
\end{equation}
with $\delta W = 1-5~GeV$ in the resonance
and threshold region. The correction factor increases rapidly
with energy, but stays below 0.9 in
the full range under consideration (Fig.\ref{fig:3.5}).

\underline{Electroweak corrections}
to the production cross section
in the continuum have been studied in \cite{been1}.  Apart from
a small region close to threshold they are negative.
Relative to the $G_F$ parametrized Born approximation they
decrease the cross section by -6.3\% to -9.3\%, if $m_t$ is
varied between 100 and 200~GeV, $m_H$ between 42 and 1000~GeV,
and $E_{cm}$ fixed at 500~GeV.  QCD and electroweak
corrections are thus of equal importance (Fig.~\ref{fig:3.6}).

Close to threshold and for relatively small Higgs boson masses a rapid 
increase of these corretions is observed (Fig.~\ref{fig:3.7}) which
can be attributed  
to the attractive Yukawa potential induced by light Higgs boson exchange.  
Several GeV above threshold, and for $m_H$ around or below 100 GeV it is 
more appropriate to split these corrections into hard and soft exchange and 
incorporate the latter in an instantaneous Yukawa potential \cite{JKsim}.

\begin{figure}
\vspace{80mm}
\caption[]{Genuine electroweak
corrections to top production in $e^+e^-$ annihilation.
From \cite{been1}.}
\label{fig:3.6}
\end{figure}
\begin{figure}
\vspace{80mm}
\caption{Relative size (in percent) of electroweak corrections in the
threshold region for $m_t=200$ GeV and different Higgs masses (from
\protect\cite{BH}).}
\label{fig:3.7}
\end{figure}

\subsubsection{Longitudinal polarization}

\begin{figure}[ht]
  \begin{center}
    \leavevmode
    \epsfxsize=12.cm
    \epsffile[85 280 510 565]{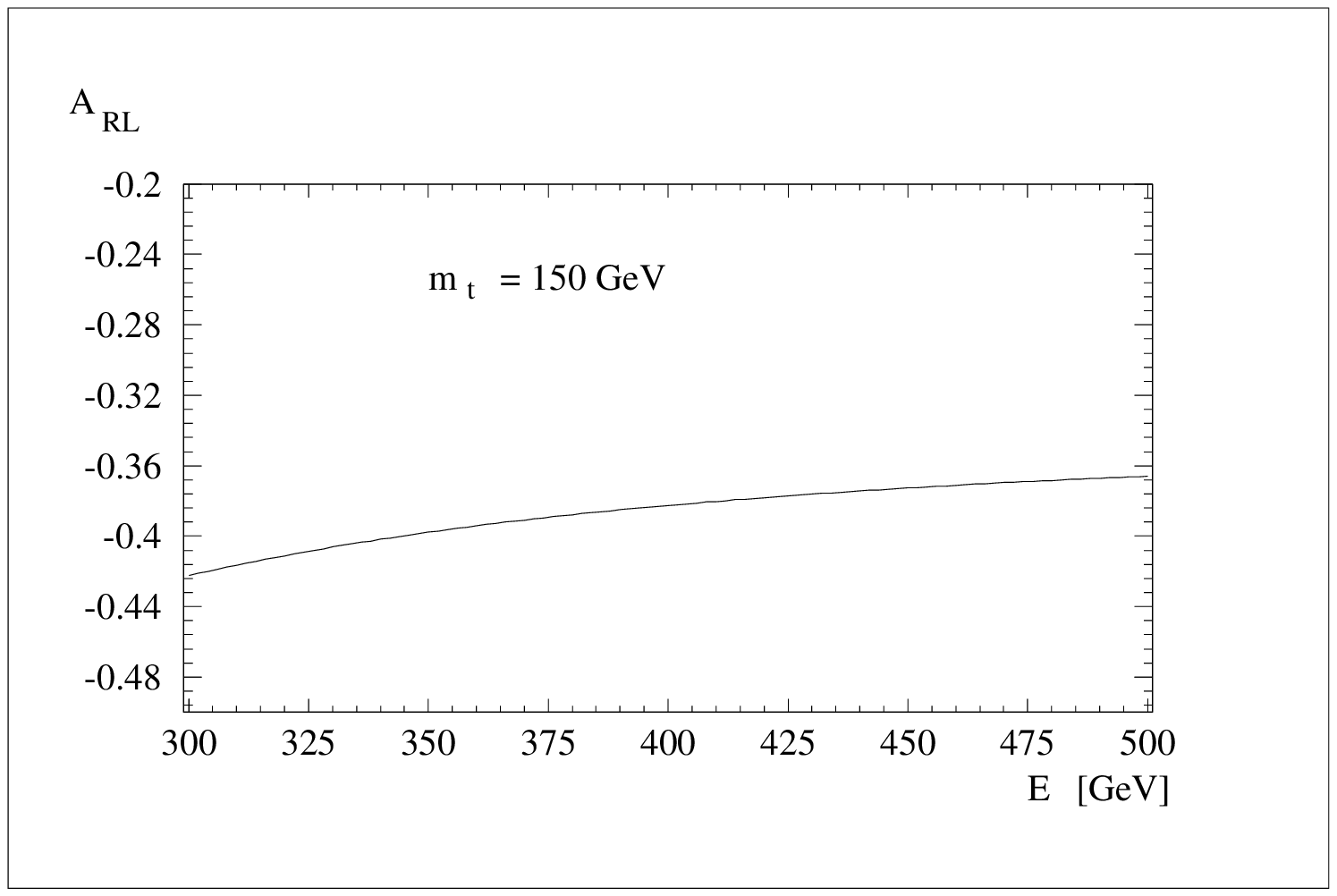}
    \hfill
\caption{Right/left asymmetry as function
of $E_{cm}$ for $m_t = 150~GeV$.}
\label{fig:RL}
  \end{center}
\end{figure}

It should be mentioned that linear colliders might well operate to
a large extent with polarized (electron) beams.  The cross section for
this case can be derived from (\ref{eq:36}).  For top quarks the resulting
right/left asymmetry
\begin{equation}
A_{LR} =  (\sigma _L - \sigma _R) / (\sigma _L + \sigma _R )
\label{eq:48}
\end{equation}
is sizable (Fig.\ref{fig:RL}) and amounts to about $-\,0.4$,
reducing the production cross section with righthanded electrons.
However, selection of righthanded electron beams decreases the $W$ pair
cross section even stronger, thereby enhancing the top quark
signal even before cuts are applied.
Electroweak corrections to $A_{LR}$ in the threshold region have been
calculated in \cite{GKKS}.

\subsection{Top quark fragmentation}

The experimental analysis of charm and bottom fragmentation functions has
clearly demonstrated that heavy quark fragmentation is hard in
contrast to the fragmentation of light
quarks. This is a consequence of the inertia of heavy particles, the momentum
of which is not altered much if a light quark $\bar q$ is attached to
the heavy quark $Q$ in the fragmentation process
to form a bound state $(Q\bar q)$, see e.g.\ \cite{fr1}. At the same
time soft infrared gluon radiation is damped if the color source is
heavy.

For $m_t \ge 100$ GeV the strong fragmentation process and
the weak decay mechanism are intimately intertwined \cite{fr3}. The
lifetime 
$\tau_* < \Lambda^{-1}$ becomes so short that the mesonic $(t\bar q)$
and baryonic $(tqq)$ bound states cannot be built--up anymore.
Depending on the initial top quark energy, even remnants of the $t$
quark jet may not form anymore \cite{bigi2}. Hadrons can be created in
the string stretched between the $t$ and the $\bar t$ only if the
quarks are separated by about 1 fermi before they decay. If the flight
path $\gamma \tau_*$ is less than $1/2$ fm,
the length of  the $t-\bar t$ string is too short to form
hadrons and jets cannot develop anymore along the flight direction
of the top quarks.  For $m_t\approx 180$ GeV  top quark energies above
1 TeV are required to allow nonperturbative strings between $t$ and
$\bar t$.  
``Early'' nonperturbative production of particles from the string between 
$t$ and $\bar t$ is thus absent for all realistic experimental 
configurations.  ``Late'' production from the $b$ and $\bar b $ jets 
produced in top decays dominates.  
\begin{figure}
\vspace{155mm}
\caption{Rapidity distributions of ``early'' particles (full) and
``late'' ones (dashed) for three different top masses: 90 GeV in a)
100 GeV in b) and 120 GeV in c).  (From \protect\cite{fr3}.)}
\label{fig:SjZ}
\end{figure}
This is illustrated in Fig.~\ref{fig:SjZ}.  Early 
production dominates for $m_t = 90$ GeV, late production for $m_t=120$ GeV 
and {\it a forteriori} for the actual value around 180 GeV.

The perturbative radiation of soft gluons, too, is interrupted by the
$t$ quark decay \cite{fr2}.
The angular distribution $(\Theta)$ and the
energy distribution $(\omega)$ of the radiated gluons is
approximately given by
\begin{equation}
dP_g = \frac{4\alpha_s}{3\pi} \frac{\Theta^2 d\Theta^2}{\left[
\Theta^2 + \frac{1}{\gamma^2} \right]^2 + \left[ \frac{\Gamma}{\gamma
\omega} \right]^2 } \frac{d\omega}{\omega}
\end{equation}
\noindent for a short--lived radiation source accelerated to $\gamma =
E_t/m_t$. The gluons accumulate on the surface of a cone with
half--aperture $\Theta_c \sim \gamma^{-1}$ for a long--lived $t$, but
$\sim \gamma^{-1} \sqrt{\gamma\Gamma/\omega}$ if the particle decays
quickly. The energy spectrum rises from zero to a maximum at $\omega
\sim \gamma\Gamma$ before falling off $\sim 1/\omega$ for large $\omega$,
if the width is greater than the confinement scale $\Lambda$.

The impact of the finite width on the angular distribution of gluon 
radiation will be visible if $\omega \le \Gamma_t E_t/m_t$. For a linear 
collider with c.m.~energy of 2 TeV gluon jets with energies of 10 GeV and 
below would be affected \cite{KOS}.  The radiation pattern  is shown in
\begin{figure}
\vspace{10cm}
\caption{Sensitivity of the soft gluon distribution to the top width
for back-to-backk $t$ and $b$: $dN/d\cos(\theta_g)$ for $M$ = 140 GeV,
$\theta_b=180^\circ$, and $\Gamma$ = 0, 0.7, 5, 20 GeV, and $\infty$
as marked. $W$ = 1000 GeV. (From \protect\cite{KOS}.)}
\label{fig:KhozeOrrSt}
\end{figure}
Fig.~\ref{fig:KhozeOrrSt}  for $m_t = 140$  GeV and  
$\sqrt{s} = 600 $ GeV, with $\Gamma_t$ tuned to different values in order 
to demonstrate the sensitivity of such a measurement.  The impact of the 
conversion of gluons to hadrons has been ignored in this study.  The 
picture is, further, complicated by the interference between radiation from 
top production and decay --- a phenomen characteristic for unstable 
particles. These phenomena allow to probe the time evolution of 
hadronisation in a unique manner.  Their understanding is a necessary 
prerequisite for any top mass measurement through a kinematic analysis of 
$b$-jet-$W$ final states.

\subsection{Static $t$ parameters}
Because of the large $t$ mass, deviations from the Standard
Model may manifest themselves in the top quark sector first.
Examples in which the large mass is crucial are provided by
multi--Higgs doublet models, models of dynamical symmetry breaking and
compositeness. These effects can globally be described by form factors
parametrizing the electroweak $t\bar t$ production current $(a =
\gamma,Z)$ and the weak $(t,b)$ decay
current $(a=-)$ \cite{st1,BerOve},
\begin{equation}
j_\mu^a \sim F_{1L}^a \gamma_\mu P_L + F_{1R}^a \gamma_\mu P_R +
\frac{i\sigma_{\mu\nu} q_\nu}{2m_t} [ F_{2L}^a P_L + F_{2R}^a P_R]
\end{equation}
\noindent [$P_{L,R}$ project on the
left and right chirality components of the wave
functions.] In the \sm, $F_{1L}^- = 1$ while all other $F^-_i$ vanish;
$F_{1L}^\gamma = F_{1R}^\gamma = 1$ and $F_{2L}^\gamma = F_{2R}^\gamma
= 0$, analogously for the $Z$ current. ${\cal{CP}}$ invariance
requires $F_{2L}^{\gamma,Z} = F_{2R}^{\gamma,Z}$ in the $t\bar t$
production current, and equal phases for $F_{1L}^-$ and $F_{2R}^-$
{\it etc.}\ in the decay current. The static values of the form factors
$F_2^{\gamma,Z}$ are the anomalous magnetic and electric dipole moments
of the top quark.

The form factors are determined experimentally by measuring the
angular distribution of the $t\bar t$ decay products, $e^+e^- \ra
t\bar t, t \ra b W^+, W^+ \ra f \bar f' $ {\it etc.}
This requires the top quark to be treated as a free particle,
the polarization of which not being affected by non--perturbative
hadronic binding effects. This assumption is justified by the short
lifetime of the top quark as discussed earlier. Details of the general
helicity analysis can be found in the literature \cite{st1,st2}.

\subsection{Normal polarization of the top quarks}
A non--zero component of the $t$ polarization vector that is
normal to the production plane can be generated only by the
interference between complex helicity flip and non--flip
amplitudes. Such relative phases can arise from ${\cal{CP}}$ violation
but also from higher order loop corrections due to gluon exchange in
the final state \cite{kuehn2,st4,st1}
or electroweak corrections involving
Higgs and gauge bosons \cite{st1}. The QCD induced normal polarization
is generally less than 5\%, the electroweak normal polarization is
smaller still.
[By contrast, longitudinal and transverse polarization components within the
$t\bar{t}$ production plane are generated already at the tree level of
the electroweak interactions and they are large in general; see
\cite{kuehn2} for the discussion of details.]

\subsection{Angular correlations of $t\bar t$ decay products}
As stated in the previous chapter, top quarks are produced through the
virtual photon and $Z$. In the threshold region they are polarized to a
degree
\begin{equation}
P_t= A_{RL}\approx - 0.4
\end{equation}
Assuming for the distribution of leptons from the decay of polarized
top quarks
\begin{equation}
\frac{dN}{dx\,d\cos\theta}=f(x)+ g(x) P_t\cos\theta
\end{equation}
(with $g(x)=f(x)$ in the \sm, see eq.\ref{eq:15})
the angular distribution 
allows to test for the chirality of the $tb$
current. Implicitly it was assumed that
hadronization does not affect
the top spin degrees of freedom \cite{acta,jk41}. This assumption can
be tested independently through the study of correlations between $t$ and
$\bar t$ decay products. In the threshold region the spins are
correlated
$\propto (1+\frac{1}{3} \vec{s_+}\cdot\vec{s_-})$. This leads to the
following correlated $\ell_+\ell_-$ distribution:
\begin{equation}
\frac{dN}{dx_+dx_-d\cos\theta_{+-}}= f_+(x_+)f_-(x_-) +
                      \frac{1}{3} g_+(x_+)g_-(x_-)\cos\theta_{+-}
\end{equation}
where $f_+=f_-$ and $g_+= -g_-$. $\theta_{+-}$ denotes the angle between
$\ell_+$ and $\ell_-$. After averaging the lepton energies,
\begin{equation}
\frac{dN}{d\cos\theta_{+-}}=1+\frac{1}{3} h_+ h_- \cos\theta_{+-}
\end{equation}
Note that the coefficient of the correlation term is $-h^2_+/3$ and
hence always negative (assuming CP conservation). Since
$|h_+|\leq1$ it ranges between 0 and $-1/3$. This limiting value is
assumed in the \sm. A detailed discussion with illustrative examples is
given in \cite{PIK}.

\subsection{Testing the Yukawa Coupling}

With its relatively large Yukawa coupling $g_Y=\sqrt{2} m_t/v\approx
1$  
the top quark 
is uniquely suited to test one of the basic ingredients of the Standard 
Model, the coupling between top quarks and the Higgs boson.  The 
verification of this crucial prediction would confirm the mechanism for the 
generation of fermion masses and hence complete the measurement and 
analysis of basic couplings.  Alternatively, any deviation would provide 
unambiguous proof for new physics.

Different strategies are at hand at an \epem collider which are closely 
tied to the cms energy available and to the mass of the Higgs boson.  For 
relatively light bosons a variety of possibilities appear to be promising: 
vertex corrections affect the cross section for $t\bar t$ production in the 
threshold region.  For a collider in its early stage with an energy around 
500 GeV it may well be the only option available and will be discussed more 
thoroughly in section 3.2.  For higher energies, say around  1 TeV, a 
promising choice is the radiation of Higgs 
bosons from $t\bar t$ (see Fig.~\ref{fig:3.2}) 
with a cross section around 1 fb 
\begin{figure}[htb]
\psfig{figure=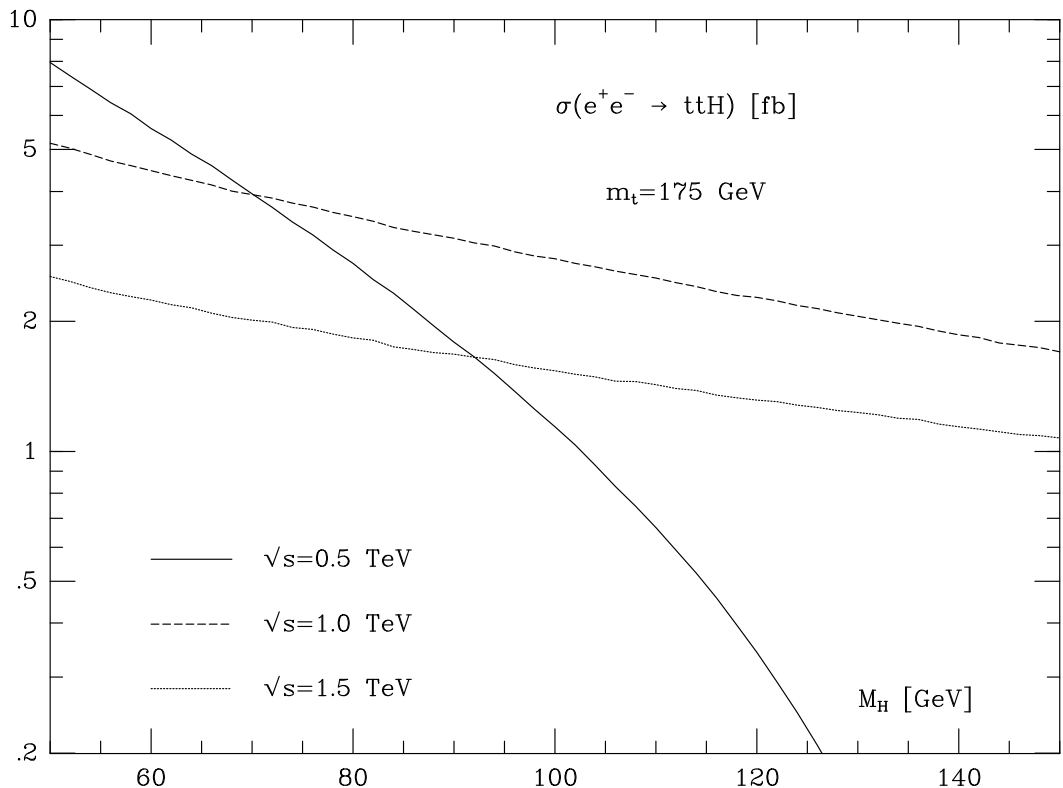,angle=-90,width=60mm,bbllx=60pt,bblly=60pt,%
bburx=510pt,bbury=340pt}
\caption{The cross section $\sigma(e^+e^- \to t\bar t H)$ (from
\protect\cite{djo}).} 
\label{fig:sigma}
\end{figure}
(Fig.~\ref{fig:sigma}) \cite{djo}. Alternatively one 
may analyse the top quark final states in conjunction with a $Z$ boson.  
This reaction receives important contributions from the left one of
the diagrams depicted in 
Fig.~\ref{fig:3.2}
if the Higgs mass happens to be relatively close to 2$m_t$
\cite{hagi}.
  Again, for 
simple kinematical reasons, high energies are crucial for the reaction to 
be accessible (Fig.~\ref{fig:Hagi}). 
\begin{figure}
\vspace{15cm}
\caption{The Higgs mass ($m_H$) dependences of the total cross
sections of $e^+e^- \to t\bar t Z$ for various top quark masses $m_t$.
The c.m.~energy $\protect\sqrt{s}$ is set to be 500 GeV (a) and 1 TeV
(b). (From \protect\cite{hagi}.)}
\label{fig:Hagi}
\end{figure}

\section{Threshold behaviour}
The previous section dealt with top quark production sufficiently far
above threshold for the reaction to be well described by the Born
cross section, modified slightly by QCD and electroweak corrections.
This is in contrast to the situation in the threshold region, where
QCD plays an important role and controls the cross section.  Strong forces
modify the Born prediction. They compensate the phase space suppression and
enhance the production rate significantly, leading to a step function like
behaviour at threshold.  The large top decay rate also plays an important
role. Quarkonium resonances cease to exist and merge into a structureless
excitation curve which joins smoothly with the continuum prediction above
the nominal threshold.

This sharply rising cross section allows to study top quarks in a
particularly clean environment and with large rates. The following physics
questions can be addressed:
\begin{itemize}
\item The QCD potential can be scrutinized at short distances, with the
non perturbative tail cut off by the top decay. As a result
$\Lambda_{QCD}$ or $\alpha_s$ could be determined accurately.
\item The top quark mass can be measured with a precision which is only
limited by the theoretical understanding of the excitation curve, but in
any case better than 500 MeV.
\item Top quarks are strongly polarized (about 40\%) even for unpolarized
beams; and longitudinal beam polarisation will enhance this value even
further. Detailed studies of top decays, in particular of the $V-A$ structure
of the $tbW$ coupling are therefore feasible.
\item The interquark potential is --- slightly --- modified by the Yukawa
potential induced by Higgs exchange.  The excitation curve and the top
quark momentum distribution may therefore lead to an indirect measurement
of the Yukawa coupling.
\item The large number of top quarks in combination with the constrained
kinematics at threshold, could facilitate the search for new decay modes
expected in extensions of the SM.
 \end{itemize}
With this motivation in mind the following points will be discussed: After
a brief review of qualitative features of threshold production (section
3.2.1), the present status of our theoretical understanding of the total
cross section will be presented in section 3.2.2.

The momentum distribution of top quarks and their decay products offers an
alternative and complementary route to probe the interquark potential, as
shown in section 3.2.3.  Spin effects and angular distributions are
sensitive towards the small $P$-wave contribution induced by the axial part
of the neutral current.  The theoretical framework and the resulting
predictions are collected in section 3.2.4.  Rescattering, relativistic
corrections and other terms of order $\alpha_s^2$ will be touched upon in
section 3.2.5.

\subsection{Introductory remarks}

For a qualitative understanding it is illustrative to compare the different 
scales which govern top production close to threshold.  The quarks are 
produced at a scale comparable to their Compton wave length
\be
d_{\rm prod} \sim 1/m_t
\ee
Electroweak vertex corrections do not alter this behaviour significantly, 
since $Z$- or $W$-boson exchange proceeds at a distance $\approx 1/m_Z$ 
which is still short compared to scales characteristic for the bound state 
dynamics. For the QCD potential 
\be
V_{\rm QCD} = -{4\over 3}{\alpha_s\over r}
\ee
one anticipates an effective coupling constant ${4\over 3}\alpha_s \approx 
0.2$, if $\alpha_s$ is evaluated at the scale of the Bohr momentum
\be
k_B \approx    {4\over 3}\alpha_s {m_t\over 2} \approx 20 GeV
\ee
The resulting Bohr radius 
\be
r_{\rm Bohr}= 1/k_B
\ee
is small compared to hadronic scales. The binding energy of the 1S level 
\be
E_B =  \left( {4\over 3}\alpha_s\right)^2 {m_t\over 4}\approx 2 
GeV
\ee
and, quite generally, the separation between different resonances, is 
smaller than the decay rate 
\be
2\Gamma_t \approx 3 GeV
\ee
whence all resonances will merge and join smoothly with the continuum.  

The coupling strength $\kappa$ of the Yukawa potential 
\be
V_Y = -{\kappa\over r}e^{-m_Hr}\quad {\rm with } \quad 
\kappa = \sqrt{2}G_F {m_t^2\over 4\pi} = 0.042
\ee
is comparable to the QCD coupling $4/3\alpha_s=0.2$ in magnitude. The 
exponential damping, however, with a cutoff $1/m_H \ll r_{\rm Bohr}$ and a 
lower limit $m_H > 65 $ GeV, reduces the impact of the Yukawa potential 
quite drastically.  (The  situation may be different in multi-Higgs-models: 
the couplings could be enhanced and, even more important, the Higgs might be 
lighter!)  Furthermore, the nonrelativistic treatment is no longer adequate 
and retardation effects must be taken into consideration.

The large top quark width plays a crucial role for the threshold behaviour, 
which is best understood in the framework of (nonrelativistic) Green's 
function techniques. The production of $t\bar t$ from a pointlike source 
(actually of extension $1/m_t$) at $x'$ with frequency $\omega = E$ is 
characterized by the time dependent Green's function  $G(\vec r,\vec r',t)$ 
which is a solution of the time dependent Schr\"odinger equation with a 
pointlike source term
\be
(i\partial_t-H) G(\vec r,\vec r',t) =  \delta(\vec r - \vec r') e^{-iEt}
\ee
with
\be
H={p^2\over 2m} + V(\vec r)
\ee
In the problem at hand $m=m_t/2$ is the reduced mass, $\vec r$ the relative 
distance between $t$ and $\bar t$, and the width $\Gamma=2\Gamma_t$.  The 
location of the source is at the origin
 $\vec r'=0$ by convention, and the second argument 
of $G$  will be suppressed, $G(\vec r,t)\equiv G(\vec r,0,t)$.  For a 
qualitative discussion one may ignore the potential and obtains for a
stable quark
\be
G(\vec r,t)&=& -{m\over 2\pi r} e^{i\sqrt{2mE}r} e^{-iEt}
\nonumber\\
\widetilde G(\vec p,t) &=& {1\over E-{p^2\over 2m}} e^{-iEt}
\label{eq:Green}
\ee
The corresponding current is flowing in radial direction from the source
\be
\vec j = {i\over 2m} \left( G\vec\nabla G^* -G^*\vec\nabla G \right)
=\vec e_r {m^2v\over 4\pi^2r^2}
\ee
with a constant flux through a sphere around the origin, reflecting the 
conservation of probability.

The width $\Gamma = 2\Gamma_t$ is introduced in the Schr\"odinger equation 
through the replacement
\be
H\to H-i\Gamma/2
\ee
and, consequently, through the substitution $E\to E+i\Gamma/2$ in 
eq.~(\ref{eq:Green}). The exponential damping of the flux in radial 
direction
\be
\vec j= \vec e_r {m^2 v\over 4 \pi^2 r^2} e^{i(\sqrt{2m(E+i\Gamma/2)}
-\sqrt{2m(E-i\Gamma/2)})r}
\ee
is most easily interpreted in two limiting cases.  For $E\ll \Gamma$ the 
decrease 
\be
\vec j\sim  {\vec e_r\over r^2} e^{-\sqrt{2m\Gamma}r}
\ee
is solely driven by the large width, with a cutoff $x_c^{-1} = 
\sqrt{2m\Gamma} \approx 2.4$ GeV for realistic parameters.  For $E\gg 
\Gamma$, on the other hand, the current decreases like
\be
\vec j\sim  {\vec e_r\over r^2} e^{-\Gamma r/v}
\ee
In this latter case top quarks may travel appreciable distances, up to 
$x_{\rm cut} \approx v/\Gamma$. However, for realisitic beam energies
they hardly propagate beyond the perturbative region.


\subsubsection{Predictions for  the Coulomb potential}
The large top decay rate restricts the range of sensitivity to the short 
distance part of the potential
\be
\widetilde V(q) = -{16\pi\over 3} {\alpha_V(q)\over q^2}
\ee
which is approximately Coulombic, with a logarithmic variation of 
$\alpha_V$. Most of the qualitative features of top quark threshold physics 
can be understood even on the basis of the results for constant $\alpha_V$, 
which are available in analytical form. 

A remarkable feature of heavy quark production is the sharp rise of the 
cross section at threshold, a consequence of the attractive Coulomb force.  
The step function joins smoothly with the smeared resonances. 

Let us try to quantify this aspect with the help of simple nonrelativistic 
quantum mechanics ($\alpha={4\over 3}\alpha_s$, $m=m_t/2$).  The narrow 
resonances below the nominal threshold ($E$=0) are located at $E_n = 
-E_{Ryd}/n^2$ with $E_{Ryd}=k_B^2/(2m)=\alpha^2m/2$;

\begin{figure}[htb]
\psfig{figure=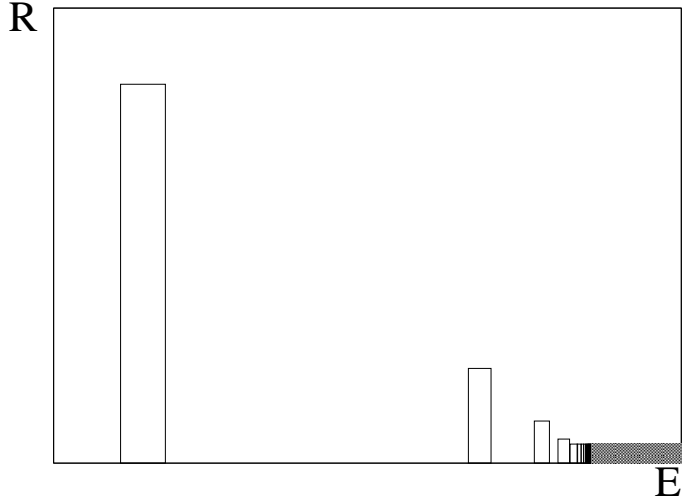,width=120mm,bbllx=0pt,bblly=270pt,%
bburx=450pt,bbury=550pt}
    \caption{Schematic representation of resonances (wiht the area of
the boxes adjusted to represent the weights of the respective delta
functions) and the continuum cross section.}
\label{fig:S18}
\end{figure}

The production amplitude from a pointlike source with frequency $\omega=E$ 
is proportional to
\be
\langle \vec r=0 | \psi_n\rangle = \psi_n(0),
\ee
the rate correspondingly to
\be
R\sim |\psi_n(0)|^2\delta(E-E_n)
\ee
The wave function at the origin decreases with the third power of the 
radial quantum number,
\be
 |\psi_n(0)|^2 \sim 1/n^3
\ee
their spacing becomes increasingly dense
\be
\Delta E = E_{n+1}-E_n = E_{Ryd} \left({1\over n^2} -  {1\over 
(n+1)^2} \right)\sim E_{Ryd} {2\over n^3}
\ee
such that their average contribution to the cross section approaches a 
constant value (Fig.~\ref{fig:S18}).
  Above threshold one has to project the state $\langle \vec 
r=0|$ onto the Coulomb wave functions in the continuum $\psi_E(\vec r)$.
These replace the plane waves which are appropriate for the case where
final state interaction is absent.

The production amplitude is thus governed by
\be
\langle\vec r=0| \psi_E\rangle = \psi_E(0)
\ee
and the rate $\sim |\psi_E(0)|^2$. The threshold phase space factor $v$ 
is thus compensated by the $1/v$ singularity in $ |\psi_E(0)|^2$, and the 
cross section approaches a constant value for $E\to 0$. 

The explicit calculation yields 
\be 
R(E) = {9\pi\over 2\alpha^2} \Gamma_e \delta(E-E_n) = 3Q_t^2{3\over 2} 
\sum_n {k_B^2\over m^2} {1\over n^3} \pi \delta\left(E-E_{Ryd}{1\over 
n^2}\right)
\label{eq:RE}
\ee
   for energies below threshold, and
\be
R(E)=3Q^2_t {3\over 2} \beta {x\over 1-e^{-x}} 
\label{eq:347}
\ee
with $x=k_B/k = \pi\alpha/\beta$
for energies above threshold.\footnote{For a textbook discussion of Coulomb 
scattering states and a derivation of this result see 
e.g.~\cite{Landau,Messiah}}

The perturbative expansion breaks down in the limit $\beta\to 0$.  The 
first term of this formal series
\be
\beta   {x\over 1-e^{-x}}  = \beta \left( {x\over 2}-\ldots\right)
\ee
underestimates the exact result by a factor two.

Eq.~(\ref{eq:RE}) allows to connect the formalism based on narrow individual 
resonances with a formulation which is tailored to the situation at hand, 
namely wide overlapping resonances which merge into a smooth continuum.

Instead of summing 
the contributions from a large number of high radial excitations
one may directly calculate the imaginary part of the Greens
function for complex energy
\begin{eqnarray}
\sigma (e^+e^- \to t\bar{t})
      = \frac{24\pi^2\alpha^2}{s} \frac{\rho_v(s)}{m^2_t}
       \left( 1 - \frac{16}{3} \frac{\alpha_s}{\pi} \right)
       \sum_{n} \left| \psi_n (0) \right|^2
       \frac{\Gamma_t}{(E_n - E)^2 + \Gamma^2_t} \nonumber\\
      = \frac{24\pi^2\alpha^2}{s} \frac{\rho_v(s)}{m^2_t}
       \left( 1 - \frac{16}{3} \frac{\alpha_s}{\pi} \right)
       \sum_{n} Im \frac{\psi^{\*}_n (0) \psi^*_n (0)}
       {E_n - E - i \Gamma_t} \nonumber\\
      = -\frac{24\pi^2\alpha^2}{s} \frac{\rho_v(s)}{m^2_t}
       \left( 1 - \frac{16}{3} \frac{\alpha_s}{\pi} \right)
       Im G(0,0,E+i\Gamma)
\label{eq:60}
\end{eqnarray}
The factor  $\rho_v(s)$  incorporates the contributions from the
intermediate photon and $Z$ and is given by 
\begin{eqnarray}
\rho^{Born}_v(s) = \left| e_t e_e + \frac{1}{y^2}
                \frac{\upsilon_t \upsilon_e M^2_\theta}
                     {s - M^2_Z + iM_Z \Gamma_Z} \right|^2
              + \left| \frac{1}{y^2}
                \frac{\upsilon_t a_e M^2_G}
                     {s - M^2_Z + iM_Z \Gamma_Z} \right|
\label{eq:58}
\end{eqnarray}
\[ \upsilon_f = 2I_{3f} - 4e_f \qquad
   a_f = 2I_{3f} \qquad
   y = 16 \sin^2 \theta_W \cos^2\theta_W \]
($\alpha=\alpha_{\rm eff}=1/128$ has been adopted in the numerical
   evaluation. Radiative corrections to this formula have been discussed in
\cite{GKKS}.)

The problem can be solved in closed analytical form for an exact
Coulomb potential \cite{fk} 
\be
{\rm Im}G_{E+i\Gamma_t} (0,0) &=& -{m_t^2\over 4\pi}
\left[ {k_2\over m_t} + {2k_B\over m_t}\arctan {k_2\over k_1}
\right.
\nonumber \\
&&\left. 
 +\sum_{n=1}^\infty {2k_B^2 \over m_t^2\pi^4} 
{\Gamma_t k_B n + k_2 \left(n^2
\sqrt{E^2+\Gamma_t^2}+{k_B^2\over m_t}\right)
\over
\left( E+{k_B^2\over m_tn^2}\right)^2+\Gamma_t^2}\right]
\nonumber\\
k_{1,2} &=& \left[
m_t\left( \sqrt{E^2+\Gamma_t^2}\mp E\right)/2\right]^{1/2},
\nonumber\\
k_B&=& {2\over 3} \alpha_s m_t.
\ee

To arrive at a realistic prediction of the total (and, in sect 3.2.3
of the differential) cross section the Coulomb potential must be
replaced by the realistic QCD potential.

\subsection{The QCD potential}

On the basis of earlier conceptual work in \cite{apple,sus} the
asymptotic behaviour of the static potential has been derived in
\cite{fis,bil}.  In momentum space the potential reads in the
$\overline{MS}$ subtraction scheme
\begin{equation}
V\left( Q^2 , \alpha _{\overline{MS}} (Q^2)\right) = - \frac{16\pi}{3}
 \frac{\alpha _{\overline{MS}} (Q^2)}{Q^2}
  \left[ 1 + \left( \frac{31}{3} -
  \frac{10}{9} n_f \right)
   \frac{\alpha _{\overline{MS}}(Q^2)}{4\pi} \right]
\label{eq:49}
\end{equation}
The renormalization scale $\mu ^2$ has been chosen as $Q^2$, and $n_f$
refers to the number of massless quarks.  Employing standard arguments based
on the renormalization group, the $Q^2$ expansion of $\alpha _{\overline{MS}}(Q^2)$
is given by
\begin{eqnarray}
\frac{\alpha_{\overline{MS}}(Q^2)}{4\pi } &=&
\frac{1}{b_0 \log
 \left( Q^2/\Lambda^2_{\overline{MS}}
 \right)}
 \left[    1 - \frac{b_1}{b^2_0}
          \frac   {\log\log
          \left(  Q^2 / \Lambda^2_{\overline{MS}} \right) }
                  {\log\left( Q^2 / \Lambda^2_{\overline{MS}}
          \right) }
 \right]\quad
\label{eq:50}\\
b_0 &=& 11 - \frac{2}{3}n_f, \qquad
b_1 = 102 - \frac{38}{3}n_f
\nonumber
\end{eqnarray}

The leading behaviour of the potential at small distances
$[\gamma_E=0.5772\cdots]$
\begin{eqnarray}
V(r) & = & \frac{16\pi}{(33-2n_f)r \log 1 / (\Lambda _{\overline{MS}}r)^2}
\nonumber \\
& & \left[ 1 - \frac{b_1}{b^2_0}
               \frac{\log\log 1 / (\Lambda _{\overline{MS}}r)^2}
               {\log 1 / (\Lambda _{\overline{MS}}r)^2}
             + \frac{\left( \frac{31}{3} - \frac{10}{9} n_f \right)
               / b_0 + 2 \gamma_E }
               {\log 1 / (\Lambda _{\overline{MS}}r)^2 } + \cdots \right]
\label{eq:51}
\end{eqnarray}
is thus directly given in terms of the QCD scale parameter $\Lambda$.
The exploration of $V(r)$ for small distances could thus lead to a
direct determination of $\Lambda$.  For quark masses above 50-100~GeV the
ground state properties become independent of the potential in the
nonperturbative region.  As discussed in the previous section the
large decay rate acts as a cutoff and the predictions are fairly
insensitive to the actual regularisation.  However, an additional
constant which can be traded against a shift in $m_t$ must be
carefully calibrated.

In practice one connects the theoretically predicted short distance part
smoothly with the empirically determined potential above $\sim$0.1~fermi.
The asymptotic form given in (\ref{eq:49})
is based on the assumption that $n_f$ species of light quarks, taken
as massless, contribute to the vacuum polarization, and heavier ones
are ignored.  The value of $\Lambda_{\overline{MS}}$ in (\ref{eq:50})
must be properly related to $\Lambda_{\overline{MS}}$ as determined
from other experiments with a different number of effective light
flavors \cite{bern,marc}.  For the momentum range of around 15~GeV explored
by the $t\bar{t}$ system $n_f=$5 seems adequate. 

\begin{figure}[htb]
  \begin{center}
    \leavevmode
    \epsfxsize=60mm
    \epsffile[130 600 390 610]{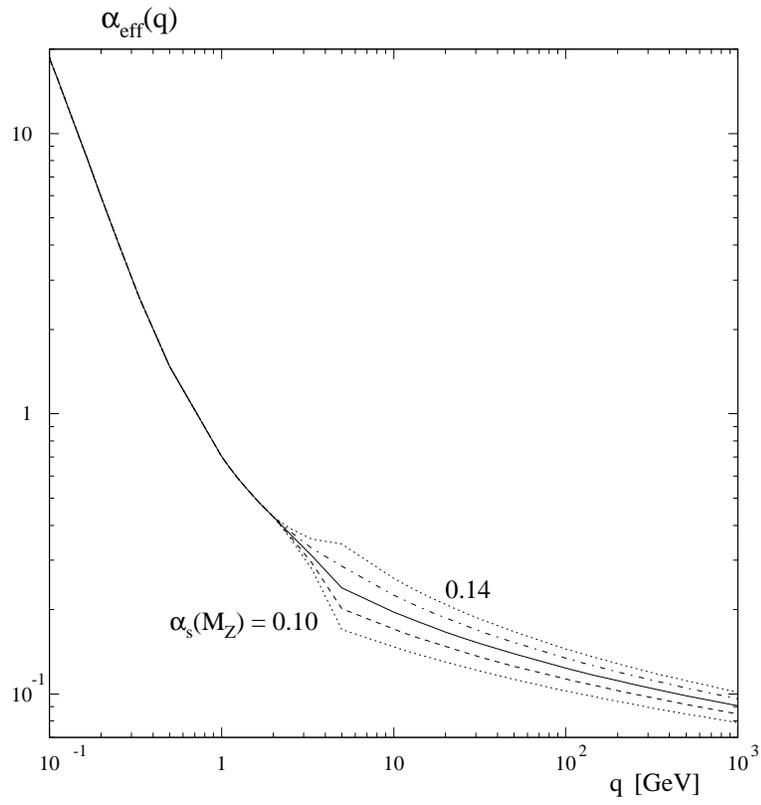}
\vspace*{110mm}
    \hfill
\caption{$\alpha_{\rm eff}$ for different values of $\alpha_s(M_Z)$:
solid: 0.12, dashed: 0.11, dashed-dotted: 0.13, dotted: 0.10 and 0.14.}
\label{fig:AEFF}
  \end{center}
\end{figure}

\begin{figure}[ht]
  \begin{center}
    \leavevmode
    \epsfxsize=12.cm
    \epsffile[70 135 515 710]{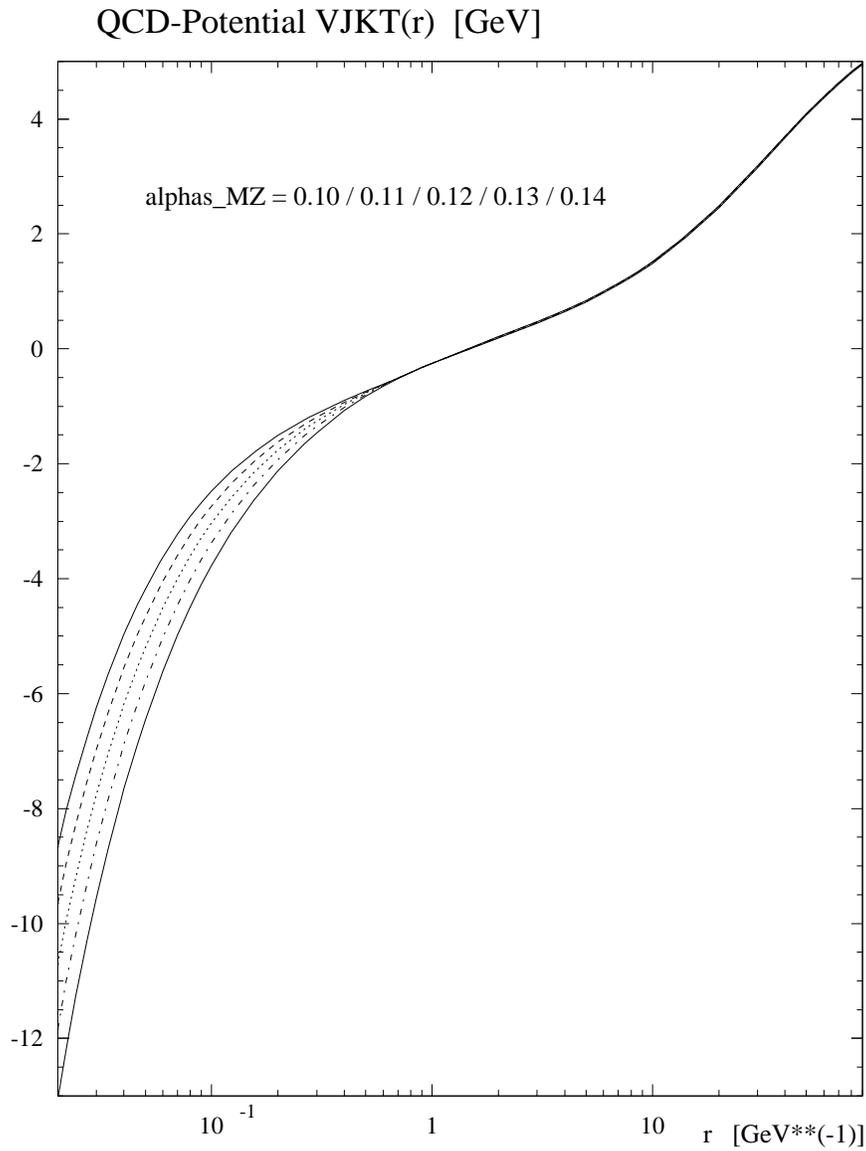}
    \hfill
\caption{QCD potential in the position space $V_{JKT}$
 for different values of $\alpha_s(M_Z)$:
solid: 0.12, dashed: 0.11, dashed-dotted: 0.13, dotted: 0.10 and 0.14.}
\label{fig:COOR}
  \end{center}
\end{figure}

In the subsequent discussion the Green's function will be calculated in 
momentum space with the help of the Lippman-Schwinger integral
equation \cite{JKT}.  The 
representation of the QCD potential in momentum space with the large $Q^2$ 
behaviour given by eqs.~\ref{eq:49} and \ref{eq:50} will be employed.
The intermediate  
and small momentum dependence will be based on Richardson's potential. This 
choice allows to vary $\alpha_{\overline{MS}}(M_Z^2)$ (or equivalently 
$\Lambda_{\overline{MS}}$) between 0.11 and 0.13, while maintaining a 
smooth $Q^2$ dependence of $\alpha_V(Q^2)$ (Fig.~\ref{fig:AEFF}).
  An additive constant 
in coordinate space (corresponding to a $\delta$-function in momentum 
space)  is adjusted to fix $V(r=1 GeV^{-1})=-1/4$ GeV for arbitrary  
$\alpha_{\overline{MS}}$. This constraint avoids the unmotivated and 
uncontroled variation of the long distance part of $V(r)$ with a change in 
$\alpha_{\overline{MS}}$. The potential in coordinate space is shown in 
Fig.~\ref{fig:COOR}.

%% file: part4.tex
\subsection{Realistic predictions for $\sigma_{t\bar t}$.}

For a realistic QCD potential the Green's function can only be calculated 
with numerical methods. An elegant algorithm for a solution in coordinate 
space has been suggested in \cite{pesk}. As a consequence of the 
optical theorem (see also eqn.~(\ref{eq:60})) only the imaginary part
of 
 $G(\vec r=0,\vec r'
=0,E+i\Gamma_t)$  
is needed to predict the total cross section.  The differential equation 
for the Green's function
\be
\left[ \left(E+i\Gamma_t\right)-\left( -{\nabla^2\over m_t} + V(\vec 
r)\right)\right] G(\vec r, \vec r'=0,E+i\Gamma_t) = \delta(\vec r)
\ee
is solved in a way which provides direct access to   Im$G(\vec r=0,\vec r' 
=0,E+i\Gamma_t)$ without the need to calculate the full $\vec r$ dependence. 
Alternatively, in \cite{JKT,TJ} the Green's function in momentum space was 
obtained from the Lippmann-Schwinger equation
\be
G(\vec p,E+i\Gamma_t)& =& G_0(\vec p,E+i\Gamma_t)  + G_0(\vec p,E+i\Gamma_t)  
\nonumber \\
&&\times \int{d\vec q\over (2\pi)^3} \widetilde V(\vec p - \vec q)
 G(\vec q,E+i\Gamma_t)  \nonumber\\
 G_0(\vec p,E+i\Gamma_t)  &=& {1\over E-p^2/m_t+i\Gamma_t}
\label{eq:355}
\ee
The total cross section is in this case obtained from the integral over the 
differential distribution
\be
{d\sigma\over d^3p} = {3\alpha^2\over \pi s m_t^2} \rho_v(s) \Gamma_t 
|G(\vec p,E+i\Gamma_t)|^2
\ee
This second formulation is particularly suited to introducing a momentum 
and energy dependent width $\Gamma(p,E)$ which allows to incorporate the 
phase space suppression and certain $\alpha_s^2$ rescattering corrections 
to be discussed below in section 3.2.6.

\begin{figure}[ht]
  \begin{center}
    \leavevmode
    \epsfxsize=12.cm
    \epsffile[110 265 465 560]{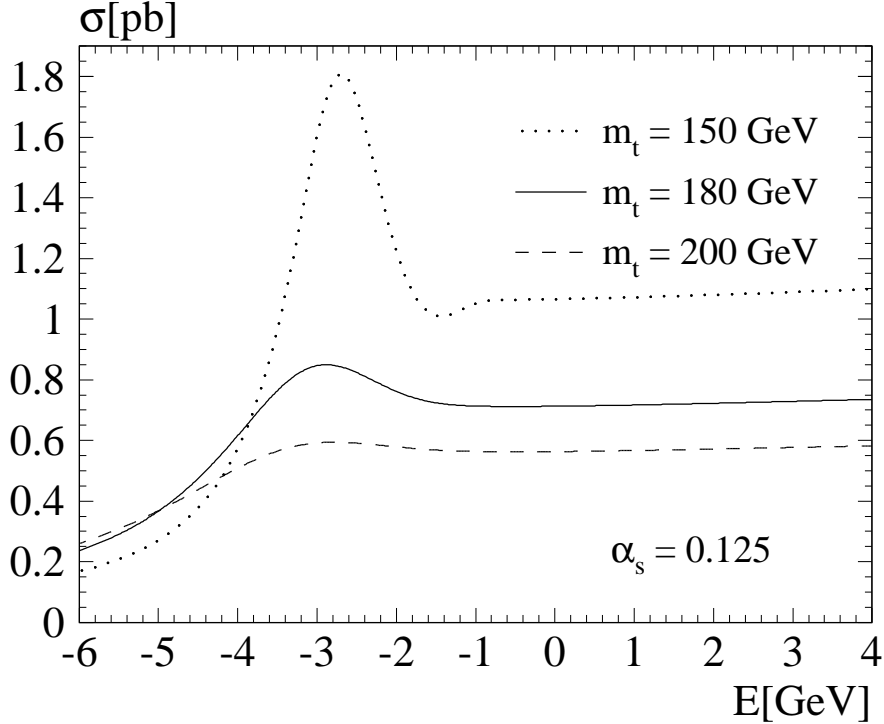}
    \hfill
\caption{Total cross section as function of $E=\protect\sqrt{s}-2m_t$
for three values of the top quark mass.}
\label{fig:310}
  \end{center}
\end{figure}

\begin{figure}[ht]
  \begin{center}
    \leavevmode
    \epsfxsize=12.cm
    \epsffile[110 265 465 560]{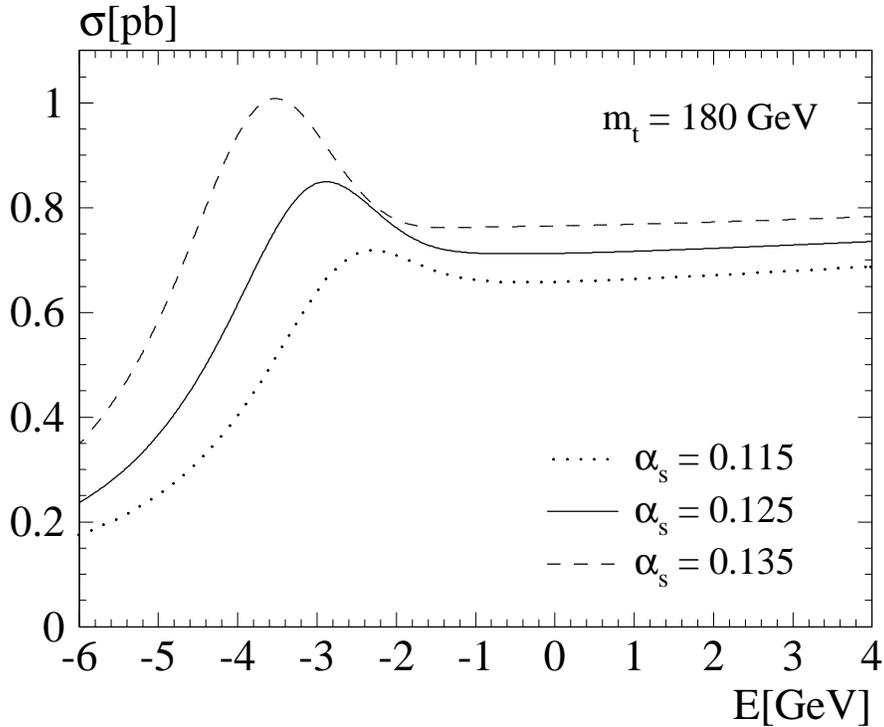}
    \hfill
\caption{Total cross section as function of $E=\protect\sqrt{s}-2m_t$
for three values of $\alpha_s$. }
\label{fig:311}
  \end{center}
\end{figure}

\begin{figure}[ht]
  \begin{center}
    \leavevmode
    \epsfxsize=12.cm
    \epsffile[110 280 465 560]{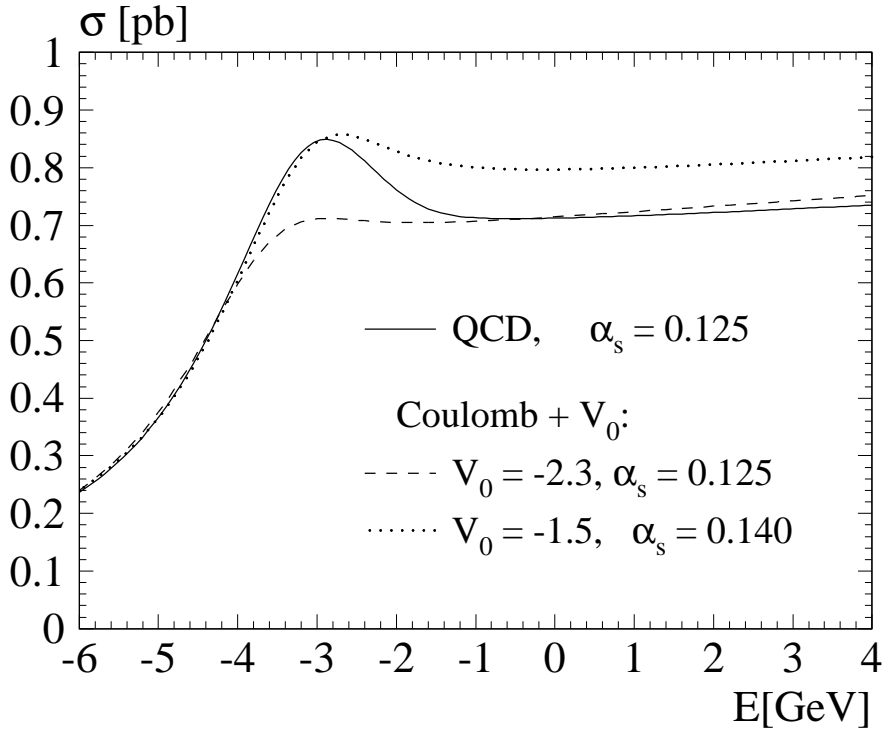}
    \hfill
\caption{Comparison between the predicted cross section for constant
(dashed and dotted lines) and running (solid lines) $\alpha_s$.}
\label{fig:312}
  \end{center}
\end{figure}

\begin{figure}[ht]
  \begin{center}
    \leavevmode
    \epsfxsize=10.cm
    \epsffile[110 260 465 595]{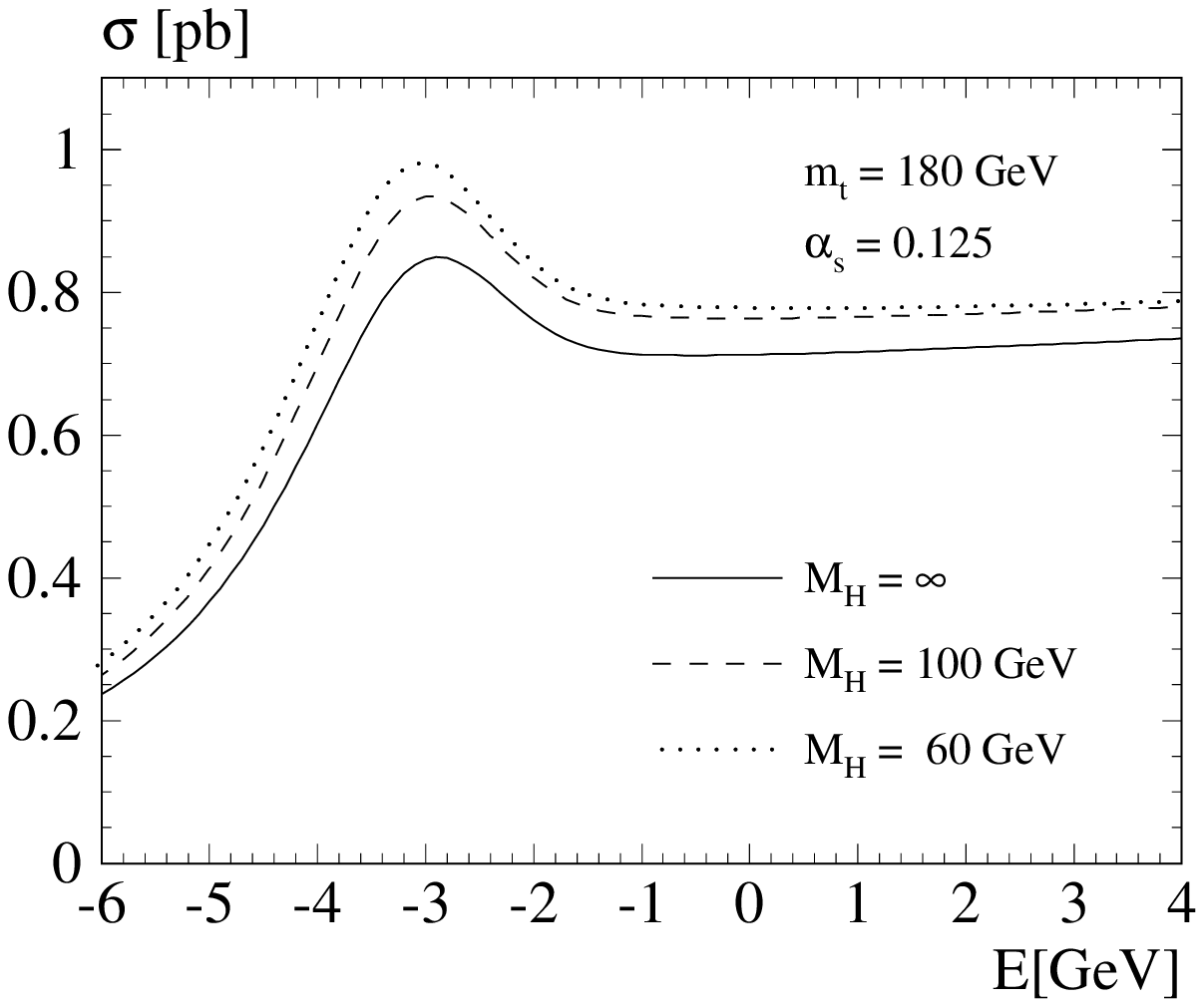}
    \hfill
\caption{Total cross section as function of $E=\protect\sqrt{s}-2m_t$
for different values of the Higgs mass.}
\label{fig:313}
  \end{center}
\end{figure}

It is well known that the coupling of the virtual photon to the quarkonium 
boundstate is modified by ``hard'' gluon exchange.  The vertex correction 
to the vector current produces an additional factor $\left(1-{16\over 
3}{\alpha_s\over \pi}\right)$ for the quarkonium decay rate into
{\epem} 
through the virtual photon or $Z$.  This factor can be calculated by 
separating the gluon exchange \cite{Buchm} correction to the vertex into 
the instanteneous potential piece and a remainder which  is attributed to 
gluons with high virtualities of order $m_t$.  A similar approach has been 
developed in \cite{JKsim} for Higgs exchange.  The vertex correction is 
again decomposed into a part which is given by the instantaneous Yukawa 
potential
\be
V_{Yuk}(r) = -\kappa {e^{-m_Hr}\over r}
\ee
with $\kappa=\sqrt{2} Gm_t^2/4\pi$ and a remainder which is dominated by 
highly virtual Higgs exchange.  The rapid increase of the correction in the  
threshold region (cf.~sect. 3.1.2) is driven by the potential; the 
remainder, the hard vertex correction, is fairly energy independent.  The
total cross section is thus sensitive to the top mass, the width (which in 
the SM is uniquely determined by $m_t$), the strong coupling constant 
$\alpha_s$ and the mass of the Higgs boson.  This dependence is illustrated 
in Figs.~\ref{fig:310}-\ref{fig:313}.  Apart from the trivial shift
of the threshold due to a  
change in $m_t$ the shape of $\sigma$ is affected by the rapidly increasing 
width of the top quark which amounts to 0.81 GeV, 1.57 GeV and 2.24 GeV for 
$m_t =$ 150 GeV, 180 GeV, and 200 GeV respectively. This is demonstrated in 
Fig.~\ref{fig:310}: A fairly 
pronounced 1S peak is still visible for $m_t$ = 150 GeV, for $m_t=200$ GeV, 
however, only a smooth shoulder is predicted.  The behaviour is 
qualitatively very similar, if we keep $m_t$ fixed say at 180 GeV and 
decrease or increase $\Gamma_t$ be the corresponding amount.  The shape of 
the cross section will therefore allow to determine the width of the top 
quark.  A qualitatively very different response is observed towards a 
change in $\alpha_s$ (Fig.~\ref{fig:311}). The binding energy increases with 
$\alpha_s$, the aparent threshold is thus lowered (This is the reason for 
the strong correlation between $\alpha_s$ and $m_t$ in the experimental 
analysis based on $\sigma_{tot}$ only \cite{Fujii,Igo}.) and the height of the 
``would-be resonance'' is increased.  Even several GeV above threshold one 
observes a slight increase of the cross section with $\alpha_s$, a 
consequence of the enhanced attraction between $t$ and $\bar t$
(cf.~eq.~\ref{eq:347}).   The 
impact of the running of $\alpha_s$ on the shape of the cross section is 
evident from Fig.~3.15.  The full QCD prediction with running
$\alpha_s$  (for  
$\alpha_{\overline{MS}}(M_Z^2)$ = 0.125) is compared to the prediction for 
a Coulomb potential with $\alpha_s$ fixed. It is impossible to describe the 
height of the peak and the continuum above with the same value of 
$\alpha_s$, even allowing for an
arbitrary additive constant $V_0$.  The influence of a variation in $m_H$ 
is shown in Fig.~\ref{fig:313}.  Cross section measurements with a
precision better  
than 10\% will become sensitive to the effect of a light Higgs boson. 

Up to this point the amplitude induced by virtual $Z$ and $\gamma$ are 
included in Born approximation only.  Electroweak corrections and 
initial state radiation are neglected.  A detailed discussion of 
electroweak corrections to the cross section and the left right asymmetry 
in the context of the SM can be found in \cite{GKKS}.     The corresponding 
discussion for the two-Higgs-doublet model is presented in \cite{GuthK}. In 
this model one might encounter enhanced Yukawa couplings which would 
amplify the effect under discussion. 

\begin{figure}
\psfig{figure=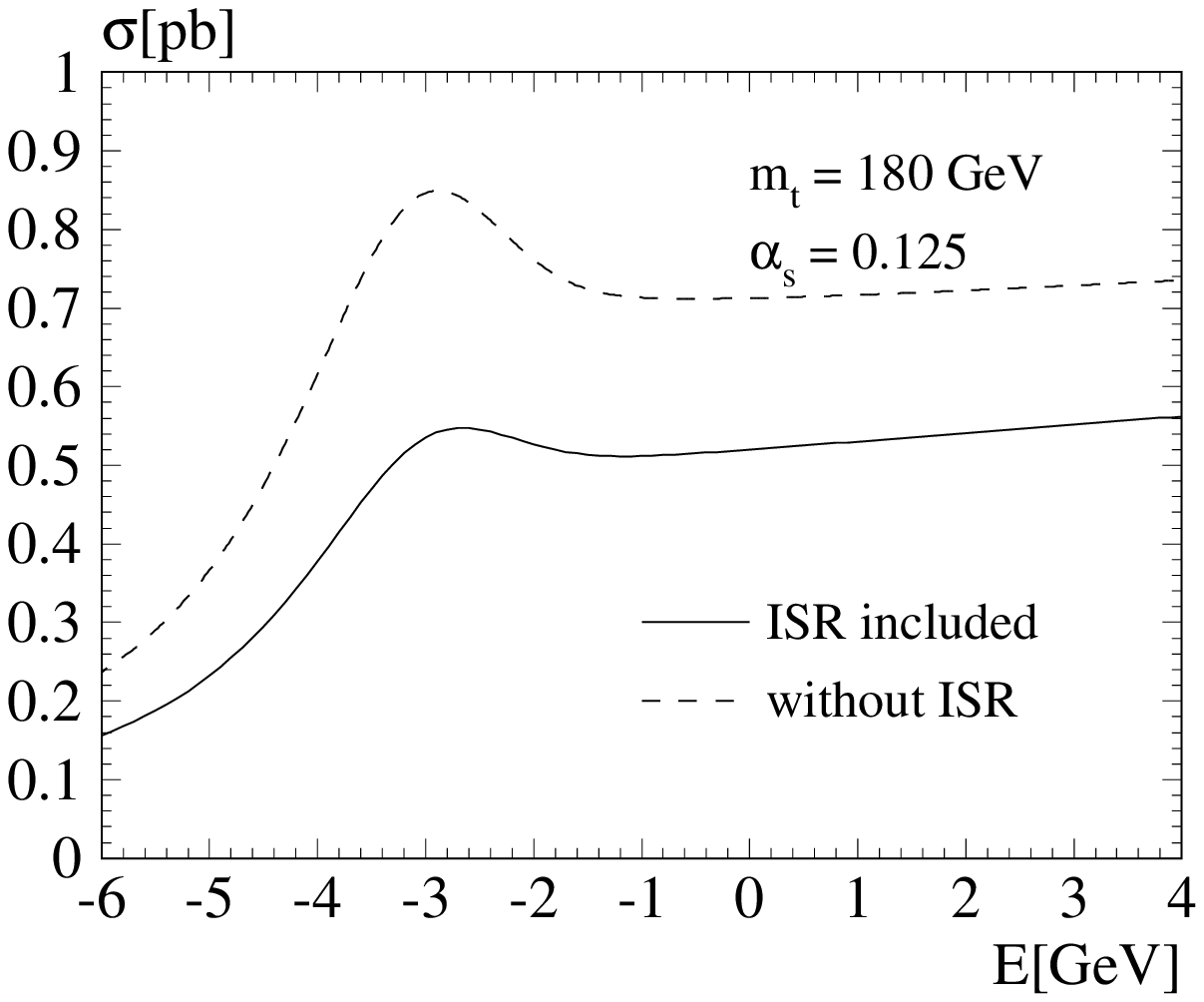,width=130mm,bbllx=40pt,bblly=270pt,%
bburx=450pt,bbury=550pt}
\caption{Comparison of the $t\bar t$ production cross section without
(dashed) and with (solid line) initial state radiation.}
\label{fig:initstate}
\end{figure}

Initial state radiation leads to a fairly drastic distortion of the shape of 
the cross section, in particular to a smearing of any pronounced structure. 
This is illustrated in Fig.~\ref{fig:initstate}
where the predictions with and without initial state radiation are compared 
for otherwise identical parameters.

Beamstrahlung and the energy spread of the beam lead to a further smearing 
of the apparent cross section.  These accelerator dependent issues are 
treated in more detail in \cite{Fujii}. (For a related discussion see also 
\cite{Igo}.)

\subsection{Momentum distributions of top quarks}

The Green's function in momentum space and the momentum distribution of top 
quarks (and thus their decay products) are intimately related. For a narrow 
quarkonium resonance with orbital quantum number $n$ the quarks' momentum 
distribution is evidently given by the wave function in momentum space
\be
{dN\over d\vec p} = {|\widetilde \psi_n(\vec p)|^2 \over (2\pi)^3}
\ee
For $J/\psi$ or $\Upsilon$ this distribution is not directly accessible to 
experiment since these states decay through $Q\bar Q$ annihilation only.  
For toponium, however, which is dominated by single quark decay, the decay 
products carry the information of their parent momentum and hence allow for 
the reconstruction of the original quark momentum distribution \cite{JKm}.

\begin{figure}
  \begin{center}
    \leavevmode
    \epsfxsize=12.cm
    \epsffile{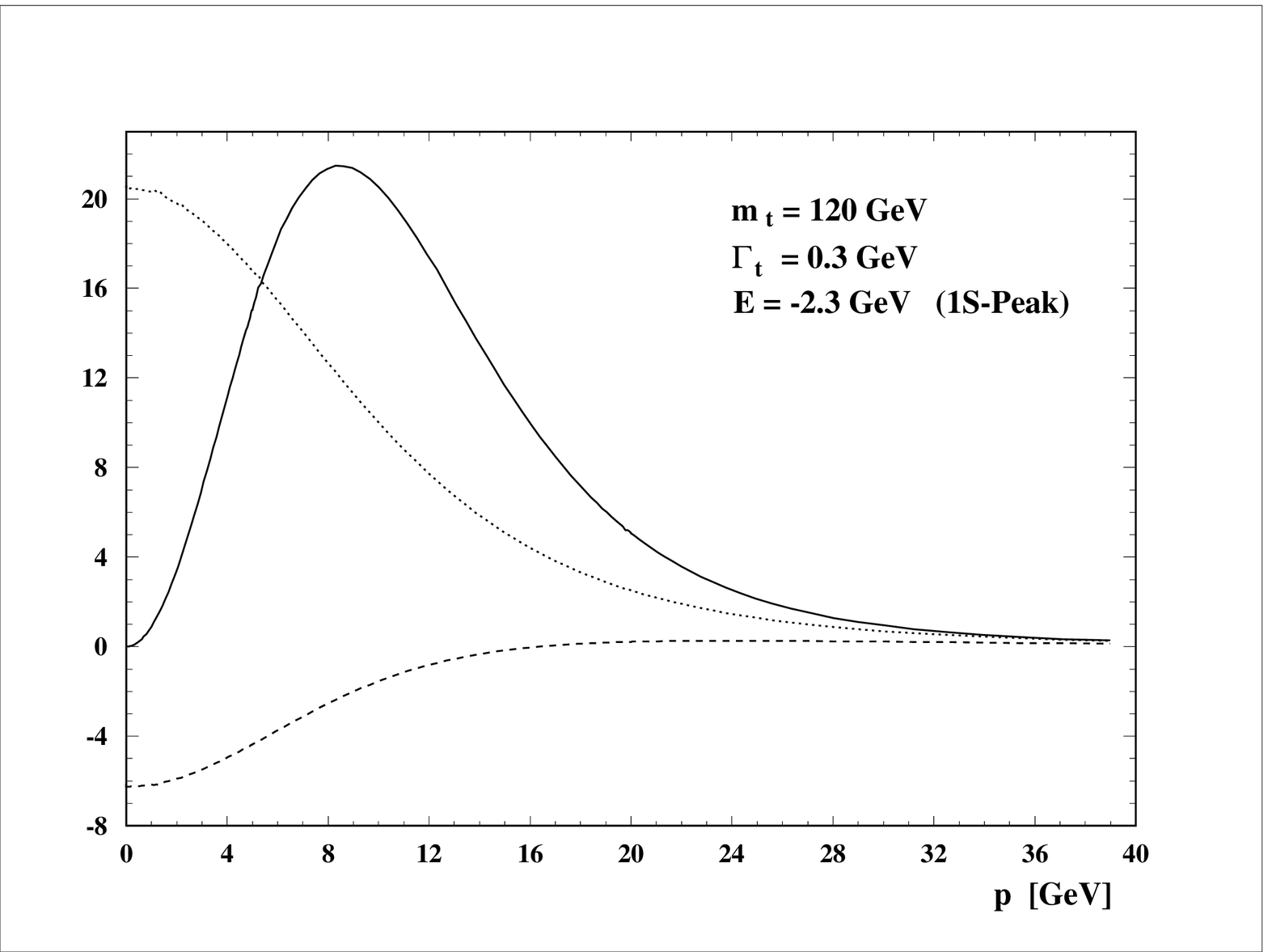}

\rule{0mm}{10mm}

\rule{-1mm}{0mm}
    \epsfxsize=12.cm
    \epsffile{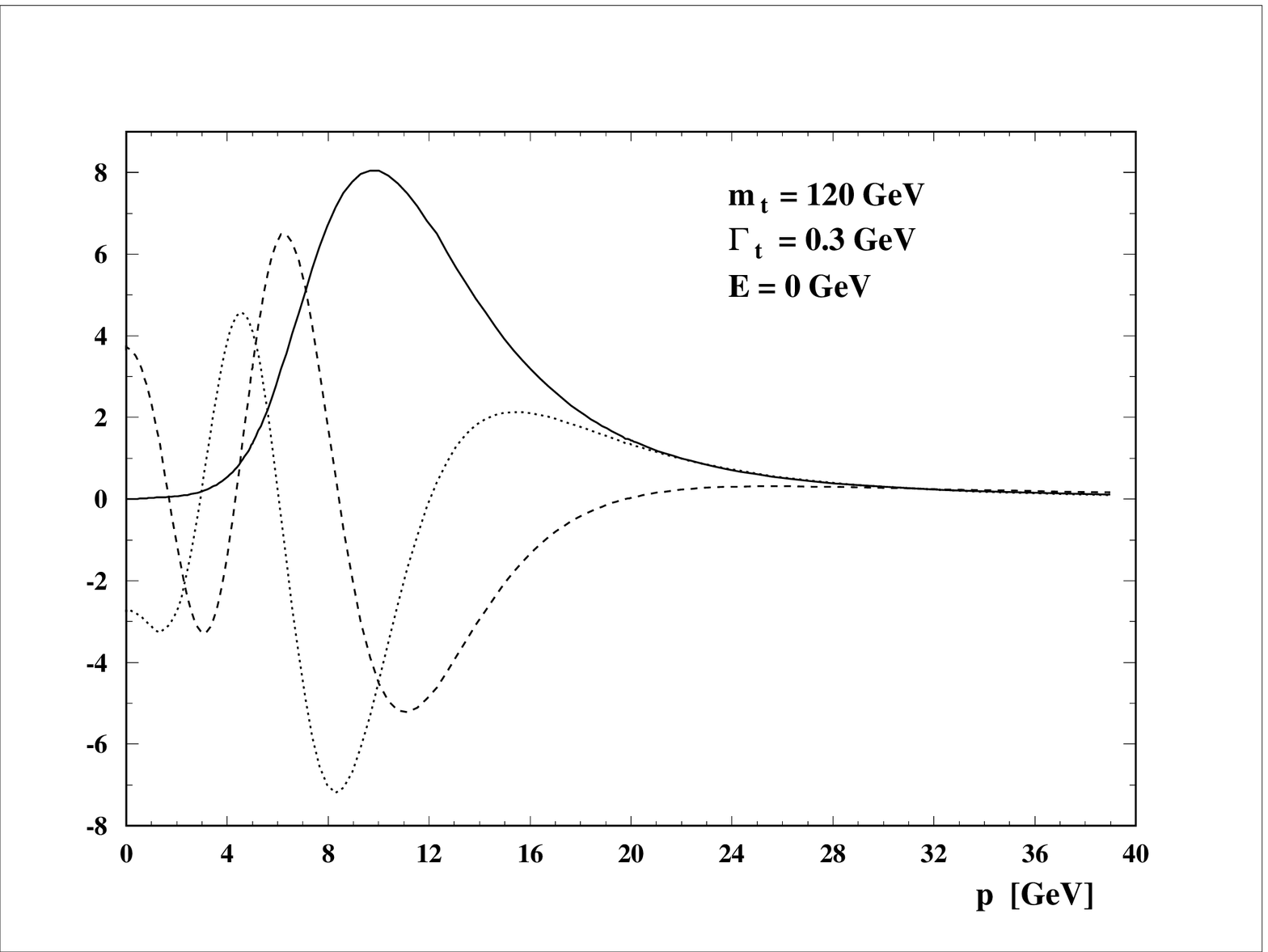}

\rule{0mm}{1mm}
\caption{Real (dashed) and imaginary (dotted) 
 parts of the Green's function for an
energy corresponding to the 1S peak (upper figure) and for $E=0$
(lower figure). Solid curve: $|pG(p)|^2\cdot 0.002.$}
\label{fig:2}
  \end{center}
\end{figure}

\begin{figure}[ht]
  \begin{center}
    \leavevmode
    \epsfxsize=12.cm
    \epsffile[15 210 580 635]{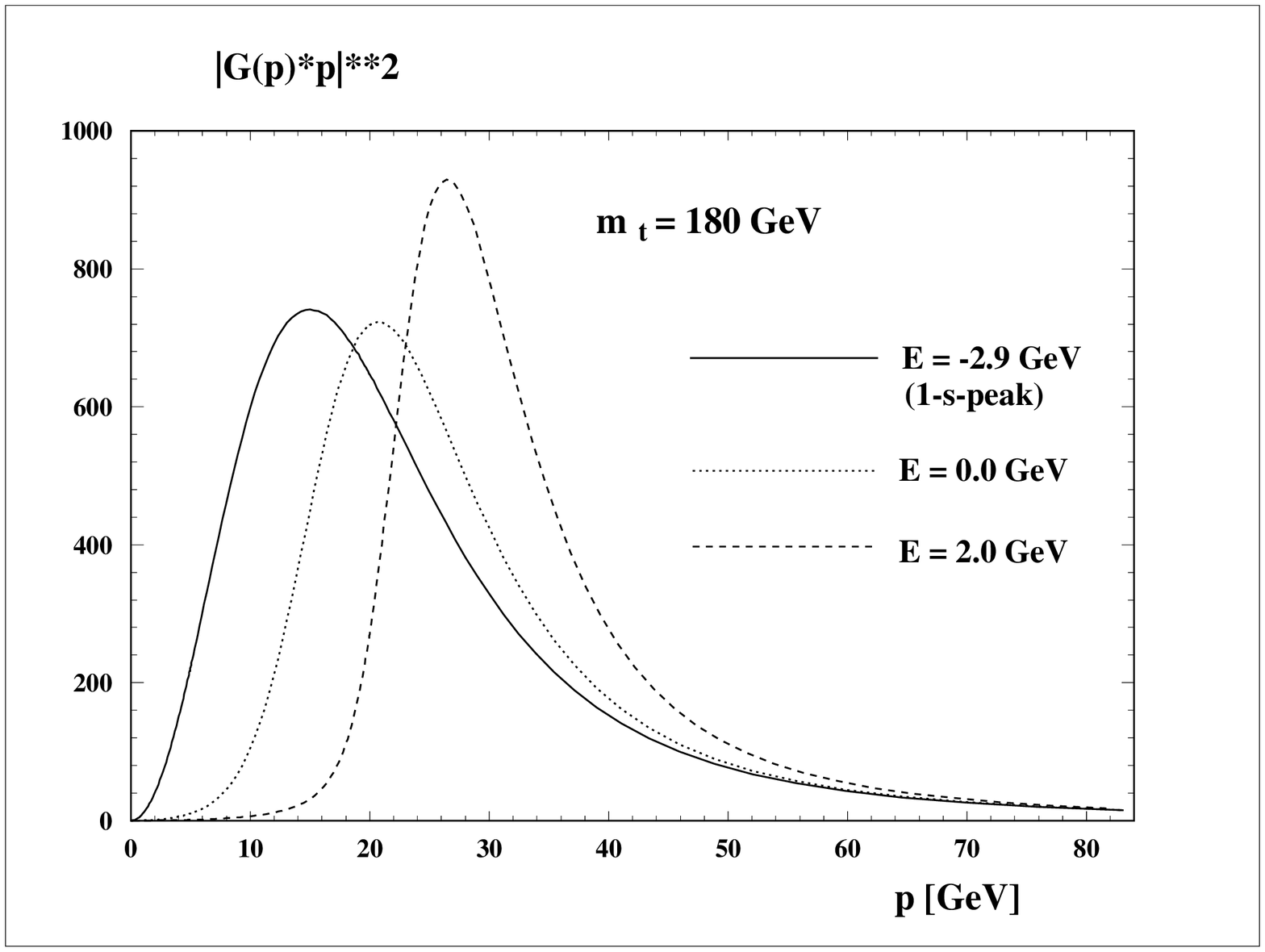}
    \hfill
\caption{Momentum distribution of top quarks for three different cms
energies.}
\label{fig:3}
  \end{center}
\end{figure}

For one individual resonance this leads to the differential $t\bar t$ 
cross section (without $Z$ contribution and transverse gluon correction).  
\be
{d \sigma_n\over d\vec p} (\vec p,E) = {3\alpha^2 Q_t^2\over
\pi sm_t^2} |\psi_n(\vec r =0)|^2 
{\Gamma_t\over (E-E_n)^2+\Gamma_t^2}  |\widetilde \psi_n(\vec p)|^2
\ee
Once $\Gamma_t$ is sufficiently large, interferences between different 
radial excitations become important and the right-hand side of this 
equation has to be replaced by the square of the Green's function 
\cite{Sumino,JKT}
\be
{d \sigma_n\over d\vec p} (\vec p,E) = 
{3\alpha^2 Q_t^2\over \pi sm_t^2}
\Gamma_t 
\left|G(\vec p,E+i\Gamma_t) \right|^2
\ee
with
\be
G(\vec p,E+i\Gamma_t) = \int d\vec r e^{i\vec p \vec r} 
G(\vec r,\vec r'=0,E+i\Gamma_t)
\ee
As discussed in sect.~3.2.2, the Green's function can be obtained in 
momentum space as a solution of the Lippmann-Schwinger equation.  For an 
energy close to the 1S peak it exhibits a  fairly smooth 
behaviour reminiscent of the 1S wave function in momentum space
(Fig.~\ref{fig:2}).
With increasing energy an oscillatory pattern of the amplitude is observed, 
and a shift towards larger momenta (Fig.~\ref{fig:2}).  These results are 
intentionally displayed for $m_t=120$ GeV, where the oscillations are still 
clearly visible, in contrast to $m_t$=180 GeV where all oscillations are 
smeared by the large width $\Gamma_t$.  The corresponding predictions for the 
distributions at $m_t = 180$ GeV are displayed in Fig.~\ref{fig:3}.   
\begin{figure}[ht]
  \begin{center}
    \leavevmode
    \epsfxsize=12.cm
    \epsffile[15 210 580 635]{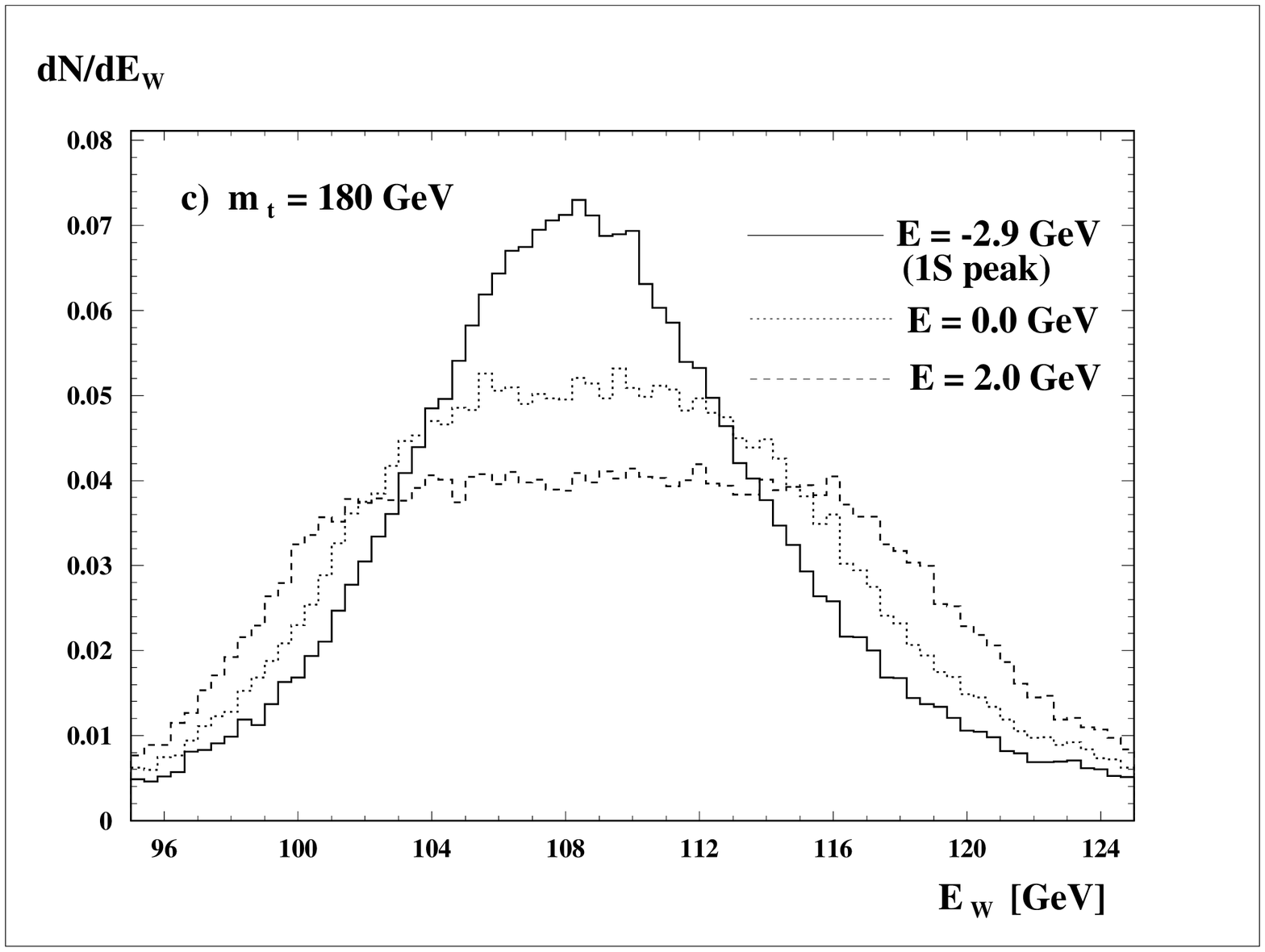}
    \hfill
\caption{Energy distribution of $W$'s from top quark decay for three
different cms energies.}
\label{fig:4}
  \end{center}
\end{figure}
The transition 
from a wide distribution below the nominal threshold to a narrow one with 
the location of the peak determined by trivial kinematics is clearly 
visible.  The impact on the energy distribution of the $W$'s from top decay 
is shown in Fig.~\ref{fig:4}. 

 To characterize the momentum distribution by a single parameter, one may 
either choose its peak value, or the expectation value of the modulus of 
the momentum $\langle p\rangle$, the latter being well adopted to the 
experimental analysis.  In the situation at hand the definition of 
$\langle p\rangle$ has to be introduced with some care. The free Green's 
function $G_0$ (see eq.~\ref{eq:355})
 drops $\sim p^{-2}$ for large momenta and this 
behaviour is recovered also in the presence of interaction. The expectation 
value $\langle p\rangle$ diverges logarithmically with the cutoff. In the 
narrow width approximation one finds for the leading terms
\be
\langle p\rangle &=& 
{ \int_0^{p_m} d\vec p p \left| G_0(\vec p,E+i\Gamma) \right|^2 
   \over
 \int_0^{p_m} d\vec p  \left| G_0(\vec p,E+i\Gamma) \right|^2 
}
\nonumber \\ &=&
\sqrt{m_tE}\left( 1+ {\Gamma\over E}{\ln p_m^2/
\left(\sqrt{E^2+\Gamma^2} m_t\right) \over \pi}\right)
\ee
where a cutoff $p_m$ has been introduced. As a consequence of the small 
numerical prefactor of the divergent term and its
 logarithmic cutoff dependence 
the result is fairly insensitive to the exact value of the cutoff for $p_m$ 
of order $m_t$.  Alternatively one may replace the phase space element 
$d\vec p/m_t$ by the relativistic version   $d\vec p/E = d\vec 
p/\sqrt{m_t^2+p^2}$ to obtain a convergent result. In future measurements the 
cutoff will presumably be provided by the experimental analysis.

In order to study the dependence of $\langle p\rangle$ on the strong 
coupling constant, consider for the moment the predictions for a stable 
quark.
Some
intuition and qualitative understanding can already be gained from the
predictions based on a pure Coulomb potential \cite{JKTZer}.

For a stable top quark of fixed mass the ``effective threshold''
can be associated with the location of the $1S$ resonance
$\sqrt{s_{thr}} = 2 m_t + E_{1S}$ with $E_{1S}=-E_{Ryd}= - \alpha^2m_t/4$
which decreases with increasing $\alpha$. The height of the resonance
cross
section is proportional to the square of the wave function at the origin
and hence proportional to $\alpha^3$, as long as the resonances are
reasonably well separated. In the limit of large $\Gamma_t$, i.e.~far
larger than $E_{Ryd}$, the overlapping $1S$, $2S$ $\dots$ resonances have
to fill the gaps between the peaks. Since these gaps themselves
increase proportional to $\alpha^2$, one is left in the extreme case of large
width with a cross section linear in $\alpha$. Note that this
corresponds to the behaviour of the cross section close to but slightly
above the threshold which is also proportional to $\alpha$.

For realistic top masses one thus observes a dependence
of the peak cross section linear in $\alpha$.
Since the location of the peak itself depends on $\alpha$, only the
analysis of the full shape allows to extract the relevant information.

In a next step also the momentum distribution of top quarks
has to be
exploited to obtain further information. The discussion is again
particularly simple for the Coulomb potential $V(r)=-\alpha/r$ and
provides a nice exercise in nonrelativistic quantum mechanics.
The average momentum,
in units of the Bohr momentum $\alpha m_t/2$, can be written in terms of
a function $f(\epsilon)$ which depends only on one variable $\epsilon=
E/E_{Ryd}$ if
the energy $E=\sqrt{s}- 2m_t$ is measured in terms of the Rydberg
energy.

\newcommand{\pav}{\langle p\rangle}
\begin{equation}\label{paveq}
\pav = \frac{\alpha m_t}{2} f(\epsilon)
\end{equation}
For positive arguments  the function $f$ can be derived
from obvious kinematical considerations.

\begin{equation}
f(\epsilon)=\sqrt{\epsilon} \qquad {\rm for} \qquad \epsilon \ge0
\end{equation}

For the discrete negative
arguments $\epsilon_n=-1/n^2$, corresponding to the
locations of the bound states,
the radial wave functions in momentum space are given 
in terms  of the Gegenbauer polynomials $C^m_n$

\begin{equation}
\psi(\vec p\,)= \frac{16\pi n^{3/2}}{(1+n^2 p^2)^2} C^1_{n-1}
\left(\frac{n^2 p^2 -1} {n^2 p^2 + 1} \right) Y_0^0\left(\theta,\varphi\right)
\end{equation}
with
\begin{equation}
\int\frac {d\vec p}{(2\pi)^3} |\psi(\vec p\,)|^2 = 1\ .
\end{equation}
Using the explicit forms of $C^m_n$
\begin{equation}
C^1_0(z)=1,\qquad C^1_1(z)=2z,\qquad C^1_2(z)=4z^2-1
\end{equation}
one obtains through straightforward calculation

\begin{equation}
f(-1)=\frac{8}{3\pi},\qquad f(-1/4)=\frac{16}{15\pi},
\qquad f(-1/9)=\frac{24}{35\pi}\ .
\end{equation}
For arbitrary $n$ one derives the general result
\begin{equation}
f\left(-\frac{1}{n^2}\right)=\frac{8n}{(2n-1)(2n+1)\pi}
\label{eq8}
\end{equation}
with the asymptotic behaviour
\begin{equation}\label{asym}
f\left(-\frac{1}{n^2}\right)\to \frac{2}{n \pi}\ .
\end{equation}

This is in accord with the result expected from classical mechanics:
For the average momentum of a particle on a closed orbit
in the Coulomb potential one derives
\begin{equation}
\langle p^{2\beta }\rangle=\left(\frac{\alpha m_t}{2}\right)^{2\beta }
\left(\frac{-E}{E_{Ryd}}\right)^\beta 
\frac{1}{2\pi}
\int_0^{2\pi} d\xi
\frac{(1-e^2\cos^2\xi)^\beta}{(1-e\cos\xi)^{2\beta -1} }\ .
\end{equation}
Quantum mechanical orbits with angular momentum zero and
high radial quantum numbers correspond to classical motions
with excentricity $e=1$ (i.e.~straight lines). 
In this limiting case the classical
expectation value is easily evaluated, and for $\beta=1/2$ one finds
agreement with the quantum mechanical result.
For small negative energies one
therefore obtains the behaviour $f(\epsilon) = 2\sqrt{-\epsilon}/\pi$.
Significantly below threshold, however, the average momentum
obtained from the Green's function
 increases
more rapidly with decreasing energy and between the $1S$ and the $2S$
state one observes an approximately
linear dependence on the energy.

From these considerations the dependence of the average momentum
on $\alpha$ (with $E$ fixed) is easily understandable, in particular
the seemingly surprising observation that well below threshold
$\pav$ decreases with
increasing $\alpha$.
From (\ref{paveq}) one derives for
a shift in $\alpha$ (keeping the energy $E$ fixed) the following
shift in $\pav$
\begin{equation}
\frac{\delta\pav}{\pav}=
\left(1-2\frac{f'(\epsilon)}{f(\epsilon)} \epsilon\right)
\frac{\delta\alpha}{\alpha}\ .
\end{equation}
Above threshold as well as close to but below threshold $f\propto
\sqrt{|\epsilon|}$.
Hence $\epsilon f'/f = 1/2$ and the average momentum remains
unaffected.
 The location of the minimum is thus an ideal place to fix the mass of
 the top quark.
Significantly below threshold, however, $\epsilon f'/f \approx 1$
and the factor in front of $\delta\alpha/\alpha$ becomes negative.
This explains the decrease of $\pav$  with increasing $\alpha$.

\begin{figure}
\epsffile{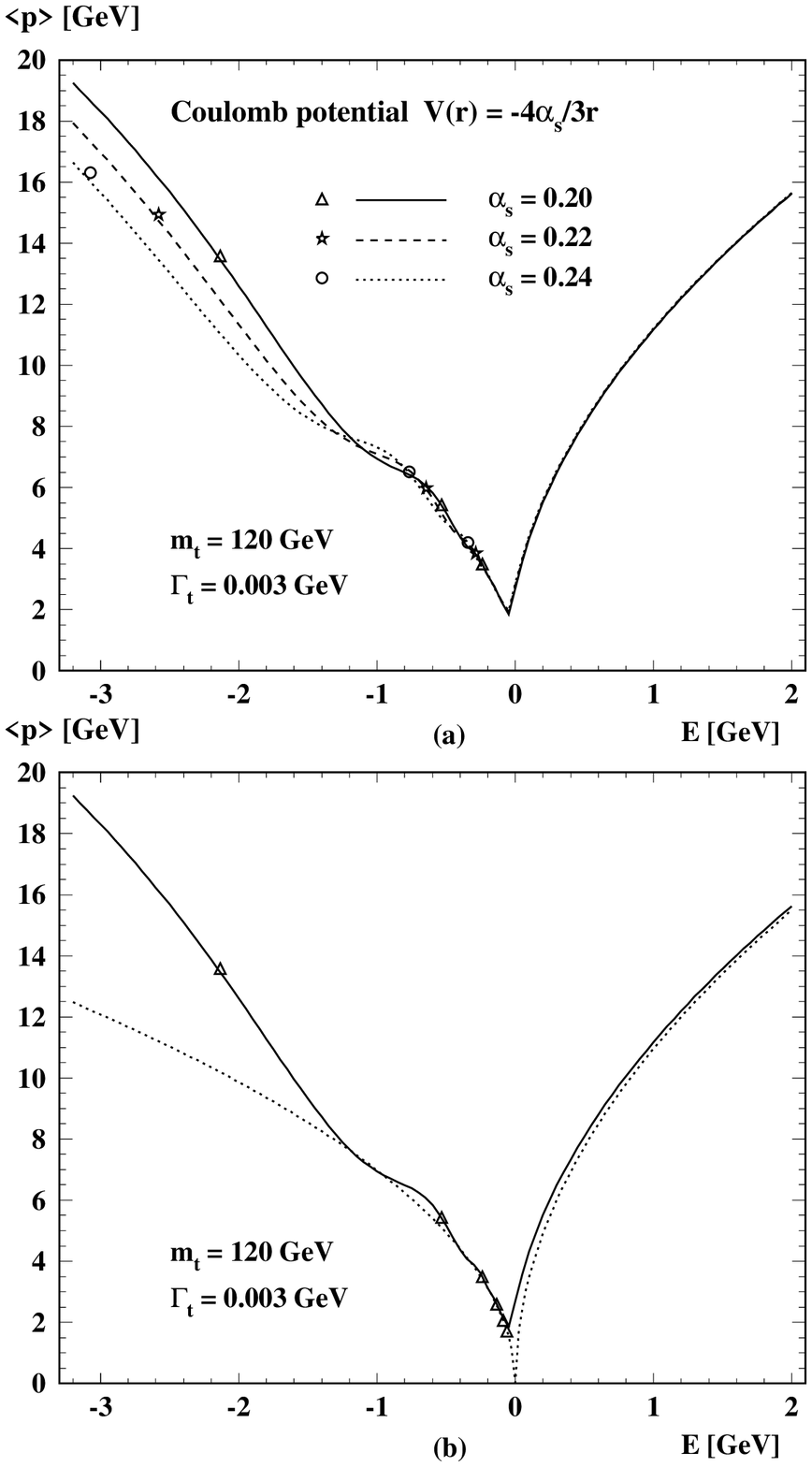}
\vspace*{-15mm}
\caption{a) Average momentum as a function of $E$ for different values
of $\alpha_s$. The markers show the results of the analytical
calculation at $1S$, $2S$, $3S$ energies.
b) Comparison with the analytical result for discrete
energies and with the square--root dependence close to threshold.}
\label{fig1}
\end{figure}

These results are illustrated in Fig.\ref{fig1}.
In Fig.\ref{fig1}b we demonstrate that $\pav$ as evaluated with the
program for the Green function (solid line) coincides perfectly
well with the values calculated from the analytical formula on
resonance, indicated
by the triangles. The prediction from classical mechanics, namely
$\pav \propto \sqrt{|\epsilon|}$ is shown by the dotted line
and agrees nicely for positive and negative energies. In Fig.\ref{fig1}a
$\alpha_s$ is increased from 0.20 to 0.24 and $\pav$ changes in accord
with the previous discussion.

For definiteness we have
chosen $m=m_t/2=60$ GeV for the reduced mass and
$\alpha=4\alpha_s/3$ with $\alpha_s$ varying between 0.20 and 0.24.
The curves demonstrate the decrease of $\pav$ by
about 10\%  for the corresponding increase in $\alpha$. The triangles
mark the locations of the resonances and the expectation values
for the momentum as calculated from (\ref{eq8}).

\begin{figure}
\epsffile{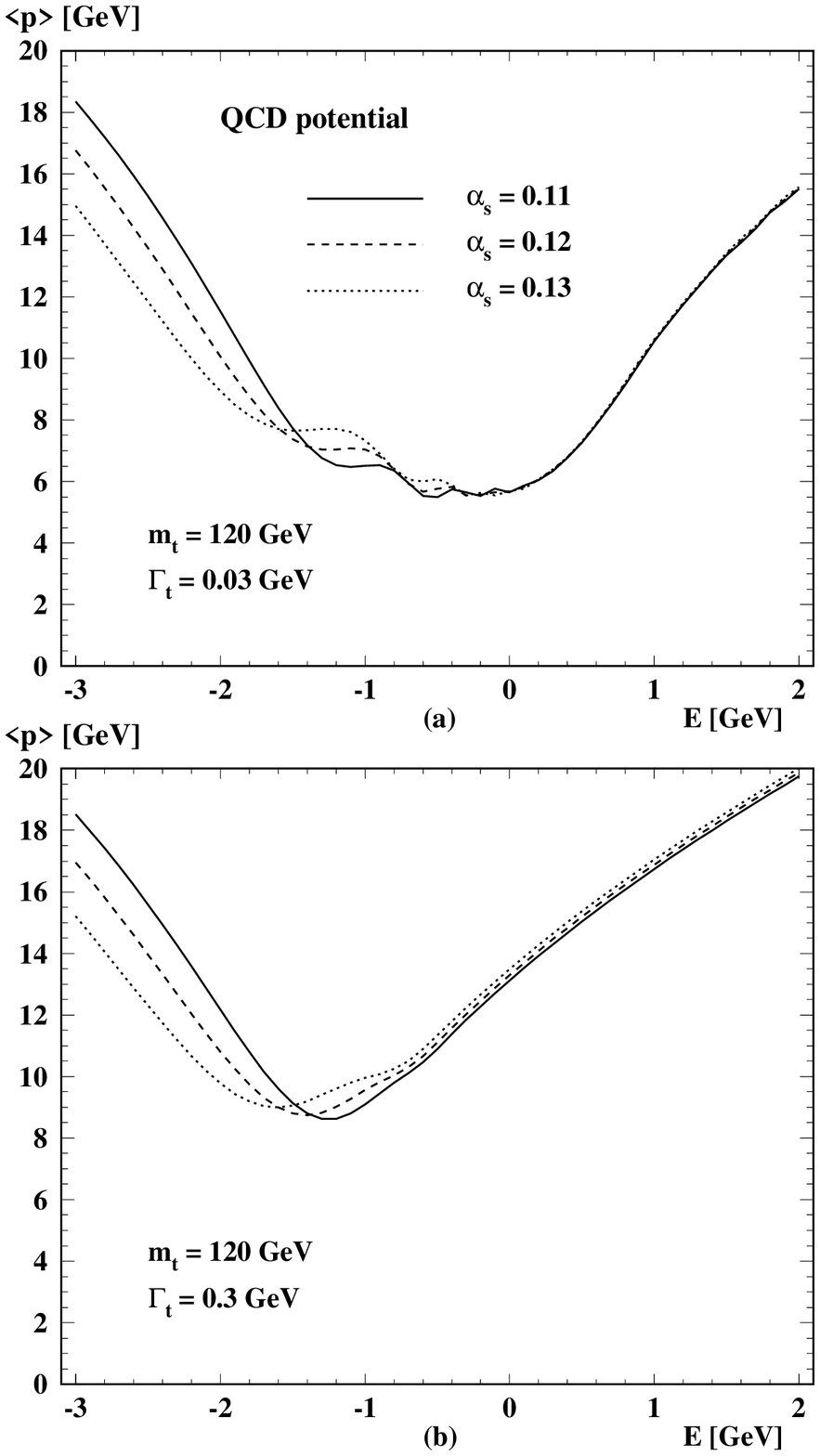}
\vspace*{-15mm}
\caption{Energy dependence of $\langle p\rangle$, the
average t quark momentum
for $\alpha_s = 0.13$ (dotted)
$0.12$ (dashed) and $0.11$ (solid) line for
$m_t=120$ GeV. a) $\Gamma_t=0.03$ GeV and b) $\Gamma_t=0.3$ GeV.}
\label{fig2}
\end{figure}

The qualitative behaviour remains unchanged for realistic QCD
potentials  corresponding to different values of $\alpha_s(M_Z)$.
Qualitatively the same behaviour is observed as in Fig.\ref{fig1}.
In Fig.\ref{fig2}a the top quark width has been set to an artificially
small value of $0.03$ GeV, in Fig.\ref{fig2}b the realistic value of
$0.3$ GeV has been adopted. The finite width leads to an additional
contribution to the momentum of order $\sqrt{\Gamma m}$.

An important feature is evident from Fig.\ref{fig2}: The momentum
calculated for
positive energy is nearly independent from $\alpha_s$ and reflects
merely the kinematic behaviour, just as in the case of the Coulomb
potential. This is characteristic for the choice of a potential \cite{TJ}
where the large distance behaviour is fixed by phenomenology and
decoupled from the short distance value of $\alpha_s$.

The different assumptions on the long distance behaviour are reflected
in differences between
the predictions of \cite{peskin,Sumino,JKT}
for the precise location of the $t\bar t$
threshold for identical values of $\alpha_s$ and $m_t$ and in
differences
in the $\alpha_s$ dependence of the momentum distributions for fixed
$m_t$ and energy (see also \cite{Igo}).
All these differences can be attributed to the
freedom in the additive constant discussed before.  The same additive
constant appears in $b\bar b$ spectroscopy, such that the mass
difference between top and bottom is independent from these
considerations.

\begin{figure}
  \begin{center}
    \leavevmode
    \epsfxsize=12.cm
    \epsffile[15 135 580 705]{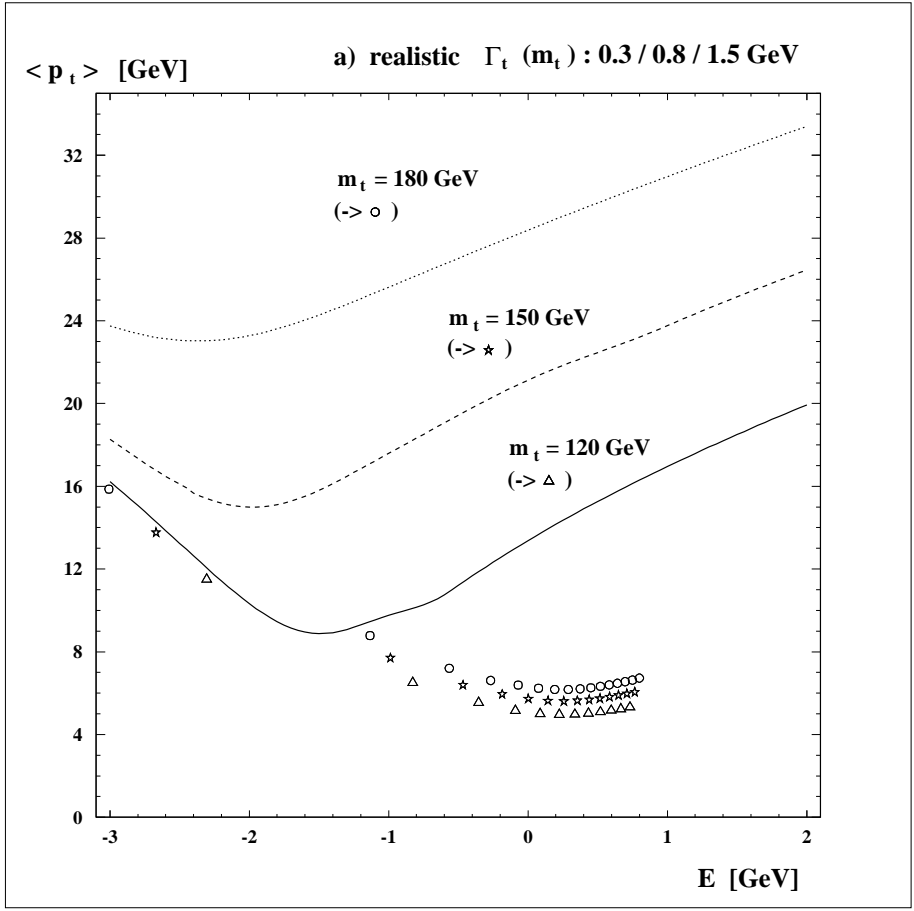}
    \hfill
\caption{Energy dependence of the average top quark momentum for $m_t$
= 120/150/180 GeV. Triangles, stars and circles correspond to $\langle
p_t \rangle$ for S-states with $\Gamma_t=0$. }
\label{fig:s30}
  \end{center}
\end{figure}

In Fig.~\ref{fig:s30} the predictions for $\langle p \rangle$  vs.~energy 
are presented for the case of a realistic QCD potential, assuming
$m_t=120/150/180$ GeV. The strong rise of $\langle p \rangle$ as a
consequence of the strong increase of $\Gamma_t$ is clearly visible. 

\subsection{Angular distributions and polarization}

Close to threshold the production amplitude is dominantly $S$-wave which 
leads to an isotropic angular distribution.  The spin of top quarks  
is alligned with the beam direction, with a degree of polarization 
determined by the electroweak couplings, the beam polarization and the mass 
of the top quark, but independent of the production dynamics, in particular 
of the potential. 

 Small, but nevertheless experimentally accessible corrections do arise 
from the small admixture of $P$-wave contributions and from rescattering of 
the top quark decay products. Let us concentrate for the moment on the 
first mechanism.  $P$-wave amplitudes are proportional to the top quark 
momentum.  For stable noninteracting particles the momentum vanishes at 
threshold.  However, as discussed in the previous section the expectation 
value of the quark momentum is nonzero for all energies ---  a consequence 
of the large top decay rate and the uncertainty principle.   Technically the 
$P$-wave contribution is calculated with the help of the Green's function 
technique. The generalization of the Lippman-Schwinger equation
(\ref{eq:355})  from $S$- 
to $P$-waves reads as follows
\be
{\cal F} (\rmp,E) &=& G_0(p,E) + G_0(p,E)
\int {d^3 k \over (2 \pi)^3}
        {\bfp\cdot \bfk \over \bfp^2}
        V(\bfp - \bfk) {\cal F} (k,E) 
\ee

It is then straightforward to calculate the
differential momentum distribution and the polarization of
top quarks produced in electron positron annihilation.
Let us recall the following conventions for the fermion
couplings
\begin{equation}
v_f = 2 I^3_f - 4 q_f \sin^2\theta_{\rm W} , \qquad  a_f = 2 I^3_f .
\end{equation}
$P_\pm$ denotes the longitudinal electron/positron
polarization and
$\chi=(P_+-P_-)/(1-P_+P_-)$
can be interpreted as effective longitudinal polarization of
the virtual intermediate photon or $Z$ boson.
The following abbreviations will be useful below:
\begin{eqnarray}
a_1 &=& q_e^2 q_t^2 + (v_e^2 + a_e^2) v_t^2 d^2 +
        2 q_e q_t v_e v_t d \nonumber \\
a_2 &=& 2 v_e a_e v_t^2 d^2 + 2 q_e q_t a_e v_t d \nonumber \\
a_3 &=& 4 v_e a_e v_t a_t d^2 + 2 q_e q_t a_e a_t d \label{coupl}\\
a_4 &=& 2 (v_e^2 + a_e^2) v_t a_t d^2 + 2 q_e q_t v_e a_t d \nonumber\\
d &=& {1\over 16 \sin^2\theta_{\rm W}\cos^2\theta_{\rm W}}\,{s\over s - M_Z^2}.
    \nonumber
\end{eqnarray}
The differential cross section, summed over polarizations of quarks
and including $S$-wave and $S$-$P$--interference contributions,
is thus given by
\begin{eqnarray}
{d^3\sigma \over dp^3} &=&
{3 \alpha^2 \Gamma_t \over 4 \pi m_t^4} (1-P_+P_-)
\left[ { (a_1 + \chi a_2)
        \left(1-{16 \alpha_{\rm s} \over 3 \pi} \right)
        \left|G(\rmp,E)\right|^2 + }\right. \nonumber\\
& & 
\left. {+(a_3+\chi a_4)
\left( 1-{12\alpha_{\rm s} \over 3 \pi} \right)
{\rmp \over m_t} \Re \left(\,G(\rmp,E) F^*(\rmp,E)\,\right)\,
\cos\vartheta} \right]  \label{dsig_d3p} .
\end{eqnarray}
The vertex corrections from hard gluon exchange for $S$-wave
and $P$-wave  amplitudes are included in this
formula. It leads to the following forward-backward asymmetry
\cite{MurSum2,hjkt}
\begin{equation}\label{afb}
{\cal A}_{\rm FB}(\rmp,E) = C_{\rm FB}(\chi)\, \varphi_{\rm _R}(\rmp,E),
\end{equation}
with
\begin{equation}
   C_{\rm FB}(\chi) = {1 \over 2}\, {a_3
    + \chi a_4 \over a_1 + \chi a_2} ,
\end{equation}
$\varphi_{\rm _R} = \Re\,\varphi$,  and
\begin{equation}\label{phi}
\varphi(\rmp,E) =
{(1-{4 \alpha_{\rm s}/3 \pi})\over (1-{8 \alpha_{\rm s}/3 \pi})}\,
        {\rmp \over m_t}\,
        {F^* \!(\rmp,E) \over G^* \!(\rmp,E)}  .
\end{equation}
This result is still differential in the top quark momentum.
Replacing $\varphi(\rmp,E)$ by
\begin{equation}\label{cap_phi}
\Phi(E) =
{(1-{4 \alpha_{\rm s}/3 \pi})\over (1-{8 \alpha_{\rm s}/3 \pi})}\,
{\int_0^{\rmp_m} d\rmp\,
{\rmp^3 \over m_t}\, F^*(\rmp,E)G(\rmp,E) \over
\int_0^{\rmp_m} d\rmp \, \rmp^2 \left|G(\rmp,E)\right|^2}  .
\end{equation}
one obtains the integrated forward-backward
asymmetry again.
Again, the cutoff $\rmp_m$ must be introduced to eliminate the
logarithmic divergence of the integral.

\begin{figure}[ht]
\begin{flushleft}
\leavevmode
\epsfxsize=8cm
\epsffile[100 370 500 520]{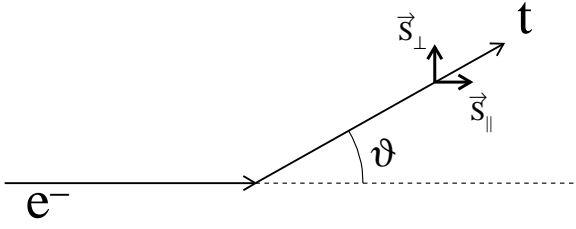}\\[-1.5cm]
\hfill
\parbox{6.cm}{\small
\caption[]{\label{dyn.ps}\sloppy Definition of the spin directions.
        The normal component ${\bf s}_{\rm N}$ points out of the plane.}}
\end{flushleft}
\end{figure}

{\em Polarization \cite{hjkt,ttp48}}\\
To describe top quark polarization in the threshold region it
is convenient to align the reference system with the beam
direction (Fig.~\ref{dyn.ps}) and to define
\begin{eqnarray}
{\bf s}_{\|} = {\bf n}_{e^-}, \quad
{\bf s}_{\rm N} = {{\bf n}_{e^-} \times {\bf n}_t \over
        |{\bf n}_{e^-} \times {\bf n}_t|},               \label{basis}
\quad
{\bf s}_\bot = {\bf s}_{\rm N} \times {\bf s}_{\|} . 
\end{eqnarray}
In the limit of small $\beta$ the quark spin is essentially
aligned with the beam direction apart from small
corrections
proportional to $\beta$, which depend on the production
angle. A system of reference with ${\bf s}_\|$ defined with respect
to the top quark momentum \cite{krz}
is convenient in the high energy limit but evidently becomes
less convenient close to threshold.

Including the QCD potential one obtains for the three components
of the polarization
\begin{eqnarray}
{\cal P}_\|(\bfp,E,\chi) &=& C_\|^0(\chi)
+ C_\|^1(\chi)\, \varphi_{\rm _R}(\rmp,E)\,\cos\vartheta\,
 \label{thr_long}\\
{\cal P}_\bot(\bfp,E,\chi) &=& C_\bot(\chi)\,
\varphi_{\rm _R}(\rmp,E)\,
\sin\vartheta\,
\label{thr_perp}\\
{\cal P}_{\rm N}(\bfp,E,\chi) &=& C_{\rm N}(\chi)
\varphi_{\rm _I}(\rmp,E)
\sin\vartheta\,
        \label{thr_norm} ,
\end{eqnarray}\\
\parbox{75.ex}{
\begin{eqnarray*}
& &\hspace{5.ex}C_\|^0 (\chi) =
-{a_2 + \chi a_1 \over a_1 + \chi a_2} ,\hspace{6.9ex}
  C_\|^1 (\chi) = \left( 1-\chi^2 \right) {a_2 a_3 - a_1 a_4   \over
        \left(a_1 + \chi a_2 \right)^2} ,\\
& &\hspace{5.ex}C_\bot(\chi)  = -{1\over 2} \,
{a_4 + \chi a_3 \over a_1 + \chi a_2} ,
    \qquad C_{\rm N}(\chi) =-{1 \over 2}\, {a_3
    + \chi a_4 \over a_1 + \chi a_2}\, =\, - C_{\rm FB}(\chi) ,
\end{eqnarray*}}
\hfill
\parbox{5.ex}{
\begin{eqnarray} \label{coefs} \end{eqnarray} }
with $\varphi_{\rm _I} = \Im\,\varphi$, and $\varphi(\rmp,E)$ as defined
in (\ref{phi}).  The momentum integrated quantities are obtained by
the replacement $\varphi(\rmp,E) \to \Phi(E)$. The case of
non-interacting stable quarks is recovered by the replacement
$\Phi\to\beta$, an obvious consequence of (\ref{cap_phi}).
\par
\begin{figure}[ht]
 \begin{center}
  \leavevmode
  \epsfxsize=14cm
  \epsffile[40 290 535 525]{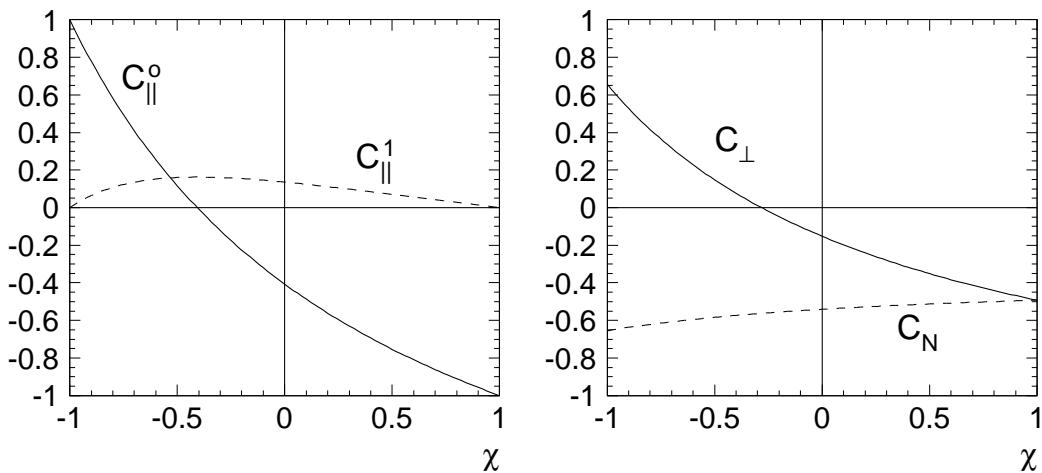}
  \caption[]{\label{pol_coefs.ps}\sloppy
        The coefficients (\ref{coefs}) for $\sqrt{s}/2 = m_t=180$ GeV.}
 \end{center}
\end{figure}

\noindent
Let us emphasize the main qualitative features of the result:
\begin{itemize}
\item Top quarks in the threshold region are highly polarized.  Even
  for unpolarized beams the longitudinal polarization amounts to about
  $-0.41$ and reaches $\pm1$ for fully polarized electron beams. This
  later feature is of purely kinematical origin and independent of the
  structure of top quark couplings.  Precision studies of polarized
  top decays are therefore feasible.
\item Corrections to this idealized picture arise from the small
  admixture of $P$-waves. The transverse and the normal components of
  the polarization are of order 10\%. The angular dependent part of
  the parallel polarization is even more suppressed.  Moreover, as a
  consequence of the angular dependence its contribution vanishes upon
  angular integration.
\item The QCD dynamics is solely contained in the functions $\varphi$
  or $\Phi$ which is the same for the angular distribution and the
  various components of the polarization. (However, this
  ``universality'' is affected by the rescattering corrections.)
  These functions which evidently
  depend on QCD dynamics can thus be studied in a variety of ways.
\item The relative importance of $P$-waves increases with energy,
  $\Phi\sim\sqrt{E/m_t}$.  This is expected from the close analogy
  between $\Phi_{\rm R}=\Re\,\Phi$ and $\beta$. 
  In fact, the order of magnitude of
  the various components of the polarization above, but close to
  threshold, can be estimated by replacing $\Phi_{\rm R}\to \rmp/m_t$.
\end{itemize}
The $C_i$ are displayed in Fig.~\ref{pol_coefs.ps} as functions of the
variable $\chi$ ($\sin^2\!\theta_{\rm W}= 0.2317$, $m_t=180$ GeV).
As discussed before, $C_\|^0$ assumes its maximal value $\pm 1$ for
$\chi=\mp 1$ and the coefficient $C_\|^1$ is small throughout. The
coefficient $C_\bot$ varies between $+0.7$ and $-0.5$ whereas $C_{\rm
  N}$ is typically around $-0.5$.  The dynamical factors $\Phi$ are
around $0.1$ or larger, such that the $P$-wave induced effects should
be observable experimentally.
\par\noindent
The normal component of the polarization which is proportional to
$\varphi_{\rm _I}$ has been predicted for stable quarks in the framework of
perturbative QCD \cite{dev,krz}. In the threshold region the phase can be
traced to the $t\bar t$ rescattering by the QCD potential. For 
stable quarks, assuming a pure Coulomb potential $V=-4\alpha_{\rm s}/3r$, 
the nonrelativistic problem can be solved analytically \cite{FKotsky} and
one finds
\begin{eqnarray}
\lim_{\Gamma_t \rightarrow 0 \atop E\rightarrow \bfp^2/m_t} 
        \left(E - {\bfp^2\over m_t} + i\Gamma_t \right) G(\rmp,E) &=&
        \exp\left(\pi {\bar k} \over 2 \rmp\right)\,
        \Gamma(1+{i{\bar k}/\rmp}) \label{lim1_gt0}\\
\lim_{\Gamma_t \rightarrow 0 \atop E\rightarrow \bfp^2/m_t} 
        \left(E - {\bfp^2\over m_t} + i\Gamma_t \right) F(\rmp,E) &=&
        \left(1 - i{{\bar k} \over \rmp}\right) 
        \exp\left(\pi {\bar k} \over 2 \rmp\right)\,
        \Gamma(1+{i{\bar k}/\rmp}) \label{lim2_gt0},\ \ \ \ \
\end{eqnarray}
with ${\bar k} = 2m_t\alpha_s/3$
and hence
\begin{eqnarray}
\varphi_{\rm _I}(\rmp,E) &\rightarrow&
{2\over 3}\alpha_{\rm s}{1-4\alpha_{\rm s}/3\pi\over
        1-8\alpha_{\rm s}/3\pi} \label{phi_gt0}\\
\Phi_{\rm I}(E) &\rightarrow&
{2\over 3}\alpha_{\rm s}{1-4\alpha_{\rm s}/3\pi\over
        1-8\alpha_{\rm s}/3\pi} . \label{cap_phi_gt0}
\end{eqnarray}
The component of the polarization normal to
the production plane is thus
approximately independent of $E$ and essentially measures the strong
coupling constant. In fact one can argue that this is a unique way to
get a handle on the scattering of heavy quarks through the QCD
potential.

\begin{figure}
 \begin{center}
  \leavevmode
  \epsfxsize=135mm
  \epsffile[40 185 535 630]{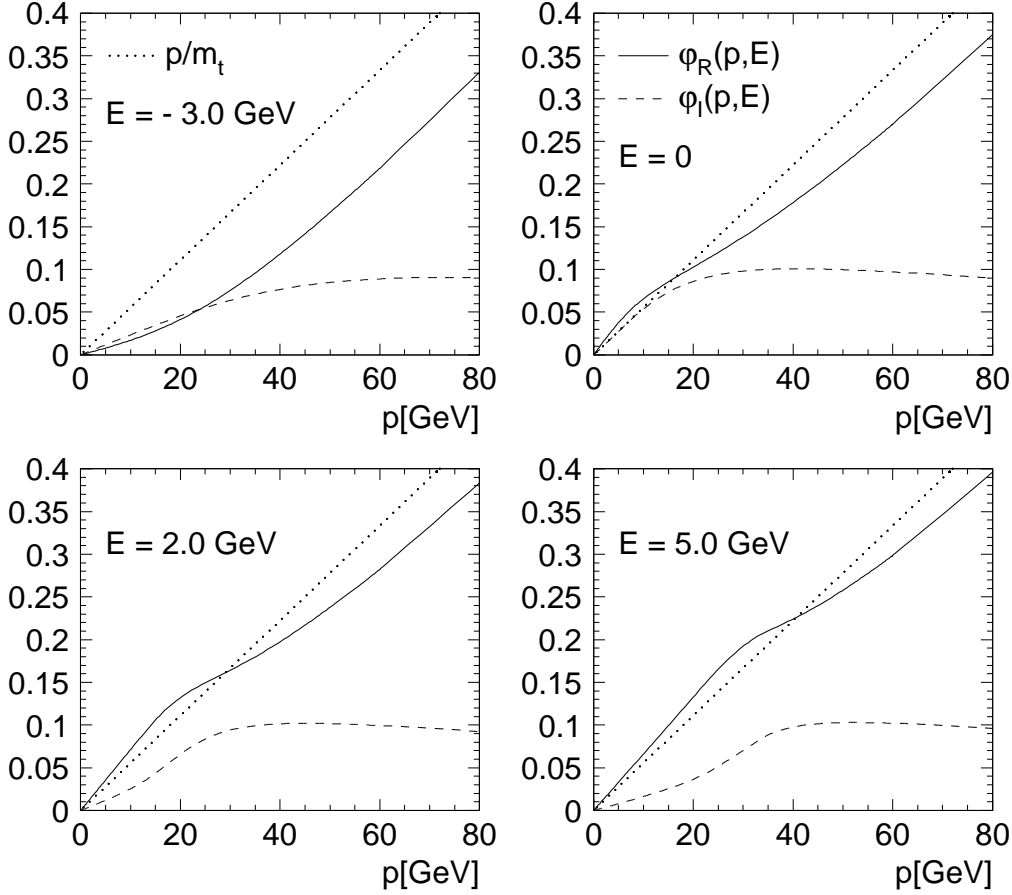}
  \caption[]{\label{phi.ps}\sloppy 
        Real (solid) and imaginary (dashed) part of the function
        $\varphi(\rmp,E)$ for $m_t = 180$ GeV, $\alpha_{\rm s}=0.125$ 
        and four
        different energies. The dotted line shows the free particle case
        $\Re\,\varphi = \mbox{p}/m_t$ (from \protect\cite{ttp48}).}
 \end{center}
\end{figure}

Predictions for real and imaginary parts of the function $\varphi$ are
displayed in Fig.~\ref{phi.ps} for four different energies.  

\begin{figure}
 \begin{center}
  \leavevmode
  \epsfxsize=135mm
  \epsffile[70 100 510 735]{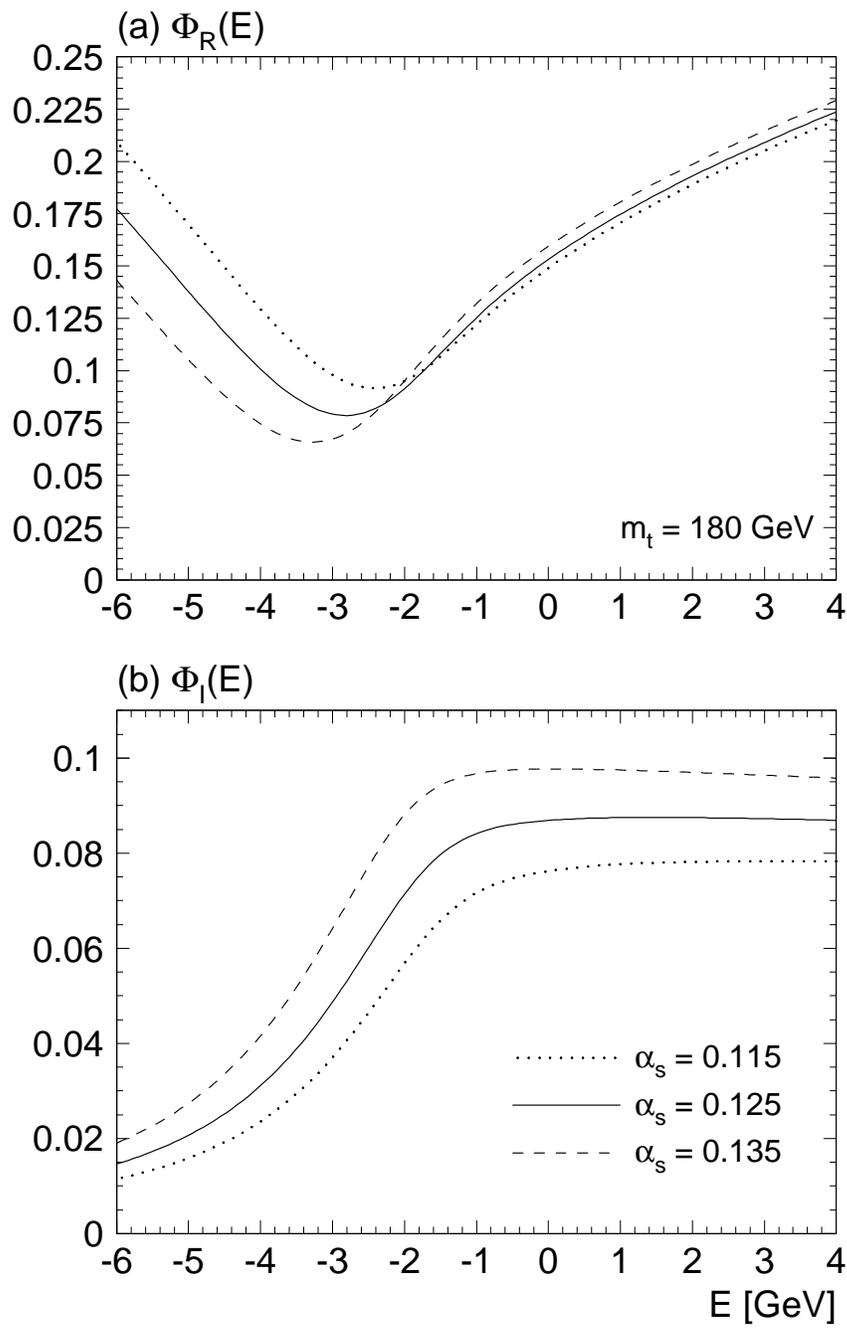}
  \caption[]{\label{alph_dep.ps}\sloppy Real and imaginary part of $\Phi(E)$
        for three different values of $\alpha_{\rm s}$ (from
\protect\cite{ttp48}). }
 \end{center}
\end{figure}

The momentum integrated functions $\Phi(E)$ are shown in
Fig.~\ref{alph_dep.ps}.  From this figure, in combination with
Fig.~\ref{pol_coefs.ps}, it is clear that the contribution of $P$-wave
amplitudes to the quark polarization will amount to 10\% at most and
by construction vanishes upon angular integration.  

\subsection{Rescattering}
\begin{figure}[h]\begin{center}
\begin{tabular}{cc}
  \epsfxsize 63mm \mbox{\epsffile[25 570 200 650]{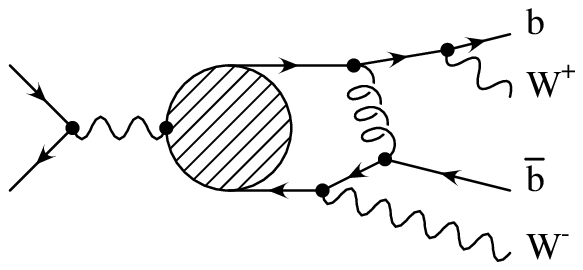}} &
  \epsfxsize 63mm \mbox{\epsffile[25 570 200 650]{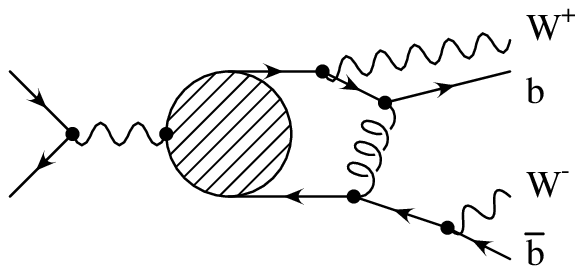}} \\
  a) & b)
\end{tabular}\end{center}
\caption{\label{tbbar}Lowest order rescattering diagrams.}
\end{figure}

For a particle with a very small decay rate production and decay amplitudes 
can be clearly separated.  This is fairly evident from the space-time 
picture of such a sequence.  Prior to its decay the particle travels away 
from the production point and any coherence is lost between the two 
reactions. The situation is different for the case under discussion, an 
unstable top quark which decays within the range of interaction between $t$ 
and $\bar t$.  In such a situation the decay products from $t$ are still 
affected by the force originating from $\bar t$ and vice versa 
(Fig.~\ref{tbbar}).

In ref.~\cite{MelYak1,sumph} it has been demonstrated that the total 
cross section remains unaffected by rescattering in order $\alpha_s$.  This 
result had been anticipated in \cite{JKT} on the basis of earlier work 
which considered the decay rate of a muon bound in the strong field of a 
nucleus \cite{Huff}.  In contrast momentum and angular distributions 
\cite{MurSum2,sumph,FMY94} as well as the top quark polarization 
\cite{ttp48} are affected by rescattering.  For example the momentum 
distribution has to be corrected by a factor $(1+\psi_1(p,E)) $ with 
\be
   \psi_1(\rmp,E) = 2\,\Im\int\!\frac{d^3k}{(2\pi)^3}V(|{\bf k}-{\bf p}|)
        \frac{G(\rmk,E)}{G(\rmp,E)}\frac{\arctan{\frac{|{\bf k}-{\bf p}|}
        {\Gamma_t}}}{|{\bf k}-{\bf p}|} \\
\ee
\begin{figure}
\begin{tabular}{cc}
  \epsfxsize 70mm \mbox{\epsffile[0 0 567 454]{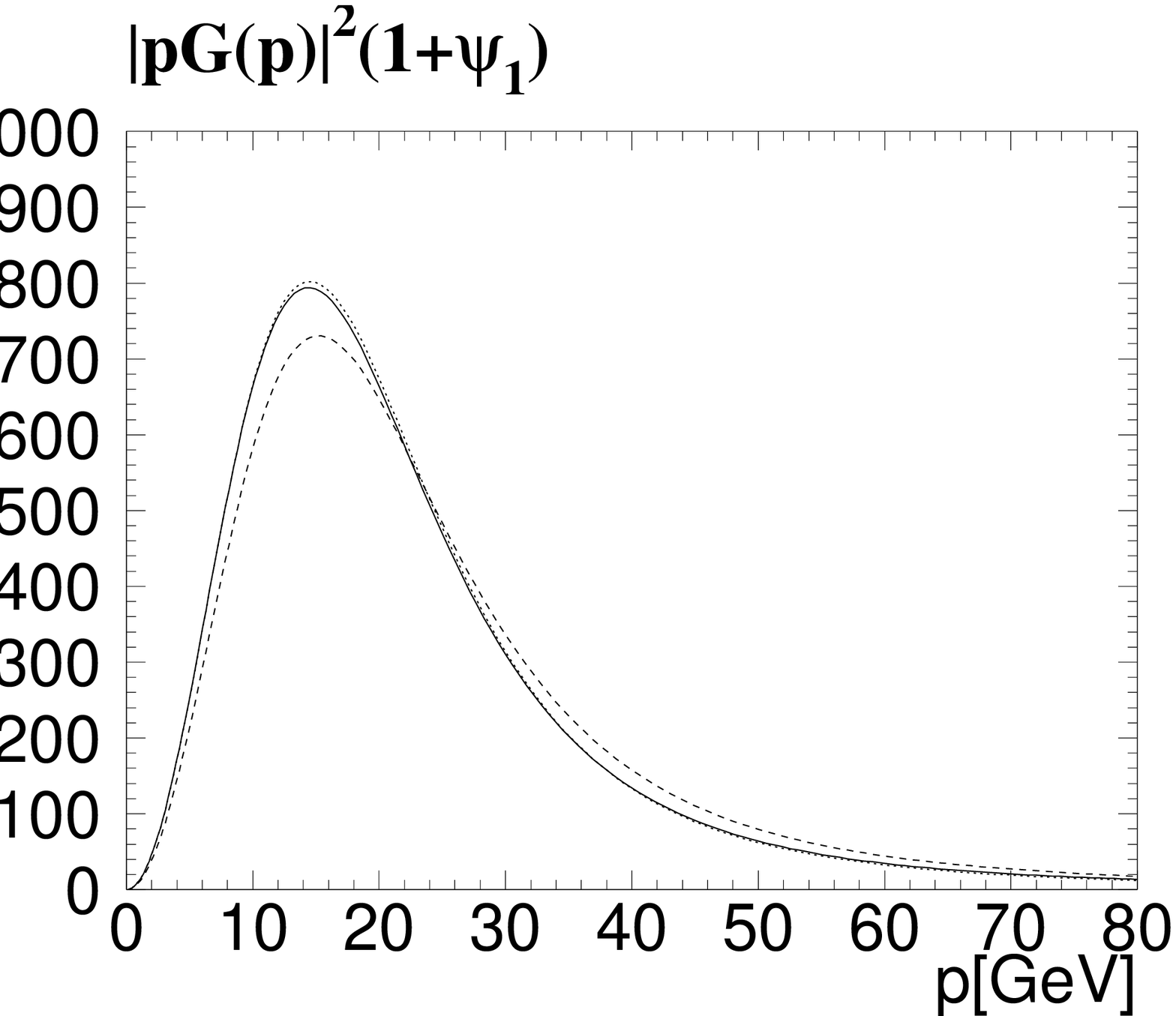}} & 
  \epsfxsize 70mm \mbox{\epsffile[0 0 567 454]{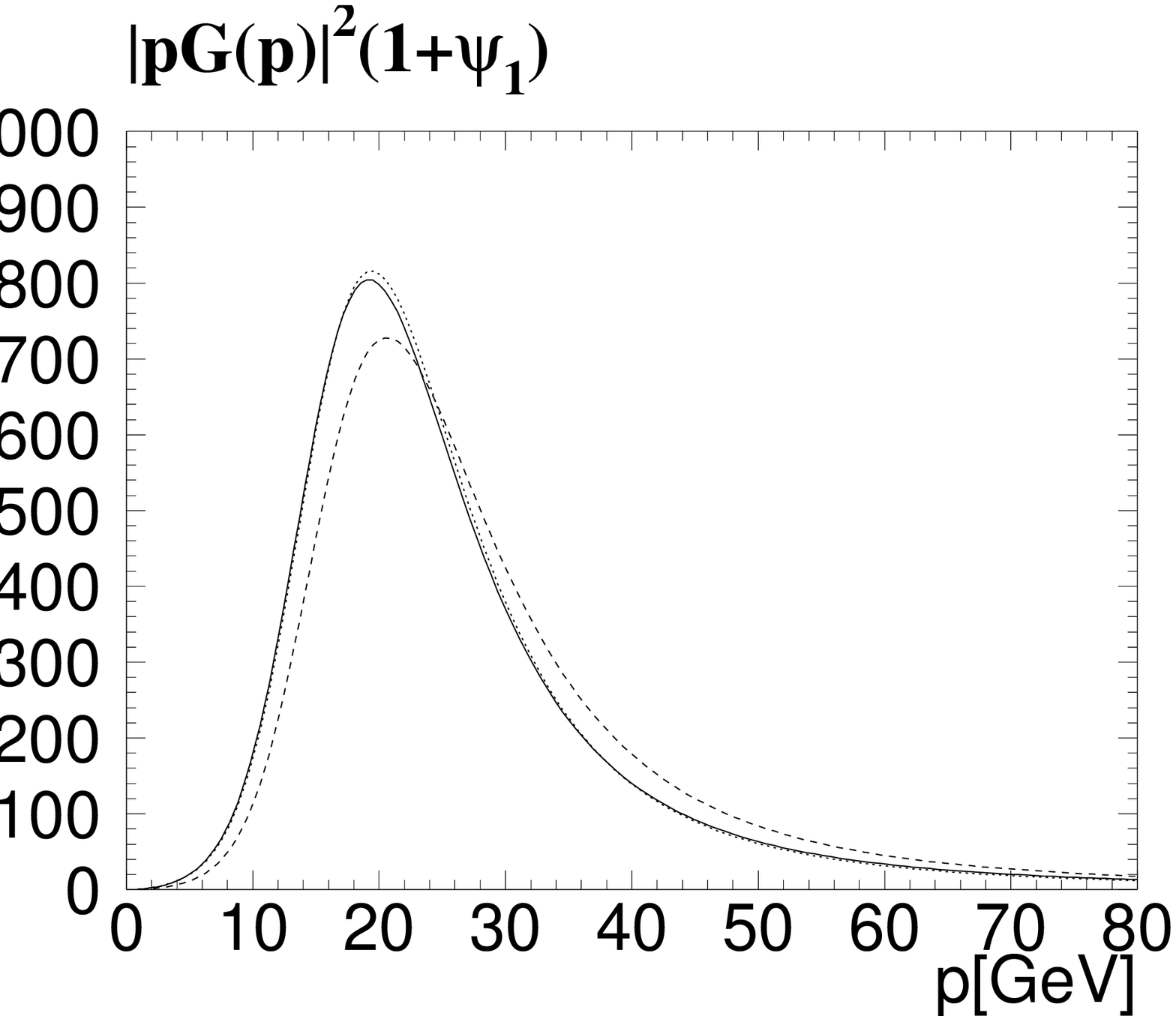}} \\  
    a) E=-3GeV & b) E=0GeV \\
  \epsfxsize 70mm \mbox{\epsffile[0 0 567 454]{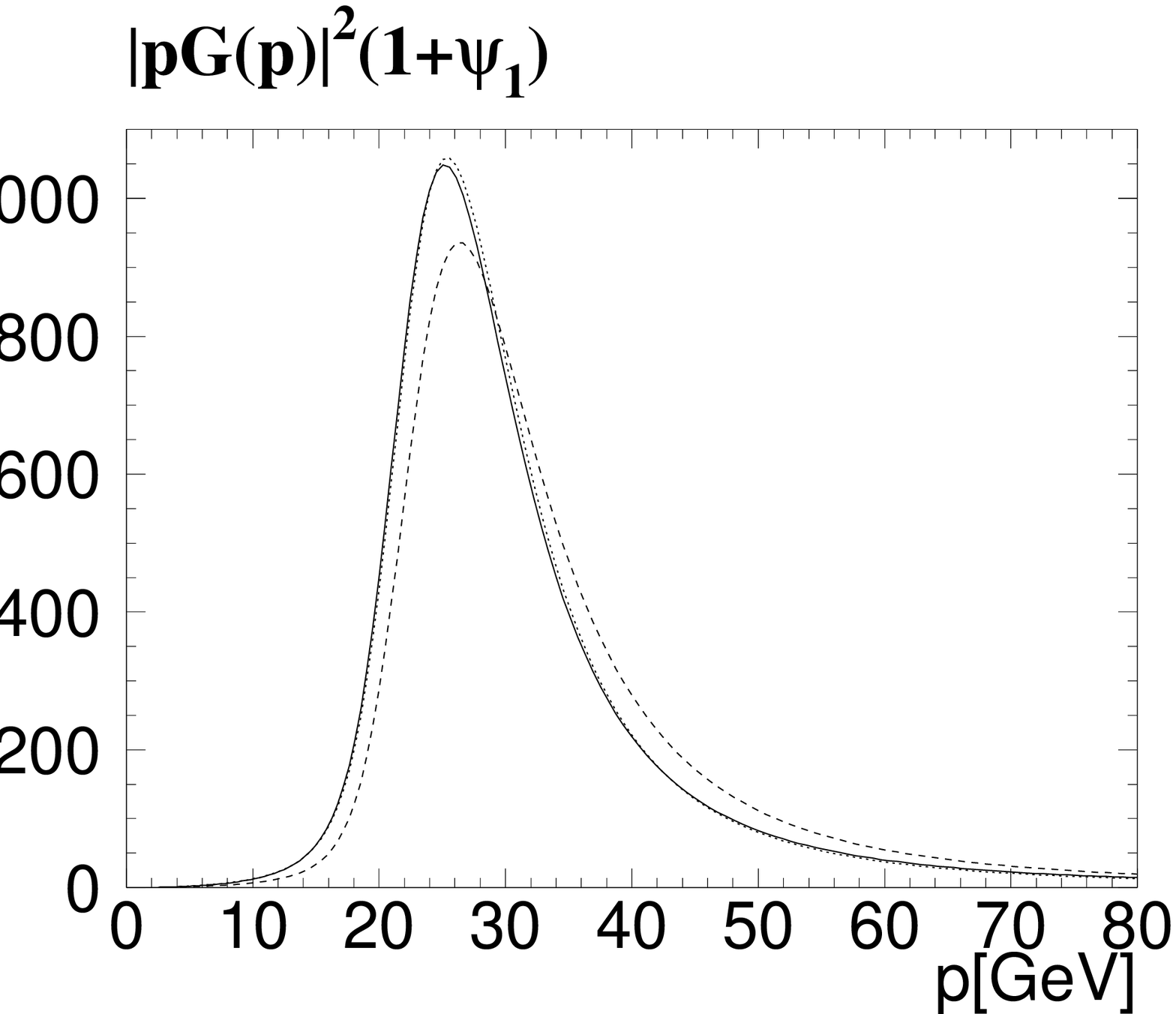}} & 
  \epsfxsize 70mm \mbox{\epsffile[0 0 567 454]{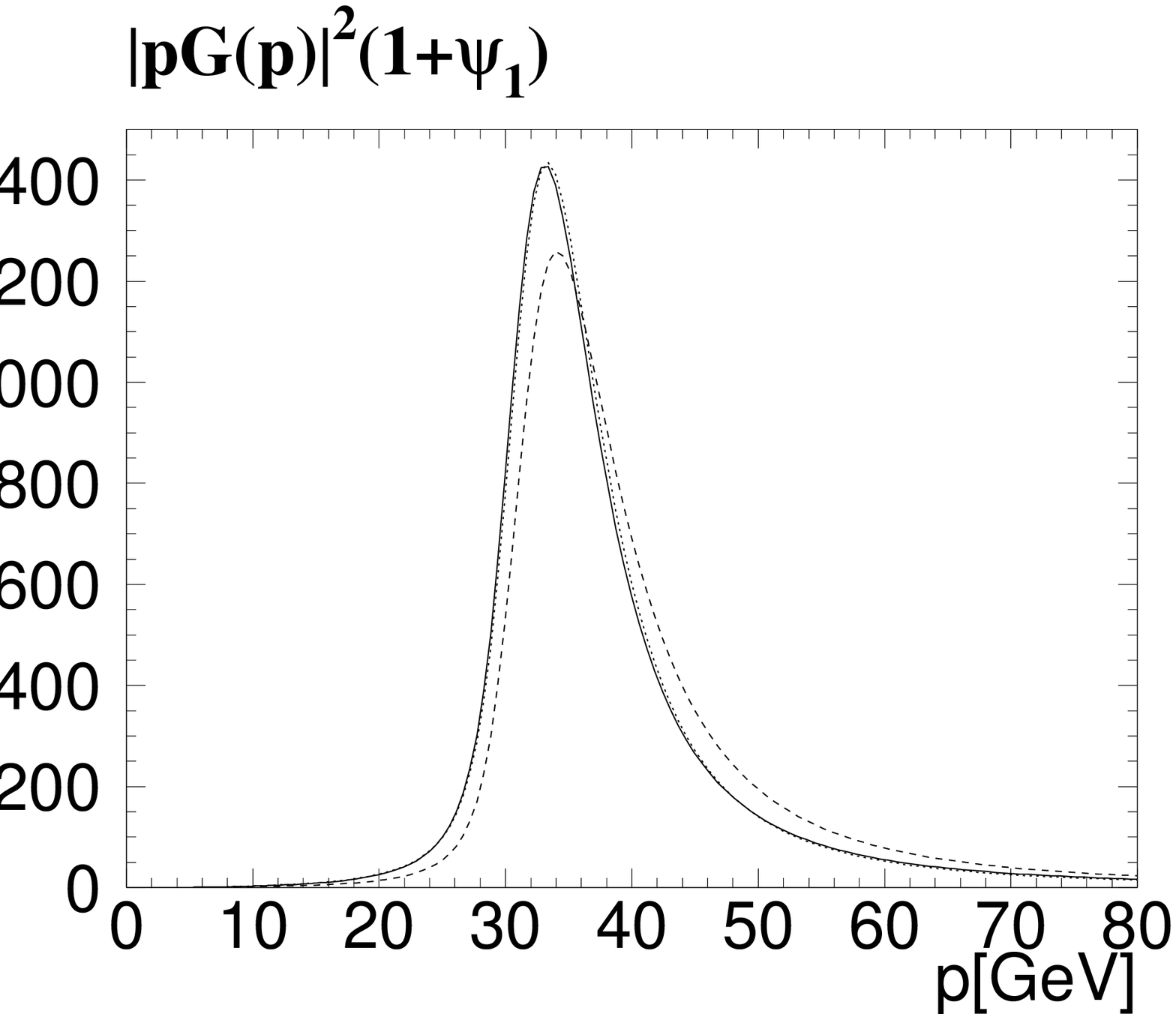}} \\
    c) E=2GeV  & d) E=5GeV
\end{tabular}
\caption{Modification of the momentum distribution through rescattering.
Dashed line: no rescattering corrections included; Solid line: rescattering
contribution with full potential included; dotted line: rescattering
contribution with pure Coulomb potential and $\alpha_{\rm s}=0.187$
included (from \protect\cite{ttp48}).
\label{rescdsig}}
\end{figure}
The distribution is shifted towards smaller momenta by about 5\% 
(Fig.~\ref{rescdsig}), an effect that could become relevant in precision 
experiments.  The influence on the forward-backward asymmetry and the 
polarization is even more pronounced
\cite{ttp48}, as far as the 
$S$-$P$-wave interference terms are concerned which are  thus
intrinsically of  
order $\beta$.  A detailed discussion of these effects is beyond the scope 
of these lectures and can be found in \cite{ttp48}.

\subsection{Relativistic corrections}
In  ${\cal O}(\alpha^2_s)$ one anticipates effects from
relativistic corrections, from the reduction of the phase space
through the binding energy and from the Coulomb wave function of the
$b$ quark. Individually these effects are large.  For the sake of the
argument, let us adopt a pure Coulomb potential and a binding energy
of -2.5 GeV.  From the virial theorem one derives a potential energy
of -5 GeV.  The phase space of the quark decaying first is therefore
reduced by this same amount.  Assuming $m_t = 180 $ GeV one would
arrive at a reduction of $\Gamma_t$ by about 10\%.  A full calculation
of all ${\cal O}(\alpha^2_s)$ effects is not available at present and
one has to resort to models and analogies \cite{JKT,Sumino,TJ}. For
example, it has been shown \cite{Huff,Uber} that the decay rate of a
muon bound in the field of a nucleus is given by
\be
\Gamma = \Gamma_{\rm free} 
\left[1-5(Z\alpha)^2\right]
\left[1+5(Z\alpha)^2\right]
\left[1-{(Z\alpha)^2\over 2}\right],
\ee
where the first correction factor originates from the phase space
suppression, the second from the Coulomb enhancement, and the third
from time dilatation. Thus there is no first order correction to the
total rate from rescattering in the nucleus potential, similar to the
$t\bar t$ case discussed above.  The second order contributions
evidently compensate to a large extent.  In a model calculation where
these features are implemented \cite{TJ} through a momentum dependent
width, it is found that the total cross section as well as the
momentum distribution are hardly affected.  These considerations have
recently been confirmed in a more formal approach \cite{MKum}.

%% file: ack.tex
\addcontentsline{toc}{chapter}{Acknowledgments}
\subsection*{Acknowledgments}
I would like to thank A. Czarnecki, R. Harlander, A. Hoang,
M. Je\.zabek, E. Mirkes, M. Peter, T. Teubner, and P. Zerwas for
pleasant collaborations and many helpful discussions and comments on
the TOPics presented in this review.  I am particularly grateful to
A. Czarnecki.  Without his {\TeX}nical help this paper would never
have been completed.  The hospitality of the SLAC theory group during
my sabbatical in the summer term 1995 is gratefully acknowledged.